\documentclass[UTF8,fontset=fandol,zihao=-4,openany,oneside]{ctexbook}
\usepackage[paperwidth=176mm,paperheight=250mm,top=21mm,bottom=21mm,left=21mm,right=19mm,headheight=15pt,headsep=6mm]{geometry}
\usepackage{amsmath,amssymb,bm,mathtools}
\usepackage{graphicx,booktabs,tabularx,array,longtable}
\usepackage{tikz,pgfplots}
\pgfplotsset{compat=1.18}
\usetikzlibrary{arrows.meta,positioning,calc}
\usepackage{caption,float,placeins}
\usepackage{enumitem,fancyhdr}
\usepackage[unicode,colorlinks=true,linkcolor=black,citecolor=black,urlcolor=black,bookmarksnumbered=true]{hyperref}
\hypersetup{pdftitle={A Concise Introduction to Quantum Electromagnetics},pdfauthor={Wei E. I. Sha}}
\makeatletter
\renewcommand{\@pnumwidth}{2em}
\renewcommand{\@tocrmarg}{3.5em}
\makeatother
\ctexset{chapter={format=\Large\normalfont,beforeskip=3pt,afterskip=22pt},section={format=\large\normalfont,beforeskip=19pt,afterskip=10pt},subsection={format=\normalsize\normalfont,beforeskip=13pt,afterskip=7pt}}
\setlist[enumerate]{leftmargin=2em,itemsep=4pt,topsep=5pt}
\renewcommand{\headrulewidth}{0.35pt}
\fancypagestyle{plain}{\fancyhf{}\fancyfoot[C]{\small\thepage}\renewcommand{\headrulewidth}{0pt}}
\newcommand{\ket}[1]{\lvert #1\rangle}
\newcommand{\bra}[1]{\langle #1\rvert}
\newcommand{\braket}[2]{\langle #1\vert #2\rangle}
\newcommand{\avg}[1]{\langle #1\rangle}
\newcommand{\Tr}{\operatorname{Tr}}
\newcommand{\Var}{\operatorname{Var}}
\newcommand{\ee}{\mathrm e}
\newcommand{\ii}{\mathrm i}
\newcommand{\dd}{\mathrm d}
\newcolumntype{Y}{>{\raggedright\arraybackslash}X}
\newcommand{\Diss}[1]{\mathcal D[#1]}
\definecolor{inkblue}{RGB}{38,67,88}
\definecolor{lightblue}{RGB}{239,244,247}
\definecolor{softgray}{RGB}{245,245,245}
\renewcommand{\Re}{\operatorname{Re}}
\renewcommand{\Im}{\operatorname{Im}}
\ctexset{chapter/name={Chapter\space,},chapter/number=\arabic{chapter},contentsname={Contents},figurename={Figure},tablename={Table},bibname={References},appendixname={Appendix}}
\begin{document}
\frontmatter
\def\BookFrontMatterEmbedded{1}

\ifdefined\BookFrontMatterEmbedded
  \let\BookFrontBegin\relax
  \let\BookFrontEnd\relax
  \let\BookFrontNumbering\relax
\else
  \documentclass[UTF8,fontset=fandol,zihao=-4,openany,oneside]{ctexbook}
  \usepackage[
    paperwidth=176mm, paperheight=250mm,
    top=21mm, bottom=21mm, left=21mm, right=19mm,
    headheight=15pt, headsep=6mm
  ]{geometry}
  \usepackage{amsmath,tikz,fancyhdr}
  \usetikzlibrary{arrows.meta}
  \usepackage[unicode,hidelinks]{hyperref}
  \hypersetup{pdftitle={Quantum Electromagnetics: Cover and Preface},pdfauthor={Wei E. I. Sha},pdfsubject={An introduction for engineering electromagnetics researchers}}
  \ctexset{chapter={format=\Large\normalfont,beforeskip=3pt,afterskip=22pt}}
  \setlength{\parindent}{2em}
  \setlength{\parskip}{2pt}
  \linespread{1.22}
  \pagestyle{fancy}
  \fancyhf{}
  \fancyhead[L]{\small Quantum Electromagnetics}
  \fancyhead[R]{\small Preface}
  \fancyfoot[C]{\small\thepage}
  \renewcommand{\headrulewidth}{0.35pt}
  \fancypagestyle{plain}{%
    \fancyhf{}
    \fancyfoot[C]{\small\thepage}
    \renewcommand{\headrulewidth}{0pt}
  }
  \def\BookFrontBegin{\begin{document}}
  \def\BookFrontEnd{\end{document}}
  \def\BookFrontNumbering{\pagenumbering{roman}}
\fi
\BookFrontBegin
\begingroup
\definecolor{QEMpaper}{HTML}{F7F6F1}
\definecolor{QEMink}{HTML}{17394C}
\definecolor{QEMblue}{HTML}{3C718A}
\definecolor{QEMlight}{HTML}{DBE5E7}
\definecolor{QEMgray}{HTML}{6C7C82}
\definecolor{QEMgold}{HTML}{A77B36}
\begin{titlepage}
\thispagestyle{empty}\mbox{}
\begin{tikzpicture}[remember picture,overlay]
\begin{scope}[shift={(current page.south west)},x=1mm,y=1mm,
 line cap=round,line join=round,>=Stealth]
  \fill[QEMpaper] (0,0) rectangle (176,250);
  \fill[QEMink] (0,0) rectangle (3,250);
  \draw[QEMgold,line width=.65pt] (22,230)--(48,230);
  \node[anchor=west,text=QEMink,inner sep=0pt,font=\songti\fontsize{18}{24}\selectfont] at (21,215) {A Concise Introduction to};
  \node[anchor=west,text=QEMink,inner sep=0pt,font=\songti\fontsize{28}{34}\selectfont] at (21,195) {Quantum Electromagnetics};
  \node[anchor=west,text=QEMgray,inner sep=0pt,font=\sffamily\fontsize{10}{13}\selectfont] at (22,177) {For researchers and engineers in electromagnetics};

  \node[text=QEMgray,rotate=90,align=center,font=\fontsize{8}{11}\selectfont] at (12,101) {Description of the field};
  \node[text=QEMgray,font=\fontsize{8}{11}\selectfont] at (97,163) {Description of matter};
  \node[text=QEMink,font=\fontsize{10}{13}\selectfont] at (67,154) {Macroscopic media};
  \node[text=QEMink,font=\fontsize{10}{13}\selectfont] at (127,154) {Quantum matter};
  \node[text=QEMink,align=center,font=\fontsize{10}{14}\selectfont] at (23,127) {Classical\\field};
  \node[text=QEMink,align=center,font=\fontsize{10}{14}\selectfont] at (23,79) {Quantum\\field};
  \fill[white,opacity=.6] (37,53) rectangle (157,149);
  \draw[QEMblue!30,line width=.45pt] (37,53) rectangle (157,149);
  \draw[QEMblue!45,line width=.45pt] (97,53)--(97,149) (37,101)--(157,101);

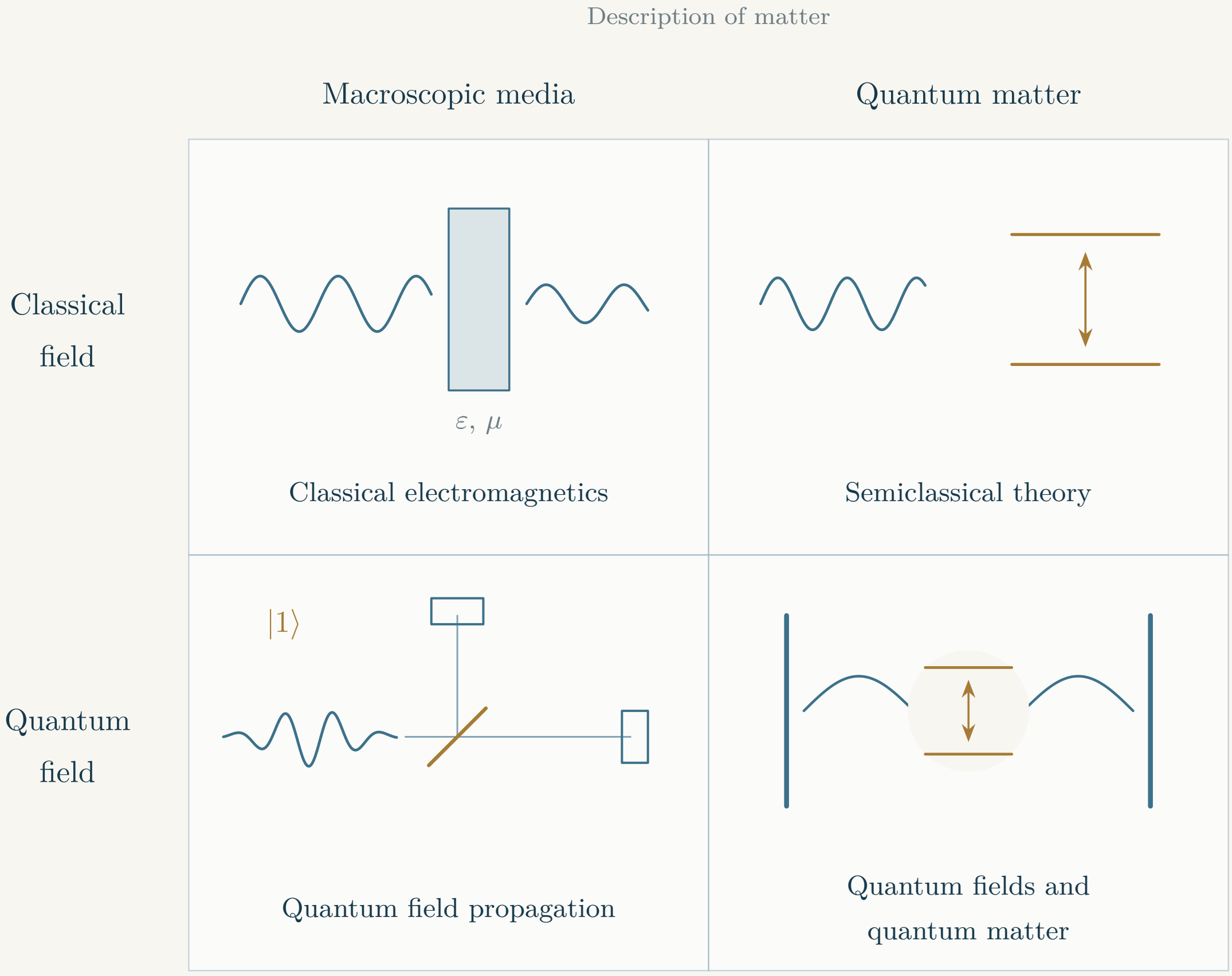
  \filldraw[fill=QEMlight,draw=QEMblue,line width=.65pt] (67,120) rectangle (74,141);
  \draw[QEMblue,line width=.9pt,smooth,domain=43:65,samples=80]
      plot (\x,{130+3.2*sin((\x-43)*40)});
  \draw[QEMblue,line width=.9pt,smooth,domain=76:90,samples=60]
      plot (\x,{130+2.2*sin((\x-76)*40)});
  \node[text=QEMgray,font=\fontsize{9}{11}\selectfont] at (70.5,116) {$\varepsilon,\,\mu$};
  \node[text=QEMink,font=\fontsize{9}{12}\selectfont] at (67,108) {Classical electromagnetics};

  \draw[QEMblue,line width=.9pt,smooth,domain=103:122,samples=80]
      plot (\x,{130+3*sin((\x-103)*45)});
  \draw[QEMgold,line width=1pt] (132,123)--(149,123) (132,138)--(149,138);
  \draw[QEMgold,<->,line width=.7pt] (140.5,125)--(140.5,136);
  \node[text=QEMink,font=\fontsize{9}{12}\selectfont] at (127,108) {Semiclassical theory};

  \node[text=QEMgold,font=\fontsize{10}{12}\selectfont] at (48,93) {$|1\rangle$};
  \draw[QEMblue,line width=.9pt,smooth,domain=41:61,samples=90]
      plot (\x,{80+3.4*exp(-((\x-51)/6)^2)*sin((\x-41)*64)});
  \draw[QEMblue!65,line width=.55pt] (62,80)--(68,80) (68,80)--(88,80) (68,80)--(68,94);
  \draw[QEMgold,line width=1.2pt] (64.7,76.7)--(71.3,83.3);
  \draw[QEMblue,line width=.75pt] (87,77) rectangle (90,83) (65,93) rectangle (71,96);
  \node[text=QEMink,font=\fontsize{9}{12}\selectfont] at (67,60) {Quantum field propagation};

  \draw[QEMblue,line width=1.6pt] (106,72)--(106,94) (148,72)--(148,94);
  \draw[QEMblue,line width=.85pt,smooth,domain=108:146,samples=100]
      plot (\x,{83+4*sin((\x-108)*14.210526)});
  \fill[QEMpaper] (127,83) circle (7);
  \draw[QEMgold,line width=.9pt] (122,78)--(132,78) (122,88)--(132,88);
  \draw[QEMgold,<->,line width=.65pt] (127,79.4)--(127,86.6);
  \node[text=QEMink,align=center,font=\fontsize{9}{12}\selectfont] at (127,60) {Quantum fields and\\quantum matter};

  \draw[QEMblue!35,line width=.4pt] (22,36)--(157,36);
  \node[anchor=west,text=QEMink,inner sep=0pt,font=\songti\fontsize{16}{20}\selectfont] at (22,25) {Wei E. I. Sha};
  \node[anchor=west,text=QEMgray,inner sep=0pt,font=\fontsize{11}{14}\selectfont] at (22,15) {Zhejiang University};
\end{scope}
\end{tikzpicture}
\end{titlepage}
\endgroup

\BookFrontNumbering
\begingroup
\setlength{\emergencystretch}{2em}
\clubpenalty=10000\widowpenalty=10000\raggedbottom
\chapter*{Preface}
\addcontentsline{toc}{chapter}{Preface}
\markboth{Preface}{Preface}
\ifdefined\fancyhead\fancyhead[R]{\small Preface}\fi

In engineering electromagnetics, we often begin with a concrete question. Can an antenna be made smaller, receiver sensitivity improved by a few decibels, or insertion loss reduced a little further? We solve Maxwell's equations, analyze field distributions and energy flow, and repeatedly adjust structures, materials, and feeds. Fabrication and measurement then test the design. Progress in wireless communications, radar sensing, high-speed interconnects, and optoelectronic devices has accumulated through such efforts. Generations of researchers and engineers have turned electromagnetics into useful technology, changing how we communicate, understand the world, and use energy.

With experience, another side becomes apparent. In some mature areas, performance gains become increasingly difficult. More complex structures, more design variables, and more accurate simulations do not always bring proportionate benefits. The question then extends beyond how to optimize further: where does the limitation come from? Is there room left in the design and fabrication, or do physical constraints already bound the performance? Resolving this question can be more valuable than obtaining another slightly improved curve.

For electrically small, linear passive antennas, the Chu limit relates size, stored energy, and radiation quality factor. The diffraction limit in conventional far-field imaging depends on wavelength, aperture, and the measurement scheme. Receivers must also account for thermal noise. Under impedance matching and conditions such as $hf\ll k_{\mathrm B}T$, the available thermal-noise power in bandwidth $B$ is approximately $k_{\mathrm B}TB$. Each result has its own conditions of validity. To understand a limit, we must first understand those conditions. When discussing a new technology, we should state which conditions change, what resources are added, and what costs follow. Quantum electromagnetics must meet the same standard.

Engineering electromagnetics increasingly emphasizes multifunctional integration, coordinated system design, and validation under realistic operating conditions. These are important and demanding tasks. Working at a university, I often ask how integration and performance improvements along established routes can build knowledge of lasting value. Leading companies have substantial strengths in fabrication, data, and system engineering. Competing only in these areas may not make the best use of a university's strengths. We also need to study physical questions that remain poorly understood and may not yet lead to a product. New device principles and measurement methods often depend on this early exploration. Quantum electromagnetics is one direction that I believe deserves sustained effort.

The familiar interaction between field and matter provides a natural entry point. In conventional electromagnetic calculations, permittivity, permeability, conductivity, or equivalent circuit parameters summarize the material response. This lets us analyze complex structures without solving for each microscopic particle. But atomic transitions, coherence, and saturation may require more than fixed material parameters: the relevant quantum states must be evolved explicitly. Likewise, average fields and average power cannot distinguish a source that emits one photon in each trial from one that occasionally emits several. Nor do they suffice to predict coincidence counts at two detectors. These questions also require the quantum statistics of the field.

Electromagnetic design thus gains new tasks: preparing states of field and matter, controlling energy exchange between them, and extracting information through measurement. Atomic levels can provide frequency references. Rydberg-atom responses can be used to measure radio-frequency electric fields. Coupling emitters to resonant structures can control single-photon emission. Photon correlations reveal information beyond average intensity. Electromagnetic structure design is thereby connected to quantum-state preparation, evolution, and readout. Any improvement from quantum methods must be assessed for a specific task, including loss, noise, and the conditions required for implementation.

The knowledge that electromagnetic researchers already possess remains highly useful. The drive experienced by an atom depends on its local electric field. Emitter--cavity coupling requires knowledge of eigenmodes and field normalization. Green functions and scattering analysis determine the channels into which radiation is emitted and how it reaches a receiver. Practical devices also involve loss, packaging, and measurement. In many cases, classical electromagnetic calculations still do most of the spatial field solving, while quantum models add material-state dynamics and field-measurement statistics. Understanding this division of work provides a starting point for further study.

I wrote this book to make these connections clearer. There are many excellent texts on quantum mechanics and quantum optics, including books for engineering readers. Yet readers accustomed to starting from Maxwell's equations, constitutive relations, and boundary conditions still encounter difficult transitions. How do familiar electromagnetic modes and complex amplitudes relate to photons, operators, and quantum states? How do classically calculated field distributions and scattering matrices enter quantum models? Why can classical results sometimes describe an average response, while fluctuations and correlations require a different description? Explaining these connections early makes many new concepts less unfamiliar.

This book is intended for researchers and engineers with a background in classical and engineering electromagnetics who have not systematically studied quantum mechanics. Field--matter interaction is the central theme. The book distinguishes classical and quantum descriptions of the field, and macroscopic parameters and explicit quantum states for matter, giving four classes of model. At each added level of quantum description, we compare the new unknowns, the information retained, and the questions that become accessible. The same device may require different models for average transmittance, atomic populations, and photon coincidence counts. The quantity being sought should determine the model.

Chapter 1 explains why we move beyond classical electromagnetics and introduces energy levels, photons, and quantum states. Chapter 2 retains classical driving fields and starts with atoms having a few relevant levels. It introduces probability amplitudes, density matrices, and dissipation, then applies them to atomic clocks and Rydberg-atom reception. Chapter 3 develops field quantization from LC circuits and electromagnetic modes. It covers photon statistics, photodetection, propagation and loss in linear media, and two-photon scattering correlations. Chapter 4 places field and matter in a joint quantum model to study energy exchange, spontaneous emission, cavity enhancement, and related photonic devices.

Chapter 5 returns to engineering analysis. It discusses model selection, the connection between classical and quantum calculations, and the assessment of sensitivity and quantum signatures. These discussions turn the preceding concepts into practical calculations and measurements.

``Concise'' in the title refers chiefly to the selection of material. The main text focuses on basic concepts, essential equations, and physical meaning, spelling out derivation steps that beginners may find difficult. Where an analytical derivation aids understanding, it is developed explicitly. Where the algebra becomes too lengthy, the text explains how to formulate the equations, solve them numerically, and check the results. Material requiring more mathematical preparation is placed in Further Reading sections. Understanding a model requires knowing both how to calculate with it and what it neglects, including its conditions of validity.

Writing this book has also been an experiment in human--AI collaboration. Artificial intelligence assisted with organizing material, drafting, expanding derivations, and revising the prose. Repeated discussion and revision helped compare explanations. They also reminded me that fluent prose does not guarantee a correct concept, and a complete-looking equation may still confuse operator ordering or omit an approximation. I therefore checked the derivations, physical interpretations, and model assumptions individually, seeking to finalize material that I understood clearly myself. I take responsibility for the selection of material and for the final content.

I expect human--AI collaboration to become increasingly common. Tools make answers easier to obtain, but researchers must still judge why an answer holds and under what conditions. I hope this book helps readers develop that judgment.

During my work on quantum electromagnetics, I have benefited from the guidance, help, and support of Xianliang Wu, Weng Cho Chew, Wallace C. H. Choy, \mbox{Nicolae Panoiu}, and Er-Ping Li. I warmly thank them here. I also thank Professor Jin Pan, who suggested that I write a short book on quantum electromagnetics.

I hope this short book helps colleagues and students in engineering electromagnetics get started. Its main aim will be met if readers can identify when classical analysis is sufficient, when a quantum description is needed, and how to build a model whose results can be calculated, tested, and explained. I am still learning, and my knowledge and experience are limited. Errors and omissions are inevitable; I welcome readers' corrections and comments.

\vspace{1em}
\begin{flushright}
Wei E. I. Sha\\[4pt]
College of Information Science and Electronic Engineering\\
Zhejiang University\\[4pt]
November 2026\\
Email: weisha@zju.edu.cn
\end{flushright}
\clearpage
\endgroup
\BookFrontEnd


\clearpage
\fancyhead[R]{\small Contents}
\begingroup
\sloppy
\setlength{\emergencystretch}{4em}
\tableofcontents
\endgroup
\mainmatter
\clearpage
\fancyhead[R]{\small Chapter 1}
\ctexset{chapter/format=\Large\normalfont,section/format=\large\normalfont,subsection/format=\normalsize\normalfont}
\captionsetup{labelfont=normalfont}
\chapter{From Classical to Quantum Electromagnetics}
\label{ch:introduction}

Readers familiar with engineering electromagnetics already know a basic method of analysis. Given electromagnetic sources, material parameters, and boundary conditions, solve Maxwell's equations, then calculate power, impedance, reflection, transmission, and other quantities from the fields. Antenna radiation, waveguide transmission, and cavity design usually follow this approach.

Quantum electromagnetics still needs this knowledge. Electromagnetic analysis determines where to place an atom, which modes the surrounding structure supports, and where energy enters and leaves. This book adds a description of energy changes within matter and of the quantum properties of electromagnetic fields during energy exchange and measurement.

This chapter introduces these ideas through familiar electromagnetic problems. No prior knowledge of quantum mechanics is needed at this stage.

\section{What Does Classical Electromagnetics Leave Out?}

To calculate the transmittance of a glass slab, its thickness and refractive index are usually enough. Glass contains many atoms, but we need not solve for the motion and state of each one. Their average effect on the field is already included in the material parameters. Representing complex microscopic processes with a few macroscopic parameters is one reason engineering electromagnetics can handle practical devices.

Some problems, however, concern precisely the details that these parameters leave out.

For example, why does an atomic gas absorb incident light strongly only near certain frequencies? How does the energy inside an atom change while an external field continues to act? If these changes alter the absorption of the gas, a fixed refractive index is no longer sufficient. We must describe the allowed energy states of the atom and how the external field changes them.

Another example is weak-light detection. A power meter reports the average power over a time interval. It does not specify how much light arrives in each experimental trial. A source that emits one photon every time can have the same average power as a source that sometimes emits nothing and sometimes emits several photons. Yet they need not perform equally well when used to transmit information. Besides the spatial field distribution, we must consider the possible outcomes of each measurement and their probabilities. The next section explains what a photon is.

We can also place an atom that has already acquired energy inside a cavity. After the external excitation stops, the atom may still release energy by emitting light. This process is called spontaneous emission. For an antenna, we usually prescribe a current and then calculate its radiation. For the atom, we must also explain how radiation arises and how energy passes from the atom to the field. The surrounding cavity participates in this process, so the emission rate may differ from its free-space value.

These problems have two aspects: how matter stores and exchanges energy, and how the electromagnetic field is excited and measured. Classical electromagnetics provides the basis for propagation and spatial field distributions. Quantum theory supplies the additional description needed for these processes.

\section{Energy Levels and Photons}

\subsection{Why Describe Atomic Energy in Terms of Levels?}

In a classical LC circuit, the capacitor voltage can vary continuously, and so can the stored energy. The bound electronic system inside an atom behaves differently: its allowed bound-state energies are discrete. These allowed values are called energy levels.

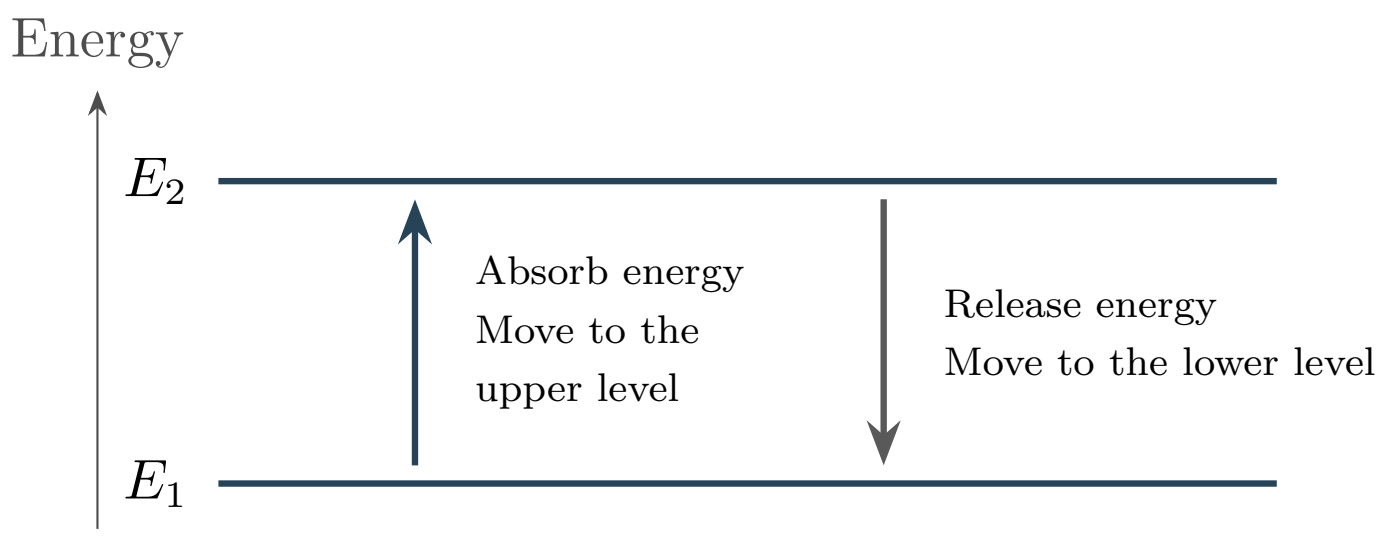
\begin{figure}[htbp]
\centering
\begin{tikzpicture}[font=\small,>=Stealth]
\draw[->,black!70] (0,-0.3)--(0,2.6) node[above]{Energy};
\draw[very thick,inkblue] (0.8,0)--(7.8,0);
\draw[very thick,inkblue] (0.8,2)--(7.8,2);
\node[anchor=east] at (0.75,0) {$E_1$};
\node[anchor=east] at (0.75,2) {$E_2$};
\draw[->,very thick,inkblue] (2.1,0.12)--(2.1,1.88);
\node[anchor=west,align=left,font=\scriptsize] at (2.35,1) {Absorb energy\\Move to the\\upper level};
\draw[->,very thick,black!65] (5.2,1.88)--(5.2,0.12);
\node[anchor=west,align=left,font=\scriptsize] at (5.45,1) {Release energy\\Move to the lower level};
\end{tikzpicture}
\caption{Two atomic energy levels and transitions between them. The upward arrow represents energy absorption; the downward arrow represents energy release. Vertical separation indicates only the energy difference.}
\label{fig:levels}
\end{figure}

The lowest energy level corresponds to the ground state, and higher levels correspond to excited states. Figure~\ref{fig:levels} represents two such levels by horizontal lines. Their heights indicate energy, not the spatial positions of electrons or their orbits around the nucleus.

A change from one energy level to another is called a transition. If the lower level has energy $E_1$ and the upper level has energy $E_2$, an upward transition requires energy $E_2-E_1$. The transition frequency $f$ is related to the energy difference by
\begin{equation}
E_2-E_1=hf.
\label{eq:transition}
\end{equation}
Here, $h$ is Planck's constant, which converts frequency into energy and has units of joule seconds. In terms of angular frequency $\omega = 2\pi f$, the relation is often written as $E_2 - E_1 = \hbar\omega$, where $\hbar = h/2\pi$. An incident electromagnetic wave near the transition frequency may cause appreciable absorption or stimulated emission, driving transitions between the two levels. This is the starting point for understanding atomic spectral lines, coherent control, and atomic frequency references.

A transition also depends on the electric-field direction and the properties of the atom. Matching the frequency alone does not guarantee that it occurs. Spectral lines also have finite widths. We will discuss transition strengths and linewidths when developing the equations for atom--field interaction.

Two kinds of discreteness must be distinguished here. A classical cavity supports only certain eigenfrequencies because its boundary conditions select the modes. Even after one mode has been selected, classical theory allows its amplitude and stored energy to vary continuously. Discrete atomic levels instead specify which energy states matter can occupy. These are different statements.

\subsection{How Are Photons Related to Electromagnetic Modes?}

Now consider the electromagnetic field. For an ideal mode of frequency $f$, quantum theory states that its excitation energy increases or decreases in units of $hf$. Each unit of excitation is called a photon. Adding one photon increases the mode energy by $hf$; removing one decreases it by $hf$. Even at zero photon number, irreducible quantum fluctuations remain. These are vacuum fluctuations.

A photon and a mode therefore describe different aspects of the field. A mode specifies its spatial distribution; the photon number specifies how strongly that mode is excited. In a linear cavity, increasing the photon number from one to two does not give the mode a new spatial shape.

Nor should a photon be pictured simply as a small ball following a definite trajectory. Light diffracts at a slit and interferes between two paths. These propagation properties remain present in single-photon experiments. Each detection records one localized event. The distribution of detection positions emerges only after many trials under the same conditions. The field distribution set by waveguides, lenses, and scatterers still helps determine these detection probabilities.

This does not mean that every weak-light problem requires a complicated quantum calculation. Classical fields often suffice for the average transmission of a linear device. But average power alone cannot tell us whether a pulse contains exactly one photon or whether two detectors can register signals together.

\section{What Else Does a Quantum State Specify?}

Having introduced energy levels and photons, we now need the word ``state.'' A quantum state describes the possible measurement outcomes of a system and the probability of each outcome. It also determines how the system evolves under an external field.

Consider an atom with two relevant energy levels. It may occupy the lower or the upper level. Suitable electromagnetic control can also prepare a coherent superposition of the two states. An energy measurement on this superposition still gives one of the allowed energies in each trial. Repeated preparation and measurement reveal the probabilities of the two outcomes.

A coherent superposition contains more information than the probabilities of occupying the two levels: it also has a relative phase. In engineering electromagnetics, changing the relative phase between two waves changes their interference even if their individual powers remain fixed. Superpositions of quantum states also involve phase, but the quantities being added are probability amplitudes for different states. A probability amplitude is a complex number whose squared magnitude gives the probability of the corresponding outcome. Probabilities alone discard the phase information.

A concrete comparison makes the distinction clear. Prepare one group of atoms with identical coherent control, giving equal probabilities for the lower and upper energy outcomes. Prepare a second group by randomly placing half the atoms in the lower level and half in the upper level. Direct energy measurements give the same statistics for both groups. If suitable coherent control is applied before the energy measurement, however, the first group responds to the relative phase established during preparation. The second group has no such phase dependence.

Recording only the fractions of atoms in each level can therefore miss information that affects their later response. These fractions are called populations; the phase relationships are described by the coherence between levels. Physically, populations describe the stored atomic energy, whereas coherence directly determines the oscillating material polarization. Chapter 2 places populations and coherence in one set of equations and explains how they determine absorption and polarization.

\begin{figure}[htbp]
\centering
\begin{tikzpicture}[font=\small,>=Stealth,x=1cm,y=1cm]
\foreach \x/\title in {0/{(a) One photon every time},6.4/{(b) Sometimes zero, sometimes two}}{
  \begin{scope}[xshift=\x cm]
  \node[font=\footnotesize] at (2,2.65) {\title};
  \draw[->] (0,0)--(4.2,0) node[above=2pt,align=center,font=\footnotesize]{Photon\\number};
  \draw[->] (0,0)--(0,2.35);
  \node[rotate=90] at (-1.1,1.2) {Probability};
  \draw (-0.06,1)--(0.06,1) node[left=3pt]{\scriptsize $0.5$};
  \draw (-0.06,2)--(0.06,2) node[left=3pt]{\scriptsize $1$};
  \foreach \u/\n in {0.8/0,2/1,3.2/2}{\node[below] at (\u,0){$\n$};}
  \node at (2,-0.9) {Mean photon number: $1$};
  \end{scope}
}
\draw[fill=black!18] (1.7,0) rectangle (2.3,2);
\draw[fill=black!18] (6.9,0) rectangle (7.5,1);
\draw[fill=black!18] (9.3,0) rectangle (9.9,1);
\end{tikzpicture}
\caption{Equal average energy need not imply equal photon statistics. Two sources emit pulses at the same frequency and repetition period. In (a), every pulse contains exactly one photon. In (b), half the pulses are empty and half contain two photons. The average powers are equal, but individual detection outcomes differ. The probabilities are obtained from many repeated measurements.}
\label{fig:ch1:statistics}
\end{figure}
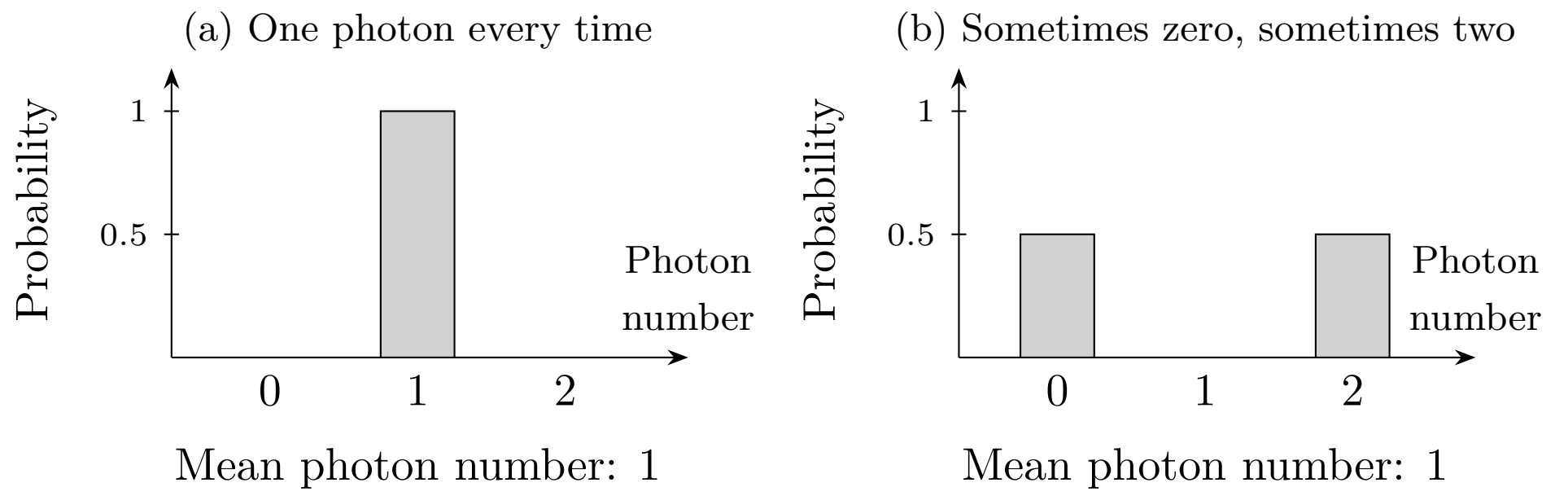

The field has a quantum state of its own. The same mode may contain exactly one photon in every trial, or have an uncertain photon number, as in the field from a conventional coherent source. States with the same average energy can still produce different measurement statistics. Chapter 3 develops their mathematical description. For now, remember that specifying a mode and its average energy does not specify all the quantum information in the field. Figure~\ref{fig:ch1:statistics} compares two photon-counting distributions.

\section{Choosing a Description of Field and Matter}

A device need not use the same description for every component. When studying atoms in a vapor cell, for example, we usually do not solve individually for the atoms in its glass walls. A permittivity still describes the glass. The states that require detailed treatment are the atomic levels relevant to the measurement and their evolution.

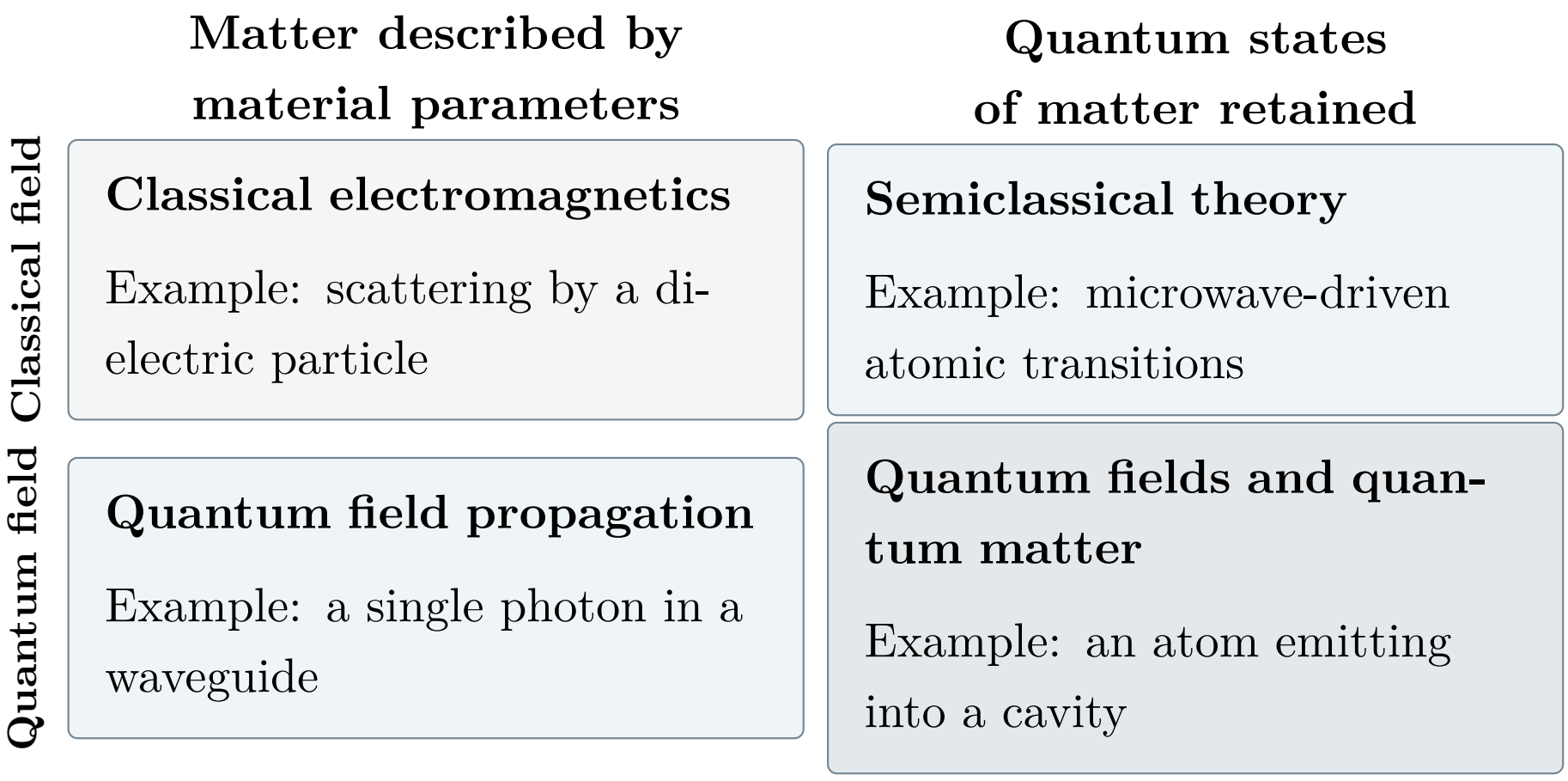
\begin{figure}[htbp]
\centering
\begin{tikzpicture}[font=\small,box/.style={draw=inkblue!65,rounded corners=2pt,text width=5.1cm,minimum height=2.0cm,align=left,inner sep=8pt}]
\node[box,fill=softgray] (a) at (0,0) {\textbf{Classical electromagnetics}\\[5pt]Example: scattering by a dielectric particle};
\node[box,fill=lightblue] (b) at (5.85,0) {\textbf{Semiclassical theory}\\[5pt]Example: microwave-driven atomic transitions};
\node[box,fill=lightblue] (c) at (0,-2.45) {\textbf{Quantum field propagation}\\[5pt]Example: a single photon in a waveguide};
\node[box,fill=inkblue!12] (d) at (5.85,-2.45) {\textbf{Quantum fields and quantum matter}\\[5pt]Example: an atom emitting into a cavity};
\node[align=center,font=\small\bfseries] at (0,1.6) {Matter described by\\material parameters};
\node[align=center,font=\small\bfseries] at (5.85,1.6) {Quantum states\\of matter retained};
\node[rotate=90,font=\footnotesize\bfseries] at (-3.16,0) {Classical field};
\node[rotate=90,font=\footnotesize\bfseries] at (-3.16,-2.45) {Quantum field};
\end{tikzpicture}
\caption{Four descriptions of field and matter, classified by the information retained in the calculation. Different questions about the same device may call for different combinations.}
\label{fig:frameworks}
\end{figure}

Considering field and matter separately gives the four combinations in Figure~\ref{fig:frameworks}. ``Material parameters'' means that quantities such as permittivity summarize the material response. ``Quantum states of matter'' means explicitly calculating the populations and coherence of the relevant levels. Likewise, the field can be treated classically or described by its quantum state, depending on the problem.

\subsection{Classical Electromagnetics}

When macroscopic parameters describe matter and classical variables describe the field, we recover familiar engineering electromagnetics. Examples include calculating scattering by a dielectric particle with a given permittivity, or antenna impedance for a given conductor geometry. Such calculations do not explicitly track individual quantum states within the material.

Material parameters can incorporate quantum effects of atoms and electrons. Using them in a classical calculation means that these microscopic processes have been summarized in the material response, not that the processes are absent.

\subsection{Classical Fields Driving Quantum Matter: Semiclassical Theory}

Semiclassical theory applies when we need to study atomic level dynamics while the driving field can still be represented as a classical wave. For example, we can first solve for a microwave field in a cavity. We then calculate how that field changes the probabilities of occupying the atomic levels.

If the atoms appreciably affect the field, their average polarization must be fed back into Maxwell's equations as a source, and the field recalculated. This resembles the self-consistent treatment of fields and material response in engineering electromagnetics. Here, however, the material response follows from changes in atomic states.

Many basic processes in atomic clocks and the average response of Rydberg-atom receivers can be studied at this level. Using quantum atoms does not require quantizing every driving field.

\subsection{Quantum-Field Propagation in a Given Structure}

Sometimes the question is how one or a few photons pass through a waveguide, beam splitter, or scatterer. We then need the quantum state of the field, while parameters such as permittivity can still describe the device materials. Classical electromagnetic calculations determine the effect of the structure on propagation; quantum theory supplies the probabilities of detection outcomes.

This treatment belongs to macroscopic quantum electrodynamics. ``Macroscopic'' means that the background material is described by its collective response rather than by tracking its atoms individually. If it absorbs, we must include both the associated energy exchange and the field fluctuations accompanying loss, or quantum noise. Chapter 3 starts with lossless modes and gradually develops these ideas.

\subsection{Joint Evolution of Quantum Fields and Quantum Matter}

To study how an atom emits a photon without external driving, or repeatedly exchanges energy with one or a few cavity photons, we must describe both atomic and field states. The field cannot always remain a prescribed drive. Nor can the atom be replaced by a fixed material parameter, or even by self-consistent feedback between its average polarization and a classical field.

This is the fully quantum framework used in this book. It concerns only the atoms and fields relevant to the problem, not every particle in the device. The surrounding substrate, package, and much of the dielectric environment can usually retain a macroscopic electromagnetic description.

Model selection should begin with a precise statement of what is to be calculated. A resonance frequency and an atomic transition probability require different information. So do average power and a single-photon detection probability. Small size, high frequency, or weak signals alone do not determine the model. A suitable model retains the physical processes that control the quantity of interest.

\section{Applications and the Chapters Ahead}

Engineering applications of quantum electromagnetics can be viewed through measurement, photonic devices, and quantum circuits.

An atomic clock uses an atomic transition as a frequency reference. The field from a local oscillator interacts with the atoms. Measurements of their state changes reveal frequency errors and provide a feedback signal. The atoms supply the reference, while the electromagnetic system controls the field frequency, amplitude, polarization, and phase. Field uniformity across the interaction region also affects the measurement.

Rydberg atoms occupy highly excited levels, and some transitions between these levels are sensitive to radio-frequency electric fields. A receiver uses the radio-frequency field to change the atomic state, converts that change into a change in probe-light transmission, and infers the electric field from the optical signal. Atomic levels explain only part of the response. Classical electromagnetic analysis is still needed to determine how the cell and surrounding structure modify the local field.

Single-photon devices concern how many photons are generated in each trial, which propagation mode they enter, and how likely they are to be detected. Making a source very weak does not ensure exactly one photon per trial. Useful output requires understanding the emitter's levels, its interaction with the surrounding structure, and losses during propagation and absorption. These questions directly connect atomic physics with wave\-guide, cavity, and scattering design.

Superconducting quantum circuits use nonlinear circuit elements to create controllable, unequally spaced discrete energy levels. Two states are selected as a quantum bit, or qubit, the basic unit of quantum information. Microwave fields control these states, while resonators and transmission lines provide coupling and readout. Modes, loss, crosstalk, and package parasitics therefore remain central design problems.

These applications use quantum properties, but performance must still be assessed for the actual task. Field sensing needs both a sufficiently large response and controlled noise. A single-photon source must suppress multiphoton events while maintaining useful output efficiency. Quantum models help us calculate these processes and their limits.

The following chapters develop the central theme of field--matter interaction. Chapter 2 combines classical fields with quantum matter. It retains populations and coherence in the relevant atomic levels, treats driving fields classically, and, where needed, solves self-consistently with Maxwell's equations through the average polarization. Chapter 3 combines quantum fields with linear media. It retains the field state, describes the background through macroscopic material response, and includes environmental fluctuations associated with loss. Chapter 4 treats quantum fields and quantum matter together, examining excitation exchange, correlations, and photon generation and detection.

These combinations answer different questions. For the same device, average absorption may be accessible with the equations of Chapter 2. Single-photon propagation or emitter--field correlations require the corresponding field-state information. Chapter 5 draws these models together, showing how to select degrees of freedom, connect electromagnetic and quantum calculations, and assess results for a specific task. Numerical methods are introduced only to the extent needed to obtain the results.

For readers trained in engineering electromagnetics, familiar fields and boundary conditions remain the starting point. As the questions extend from propagation to energy exchange and measurement, we add energy levels, quantum states, and their equations of motion. This progression makes the role of each part of the theory clear in device analysis.

\clearpage
\fancyhead[R]{\small Chapter 2}
\ctexset{chapter/format=\Large\normalfont,section/format=\large\normalfont,subsection/format=\normalsize\normalfont}
\captionsetup{labelfont=normalfont}

\chapter{Classical Fields and Quantum Matter}
\label{chap:semiclassical}

In engineering electromagnetics, specifying a material's permittivity allows us to solve for field propagation and scattering. But permittivity itself has a physical origin. An applied electric field drives charges within matter, producing polarization; that polarization then acts as a source that affects the electromagnetic field. For an atomic gas, resolving the absorption, phase response, and saturation of a particular transition requires an explicit calculation of this part of the material response.

Chapter 1 explained that energy-level populations represent the energy stored in an atom, while coherence between levels corresponds to polarization that oscillates in time. This chapter expresses that statement in equations. The electromagnetic field remains classical, while the atom is described by a few relevant energy levels and their probability amplitudes. We begin with a two-level atom, then consider atomic clocks and Rydberg-atom receivers. The quantum mechanics needed for each model is introduced along the way.

\section{From Energy Levels to State Equations}

\subsection{Probability Amplitudes and State Vectors}

Suppose that only two atomic energy levels participate in the process of interest. Denote the lower level by $g$ and the upper level by $e$. Set the lower-level energy to zero and the upper-level energy to $\hbar\omega_0$. Here {$\hbar=h/(2\pi)$ is the reduced Planck constant, and }$\omega_0=2\pi f_0$ is the atomic transition angular frequency{ (so $\hbar\omega_0=hf_0$, consistent with Chapter 1)}. The choice of energy zero does not affect the energy difference between the levels.

Represent the atomic state by two complex numbers, $c_g(t)$ and $c_e(t)$: an energy measurement finds the lower level with probability $|c_g|^2$ and the upper level with probability $|c_e|^2$. These complex numbers are probability amplitudes. When only these two levels are considered, the total probability is one, so
\begin{equation}
 |c_g|^2+|c_e|^2=1.
 \label{eq:normalization}
\end{equation}
The probability amplitudes can be assembled into a column vector. Quantum mechanics commonly uses the following notation:
\begin{equation}
 \ket{\psi}=c_g\ket{g}+c_e\ket{e}
 \quad\longleftrightarrow\quad
 \begin{pmatrix}c_g\\c_e\end{pmatrix},
 \qquad
 \ket{g}=\begin{pmatrix}1\\0\end{pmatrix},\quad
 \ket{e}=\begin{pmatrix}0\\1\end{pmatrix}.
 \label{eq:statevector}
\end{equation}
The symbol $\ket{\psi}$ is called a state vector. For this model, it is simply a column vector containing the two probability amplitudes; $\ket{g}$ and $\ket{e}$ are the chosen basis vectors. Additional levels require only additional components.

Transposing the column vector and taking the complex conjugate of each component gives the row vector $\bra{\psi}=(c_g^*,c_e^*)$. The asterisk denotes complex conjugation. The ordinary vector inner product can therefore be written as
\[
 \braket{\psi}{\psi}=(c_g^*,c_e^*)
 \begin{pmatrix}c_g\\c_e\end{pmatrix}=|c_g|^2+|c_e|^2=1.
\]
Similarly, $\braket{g}{\psi}=c_g$ and $\braket{e}{\psi}=c_e$ extract the corresponding components of the state. Most of the angle-bracket expressions below can be understood as row and column vector operations{. A bra on the left and a ket on the right, as in $\braket{\phi}{\psi}$, form a row-times-column inner product, whose result is a complex number. A ket on the left and a bra on the right, as in $\ket{\psi}\bra{\psi}$ in Eq.~\eqref{eq:densitydefinition} below, form a column-times-row outer product, whose result is a matrix}. The rule that probability equals the squared magnitude of the corresponding amplitude is a basic probabilistic postulate of quantum mechanics; it cannot be derived solely from classical electromagnetic energy formulas.

In engineering electromagnetics, the complex amplitudes of several modes have both magnitudes and phases. Complex information must also be retained here, but the squared magnitudes of $c_g$ and $c_e$ give measurement probabilities and cannot be identified directly with electric-field energy. Multiplying both probability amplitudes by the same phase factor does not change measurement results; their relative phase, however, affects subsequent driven evolution.

\subsection{How the Hamiltonian Matrix Determines Evolution}

Neglecting dissipation for now, the probability amplitudes obey the Schr\"odinger equation:
\begin{equation}
 \ii\hbar\frac{\dd}{\dd t}\ket{\psi}=H\ket{\psi},
 \qquad
 \ii\hbar\frac{\dd}{\dd t}
 \begin{pmatrix}c_g\\c_e\end{pmatrix}
 =
 \begin{pmatrix}H_{gg}&H_{ge}\\H_{eg}&H_{ee}\end{pmatrix}
 \begin{pmatrix}c_g\\c_e\end{pmatrix}.
 \label{eq:schrodinger}
\end{equation}
Here $\ii$ is the imaginary unit, and $H$ is the Hamiltonian matrix; every matrix element has units of energy. In the chosen energy-level basis, its diagonal elements in the absence of an applied field give the level energies, while off-diagonal elements introduced by the field describe coupling between states. Equation~\eqref{eq:schrodinger} is the fundamental law of quantum evolution used in this chapter. It cannot be derived from Maxwell's equations alone; once it is given, the few-level equations below follow by algebra.

Expanding the matrix product gives
\[
 \ii\hbar\dot c_g=H_{gg}c_g+H_{ge}c_e,\qquad
 \ii\hbar\dot c_e=H_{eg}c_g+H_{ee}c_e.
\]
For example, $H_{eg}c_g$ in the second equation is the contribution of the lower-level component to the change in the upper-level probability amplitude. For evolution without dissipation, $H$ equals its conjugate transpose, $H=H^\dagger$; thus its diagonal elements are real and $H_{ge}=H_{eg}^*$. Substituting the state equation{ $\dot{\ket{\psi}}=-(\ii/\hbar)H\ket{\psi}$ }and its conjugate transpose{ $\dot{\bra{\psi}}=(\ii/\hbar)\bra{\psi}H^\dagger$ (complex conjugation reverses the sign of the imaginary unit, and transposition reverses the order of a matrix product)} into the time derivative of the inner product gives
\[
 \frac{\dd}{\dd t}\braket{\psi}{\psi}
 =\frac{\ii}{\hbar}\bra{\psi}(H^\dagger-H)\ket{\psi}=0.
\]
Thus population can be exchanged between the levels while the total probability remains one.

Without an applied field, $H=\operatorname{diag}(0,\hbar\omega_0)$. Substitution into Eq.~\eqref{eq:schrodinger} gives $c_g(t)=c_g(0)$ and $c_e(t)=c_e(0)\ee^{-\ii\omega_0t}$. The populations remain unchanged, but the relative phase of the upper- and lower-level probability amplitudes evolves in time. Equations for populations alone would omit this information.

An applied field generally introduces off-diagonal elements into $H$, coupling the two probability amplitudes. The solution procedure resembles that for coupled-mode equations: given the initial state and coupling coefficients, integrate a system of ordinary differential equations with complex coefficients. What distinguishes the quantum model is that these coefficients correspond to energy levels and transitions, and the resulting amplitudes determine measurement probabilities.

\subsection{Where Energy Levels and Coupling Coefficients Come From}

The two-level model has not yet described the atom's internal spatial structure. Starting from the motion of charged particles requires a wavefunction $\psi(\mathbf r,t)$. The quantity $|\psi(\mathbf r,t)|^2\dd^3r$ is the probability of finding the particle in a small volume near position $\mathbf r$. The wavefunction is neither the actual electromagnetic field nor the particle's trajectory.

Consider a nonrelativistic particle of mass $m$, confined by a potential energy $V(\mathbf r)$. In the absence of an applied field, its stationary states satisfy
\begin{equation}
 \left[-\frac{\hbar^2}{2m}\nabla^2+V(\mathbf r)\right]
 \varphi_n(\mathbf r)=E_n\varphi_n(\mathbf r).
 \label{eq:eigenstates}
\end{equation}
This is an eigenvalue problem{. The first term on the left corresponds to the classical kinetic energy $p^2/(2m)$, obtained by squaring the momentum operator $\hat{\mathbf p}=-\ii\hbar\nabla$ introduced below; the second term is the confining potential energy. Its form is analogous to the Helmholtz equation used to find cavity eigenmodes in electromagnetics}. Here $E_n$ is an allowed energy and $\varphi_n$ the corresponding spatial wavefunction. The wavefunctions can be chosen normalized and mutually orthogonal, so that $\int\varphi_m^*\varphi_n\dd^3r=\delta_{mn}$, where $m=n$ gives $\delta_{mn}=1$ and otherwise the result is zero.

An operator represents a linear operation on a state. For example, the momentum operator $\hat{\mathbf p}=-\ii\hbar\nabla$ differentiates the wavefunction. Once a basis is chosen, an operator can be represented by a matrix. Below, a hat emphasizes an operator, whereas $H$ directly denotes its matrix in the chosen energy-level basis.

To obtain equations for a finite set of probability amplitudes, expand the wavefunction as $\psi(\mathbf r,t)=\sum_n c_n(t)\varphi_n(\mathbf r)$. {Denote the field-free energy operator on the left of Eq.~\eqref{eq:eigenstates} by $\hat H_0$, and }the applied-field interaction by $\hat H_{\mathrm{int}}$. Substitute the expansion into the spatial form of the {time-dependent} Schr\"odinger equation{ $\ii\hbar\partial_t\psi=(\hat H_0+\hat H_{\mathrm{int}})\psi$}, multiply from the left by $\varphi_m^*$, and integrate over space. Using orthogonality gives
\begin{equation}
 \ii\hbar\dot c_m=E_mc_m+\sum_n V_{mn}(t)c_n,
 \qquad V_{mn}=\int\varphi_m^*\hat H_{\mathrm{int}}\varphi_n\,\dd^3r.
 \label{eq:modeprojection}
\end{equation}
The spatial problem has now become a problem of probability-amplitude evolution: $E_m$ comes from the atomic eigenstates and $V_{mn}$ from coupling induced by the applied field. This parallels the electromagnetic procedure of first finding eigenmodes and then projecting onto them to obtain coupled-mode equations. If other levels are barely excited at the frequencies and field strengths used, retaining only the relevant states $g$ and $e$ returns the two-by-two matrix form of Eq.~\eqref{eq:schrodinger}.

Real atoms usually contain several electrons, and their levels may be further split. This chapter does not calculate these levels from first principles, but takes transition frequencies and transition dipole moments as known parameters. They can be obtained from atomic-structure calculations or spectroscopic data. For device analysis, the key is to select the levels that participate in the interaction and use parameters appropriate to the actual levels and field polarization.

\section{How Classical Electromagnetic Fields Enter Quantum Equations}

\subsection{Electromagnetic Potentials and Minimal Coupling}

Engineering electromagnetics usually works directly with $\mathbf E$ and $\mathbf B$. In quantum equations for a charged particle, it is more convenient to use the vector potential $\mathbf A$ and scalar potential $\phi$:
\begin{equation}
 \mathbf E=-\nabla\phi-\frac{\partial\mathbf A}{\partial t},
 \qquad \mathbf B=\nabla\times\mathbf A.
 \label{eq:potentials}
\end{equation}
This introduces no additional independent electromagnetic fields. In vacuum, imposing the Lorenz gauge $\nabla\cdot\mathbf A+c^{-2}\partial_t\phi=0$ allows Maxwell's equations to be written as
\begin{equation}
 \left(\nabla^2-\frac{1}{c^2}\frac{\partial^2}{\partial t^2}\right)
 \mathbf A=-\mu_0\mathbf J,
 \qquad
 \left(\nabla^2-\frac{1}{c^2}\frac{\partial^2}{\partial t^2}\right)
 \phi=-\frac{\rho_{\mathrm c}}{\varepsilon_0}.
 \label{eq:potentialwave}
\end{equation}
Here $\rho_{\mathrm c}$ is the charge density, $\mathbf J$ the current density, and $c$ the speed of light in vacuum. Material interfaces and boundary conditions must still be treated according to the particular electromagnetic problem.

Let the particle charge be $q$. For an electron, $q=-q_e$, where $q_e>0$ is the elementary charge. The applied electromagnetic field enters its energy operator through
\begin{equation}
 \hat H=
 \frac{1}{2m}\bigl[-\ii\hbar\nabla-q\mathbf A\bigr]^2
 +q\phi+V.
 \label{eq:minimal}
\end{equation}
This is called minimal coupling. The original confining potential energy is already represented by $V$, while $\mathbf A$ and $\phi$ here represent the applied-field potentials. The same binding interaction must not be counted again in $q\phi$.

The combination $\hat{\mathbf p}-q\mathbf A$ has a direct classical counterpart. The canonical momentum of a charged particle is $\mathbf p=m\mathbf v+q\mathbf A$, so the mechanical momentum associated with its actual motion is $m\mathbf v=\mathbf p-q\mathbf A$, and its kinetic energy is $(\mathbf p-q\mathbf A)^2/(2m)$. Replacing canonical momentum by $-\ii\hbar\nabla$ in the quantum equation gives the kinetic-energy term in Eq.~\eqref{eq:minimal}. Canonical momentum must not be identified directly with $m\mathbf v$ here.

The kinetic-energy term in Eq.~\eqref{eq:minimal} contains terms linear in $\mathbf A$ as well as an $\mathbf A^2$ term. When $\mathbf A$ varies with position, {the operator $\hat{\mathbf p}=-\ii\hbar\nabla$ differentiates the entire product to its right, so }the order of momentum and vector-potential factors must be retained: the linear term is $-q(\hat{\mathbf p}\cdot\mathbf A+\mathbf A\cdot\hat{\mathbf p})/(2m)${ (the order can be interchanged only when $\nabla\cdot\mathbf A=0$)}. These details matter when constructing a model from spatial wavefunctions. For the near-resonant atomic transitions considered here, the interaction can be transformed into the more intuitive electric-dipole form.

\subsection{From Electromagnetic Potentials to Electric-Dipole Coupling}

An atom is usually much smaller than the wavelength of the applied electromagnetic wave, so the applied electric field can be treated as uniform within the atom. Denote the atomic center by $\mathbf r_0$ and the position of an internal charge relative to that center by $\mathbf x$. Retaining the leading interaction of the electric field with the electric dipole gives
\begin{equation}
 \hat H_{\mathrm{int}}=-\hat{\mathbf d}\cdot\mathbf E(\mathbf r_0,t),
 \qquad \hat{\mathbf d}=q\mathbf x
 \label{eq:dipoleinteraction}
\end{equation}
for a single particle; in a many-particle atom, the dipole moment is the sum of all charge contributions. This is the electric-dipole approximation. We first consider electric-dipole-allowed transitions; magnetic-dipole transitions require the corresponding magnetic-field coupling term.

Equation~\eqref{eq:dipoleinteraction} can be obtained from Eq.~\eqref{eq:minimal} by a gauge transformation. A gauge transformation changes the representation of electromagnetic potentials: for any function $\chi(\mathbf r,t)$, set
\begin{equation}
 \mathbf A'=\mathbf A+\nabla\chi,\qquad
 \phi'=\phi-\partial_t\chi,\qquad
 \psi'=\ee^{\ii q\chi/\hbar}\psi.
 \label{eq:gaugetransform}
\end{equation}
The first two relations leave $\mathbf E$ and $\mathbf B$ unchanged, while the third changes the particle description consistently. To see the relation between the two coupling forms, first represent the applied field {at the atomic center $\mathbf r_0$} by a spatially uniform $\mathbf A(t)$ within the electric-dipole approximation, and set the applied scalar potential to zero. Then $\mathbf E{(\mathbf r_0,t)}=-\dot{\mathbf A}$. Choosing $\chi=-\mathbf x\cdot\mathbf A(t)$ gives
\[
 \mathbf A'=\mathbf A-\mathbf A=0,\qquad
 \phi'=\mathbf x\cdot\dot{\mathbf A}=-\mathbf x\cdot\mathbf E.
\]
The new kinetic-energy term no longer contains the applied vector potential, while the potential-energy term becomes $q\phi'=-q\mathbf x\cdot\mathbf E=-\hat{\mathbf d}\cdot\mathbf E$. This procedure also changes the wavefunction's phase representation; it does not remove the physical action of the electric field. Describing the coupling through potentials or through the electric field must therefore give the same physical results.

Equivalence of the complete models does not guarantee equivalence after an arbitrary truncation. Retaining only two levels and then discarding interaction terms arbitrarily may break this consistency. For the weak-coupling, near-resonant problems in this chapter, the electric-dipole form can be used directly. At stronger coupling, the consistency of level truncation and gauge choice must be checked~\cite{gauge}.

After selecting the levels $g$ and $e$, the off-diagonal element of the electric-dipole operator is
\begin{equation}
 \mathbf d_{eg}=\bra{e}\hat{\mathbf d}\ket{g}
 =\int\varphi_e^*(\mathbf x)\,q\mathbf x\,
 \varphi_g(\mathbf x)\,\dd^3x.
 \label{eq:transitiondipole}
\end{equation}
The symbol $\bra{e}$ is the conjugate transpose of $\ket{e}$, and the asterisk denotes complex conjugation. The quantity $\mathbf d_{eg}$ is the transition dipole moment, measured in coulomb meters. Its dot product with the applied electric field determines the coupling strength, so polarization matters just as field strength does. If $\mathbf d_{eg}\cdot\mathbf E=0$ for the chosen polarization, this electric-dipole term cannot drive the transition even when the field frequency is resonant.

This corresponds directly to the coupling integral in Eq.~\eqref{eq:modeprojection}: because the applied field is approximately uniform within the atom, it can be taken outside the integral, giving $V_{eg}=-\mathbf d_{eg}\cdot\mathbf E$. The transition dipole moment characterizes coupling between two internal wavefunctions; it does not mean that the atom always carries a classical dipole moment of fixed magnitude and direction. {For atomic eigenstates with inversion symmetry, $|\varphi_n(\mathbf x)|^2$ is even, so its product with the odd function $q\mathbf x$ integrates to zero. Thus the diagonal elements satisfy $\mathbf d_{gg}=\mathbf d_{ee}=0$ (there is no permanent electric dipole moment); a nonzero $\mathbf d_{eg}$ is possible only between levels of opposite parity.}

\section{Response of a Two-Level Atom to a Harmonic Field}

\subsection{Detuning and Rabi Frequency}

Let the electric field at the atom be monochromatic:
\begin{equation}
 \mathbf E(t)=\frac12\left(\mathbf E_0\ee^{-\ii\omega t}
 +\mathbf E_0^*\ee^{\ii\omega t}\right),
 \qquad
 \Delta=\omega-\omega_0,\qquad
 \Omega=\frac{\mathbf d_{eg}\cdot\mathbf E_0}{\hbar}.
 \label{eq:fieldrabi}
\end{equation}
Here $\mathbf E_0$ is the complex peak electric-field amplitude, not its root-mean-square value. The detuning $\Delta$ is the difference between the driving and transition angular frequencies. The Rabi angular frequency $\Omega$ may be complex, and its phase carries the phase information of the drive. The quantities $\omega$, $\Delta$, and $\Omega$ all use the angular-frequency convention; divide by $2\pi$ to convert to hertz{. Quantum optics conventionally uses $\ee^{-\ii\omega t}$ as the positive-frequency phasor factor, opposite in sign to the $\ee^{+\mathrm j\omega t}$ commonly used in microwave circuits}.

First insert the electric-dipole coupling into the two probability-amplitude equations. For two levels without permanent electric dipole moments,
\begin{equation}
 \begin{aligned}
 \ii\dot c_g&=-\frac{\mathbf d_{ge}\cdot\mathbf E(t)}{\hbar}c_e,\\
 \ii\dot c_e&=\omega_0c_e-\frac{\mathbf d_{eg}\cdot\mathbf E(t)}{\hbar}c_g,
 \end{aligned}
 \label{eq:labamplitudes}
\end{equation}
where $\mathbf d_{ge}=\mathbf d_{eg}^*$. The applied field first changes the probability amplitudes, whose squared magnitudes then give population changes. Treating field strength directly as a population-change rate would omit phase effects.

Even without an applied field, $c_e$ contains the rapid variation $\ee^{-\ii\omega_0t}$. To separate this phase, set $c_g=b_g$ and $c_e=b_e\ee^{-\ii\omega t}$, using the drive frequency as a reference{. This is analogous to demodulation in radio-frequency engineering, where the carrier phase is removed to extract a slowly varying complex envelope}. When differentiating, the derivative of the phase factor must be retained: $\dot c_e=(\dot b_e-\ii\omega b_e)\ee^{-\ii\omega t}$. Substituting into the second of Eqs.~\eqref{eq:labamplitudes} and canceling $\ee^{-\ii\omega t}$ on both sides gives
\begin{equation}
 \ii\dot b_e=-\Delta b_e-\frac{\Omega}{2}b_g
 -\frac{\mathbf d_{eg}\cdot\mathbf E_0^*}{2\hbar}
   \ee^{2\ii\omega t}b_g.
 \label{eq:beforerwa}
\end{equation}
The diagonal term $-\Delta=\omega_0-\omega$ is precisely the frequency difference between the atom and the chosen phase reference. The near-resonant coupling $-\Omega/2$ no longer oscillates rapidly, whereas the final term still varies at angular frequency $2\omega${, much like the doubled-frequency component produced by mixing. Similarly, substituting $c_e=b_e\ee^{-\ii\omega t}$ into the first of Eqs.~\eqref{eq:labamplitudes} gives $\ii\dot b_g=-(\Omega^*/2)b_e-[(\mathbf d_{ge}\cdot\mathbf E_0)/(2\hbar)]\ee^{-2\ii\omega t}b_e$. The slowly varying coupling $-\Omega^*/2$ arises from phase cancellation between the field component $\mathbf E_0^*\ee^{\ii\omega t}$ and $\ee^{-\ii\omega t}$}.

When $|\Omega|$, $|\Delta|$, and the relevant relaxation rates are all much smaller than $\omega_0$, the amplitudes $b_g$ and $b_e$ change little within an optical cycle. Contributions from opposite phases of the rapidly varying terms largely cancel, while the near-resonant terms accumulate over longer times. Neglecting the rapid terms is called the rotating-wave approximation. It is a separation-of-timescales approximation, not an exact transformation. The probability amplitudes then satisfy
\begin{equation}
 \ii\frac{\dd}{\dd t}
 \begin{pmatrix}b_g\\b_e\end{pmatrix}
 =\frac{H_{\mathrm r}}{\hbar}
 \begin{pmatrix}b_g\\b_e\end{pmatrix},
 \qquad
 \frac{H_{\mathrm r}}{\hbar}=
 \begin{pmatrix}
 0&-\Omega^*/2\\
 -\Omega/2&-\Delta
 \end{pmatrix}.
 \label{eq:rotatinghamiltonian}
\end{equation}
The subscript $\mathrm r$ denotes the matrix in this rotating phase reference. It still describes the same atom; only the rapid phase has been separated from the unknown amplitudes. The negative signs come from $-\hat{\mathbf d}\cdot\mathbf E$. With consistent definitions, a different phase convention leaves transition probabilities unchanged.

\subsection{Rabi Oscillations and Pulse Control}

Set the drive exactly on resonance, $\Delta=0$, and choose its phase so that $\Omega$ is real and positive. Initially the atom occupies the lower level: $b_g(0)=1$ and $b_e(0)=0$. Expanding Eq.~\eqref{eq:rotatinghamiltonian} gives
\[
 \dot b_g=\frac{\ii\Omega}{2}b_e,\qquad
 \dot b_e=\frac{\ii\Omega}{2}b_g.
\]
Differentiating the first equation once more and using the second to eliminate $\dot b_e$ gives
\[
 \ddot b_g=\frac{\ii\Omega}{2}\dot b_e
 =-\frac{\Omega^2}{4}b_g.
\]
This is the familiar simple-harmonic-oscillator equation. Using $b_g(0)=1$ and $\dot b_g(0)=0$, first solve for $b_g$, then obtain $b_e$ from the first-order equation:
\begin{equation}
 b_g(t)=\cos\frac{\Omega t}{2},\qquad
 b_e(t)=\ii\sin\frac{\Omega t}{2},\qquad
 p_e(t)=|b_e|^2=\sin^2\frac{\Omega t}{2}.
 \label{eq:rabionresonance}
\end{equation}
The atom is first excited to the upper level, then driven back to the lower level by the same field, producing an oscillation of population in time. This is a Rabi oscillation. The probability amplitudes contain $\Omega/2$, whereas the population is $[1-\cos(\Omega t)]/2$, so the population oscillates at angular frequency $\Omega$. Continuous illumination does not mean continuous energy absorption: absorption and stimulated emission are both contained in the same coupled equations.

For a constant-amplitude pulse of duration $\tau$, the condition $\Omega\tau=\pi$ transfers all initial lower-level population to the upper level; this is a $\pi$ pulse. When $\Omega\tau=\pi/2$, each level has half the population while retaining a definite relative phase; this is a $\pi/2$ pulse. The latter prepares a coherent superposition, which differs from a random mixture with half the atoms in each level.

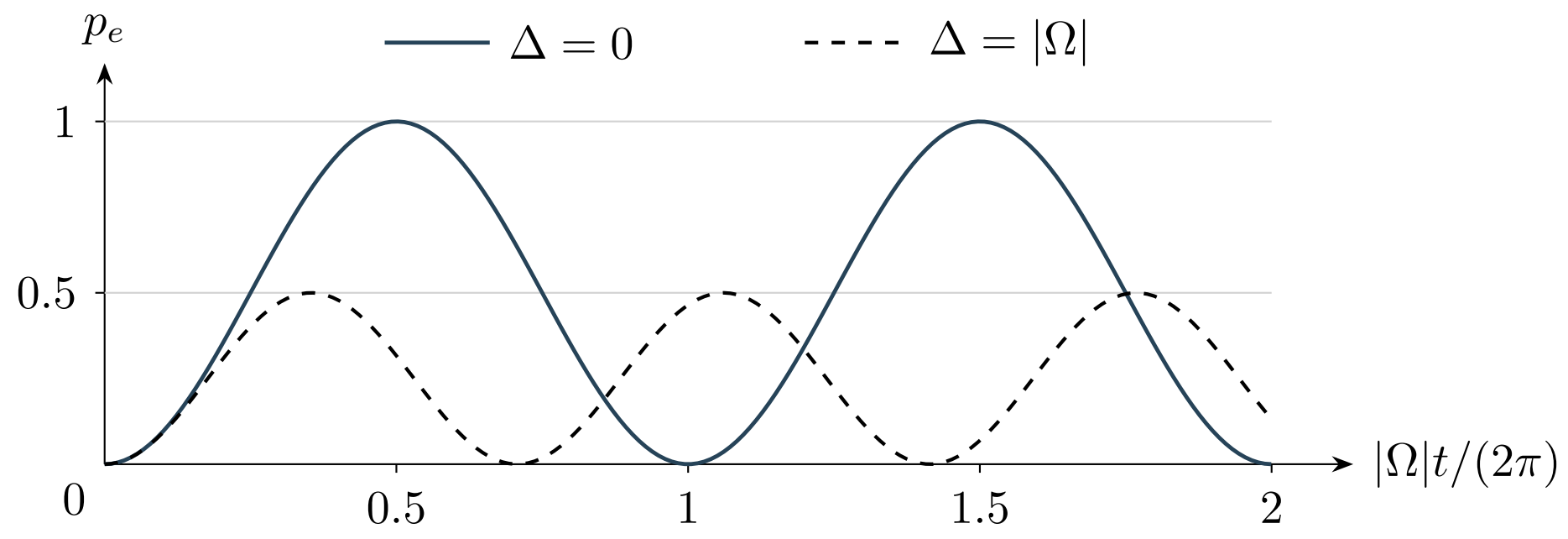
\begin{figure}[htbp]
 \centering
\begin{tikzpicture}[x=4.6cm,y=2.7cm,>=Stealth,font=\small]
 \draw[->] (0,0)--(2.14,0) node[right] {$|\Omega|t/(2\pi)$};
 \draw[->] (0,0)--(0,1.17) node[above] {$p_e$};
 \foreach \x in {0.5,1,1.5,2}{\draw (\x,0)--(\x,-.025) node[below] {\x};}
 \foreach \y in {0.5,1}{\draw (0,\y)--(-.016,\y) node[left] {\y};}
 \node[below left] at (0,0) {0};
 \draw[gray!35,thin] (0,.5)--(2,.5);
 \draw[gray!35,thin] (0,1)--(2,1);
 \draw[inkblue,line width=.9pt,domain=0:2,samples=180]
 plot (\x,{sin(180*\x)^2});
 \draw[black,dashed,line width=.8pt,domain=0:2,samples=180]
 plot (\x,{.5*sin(180*sqrt(2)*\x)^2});
 \draw[inkblue,line width=.9pt] (.48,1.23)--(.66,1.23);
 \node[right] at (.66,1.23) {$\Delta=0$};
 \draw[dashed,line width=.8pt] (1.2,1.23)--(1.38,1.23);
 \node[right] at (1.38,1.23) {$\Delta=|\Omega|$};
\end{tikzpicture}
 \caption{Rabi oscillations of a two-level atom without dissipation. The horizontal axis is normalized to the resonant Rabi period. At a detuning of $|\Omega|$, the maximum excitation probability decreases to $1/2$.}
 \label{fig:rabi}
\end{figure}

With detuning, variables can be eliminated in the same way. For constant $\Omega$ and $\Delta$, the upper-level amplitude obeys
\[
 \ddot b_e-\ii\Delta\dot b_e+\frac{|\Omega|^2}{4}b_e=0.
\]
Setting $b_e=\ee^{\ii\Delta t/2}z(t)$ eliminates the first derivative, giving $\ddot z+(|\Omega|^2+\Delta^2)z/4=0$. Define $\Omega_{\mathrm R}=\sqrt{|\Omega|^2+\Delta^2}$ and use $z(0)=0$ and $\dot z(0)=\ii\Omega/2$ to obtain
\[
 b_e(t)=\frac{\ii\Omega}{\Omega_{\mathrm R}}
 \ee^{\ii\Delta t/2}\sin\frac{\Omega_{\mathrm R}t}{2}.
\]
Taking the squared magnitude gives
\begin{equation}
 p_e(t)=\frac{|\Omega|^2}{|\Omega|^2+\Delta^2}
 \sin^2\left(\frac{\sqrt{|\Omega|^2+\Delta^2}}{2}t\right).
 \label{eq:rabidetuned}
\end{equation}
Detuning increases the oscillation angular frequency but reduces the maximum excited-state population, as shown in Figure~\ref{fig:rabi}. Faster population oscillations therefore do not necessarily imply a stronger local electric field; the detuning must also be checked.

For example, if $|\Omega|/(2\pi)=1\,\mathrm{MHz}$, a resonant $\pi$ pulse lasts $0.5\,\mu\mathrm{s}$. This interaction time is set by the field at the atom. Converting feed-source output power into $\Omega$ requires calculating the field distribution of the antenna, vapor cell, or resonant structure.

\section{Population, Coherence, and Dissipation}

\subsection{Why We Also Need a Density Matrix}

Probability amplitudes apply when the state is prepared in a definite way and environmental effects are temporarily neglected. A real atomic sample may contain different initial states, and its evolution may be affected by collisions, radiation, and field fluctuations. A single state vector is then usually insufficient: the possible states and their statistical weights must be retained together.

If the sample has probability $w_a$ of occupying state $\ket{\psi_a}$, define its density matrix by
\begin{equation}
 \rho=\sum_a w_a\ket{\psi_a}\bra{\psi_a},
 \qquad w_a\geq0,\quad\sum_a w_a=1.
 \label{eq:densitydefinition}
\end{equation}
Here $\ket{\psi_a}\bra{\psi_a}$ is the product of a column vector and its conjugate row vector. For a single definite state, the two-level density matrix is
\begin{equation}
 \rho=
 {\begin{pmatrix}\rho_{gg}&\rho_{ge}\\\rho_{eg}&\rho_{ee}\end{pmatrix}=}
 \begin{pmatrix}
 |c_g|^2&c_gc_e^*\\
 c_ec_g^*&|c_e|^2
 \end{pmatrix}.
 \label{eq:puredensity}
\end{equation}
The diagonal elements store populations, and the off-diagonal elements store coherence between levels. A density matrix does not describe a different physical system; it is a representation that accommodates both definite states and statistical mixtures.

Specifically, if $c_g=|c_g|\ee^{\ii\theta_g}$ and $c_e=|c_e|\ee^{\ii\theta_e}$, then $\rho_{eg}=|c_e||c_g|\ee^{\ii(\theta_e-\theta_g)}$, which retains the relative phase of the amplitudes. Statistical averaging over many preparations can drive the average phase factor toward zero even when $\rho_{gg}$ and $\rho_{ee}$ remain unchanged. This is the matrix meaning of equal populations with different coherences.

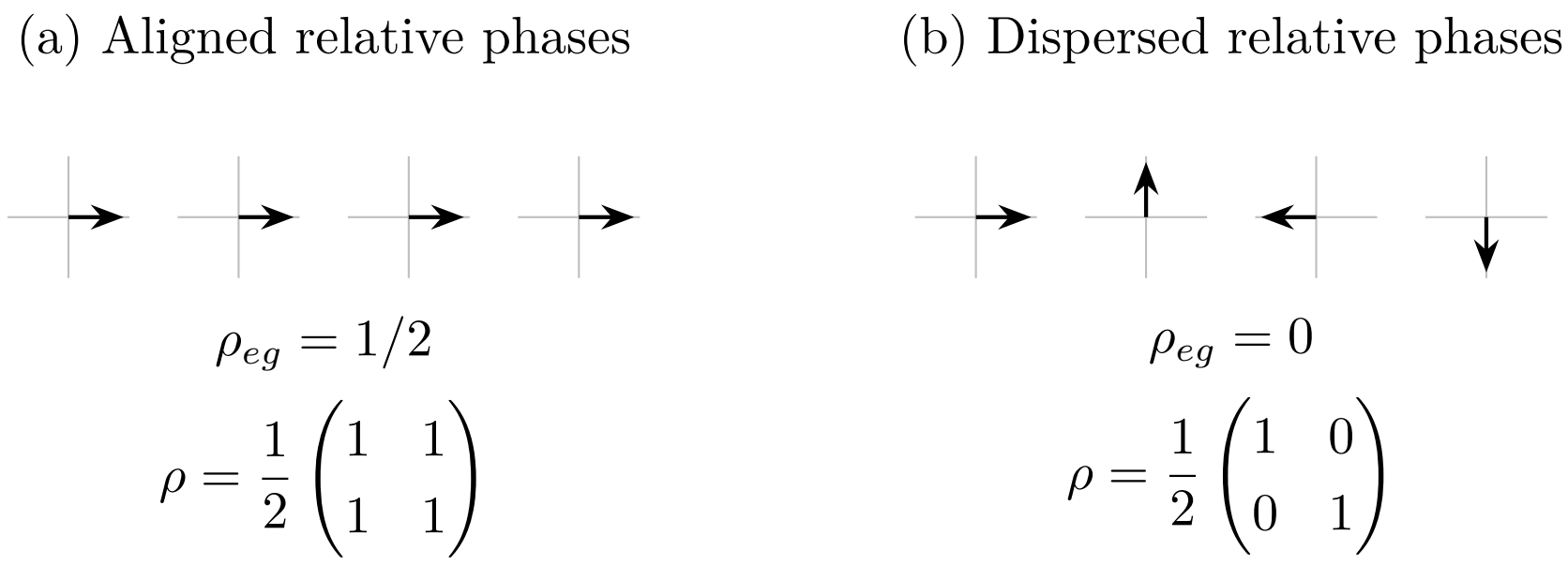
\begin{figure}[htbp]
\centering
\begin{tikzpicture}[font=\small,>=Stealth]
\foreach \shift/\title in {0/{(a) Aligned relative phases},6.4/{(b) Dispersed relative phases}}{
 \begin{scope}[xshift=\shift cm]
 \node at (2.25,1.25) {\title};
 \foreach \x in {0.45,1.65,2.85,4.05}{
  \draw[black!25] (\x-0.43,0)--(\x+0.43,0);
  \draw[black!25] (\x,-0.43)--(\x,0.43);
 }
 \end{scope}
}
\foreach \x in {0.45,1.65,2.85,4.05}{\draw[->,thick] (\x,0)--++(0.39,0);}
\foreach \x/\a in {6.85/0,8.05/90,9.25/180,10.45/270}{\draw[->,thick] (\x,0)--++(\a:0.39);}
\node[align=center] at (2.25,-1.55) {$\rho_{eg}=1/2$\\[5pt]$\displaystyle\rho=\frac12\begin{pmatrix}1&1\\1&1\end{pmatrix}$};
\node[align=center] at (8.65,-1.55) {$\rho_{eg}=0$\\[5pt]$\displaystyle\rho=\frac12\begin{pmatrix}1&0\\0&1\end{pmatrix}$};
\end{tikzpicture}
\caption{Equal populations can have different coherences. Each small coordinate system shows the phase of the complex quantity $c_ec_g^*$ in one preparation; every preparation has $|c_g|^2=|c_e|^2=1/2$. On the left, aligned phases preserve nonzero coherence after averaging. On the right, four equally probable phases $0,\pi/2,\pi,3\pi/2$ represent phase dispersion, and the averaged off-diagonal elements vanish. The density matrix on the right also equals the energy-level mixture in Eq.~\eqref{eq:superpositionmixture}: different preparations can produce the same density matrix.}
\label{fig:ch2:coherence}
\end{figure}

For example, the definite superposition $(\ket{g}+\ket{e})/\sqrt2$ and a mixture with equal probabilities of occupying $\ket{g}$ and $\ket{e}$ correspond, respectively, to
\begin{equation}
 \rho_{\mathrm{sup}}=\frac12
 \begin{pmatrix}1&1\\1&1\end{pmatrix},
 \qquad
 \rho_{\mathrm{mix}}=\frac12
 \begin{pmatrix}1&0\\0&1\end{pmatrix}.
 \label{eq:superpositionmixture}
\end{equation}
Their energy-measurement statistics are identical, but their responses to subsequent pulses with a phase reference differ. The distinction appears explicitly in the off-diagonal elements. Figure~\ref{fig:ch2:coherence} illustrates the relation between phase averaging and matrix elements.

For a definite state, quantum mechanics uses $\bra{\psi}O\ket{\psi}$ to evaluate the mean of a physical quantity $O${. When the operator is diagonal, this is the average of the level-specific measurement values weighted by the populations $|c_i|^2$; when it has off-diagonal elements, coherent cross terms also contribute}. Multiplying the row and column vectors and using $\rho_{ji}=c_jc_i^*$ gives
\[
 \bra{\psi}O\ket{\psi}
 =\sum_{i,j}c_i^*O_{ij}c_j
 =\sum_{i,j}\rho_{ji}O_{ij}=\Tr(\rho O).
\]
For a statistical mixture, take a further weighted sum with weights $w_a$. Linearity again gives
\begin{equation}
 \avg{O}=\Tr(\rho O).
 \label{eq:expectation}
\end{equation}
The symbol $\Tr$ denotes the sum of the diagonal matrix elements. For the upper-level population, for example, $O=\ket{e}\bra{e}=\operatorname{diag}(0,1)$; multiplying and taking the trace gives $\avg{O}=\rho_{ee}$. This matrix is the upper-level projection operator: it retains the upper-level component and sets the lower-level component to zero. The density matrix satisfies $\Tr\rho=1$ and $\rho=\rho^\dagger$ and has nonnegative eigenvalues, ensuring total probability, conjugate symmetry, and nonnegative probabilities, respectively.

\subsection{Relaxation and Coherence Times}

Consider two distinct decay processes. An atom in the upper level can decay to the lower level, reducing the excited-state population at a rate denoted by $\Gamma$. Even without a population change, phases accumulated by different atoms may gradually spread, causing the average coherence to decay. Denote this additional decay rate by $\gamma_\phi$, the pure-dephasing rate.

First write the density-matrix equation without dissipation. Differentiate $\rho=\ket{\psi}\bra{\psi}$ and substitute the state-vector equation and its conjugate transpose to obtain
\[
 \dot\rho=\dot{\ket{\psi}}\bra{\psi}
 +\ket{\psi}\dot{\bra{\psi}}
 =-\frac{\ii}{\hbar}H\rho+\frac{\ii}{\hbar}\rho H.
\]
Weighting by the preparation probabilities shows that a statistical mixture satisfies the same equation. Dissipation is commonly included through the Lindblad master equation:
\begin{equation}
 \frac{\dd\rho}{\dd t}
 =-\frac{\ii}{\hbar}[H,\rho]
 +\sum_k\left(
 L_k\rho L_k^\dagger
 -\frac12 L_k^\dagger L_k\rho
 -\frac12\rho L_k^\dagger L_k\right).
 \label{eq:lindblad}
\end{equation}
The commutator is $[H,\rho]=H\rho-\rho H$. The first term describes coherent evolution due to the applied field and level energies; each subsequent group of terms describes one dissipation channel. The matrix $L_k$ is a Lindblad operator, also called a jump operator (or a transition operator below). It specifies how that channel acts on the state and has units of $\mathrm{s}^{-1/2}$. A ``jump'' need not transfer population between energy levels: pure dephasing can also be represented by such operators.

For the two-level model in this chapter, use
\begin{equation}
 L_1=\sqrt{\Gamma}\ket{g}\bra{e},\qquad
 L_\phi=\sqrt{\frac{\gamma_\phi}{2}}
 \left(\ket{e}\bra{e}-\ket{g}\bra{g}\right).
 \label{eq:collapseoperators}
\end{equation}
The action of these matrices can be calculated directly. For $L_1$, for example, in the $(g,e)$ basis the only nonzero element is the upper-right entry $\sqrt{\Gamma}$. Thus
\[
 L_1\rho L_1^\dagger=
 \begin{pmatrix}\Gamma\rho_{ee}&0\\0&0\end{pmatrix},
 \qquad L_1^\dagger L_1=
 \begin{pmatrix}0&0\\0&\Gamma\end{pmatrix}.
\]
The first term adds the population lost from the upper level to the lower level. The other two subtract upper-level population and also act on coherence. Combining the three terms for this channel into $\mathcal D_1[\rho]$ gives
\begin{equation}
 \mathcal D_1[\rho]=
 \begin{pmatrix}
 \Gamma\rho_{ee}&-\Gamma\rho_{ge}/2\\
 -\Gamma\rho_{eg}/2&-\Gamma\rho_{ee}
 \end{pmatrix},\qquad
 \mathcal D_\phi[\rho]=
 \begin{pmatrix}0&-\gamma_\phi\rho_{ge}\\
 -\gamma_\phi\rho_{eg}&0\end{pmatrix}.
 \label{eq:dissipativematrices}
\end{equation}
The second matrix follows similarly from $L_\phi${: $L_\phi\rho L_\phi^\dagger$ reverses the signs of the off-diagonal elements, and combining it with the other two terms cancels the factor $1/2$}. Both matrices have zero trace, showing that no probability is lost from the chosen closed two-level system. The term $\mathcal D_1$ causes population to decay at $\Gamma$ but coherence at only $\Gamma/2$; $\mathcal D_\phi$ leaves populations unchanged and further reduces coherence.

After the drive is switched off, therefore,
\[
 \rho_{ee}(t)=\rho_{ee}(0)\ee^{-\Gamma t},\qquad
 \rho_{eg}(t)=\rho_{eg}(0)
 \ee^{-(\Gamma/2+\gamma_\phi)t}\ee^{-\ii\omega_0t}.
\]
The times for the population and coherence amplitude to decay to $1/\ee$ of their initial values are defined as $T_1$ and $T_2$, respectively. Then
\begin{equation}
 \frac{1}{T_1}=\Gamma,\qquad
 \frac{1}{T_2}=\gamma_2=\frac{\Gamma}{2}+\gamma_\phi.
 \label{eq:t1t2}
\end{equation}
Population decay itself therefore destroys coherence, so even without pure dephasing one has only $T_2=2T_1$. In general, $T_2\leq2T_1$. With thermal excitation, particles entering or leaving, or additional levels, these times must be determined afresh in the appropriate model.

Representing environmental effects by fixed rates is appropriate when the environmental memory time is much shorter than the evolution time of interest and the corresponding weak-coupling approximation holds. If $\Gamma$ includes spontaneous emission, it is used here as a prescribed parameter. This allows calculation of the average atomic response but does not describe the state of the emitted photons. Chapter 4 considers how spontaneous emission arises from coupling between atoms and a quantum field.

Average evolution must also be distinguished from evolution conditioned on a known measurement result. The $\rho$ in Eq.~\eqref{eq:lindblad} describes a statistical ensemble without selection by environmental measurement outcomes. If an outcome is recorded, such as detection of an emission event, the state must be updated accordingly to obtain a conditional state. Averaging over all outcomes with their probabilities recovers the master-equation evolution. Not reading the outcomes does not mean that the environment has no effect. Further Reading 2A derives this relation from the possible outcomes in a short time step.

\subsection{Optical Bloch Equations}

Using the same rotating phase reference, denote the excited-state population by $p=\rho_{ee}$ and the slowly varying coherence by $s$, so that in the laboratory reference $\rho_{eg}=s\ee^{-\ii\omega t}$. The density matrix in the rotating reference is then
\begin{equation}
 \widetilde\rho=
 \begin{pmatrix}1-p&s^*\\s&p\end{pmatrix}.
 \label{eq:rotatingdensity}
\end{equation}
First calculate the part without dissipation. {As with the probability-amplitude transformation, under the rotating-wave approximation the coherent evolution of $\widetilde\rho$ is determined by the rotating-reference Hamiltonian $H_{\mathrm r}$. The dissipative terms in Eq.~\eqref{eq:dissipativematrices} are linear in each matrix element and retain their form when expressed in terms of $\widetilde\rho$.} To obtain $\dot p$, evaluate $-\ii[H_{\mathrm r},\widetilde\rho]/\hbar$ and take its $ee$ element:
\[
 \begin{aligned}
 (H_{\mathrm r}\widetilde\rho)_{ee}&=-\hbar\frac{\Omega s^*}{2}-\hbar\Delta p,\\
 (\widetilde\rho H_{\mathrm r})_{ee}&=-\hbar\frac{\Omega^*s}{2}-\hbar\Delta p.
 \end{aligned}
\]
Subtracting cancels the detuning terms. Multiplication by $-\ii/\hbar$ then gives $-\ii(\Omega^*s-\Omega s^*)/2=\operatorname{Im}(\Omega^*s)$, where $\operatorname{Im}$ denotes the imaginary part. Similarly, the $eg$ element is
\[
 (H_{\mathrm r}\widetilde\rho-\widetilde\rho H_{\mathrm r})_{eg}
 =-\hbar\Delta s-\frac{\hbar\Omega}{2}(1-2p).
\]
Finally, adding the corresponding elements from Eq.~\eqref{eq:dissipativematrices} gives
\begin{equation}
 \begin{aligned}
 \dot p&=-\Gamma p+\operatorname{Im}(\Omega^*s),\\
 \dot s&=(\ii\Delta-\gamma_2)s
       +\frac{\ii\Omega}{2}(1-2p).
 \end{aligned}
 \label{eq:bloch}
\end{equation}
These are the two-level optical Bloch equations used in this chapter. A dot denotes a time derivative, and $\operatorname{Im}$ denotes the imaginary part. The word ``optical'' does not restrict the drive frequency range; the same equations apply to suitable microwave transitions.

In the first equation, $-\Gamma p$ represents loss of excitation energy, and $\operatorname{Im}(\Omega^*s)$ represents energy exchange between the drive and atom. In the second, $\Delta$ rotates the coherence relative to the drive, $\gamma_2$ causes it to decay, and the population difference $1-2p=\rho_{gg}-\rho_{ee}$ determines how much coherent response the field can establish{. When the lower level dominates, $1-2p>0$ gives net absorption; for equal populations, $1-2p=0$ and the net response saturates to zero; when the upper level dominates, $1-2p<0$ reverses the response to stimulated amplification}. Population and coherence are thus coupled, and neither can be retained alone.

For example, suppose the atom starts in the lower level, with $p=0$ and $s=0$. Immediately after the drive is switched on, $\dot p=0$, but $\dot s=\ii\Omega/2$ is already nonzero. At short times, $s\simeq\ii\Omega t/2$, hence $\dot p\simeq|\Omega|^2t/2$; integration gives $p\simeq|\Omega|^2t^2/4$, consistent with the short-time expansion of the Rabi solution. The field first establishes coherence, then changes population through that coherence. These steps unfold continuously in the equations.

Equation~\eqref{eq:bloch} contains only one real variable $p$ and one complex variable $s$, or three real variables in total. Given $\Omega(t)$, $\Delta$, $\Gamma$, $\gamma_2$, and initial conditions, it describes pulse control, continuous-wave steady states, and small-signal response without a separate set of atomic equations of motion.

Readout-noise calculations also require correlations between fluctuations at different times; average populations and coherences alone do not provide this information. Under the corresponding Markov approximation, auxiliary matrices can be evolved with the same master equation to obtain two-time correlation functions and then noise spectra. Further Reading 2B gives the quantum-regression procedure and the one-sided spectral convention used in this book. A fluctuation auxiliary matrix is generally not a normalized conditional state and must be distinguished from a density matrix updated according to a measurement result.

\section{From Atomic Response to Material Polarization}

\subsection{How Coherence Produces Polarization}

For two stationary states without permanent electric dipole moments, the diagonal dipole elements vanish, so the dipole matrix is
\[
 \hat{\mathbf d}\ \longleftrightarrow\quad
 \begin{pmatrix}0&\mathbf d_{ge}\\\mathbf d_{eg}&0\end{pmatrix}.
\]
Multiplying by the density matrix and taking the trace gives $\avg{\mathbf d}=\rho_{eg}\mathbf d_{ge}+\rho_{ge}\mathbf d_{eg}$. Substituting $\rho_{eg}=s\ee^{-\ii\omega t}$ and $\rho_{ge}=s^*\ee^{\ii\omega t}$ gives the mean dipole moment of a single atom:
\begin{equation}
 \avg{\mathbf d}(t)
 =\mathbf d_{ge}s\ee^{-\ii\omega t}
 +\mathbf d_{eg}s^*\ee^{\ii\omega t}.
 \label{eq:dipoleaverage}
\end{equation}
Here $\mathbf d_{ge}=\mathbf d_{eg}^*$. The mean internal energy, $\hbar\omega_0p$, is determined by population, while the mean oscillating dipole moment is determined by $s$. With populations but no coherence, $s=0$, this mean oscillating dipole moment vanishes. The atom may still store excitation energy, but that energy alone does not imply a definite oscillating polarization.

Let $N$ be the number of atoms per unit volume participating in the transition, all with the same local response. Macroscopic polarization is the mean dipole moment per unit volume, so $\mathbf P=N\avg{\mathbf d}$. Its complex peak-amplitude representation is
\[
 \mathbf P(t)=\operatorname{Re}[\mathbf P_0\ee^{-\ii\omega t}]
 =\frac12\left(\mathbf P_0\ee^{-\ii\omega t}
 +\mathbf P_0^*\ee^{\ii\omega t}\right).
\]
Comparing coefficients of $\ee^{-\ii\omega t}$ in the two expressions gives
\begin{equation}
 \mathbf P_0=2N\mathbf d_{ge}s.
 \label{eq:polarization}
\end{equation}
The factor of 2 comes from the complex peak-amplitude convention and corresponds to the $1/2$ in Eq.~\eqref{eq:fieldrabi}. If the field-amplitude definition changes, the definitions of $\Omega$ and $\mathbf P_0$ must change together.

\subsection{Weak-Field Response and Saturation}

In a weak field, the excited-state population is small, so take $p\simeq0$: most atoms remain in the lower level. A continuous-wave steady state satisfies $\dot s=0$, and the second of Eqs.~\eqref{eq:bloch} gives
\[
 (\gamma_2-\ii\Delta)s=\frac{\ii\Omega}{2},\qquad
 s=\frac{\ii\Omega}{2(\gamma_2-\ii\Delta)}.
\]
Multiplying numerator and denominator by $\gamma_2+\ii\Delta$ separates the real and imaginary parts:
\begin{equation}
 s=\frac{\Omega}{2}
 \frac{-\Delta+\ii\gamma_2}{\Delta^2+\gamma_2^2}.
 \label{eq:weakcoherence}
\end{equation}
To make the material parameters explicit, choose a fixed linear polarization and take the transition dipole moment along it to be a real number $d$. From $\Omega=dE_0/\hbar$ and $P_0=2Nds$, the polarization $P_0$ is proportional to $E_0$. Comparing with the engineering-electromagnetics definition $P_0=\varepsilon_0\chi E_0$ gives
\begin{equation}
 \chi(\omega)=\frac{Nd^2}{\varepsilon_0\hbar}
 \frac{-\Delta+\ii\gamma_2}{\Delta^2+\gamma_2^2}.
 \label{eq:susceptibility}
\end{equation}
This is the familiar complex electric susceptibility: its real part describes dispersion and its imaginary part absorption. With the $\ee^{-\ii\omega t}$ convention used here, passive absorption corresponds to $\operatorname{Im}\chi>0$. The imaginary part has a Lorentzian line shape, with full width at half maximum $2\gamma_2$ in angular frequency, or $\gamma_2/\pi$ in hertz.

The connection between the imaginary part and absorption can also be seen directly from power. The polarization current density is $J_{\mathrm P}=\partial_tP$, with complex amplitude $-\ii\omega P_0$. The cycle-averaged work done by the field on this current is
\begin{equation}
 \overline{EJ_{\mathrm P}}
 =\frac12\operatorname{Re}\!\left[E_0^*(-\ii\omega P_0)\right]
 =\frac{\omega\varepsilon_0}{2}\operatorname{Im}\chi\,|E_0|^2.
 \label{eq:absorbedpower}
\end{equation}
A positive result means that the electromagnetic field continuously supplies energy to the medium. On resonance, $s$ has a phase difference of $\pi/2$ relative to the driving amplitude, giving the polarization current a component in phase with the electric field. The phase of atomic coherence is thus directly connected to the familiar concept of real power.

This result shows that, for weak fields near resonance, a quantum atomic model gives a material response similar to that of a classical driven oscillator. The quantum model also retains level populations, however. As the field grows, the upper level becomes occupied, and absorption cannot increase indefinitely according to the weak-field relation.

For stronger driving, the population difference can no longer be set to one. First set $\dot s=0$ while retaining $p$, giving
\[
 s=\frac{\Omega}{2}
 \frac{-\Delta+\ii\gamma_2}{\Delta^2+\gamma_2^2}(1-2p).
\]
Substituting into the steady-state population equation $\Gamma p=\operatorname{Im}(\Omega^*s)$ gives
\[
 \Gamma p=
 \frac{|\Omega|^2\gamma_2}{2(\Delta^2+\gamma_2^2)}(1-2p).
\]
The left side is the population-decay rate toward the lower level, and the right side the net excitation rate due to the drive. Balancing them and collecting terms in $p$ gives
\begin{equation}
 p_{\mathrm{ss}}=\frac{\mathcal S}{2(1+\mathcal S)},
 \qquad
 \mathcal S=
 \frac{|\Omega|^2\gamma_2}{\Gamma(\Delta^2+\gamma_2^2)}.
 \label{eq:saturation}
\end{equation}
Here $\mathcal S$ is a dimensionless saturation parameter, and the subscript $\mathrm{ss}$ denotes steady state. For $\mathcal S\ll1$, the weak-field approximation holds. For $\mathcal S\gg1$, both level populations approach $1/2$, the population difference decreases, and absorption saturates. The coherence $s$ is then Eq.~\eqref{eq:weakcoherence} divided by $1+\mathcal S$, so the material response depends on field strength.

This does not contradict complete excitation by a $\pi$ pulse. A $\pi$ pulse describes transient control in the approximation of negligible dissipation; Eq.~\eqref{eq:saturation} describes a continuous-wave steady state with ongoing relaxation. These operating conditions must be distinguished when comparing experiments.

\subsection{Self-Consistent Maxwell--Bloch Calculations}

For a thin sample with weak absorption and scattering, one may first calculate the applied field and then use it to drive the atoms. When atomic polarization appreciably modifies the field, Eq.~\eqref{eq:polarization} must be fed back into Maxwell's equations. For a nonmagnetic background with constant permittivity $\varepsilon_{\mathrm b}=\varepsilon_0\varepsilon_{r,\mathrm b}$ (absolute permittivity, in $\mathrm{F/m}$), the field equation is
\begin{equation}
 \nabla\times\mathbf E=-\mu_0\partial_t\mathbf H,
 \qquad
 \nabla\times\mathbf H=
 \mathbf J_{\mathrm{ext}}+\varepsilon_{\mathrm b}\partial_t\mathbf E
 +\partial_t\mathbf P.
 \label{eq:maxwellbloch}
\end{equation}
Here $\mathbf J_{\mathrm{ext}}$ is the applied source current, and $\mathbf P$ includes only the atomic transitions modeled explicitly. The background permittivity must not include those same transitions again.

The field determines the local $\Omega$; the Bloch equations determine $p$ and $s$; coherence in turn determines the polarization source. Together these form the Maxwell--Bloch model. It describes absorption, saturation, and pulse evolution during propagation. The field remains a classical mean field, however, so these equations alone do not determine the distribution of emitted photon numbers.

When using slowly varying variables in a rotating reference, reconstruct the oscillating polarization according to Eq.~\eqref{eq:dipoleaverage} before feeding it into the full-wave field equation, or rewrite Maxwell's equations in envelope form as well. The slowly varying $s$ cannot be used directly as polarization at the actual field frequency. Spatial discretization, boundary conditions, and treatment of open regions otherwise follow standard engineering electromagnetic methods.

\section{How Two Pulses Form an Atomic Clock}

An atomic clock measures the offset of a local oscillator's frequency from an atomic transition frequency. Besides scanning an absorption spectrum, this offset can be measured through phase accumulation. The latter approach is Ramsey interrogation~\cite{ramsey}.

First, a $\pi/2$ pulse prepares an atom initially in $g$ in a coherent superposition with equal populations. The drive is then switched off for a free-evolution time $T$, while the local oscillator keeps running. Because their angular frequencies differ by $\Delta$, the atom and oscillator accumulate a relative phase $\Delta T$. A second $\pi/2$ pulse converts this phase difference into a population difference, after which the upper-level population is measured{. The sequence resembles a two-arm interferometer in time: the two $\pi/2$ pulses act like microwave or optical $3\,\mathrm{dB}$ splitting and combining elements, and free evolution corresponds to phase accumulation between the arms}.

We can calculate the two-pulse sequence directly. With the preceding phase convention, the state column vector after the first $\pi/2$ pulse is $(1,\ii)^{\mathsf T}/\sqrt2$. While the drive is off, $\dot b_g=0$ and $\dot b_e=\ii\Delta b_e$ in the local oscillator's phase reference. After free evolution for $T$,
\[
 \begin{pmatrix}b_g\\b_e\end{pmatrix}_{T}
 =\frac1{\sqrt2}\begin{pmatrix}1\\\ii\ee^{\ii\Delta T}\end{pmatrix}.
\]
The level populations are still equal. The frequency-offset information is stored in the relative phase and has not yet become a population change.

\begin{figure}[htbp]
 \centering
\begin{tikzpicture}[>=Stealth,font=\small]
 \draw[->] (0,3.7)--(10.5,3.7) node[right] {$t$};
 \draw[fill=lightblue,draw=inkblue,thick] (.8,3.7) rectangle (1.4,4.6);
 \draw[fill=lightblue,draw=inkblue,thick] (8.2,3.7) rectangle (8.8,4.6);
 \node[above] at (1.1,4.6) {$\pi/2$};
 \node[above] at (8.5,4.6) {$\pi/2$};
 \draw[<->] (1.4,4.15)--(8.2,4.15)
 node[midway,fill=white,inner sep=4pt] {Free evolution $T$};
 \node[below,align=center] at (1.1,3.65) {Prepare\\superposition};
 \node[below,align=center] at (8.5,3.65) {Convert phase\\to population};
 \node[above] at (10,3.8) {Readout};
 \begin{scope}[shift={(5.3,-.7)},x=4cm,y=2.1cm]
  \draw[->] (-1.15,0)--(1.24,0) node[below right] {$(f-f_0)T$};
  \draw[->] (-1.15,0)--(-1.15,1.17) node[above] {$p_e$};
  \foreach \x/\tick in {-1/-1,-.5/-0.5,0/0,.5/0.5,1/1}{\draw (\x,0)--(\x,-.03) node[below] {$\tick$};}
  \draw (-1.15,1)--(-1.18,1) node[left] {1};
  \draw (-1.15,.5)--(-1.18,.5) node[left] {$1/2$};
  \draw[gray!35] (-1.15,.5)--(1.05,.5);
  \draw[inkblue,line width=.9pt,domain=-1.05:1.05,samples=180]
   plot (\x,{.5*(1+cos(360*\x))});
 \end{scope}
\end{tikzpicture}
 \caption{Ramsey interrogation. The phase accumulated between two $\pi/2$ pulses is converted into measurable population by the second pulse. The lower horizontal axis is the product of frequency offset and free-evolution time, with $f_0=\omega_0/(2\pi)$.}
 \label{fig:ramsey}
\end{figure}
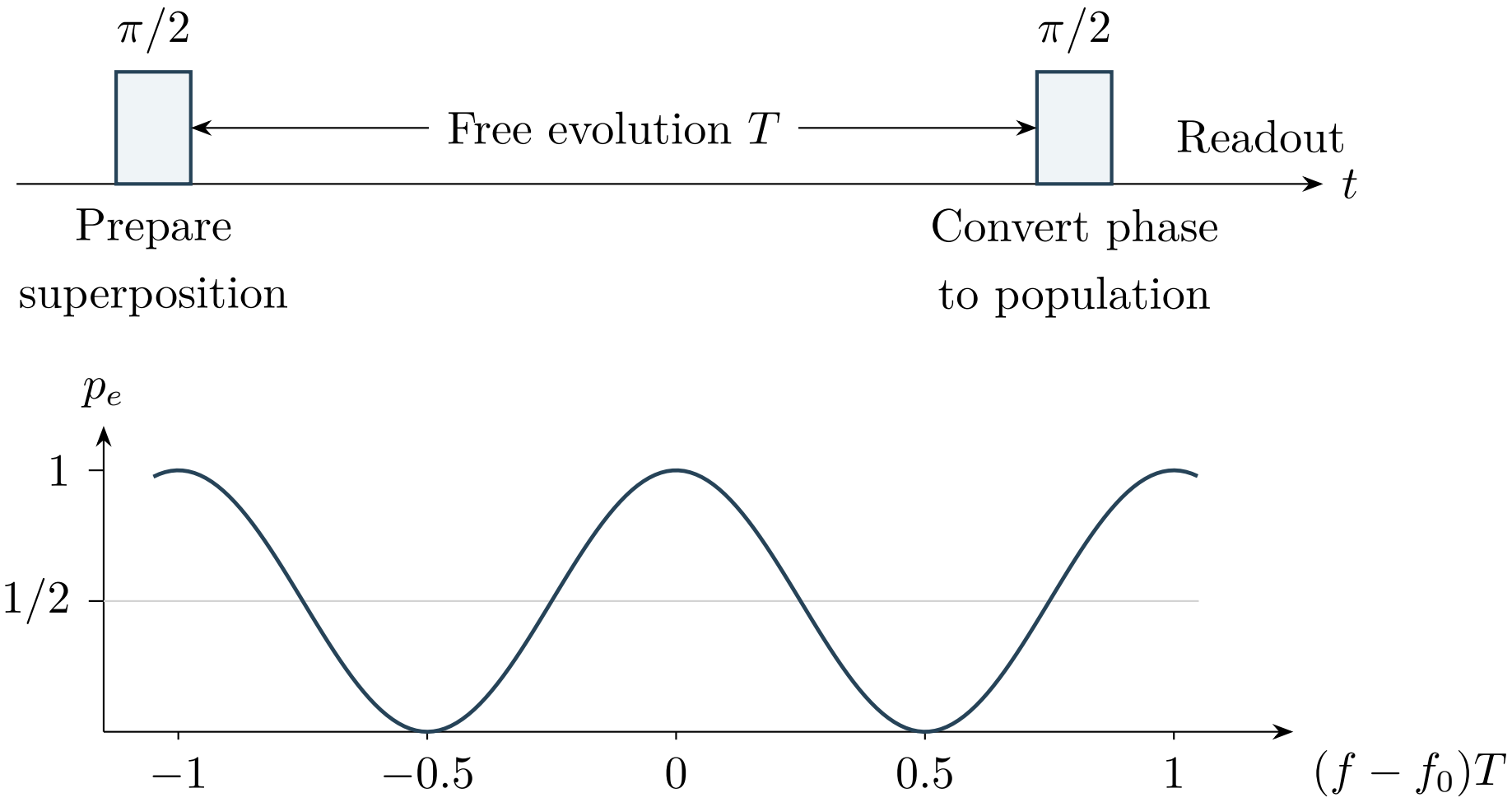

The resonant Rabi solution gives the action of a same-phase $\pi/2$ pulse on an arbitrary initial state as
\[
 U_{\pi/2}=\frac1{\sqrt2}
 \begin{pmatrix}1&\ii\\\ii&1\end{pmatrix}.
\]
Multiplying this matrix by the freely evolved state gives the outgoing upper-level amplitude $b_e^{\mathrm{out}}=\ii(1+\ee^{\ii\Delta T})/2$. Hence
\[
 p_e=\left|\frac{\ii}{2}(1+\ee^{\ii\Delta T})\right|^2
 =\frac{2+2\cos(\Delta T)}4.
\]
This also clarifies the role of the second pulse: it recombines probability amplitudes with different phases, making the phase difference observable as population. Detuning and dissipation during the pulses have been neglected. Representing coherence loss by a fringe contrast $C$, the result becomes
\begin{equation}
 p_e=\frac12\bigl[1+C\cos(\Delta T)\bigr].
 \label{eq:ramsey}
\end{equation}
Ideally, the fringe contrast is $C=1$. With only uniform coherence decay during free evolution, $C=\ee^{-\gamma_2T}$. If the second pulse {has a phase offset $\phi$ relative to the first}, the cosine changes accordingly {to $\cos(\Delta T-\phi)$. For example, alternating $\phi=\pm\pi/2$ gives $p_e=\frac12[1\pm C\sin(\Delta T)]$, placing zero detuning exactly at the points of maximum slope}. Figure~\ref{fig:ramsey} shows the pulse sequence and ideal fringes.

The fringe period in frequency is $1/T$. Increasing $T$ gives a larger phase change for the same frequency offset and thus improves frequency discrimination. But increasing $T$ also reduces contrast and increases the effects of atom loss and local-oscillator phase noise, so longer free evolution is not the only objective.

At the resonance maximum, the first derivative of population with respect to small detuning is zero, making it unsuitable for directly generating a signed locking-error signal. In practice, the phase of the second pulse can be alternated to place two measurements on opposite slopes, and their populations subtracted. Near the target frequency, this difference changes sign with detuning and provides feedback to correct the local oscillator.

Many atomic clocks use magnetic-dipole transitions. Their coupling must be constructed from $-\hat{\boldsymbol\mu}\cdot\mathbf B$, with $\Omega$ defined through the magnetic transition moment. The same two-level evolution and Ramsey phase accumulation still apply. Engineering design must also control field uniformity, pulse area, and phase distribution in the interaction region. If atoms at different positions experience different control, the ensemble fringes lose contrast.

\section{Optical Readout of Radio-Frequency Fields: Rydberg-Atom Receivers}

\subsection{First Establish a Three-Level Transparency Window}

\begin{figure}[htbp]
 \centering
\begin{tikzpicture}[>=Stealth,font=\small]
 \foreach \y/\lab in {0/1,1.6/2,3.3/3,4.2/4}{
  \draw[thick] (0,\y)--(3.3,\y);
  \node[left] at (0,\y) {$\ket{\lab}$};
 }
 \draw[<->,inkblue,line width=1pt] (1.0,.08)--(1.0,1.52);
 \draw[<->,inkblue,line width=1pt] (2.2,1.68)--(2.2,3.22);
 \draw[<->,inkblue,line width=1pt] (1.0,3.38)--(1.0,4.12);
 \node[right] at (1.12,.8) {$\Omega_{\mathrm p}$};
 \node[left] at (2.06,2.45) {$\Omega_{\mathrm c}$};
 \node[right] at (1.12,3.75) {$\Omega_{\mathrm r}$};
 \node[anchor=west] at (4,.8) {Probe: read transmission changes};
 \node[anchor=west] at (4,2.45) {Coupling field: establish transparency};
 \node[anchor=west] at (4,3.75) {RF field: couple Rydberg states};
\end{tikzpicture}
 \caption{Four-level Rydberg receiver model. The probe and coupling fields establish a three-level transparency response; a radio-frequency field couples two Rydberg states and changes probe transmission. Level spacings are not to scale, and level 4 is taken to lie above level 3.}
 \label{fig:rydberg}
\end{figure}
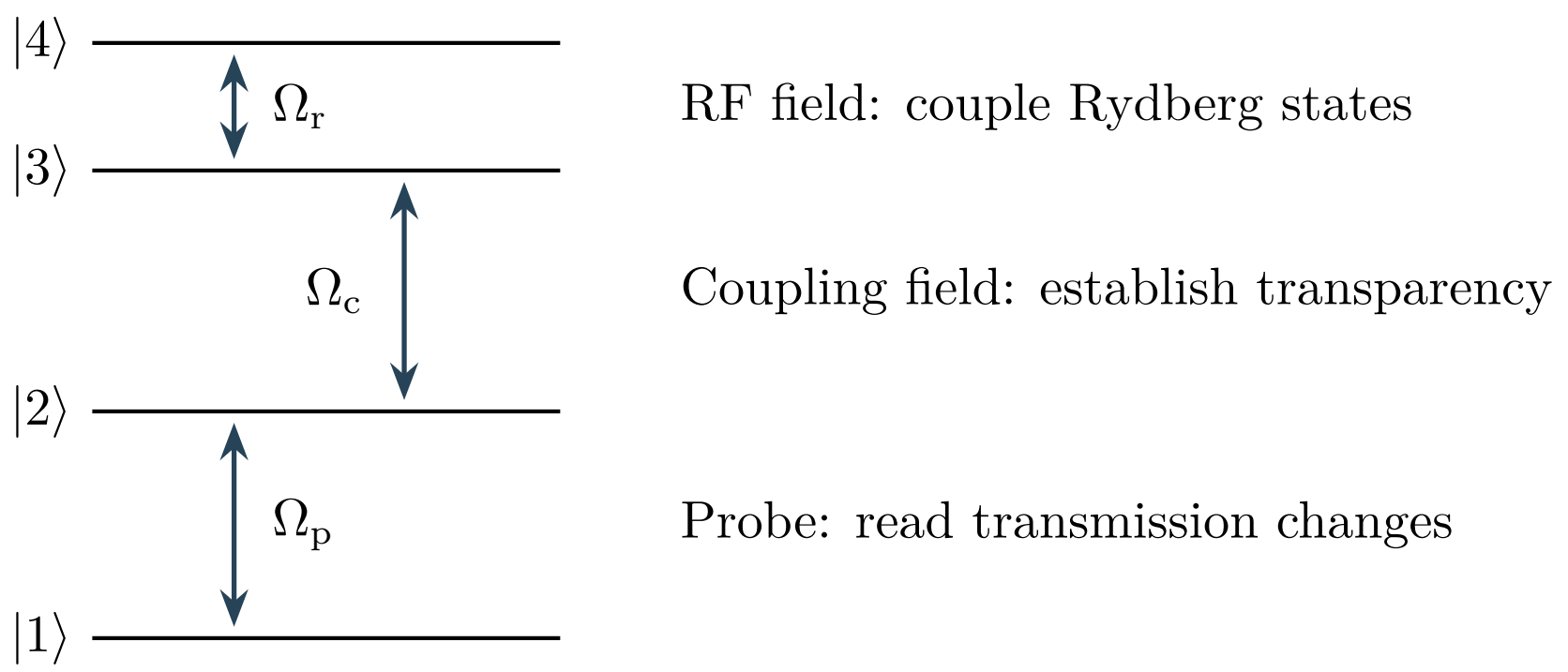

A Rydberg-atom receiver converts the action of a radio-frequency field into a change in optical transmission. Consider the ladder levels in Figure~\ref{fig:rydberg}: $\ket{1}$ is the lower level, $\ket{2}$ an intermediate excited state, and $\ket{3}$ a Rydberg state. A weak probe field couples $1\leftrightarrow2$, and a stronger coupling field couples $2\leftrightarrow3$, with Rabi angular frequencies $\Omega_{\mathrm p}$ and $\Omega_{\mathrm c}$, respectively.

With only the probe present, it is generally absorbed near the $1\leftrightarrow2$ resonance. Adding the coupling field allows response through different coherent excitation pathways. Under suitable two-photon resonance conditions, these pathways suppress excitation of the intermediate state and reduce probe absorption, producing electromagnetically induced transparency, or EIT~\cite{eit}.

Two-photon resonance means that the sum of the optical angular frequencies matches the $1\leftrightarrow3$ energy separation. Begin with the simpler case where each field is exactly resonant, and choose basis-vector phases so that $\Omega_{\mathrm p}$ and $\Omega_{\mathrm c}$ are real. In the rotating reference, the intermediate-state amplitude obeys
\[
 \ii\dot b_2=-\frac12(\Omega_{\mathrm p}b_1+\Omega_{\mathrm c}b_3).
\]
To prevent excitation of the intermediate state, both $b_2=0$ and its time derivative must vanish. Thus $\Omega_{\mathrm p}b_1+\Omega_{\mathrm c}b_3=0$, or $b_3/b_1=-\Omega_{\mathrm p}/\Omega_{\mathrm c}$. Normalizing with $|b_1|^2+|b_3|^2=1$ gives
\begin{equation}
 \ket{D}=
 \frac{\Omega_{\mathrm c}\ket{1}-\Omega_{\mathrm p}\ket{3}}
 {\sqrt{\Omega_{\mathrm c}^2+\Omega_{\mathrm p}^2}}.
 \label{eq:darkstate}
\end{equation}
In this state, the contributions from $\ket{1}$ and $\ket{3}$ to excitation of $\ket{2}$ cancel exactly. Since $b_2=0$, the other two amplitudes are no longer driven by the intermediate state either, and the state can remain unchanged. It contains no component of the readily decaying intermediate state and is called a dark state. The cancellation is between atomic excitation probability amplitudes, not between two optical fields of different frequencies directly canceling in space. With a weak probe, $|\Omega_{\mathrm p}|\ll|\Omega_{\mathrm c}|$, the dark state is mostly $\ket{1}$, but its small $\ket{3}$ component is still essential for cancellation.

The dark state explains the origin of transparency, but the window's depth and width also depend on coupling strength, coherence decay, and propagation. Rapid loss of the phase relation between states $1$ and $3$ weakens the transparency window. Knowing only the three level positions is therefore insufficient to predict the transmission spectrum.

\subsection{Add Radio-Frequency Coupling and Calculate Probe Polarization}

A radio-frequency field further couples $3\leftrightarrow4$, with Rabi angular frequency $\Omega_{\mathrm r}$. Defining $\omega_{ij}=(E_i-E_j)/\hbar$, the three field detunings are
\begin{equation}
 \Delta_{\mathrm p}=\omega_{\mathrm p}-\omega_{21},\qquad
 \Delta_{\mathrm c}=\omega_{\mathrm c}-\omega_{32},\qquad
 \Delta_{\mathrm r}=\omega_{\mathrm r}-\omega_{43}.
 \label{eq:fourdetunings}
\end{equation}
The cumulative detunings in the matrix follow the same procedure as in the two-level model. Set $E_1=0$ and transform the amplitudes as
\[
 \begin{aligned}
 c_1&=b_1, &c_2&=b_2\ee^{-\ii\omega_{\mathrm p}t},\\
 c_3&=b_3\ee^{-\ii(\omega_{\mathrm p}+\omega_{\mathrm c})t},
 &c_4&=b_4\ee^{-\ii(\omega_{\mathrm p}+\omega_{\mathrm c}+\omega_{\mathrm r})t}.
 \end{aligned}
\]
For example, the third state's energy relative to its chosen reference becomes $E_3-\hbar(\omega_{\mathrm p}+\omega_{\mathrm c})=-\hbar(\Delta_{\mathrm p}+\Delta_{\mathrm c})$. The remaining diagonal terms follow similarly, while near-resonant coupling between adjacent states remains $-\hbar\Omega_j/2$. The four-level matrix under the rotating-wave approximation is therefore
\begin{equation}
 \frac{H_{\mathrm r}}{\hbar}=
 \begin{pmatrix}
 0&-\Omega_{\mathrm p}^*/2&0&0\\
 -\Omega_{\mathrm p}/2&-\Delta_{\mathrm p}&-\Omega_{\mathrm c}^*/2&0\\
 0&-\Omega_{\mathrm c}/2&-(\Delta_{\mathrm p}+\Delta_{\mathrm c})&-\Omega_{\mathrm r}^*/2\\
 0&0&-\Omega_{\mathrm r}/2&-(\Delta_{\mathrm p}+\Delta_{\mathrm c}+\Delta_{\mathrm r})
 \end{pmatrix}.
 \label{eq:fourhamiltonian}
\end{equation}
It follows exactly the same rules as the two-level matrix: diagonal elements are negative cumulative detunings, and adjacent levels have couplings $-\Omega/2$. Level decay and coherence loss are then included through Eq.~\eqref{eq:lindblad}.

To see directly how the radio-frequency field affects the probe, take the weak-probe limit. Assume that without the probe all atoms remain in $\ket{1}$ and no other process pumps them into upper levels. With a very weak probe, $\widetilde\rho_{21}$, $\widetilde\rho_{31}$, and $\widetilde\rho_{41}$ are first order in $\Omega_{\mathrm p}$, while upper-level populations {and coherences between upper levels, such as $\widetilde\rho_{32}$,} are second order. The linear response therefore uses $\widetilde\rho_{11}\simeq1$ and approximately zero population in the other levels.

For the $21$ coherence, for example, the master equation gives
\[
 \dot{\widetilde\rho}_{21}
 =(\ii\Delta_{\mathrm p}-\gamma_{21})\widetilde\rho_{21}
 +\frac{\ii\Omega_{\mathrm p}}2(\widetilde\rho_{11}-\widetilde\rho_{22})
 +\frac{\ii\Omega_{\mathrm c}^*}2\widetilde\rho_{31}.
\]
Compared with the two-level coherence equation, this includes an additional term that feeds upper-level coherence back through the coupling field. Write $s_2=\widetilde\rho_{21}$, $s_3=\widetilde\rho_{31}$, and $s_4=\widetilde\rho_{41}$, and define
\begin{equation}
 \begin{aligned}
 D_2&=\gamma_{21}-\ii\Delta_{\mathrm p},\\
 D_3&=\gamma_{31}-\ii(\Delta_{\mathrm p}+\Delta_{\mathrm c}),\\
 D_4&=\gamma_{41}-\ii(\Delta_{\mathrm p}+\Delta_{\mathrm c}+\Delta_{\mathrm r}).
 \end{aligned}
 \label{eq:fourdenominators}
\end{equation}
Here $\gamma_{j1}$ is the corresponding coherence decay rate. The steady-state equations reduce to
\begin{equation}
 \begin{aligned}
 D_2s_2&=\frac{\ii\Omega_{\mathrm p}}2
          +\frac{\ii\Omega_{\mathrm c}^*}2s_3,\\
 D_3s_3&=\frac{\ii\Omega_{\mathrm c}}2s_2
          +\frac{\ii\Omega_{\mathrm r}^*}2s_4,\\
 D_4s_4&=\frac{\ii\Omega_{\mathrm r}}2s_3.
 \end{aligned}
 \label{eq:fourcoherences}
\end{equation}
Eliminate variables starting from the last equation. Using $s_4=\ii\Omega_{\mathrm r}s_3/(2D_4)$ first rewrites the second equation as
\[
 \left(D_3+\frac{|\Omega_{\mathrm r}|^2}{4D_4}\right)s_3
 =\frac{\ii\Omega_{\mathrm c}}2s_2.
\]
Thus $s_3$ equals the right side divided by the parenthesized coefficient. Substituting this into the first equation, the product of two factors of $\ii$ gives a minus sign; moving the term to the left gives
\[
 \left[D_2+\frac{|\Omega_{\mathrm c}|^2}
 {4\left(D_3+|\Omega_{\mathrm r}|^2/(4D_4)\right)}\right]s_2
 =\frac{\ii\Omega_{\mathrm p}}2.
\]
The final result is
\begin{equation}
 s_2=\frac{\ii\Omega_{\mathrm p}/2}
 {\displaystyle D_2+
  \frac{|\Omega_{\mathrm c}|^2}
  {\displaystyle4\left(D_3+\frac{|\Omega_{\mathrm r}|^2}{4D_4}\right)}}.
 \label{eq:fourresponse}
\end{equation}
The nested structure of the formula corresponds to the three adjacent transitions. Setting $\Omega_{\mathrm r}=0$ recovers three-level EIT; setting $\Omega_{\mathrm c}=0$ as well recovers the weak-field two-level response. The radio-frequency field does not directly drive the lower level, but changes $s_2$ through upper-level coherences and thus changes the probe polarization.

This expression also verifies the transparency condition. Turn off the radio-frequency field and put both optical fields on resonance; the denominator becomes $\gamma_{21}+|\Omega_{\mathrm c}|^2/(4\gamma_{31})$. When $\gamma_{31}$ is small and the coupling field is nonzero, {the second term $|\Omega_{\mathrm c}|^2/(4\gamma_{31})$ is large. This resembles the large reflected impedance of a secondary high-$Q$ resonator in a coupled microwave resonant circuit, which suppresses the primary-circuit response. Consequently, }$s_2$ and the associated absorption approach zero. This agrees with the dark-state derivation: coherence is preserved between the lower and Rydberg states while absorption on the probe transition is suppressed. Increasing $\gamma_{31}$ weakens this suppression.

Take the transition dipole moment along the probe polarization to be $d_{12}${, with $d_{12}=d_{21}^*$ and $\Omega_{\mathrm p}=d_{21}E_{\mathrm p}/\hbar$}, and the complex peak probe amplitude to be $E_{\mathrm p}$. Then
\begin{equation}
 \chi_{\mathrm p}=\frac{2Nd_{12}s_2}{\varepsilon_0E_{\mathrm p}}.
 \label{eq:probechi}
\end{equation}
For a uniform dilute gas with small susceptibility, the refractive index is approximately $n=\sqrt{1+\chi_{\mathrm p}}\simeq1+\chi_{\mathrm p}/2$. With this chapter's time convention, a field propagating in the positive direction contains $\ee^{\ii k_{\mathrm p}nL}$, whose magnitude decays as $\ee^{-k_{\mathrm p}L\operatorname{Im}\chi_{\mathrm p}/2}$. Intensity is proportional to the squared field magnitude, so neglecting interface reflections gives
\begin{equation}
 I_{\mathrm{out}}=I_{\mathrm{in}}
 \exp[-k_{\mathrm p}L\operatorname{Im}\chi_{\mathrm p}],
 \qquad k_{\mathrm p}=\omega_{\mathrm p}/c.
 \label{eq:transmission}
\end{equation}
The radio-frequency electric field, atomic coherence, and probe intensity are now connected within a single model.

For a sufficiently strong resonant radio-frequency drive, the coupled upper states acquire distinct eigenenergies, and a resolvable Autler--Townes splitting can appear in the spectrum. Considering only states $3$ and $4$ and subtracting their common energy gives the matrix $-\hbar\bigl(\begin{smallmatrix}0&\Omega_{\mathrm r}^*/2\\\Omega_{\mathrm r}/2&0\end{smallmatrix}\bigr)$. Its characteristic equation is $\lambda^2-\hbar^2|\Omega_{\mathrm r}|^2/4=0$, so the eigenenergies differ by $\hbar|\Omega_{\mathrm r}|$. In the ideal limit of stationary atoms, a weak probe, and suitable scanning conditions, the corresponding frequency splitting is $|\Omega_{\mathrm r}|/(2\pi)$ and can be converted into an electric-field strength~\cite{sedlacek}.

EIT emphasizes coherent cancellation of excitation pathways, whereas Autler--Townes splitting emphasizes the resolvable separation of driven levels. If the radio-frequency-induced energy separation is smaller than the linewidth, the resonances generally cannot be resolved. Spectral interpretation must compare coupling strength with linewidth rather than rely on a double-peaked shape alone.

Atoms in a thermal vapor cell have a velocity distribution. An atom with velocity $\mathbf v$ sees the detuning of field $j$ as $\Delta_j-\mathbf k_j\cdot\mathbf v$, where $\mathbf k_j$ is that field's wavevector. Calculate each velocity class's response first, then average over the velocity distribution. Counterpropagating optical fields reduce two-photon Doppler detuning, but unequal wavelengths generally prevent complete cancellation. Before converting measured peak separations directly into field strength, the scan method and such corrections must be checked.

\subsection{A Local Oscillator Enables Linear Readout of Weak Signals}

For a very weak radio-frequency field, the splitting may be smaller than the linewidth, making direct identification of two peaks ineffective. An alternative applies a strong local-oscillator field and a weak signal simultaneously to the radio-frequency transition, using the atoms to convert their beat into an optical-intensity change~\cite{jing}.

Using the local-oscillator frequency as the rotating reference and setting its phase to zero, write the radio-frequency coupling as
\begin{equation}
 \Omega_{\mathrm r}(t)=\Omega_{\mathrm{LO}}
 +\Omega_{\mathrm s}\ee^{-\ii\delta t+\ii\varphi},
 \qquad \delta=\omega_{\mathrm s}-\omega_{\mathrm{LO}}.
 \label{eq:localoscillator}
\end{equation}
Here $\Omega_{\mathrm{LO}}$ and $\Omega_{\mathrm s}$ are real and positive, and $\varphi$ is the initial signal phase relative to the local oscillator. For the same polarization component and $\Omega_{\mathrm s}\ll\Omega_{\mathrm{LO}}$,
\begin{equation}
 |\Omega_{\mathrm r}(t)|^2
 \simeq\Omega_{\mathrm{LO}}^2
 +2\Omega_{\mathrm{LO}}\Omega_{\mathrm s}
 \cos(\delta t-\varphi).
 \label{eq:beat}
\end{equation}
For this four-level model with all other parameters fixed, the steady-state transmission is a function $\mathcal T(K)$, where $K=|\Omega_{\mathrm r}|^2$. Expanding to first order about the local-oscillator-only operating point $K_0=\Omega_{\mathrm{LO}}^2$ gives
\[
 \delta I_{\mathrm{out}}(t)\simeq
 I_{\mathrm{in}}\left.\frac{\dd\mathcal T}{\dd K}\right|_{K_0}
 2\Omega_{\mathrm{LO}}\Omega_{\mathrm s}\cos(\delta t-\varphi).
\]
Provided the operating-point slope is nonzero, the optical signal contains a beat term proportional to the unknown field amplitude and retains its relative phase. Without a local oscillator, the response starts at $|\Omega_{\mathrm s}|^2$; adding one introduces the cross term responsible for first-order readout{. Here the atomic response to $|\Omega_{\mathrm r}(t)|^2$ acts as a radio-frequency mixer, downconverting the high-frequency signal into a difference-frequency beat in optical intensity. This is often called atomic superheterodyne reception}. The phase reference comes from the local oscillator; a single steady-state transmission measurement cannot yield the absolute phase of the radio-frequency field.

Equation~\eqref{eq:fourresponse} is a steady-state expression. Substituting the instantaneous $\Omega_{\mathrm r}(t)$ into it is an approximation valid only for sufficiently slow beating. When the frequency difference approaches or exceeds the atomic response bandwidth, the time-dependent master equation must be solved. Practical readout also involves optical beam profiles, atomic transit, collisions, and detection noise. The local radio-frequency field inside the vapor cell must be determined electromagnetically and cannot simply be replaced by the incident field.

\section{Interactions Between Atoms and Model Checks}

\subsection{From Dipole Interaction to Rydberg Blockade}

The preceding response assumes negligible interactions between atoms. At higher density, particularly with substantial Rydberg excitation, one atom's state can alter the transition conditions of its neighbors. Multiplying an independent-atom response by the number density is then insufficient.

Let two atoms be separated by $R$, along direction $\widehat{\mathbf R}$. When the atomic size is much smaller than $R$, and $R$ is much smaller than the relevant transition wavelength, the leading electric-dipole interaction is
\begin{equation}
 \hat V_{\mathrm{dd}}=
 \frac{\hat{\mathbf d}_A\cdot\hat{\mathbf d}_B
 -3(\hat{\mathbf d}_A\cdot\widehat{\mathbf R})
    (\hat{\mathbf d}_B\cdot\widehat{\mathbf R})}
 {4\pi\varepsilon_0R^3}.
 \label{eq:dipoledipole}
\end{equation}
This expression follows directly from the classical near field. The dipole field of atom $A$ at $B$ is approximately $\mathbf E_A=[3\widehat{\mathbf R}(\mathbf d_A\cdot\widehat{\mathbf R})-\mathbf d_A]/(4\pi\varepsilon_0R^3)$, and the interaction energy of atom $B$ is $-\mathbf d_B\cdot\mathbf E_A$. Replacing both dipole moments by their operators gives Eq.~\eqref{eq:dipoledipole}. The electromagnetic field still determines the spatial relation, while dipole matrix elements determine the allowed level couplings.

The joint state $\ket{eg}$ means that atom $A$ is in the upper level and atom $B$ in the lower level; $\ket{ge}$ denotes the reverse. For two identical resonant atoms, subtracting the common energy gives the coupling between these states as
\begin{equation}
 H_{\mathrm{pair}}=\hbar
 \begin{pmatrix}0&J\\J^*&0\end{pmatrix},
 \qquad J=\frac{\bra{eg}\hat V_{\mathrm{dd}}\ket{ge}}{\hbar}.
 \label{eq:exchange}
\end{equation}
This describes exchange of one excitation between two atoms, with a coupling strength varying as $R^{-3}$. Mathematically it is still a coupling between two states, but the basis vectors now describe joint states of two atoms.

Suppose the Rydberg pair state $\ket{rr}$ is not resonant with another pair state $\ket{ab}${. Since an individual eigenstate $\ket{r}$ has no permanent dipole moment, the first-order dipole interaction energy of $\ket{rr}$ itself is $\bra{rr}\hat V_{\mathrm{dd}}\ket{rr}=0$; the dipole operator can only couple it to other two-atom states $\ket{ab}$}. Let the energy difference be $\Delta_{\mathrm{pair}}=E_a+E_b-2E_r$ and the coupling matrix element $V{=\bra{rr}\hat V_{\mathrm{dd}}\ket{ab}\propto R^{-3}}$. Taking the uncoupled $\ket{rr}$ energy as zero gives
\[
 H_{\mathrm{pair}}=\begin{pmatrix}0&V\\V^*&\Delta_{\mathrm{pair}}\end{pmatrix},
 \qquad \lambda(\lambda-\Delta_{\mathrm{pair}})-|V|^2=0.
\]
When $|V|\ll|\Delta_{\mathrm{pair}}|$, the eigenvalue connected to the original $\ket{rr}$ remains close to zero. Neglecting $\lambda^2$ in the equation for this root gives $-\lambda\Delta_{\mathrm{pair}}\simeq|V|^2$, hence
\begin{equation}
 \delta E_{rr}\simeq-\frac{|V|^2}{\Delta_{\mathrm{pair}}}
 =\frac{C_6}{R^6}.
 \label{eq:vdw}
\end{equation}
This is the familiar van der Waals energy shift. The coefficient $C_6$ has units of $\mathrm{J\,m^6}$, and its sign depends on the level structure. The $R^{-6}$ dependence arises from two dipole couplings and must not be confused with the $R^{-3}$ dependence of resonant exchange. Near a pair resonance, this approximation fails, and the relevant joint states must be retained and solved directly~\cite{rydberg}.

If the doubly excited state's energy shift exceeds the driving energy scale and relevant linewidths, excitation of the first atom moves the second atom's excitation out of resonance. The resulting suppression of double excitation is called Rydberg blockade. It directly means that the frequency condition for double excitation has changed; it does not simply mean that atoms always experience a repulsive force.

Explicit treatment of these interactions requires two-atom or many-atom states. If a self-consistent Maxwell field is also used, interactions already included through polarization feedback must be distinguished from those represented by additional pair terms, to avoid double counting. More general problems involving radiative propagation and collective decay require the field--atom models of later chapters.

\subsection{Response Slope, Bandwidth, and Noise}

A steady-state transmission curve tells us how much the optical intensity changes with field strength, but a large slope does not automatically mean small measurement uncertainty. We must also know how rapidly the response follows changes and how strongly the output itself fluctuates.

Collect the independent real density-matrix variables into a vector $\mathbf x${, for example the three real Bloch variables $\mathbf x=(p,\,\operatorname{Re}s,\,\operatorname{Im}s)^{\mathsf T}$}. A small-signal expansion about a steady state gives
\begin{equation}
 \delta\dot{\mathbf x}=M\delta\mathbf x+\mathbf b\,\delta E,
 \qquad \delta y=\mathbf a^{\mathsf T}\delta\mathbf x.
 \label{eq:linearresponse}
\end{equation}
Here $\delta E$ is a small change in the measured electric-field component, $y$ is the readout, and $M$ is obtained by differentiating the atomic evolution equations at the operating point. The vectors $\mathbf b$ and $\mathbf a$ describe how the input drives the state and how the state produces the readout, respectively; $\mathsf T$ denotes transpose. For a small signal at modulation angular frequency $\nu$, using the $\ee^{-\ii\nu t}$ convention replaces time differentiation by $-\ii\nu$, so
\[
 (-\ii\nu I-M)\delta\mathbf x=\mathbf b\,\delta E.
\]
Solving for $\delta\mathbf x$ and substituting into the output relation gives
\begin{equation}
 G(\nu)=\frac{\delta y}{\delta E}
 =\mathbf a^{\mathsf T}(-\ii\nu I-M)^{-1}\mathbf b.
 \label{eq:transferfunction}
\end{equation}
This is the usual linear-system transfer function, with identity matrix $I$. Here $\nu$ is the angular frequency of the perturbation or envelope, distinct from the driving carrier frequency $\omega$. The value $G(0)$ gives the steady-state slope, while $G(\nu)$ gives the gain and phase at finite frequency. The atomic decay rates set the associated timescales.

Let the output noise have one-sided power spectral density $S_y(f)$, defined so that its integral over positive frequencies gives the variance. With the detection-chain response included, the equivalent input electric-field amplitude spectral density is
\begin{equation}
 S_E^{1/2}(f)=\frac{\sqrt{S_y(f)}}{|G(2\pi f)|}.
 \label{eq:equivalentnoise}
\end{equation}
Its units are usually $\mathrm{(V/m)/\sqrt{Hz}}$. A sensitivity specification should state the frequency and bandwidth or integration time. An average response without a noise model is insufficient to determine this metric.

Even identically prepared atomic states yield probabilistic fluctuations in energy-level measurements. For one atom, assign $X=1$ to an upper-level result and $X=0$ to a lower-level result, with probabilities $p$ and $1-p$. Then
\[
 \avg{X}=1\times p+0\times(1-p)=p,\qquad
 \avg{X^2}=1^2\times p+0^2\times(1-p)=p.
\]
The single-atom measurement variance is therefore $\Var(X)=\avg{X^2}-\avg{X}^2=p(1-p)$. In operator language, the projection operator $\ket{e}\bra{e}$ equals its own square, so the second moment also equals $p$. This describes the statistical fluctuation of a single energy-level measurement and does not first require a time-correlation function.

For $N_{\mathrm a}$ uncorrelated atoms with the same excitation probability, the excited-atom count is the sum of their individual $X$ values. Their variances add to $N_{\mathrm a}p(1-p)$. The measured population fraction is further divided by $N_{\mathrm a}$, so its standard deviation is
\begin{equation}
 \delta p=\sqrt{\frac{p(1-p)}{N_{\mathrm a}}}.
 \label{eq:projectionnoise}
\end{equation}
This is projection noise. At the maximum slope of a Ramsey fringe, $p=1/2$, hence $\delta p=1/(2\sqrt{N_{\mathrm a}})$. Equation~\eqref{eq:ramsey} also gives $|\partial p/\partial\Delta|=CT/2$. Dividing population uncertainty by the frequency-discrimination slope gives the single-measurement angular-frequency uncertainty
\begin{equation}
 \delta\Delta=\frac{\delta p}{|\partial p/\partial\Delta|}
 =\frac{1}{CT\sqrt{N_{\mathrm a}}}.
 \label{eq:ramseynoise}
\end{equation}
This estimate assumes uncorrelated atoms, adequate readout, and negligible other noise sources. It is an uncertainty for one measurement and cannot be used directly as a noise measure per hertz of bandwidth. Actual performance may also be limited by laser and local-oscillator noise, probe-photon counting fluctuations, and electronic readout noise. Which contribution dominates must be calculated for the experimental conditions; they cannot all be absorbed into one universal ``quantum noise'' parameter.

\subsection{How to Obtain Trustworthy Numerical Results}

Numerical solution of a few-level model is straightforward. Without dissipation, solve the probability-amplitude ordinary differential equations; with dissipation, solve the density-matrix master equation. A steady-state calculation sets time derivatives to zero and adds $\Tr\rho=1$. A four-level density matrix has only $4\times4$ elements, so most of the work lies in specifying the correct couplings, dissipation channels, and readout relation, rather than handling a large matrix.

Before calculating, make energy, angular-frequency, and ordinary-frequency units consistent, and check whether field amplitudes are peak or root-mean-square values. Then test limiting cases: turning off the coupling field should recover the two-level response; turning off dissipation should recover Rabi oscillations; a weak drive should give the line shape in Eq.~\eqref{eq:susceptibility}. Time evolution should also preserve unit trace, conjugate symmetry, and nonnegative eigenvalues of the density matrix. These checks are more reliable than judging a plausible-looking spectrum alone.

If the steady state is not unique, the result may depend on the initial state, so an arbitrary solution returned by a linear solver is insufficient. Thermal vapor cells also require velocity and spatial averaging, and substantial field variation requires coupling to propagation equations. Standard master-equation software can perform these operations; the QuTiP documentation provides interface details~\cite{qutip}, but software does not replace the choice of levels and physical processes.

Starting from field-driven level coupling, this chapter obtained populations, coherence, and polarization and connected them back to Maxwell's equations. Atomic clocks use phase accumulation to determine frequency; Rydberg receivers use multilevel coherence to convert radio-frequency fields into optical signals. In all these calculations, the drive remains classical. The next chapter turns to the field itself: how the energy of an electromagnetic mode is quantized, and why identical average powers can produce different photon-measurement statistics.

\clearpage
\section*{References (Including Further Reading)}
\addcontentsline{toc}{section}{References (Including Further Reading)}
\begingroup
\small
\linespread{1.1}\selectfont
\renewcommand{\chapter}[2]{}%

\endgroup

\clearpage
\begingroup

\section*{Further Reading 2A\quad Origin of the Lindblad Dissipator}
\markboth{Further Reading 2A\quad Origin of the Lindblad Dissipator}{Further Reading 2A\quad Origin of the Lindblad Dissipator}
\phantomsection
\addcontentsline{toc}{section}{Further Reading 2A: Origin of the Lindblad Dissipator}

\setcounter{subsection}{0}
\renewcommand{\thesubsection}{2A.\arabic{subsection}}

\setcounter{equation}{0}
\renewcommand{\theequation}{2A.\arabic{equation}}

\renewcommand{\theHsubsection}{readingA.\arabic{subsection}}
\renewcommand{\theHequation}{readingA.\arabic{equation}}

The main text uses the Lindblad master equation to describe atomic decay and dephasing. Each dissipation channel always appears as a group of terms:
\begin{equation}
 \mathcal D_L[\rho]=L\rho L^\dagger
 -\frac12L^\dagger L\rho-\frac12\rho L^\dagger L.
 \label{extA:dissipator}
\end{equation}
Why can the first term not be used alone, and why do the other two each carry $1/2$? These coefficients ensure probability conservation and positivity of the state. We now construct this structure step by step, starting with atomic decay over a very short interval.

\subsection{Population Decay Alone Is Not Enough}

Suppose the upper level of a two-level atom decays to the lower level at rate $\Gamma$. The population equations are straightforward: $\dot\rho_{ee}=-\Gamma\rho_{ee}$ and $\dot\rho_{gg}=\Gamma\rho_{ee}$. They ensure unit total probability but do not specify how coherence changes.

What goes wrong if populations decay while the off-diagonal elements are left unchanged? For the initial state $(\ket{g}+\ket{e})/\sqrt2$, this prescription gives
\[
 \rho(t)=\begin{pmatrix}
 1-\tfrac12\ee^{-\Gamma t}&\tfrac12\\
 \tfrac12&\tfrac12\ee^{-\Gamma t}
 \end{pmatrix},\qquad
 \det\rho=-\frac14(1-\ee^{-\Gamma t})^2.
\]
For any $t>0$, the determinant {(the product of the two eigenvalues)} is negative, so one eigenvalue must be negative{, violating nonnegative probability}. Such a matrix cannot represent a physical state. Conserving population-rate equations alone therefore do not guarantee a valid quantum state; dissipation of population and coherence must be mutually consistent.

\subsection{Two Outcomes Within a Short Time Step}

Consider decay alone, temporarily omitting Hamiltonian evolution. For a sufficiently short interval $\delta t$, let $\epsilon=\Gamma\delta t\ll1$. During this step, the environment may record one decay event or no event. The corresponding state changes are represented by two matrices:
\begin{equation}
 K_1=\sqrt{\epsilon}\ket{g}\bra{e}{=\begin{pmatrix}0&\sqrt{\epsilon}\\0&0\end{pmatrix}},\qquad
 K_0=\ket{g}\bra{g}+\sqrt{1-\epsilon}\ket{e}\bra{e}{=\begin{pmatrix}1&0\\0&\sqrt{1-\epsilon}\end{pmatrix}}.
 \label{extA:amplitudechannel}
\end{equation}
The operator $K_1$ transfers the upper-level component to the lower level. The probability of one decay is $\Tr(K_1\rho K_1^\dagger)=\epsilon\rho_{ee}$, so its coefficient must be the square root of the probability factor, $\sqrt\epsilon$.

The operator $K_0$ describes the part of the evolution in which this decay does not occur. It cannot simply be the identity: if the atom starts entirely in the upper level, its no-decay probability must be $1-\epsilon$, requiring multiplication of the upper-level amplitude by $\sqrt{1-\epsilon}$. The two outcome probabilities sum to one, as expressed by
\[
 K_0^\dagger K_0+K_1^\dagger K_1=I.
\]
If the environmental record is not read and only the average atomic state is needed, add the two outcomes:
\begin{equation}
 \rho(t+\delta t)=K_0\rho(t)K_0^\dagger+K_1\rho(t)K_1^\dagger.
 \label{extA:map}
\end{equation}
Each term is unnormalized; its trace{ $p_k=\Tr(K_k\rho K_k^\dagger)$ }is the probability of that outcome. Only when the outcome of an individual experiment is known do we divide the corresponding term by its own trace to obtain the conditional state{ $\rho_k=K_k\rho K_k^\dagger/p_k$. Weighting the two conditional states by their respective probabilities $p_k$, as in Eq.~\eqref{eq:densitydefinition}, cancels the denominators $p_k$ and directly gives Eq.~\eqref{extA:map}}.

Expanding the matrix products in Eq.~\eqref{extA:map} gives
\[
 \begin{aligned}
 \rho_{ee}(t+\delta t)&=(1-\epsilon)\rho_{ee}(t),\\
 \rho_{gg}(t+\delta t)&=\rho_{gg}(t)+\epsilon\rho_{ee}(t),\\
 \rho_{eg}(t+\delta t)&=\sqrt{1-\epsilon}\,\rho_{eg}(t).
 \end{aligned}
\]
Using $\sqrt{1-\epsilon}=1-\epsilon/2+O(\epsilon^2)$ gives $\dot\rho_{eg}=-\Gamma\rho_{eg}/2$. Thus population decay at $\Gamma$ and coherence decay at $\Gamma/2$ in the main text arise from the same short-time process with consistent probabilities.

The decay probability in the continuous-time solution is $1-\ee^{-\Gamma\delta t}$; the expression $\epsilon=\Gamma\delta t$ above is its first-order approximation.

\subsection{How the Two Negative Terms Arise in the General Form}

Extend this construction to several dissipation channels. If environmental memory is negligible, each short step can be determined by the current $\rho$. Let $K_k$ represent the change associated with channel $k$ during that step, and write
\begin{equation}
 \rho' =\sum_k K_k\rho K_k^\dagger,\qquad
 \sum_kK_k^\dagger K_k=I.
 \label{extA:kraus}
\end{equation}
This sum is a Kraus representation. The second condition ensures $\Tr\rho'=\Tr\rho$; every $K_k\rho K_k^\dagger$ preserves positivity, and so does their sum. Even if the atom is correlated with another system that does not participate in the evolution, this form ensures that the joint-state matrix has nonnegative eigenvalues{. This property is called complete positivity}.

In the continuous, memoryless limit, select one term close to the identity transformation and write the other, low-probability channels as
\[
 K_0=I+A\delta t+O(\delta t^2),\qquad
 K_k=\sqrt{\delta t}\,L_k\quad(k\geq1).
\]
Here $A$ has units of $\mathrm{s}^{-1}$ and $L_k$ of $\mathrm{s}^{-1/2}$. Substituting into the probability-conservation condition and retaining only first-order terms in $\delta t$ gives
\[
 A+A^\dagger+\sum_kL_k^\dagger L_k=0.
\]
Any matrix can be decomposed as
\[
 A=\frac{A+A^\dagger}{2}+\frac{A-A^\dagger}{2}.
\]
The first part equals its conjugate transpose and is called the Hermitian part. The second has a conjugate transpose equal to its negative and is called the anti-Hermitian part. Probability conservation has already fixed the first part. The second can be written $-\ii H/\hbar$, with $H=H^\dagger$, corresponding to the nondissipative evolution in the main text. Thus
\begin{equation}
 A=-\frac{\ii}{\hbar}H-\frac12\sum_kL_k^\dagger L_k.
 \label{extA:generatorA}
\end{equation}
Next expand the state change:
\[
 \rho'-\rho=\delta t\left(A\rho+\rho A^\dagger
 +\sum_kL_k\rho L_k^\dagger\right)+O(\delta t^2).
\]
Substituting Eq.~\eqref{extA:generatorA}, dividing by $\delta t$, and taking the continuous limit gives
\begin{equation}
 \dot\rho=-\frac{\ii}{\hbar}[H,\rho]
 +\sum_k\left(L_k\rho L_k^\dagger
 -\frac12L_k^\dagger L_k\rho
 -\frac12\rho L_k^\dagger L_k\right).
 \label{extA:lindblad}
\end{equation}
The two factors of $1/2$ arise from the no-jump part acting on the left and right of the density matrix. Together with the positive jump term, they conserve probability; they cannot be selected separately as three unrelated empirical corrections.

This is a short-time derivation of the structure. For a finite-level system, the Lindblad form is general under time-homogeneous, completely positive, trace-preserving evolution with the semigroup property~\cite{extA:lindblad1976,extA:gks1976}. Here ``semigroup'' means that evolution over adjacent time intervals can be composed directly without retaining the environment's past state. The derivation determines the equation's structure, not specific decay rates; those must still come from microscopic models or experiments.

\subsection{Physical Choices of Decay and Dephasing Operators}

For decay $e\to g$, compare $K_1=\sqrt{\Gamma\delta t}\ket g\bra e$ in Eq.~\eqref{extA:amplitudechannel} with $K_1=\sqrt{\delta t}L_1$ to obtain
\[
 L_1=\sqrt\Gamma\ket{g}\bra{e}.
\]
The term $L_1\rho L_1^\dagger$ adds the decayed population to the lower level. The other two terms remove upper-level population and cause the associated off-diagonal elements to decay. Retaining only the negative terms tracks only the part in which no decay jump has yet occurred, whose trace decreases; this is not the complete unconditional atomic state.

Pure dephasing admits another simple model. Define $Z=\ket{e}\bra{e}-\ket{g}\bra{g}$. It leaves both populations unchanged but adds a relative phase of $\pi$ between the components. If this phase flip occurs with probability $r\delta t$ in a short step, and no flip occurs otherwise, then
\[
 \rho'=(1-r\delta t)\rho+r\delta t\,Z\rho Z.
\]
Thus $\dot\rho=r(Z\rho Z-\rho)$. The off-diagonal element of $Z\rho Z$ is $-\rho_{eg}$, so $\dot\rho_{eg}=-2r\rho_{eg}$. To obtain the coherence decay rate $\gamma_\phi$ defined in the main text, choose $r=\gamma_\phi/2$, corresponding to
\begin{equation}
 L_\phi=\sqrt{\frac{\gamma_\phi}{2}}Z.
 \label{extA:dephasing}
\end{equation}
This explains the $1/2$ under the square root. Random phase flips provide one model that produces the same average dephasing equation; the real environment need not apply discrete $\pi$ phase shifts one at a time.

Multilevel systems follow the same procedure. For independent decay from $j$ to $i$ at rate $\Gamma_{i\leftarrow j}$, one may choose
\[
 L_{i\leftarrow j}=\sqrt{\Gamma_{i\leftarrow j}}\ket{i}\bra{j}.
\]
When channels can be treated separately, let the total outgoing rates from levels $i$ and $j$ be $\Gamma_i$ and $\Gamma_j$. Their coherence decay rate is then
\begin{equation}
 \gamma_{ij}=\frac{\Gamma_i+\Gamma_j}{2}+\gamma_{ij}^{\phi}.
 \label{extA:multilevelwidth}
\end{equation}
The linewidths in the four-level model must be consistent with its dissipation channels{. For example, if the lower level $\ket{1}$ does not decay, then $\Gamma_1=0$, and $\gamma_{j1}$ in Eq.~\eqref{eq:fourdenominators} equals $\Gamma_j/2+\gamma_{j1}^{\phi}$}. Transitions indistinguishable to the environment, shared radiation channels, or correlated phase noise may require a combined treatment.

The $L_k$ representation of a master equation is not unique either. Replacing two channels by $(L_1+L_2)/\sqrt2$ and $(L_1-L_2)/\sqrt2$, for example, cancels the cross terms and leaves the total dissipator unchanged. Operators should be selected according to energy flow and coherence loss; when environmental memory is appreciable, the applicability of this master equation must also be checked.

\endgroup

\begingroup
\section*{Further Reading 2B\quad The Standard Quantum Limit for Atomic Systems}
\markboth{{Further Reading 2B\quad The Standard Quantum Limit for Atomic Systems}}{{Further Reading 2B\quad The Standard Quantum Limit for Atomic Systems}}
\phantomsection
\addcontentsline{toc}{section}{Further Reading 2B: The Standard Quantum Limit for Atomic Systems}
\setcounter{subsection}{0}
\renewcommand{\thesubsection}{2B.\arabic{subsection}}
\setcounter{equation}{0}
\renewcommand{\theequation}{2B.\arabic{equation}}
\renewcommand{\theHsubsection}{readingB.\arabic{subsection}}
\renewcommand{\theHequation}{readingB.\arabic{equation}}

The master equation in the main text gives average populations and polarization. Measurement precision also depends on fluctuations: how widely individual results vary and how long fluctuations in two measurements remain correlated. This reading first defines these quantities, then gives a procedure for calculating noise from the master equation.

\subsection{Observables, Means, and Projection Variance}

The density matrix $\rho$ describes the atomic state, while a Hermitian matrix $O$ specifies the physical quantity to be measured. Once the basis and readout are chosen, $O$ can be a fixed matrix, and the mean readout is $\Tr(O\rho)$. In the steady state $\rho_{\mathrm{ss}}$, define
\begin{equation}
 \overline O=\Tr(O\rho_{\mathrm{ss}}),\qquad
 \delta O=O-\overline OI,\qquad
 V_O=\Tr[(\delta O)^2\rho_{\mathrm{ss}}].
 \label{extB:variance}
\end{equation}
Here $\delta O$ denotes deviation from the mean, and $V_O$ is the measurement variance of the observable. In steady state, $\rho_{\mathrm{ss}}$, $O$, and $\delta O$ are all fixed matrices; a fixed probability distribution can still produce different measurement outcomes.

For an excited-state population measurement on a two-level atom, for example, $O=P_e=\ket e\bra e$, and the outcome is either 0 or 1. Since $P_e^2=P_e$, an excitation probability $p$ gives $V_{P_e}=p-p^2$. For $N_{\mathrm a}$ independent, identically prepared atoms, the standard deviation of the average population is
\begin{equation}
 \delta p=\sqrt{\frac{p(1-p)}{N_{\mathrm a}}}.
 \label{extB:projectiongain}
\end{equation}
This is the projection noise introduced in the main text~\cite{extB:itano}. The standard quantum limit here is the precision benchmark for uncorrelated atoms under specified preparation, measurement time, and ideal readout conditions. Resources such as spin squeezing or entanglement can change this benchmark~\cite{extB:degen}. Projection variance contains no time information and cannot be treated directly as noise per hertz of bandwidth.

\subsection{Two-Time Correlations and the Origin of the Auxiliary Matrix}

The same variance can accompany either rapidly or slowly varying fluctuations. To distinguish them, calculate the correlation at a time separation $\tau$. Conventionally, the Heisenberg picture assigns time evolution to the observable operator, denoted $O_{\mathrm H}(t)$. The subscript $\mathrm H$ labels this description and has no connection to the Hamiltonian matrix $H$. Taking a time in steady state as the origin, define the symmetrized correlation function
\begin{equation}
 C_O(\tau)=\frac12\avg{\delta O_{\mathrm H}(\tau)\delta O_{\mathrm H}(0)
                  +\delta O_{\mathrm H}(0)\delta O_{\mathrm H}(\tau)}_{\mathrm{ss}}.
 \label{extB:correlation}
\end{equation}
Averaging the two multiplication orders makes $C_O$ real. In steady state it depends only on the time separation, satisfies $C_O(-\tau)=C_O(\tau)$, and has $C_O(0)=V_O$. The variance is therefore the zero-delay value of the correlation curve, not a separate noise contribution.

How does this definition become a matrix calculation? First consider nondissipative evolution, with evolution matrix $U(\tau)$. The state evolves as $\rho(\tau)=U(\tau)\rho(0)U^\dagger(\tau)$, while the Heisenberg operator is $\delta O_{\mathrm H}(\tau)=U^\dagger(\tau)\delta O U(\tau)$. Substituting into the first correlation term and using cyclic invariance of the trace gives
\[
 \begin{aligned}
 \avg{\delta O_{\mathrm H}(\tau)\delta O_{\mathrm H}(0)}_{\mathrm{ss}}
 &=\Tr[U^\dagger\delta O U\delta O\rho_{\mathrm{ss}}]\\
 &=\Tr[\delta O\,U(\delta O\rho_{\mathrm{ss}})U^\dagger].
 \end{aligned}
\]
The other term similarly becomes $\Tr[\delta O\,U(\rho_{\mathrm{ss}}\delta O)U^\dagger]$. Combining them gives
\begin{equation}
 \begin{aligned}
 \sigma(0)&=\frac{\delta O\rho_{\mathrm{ss}}+\rho_{\mathrm{ss}}\delta O}{2},\\
 \sigma(\tau)&=U(\tau)\sigma(0)U^\dagger(\tau),\\
 C_O(\tau)&=\Tr[\delta O\sigma(\tau)].
 \end{aligned}
 \label{extB:qrtinitial}
\end{equation}
This is how the auxiliary matrix $\sigma$ arises from the correlation function: a product involving time-dependent operators is rewritten as evolution from a special initial matrix. The physical state starts from $\rho_{\mathrm{ss}}$ and remains stationary; the correlation calculation starts from $\sigma(0)$ and generally changes with $\tau$. Only $\sigma(\tau)$ is updated during the calculation; $\delta O$ stays fixed.

With dissipation, the atom's evolution can no longer be written as the unitary transformation above. Under the same memoryless-environment approximation as in the main text, the quantum regression theorem allows the same Lindblad evolution operator $\mathcal L$ to be used in calculating the correlation for nonnegative delays~\cite{extB:qrt}
\begin{equation}
 \frac{\dd\sigma(\tau)}{\dd\tau}=\mathcal L[\sigma(\tau)],
 \qquad C_O(\tau)=\Tr[\delta O\sigma(\tau)].
 \label{extB:qrtoutput}
\end{equation}
For negative delays, use the even symmetry stated above; do not integrate the dissipative equation backward. The initial matrix is still given by the first of Eqs.~\eqref{extB:qrtinitial}. The auxiliary quantity $\sigma$ has zero trace and may have negative eigenvalues; it does not satisfy the probability-normalization requirement of a density matrix. Figure~\ref{fig:ch2:qrt} compares the two calculations.

\begin{figure}[htbp]
\centering
\begin{tikzpicture}[font=\small,>=Stealth,
box/.style={draw,minimum width=2.5cm,minimum height=1.1cm,align=center,inner sep=4pt}]
\node[anchor=west] at (-1.25,2.1) {Physical state: steady-state initial matrix remains unchanged};
\node[box] (r0) at (0,1) {$\rho(0)=\rho_{\mathrm{ss}}$};
\node[box] (rt) at (4.4,1) {$\rho(\tau)=\rho_{\mathrm{ss}}$};
\draw[->] (r0)--node[above]{$\mathcal L$}(rt);
\node[align=left,anchor=west] at (6.1,1) {$\mathcal L[\rho_{\mathrm{ss}}]=0$\\$\Tr\rho=1$};
\node[anchor=west] at (-1.25,-0.15) {Correlation: evolution from a special initial matrix};
\node[box] (s0) at (0,-1.2) {$\sigma(0)$\\From Eq.~\eqref{extB:qrtinitial}};
\node[box] (st) at (4.4,-1.2) {$\sigma(\tau)$\\$\Tr\sigma=0$};
\node[box] (co) at (8.8,-1.2) {$C_O(\tau)$\\$\Tr[\delta O\sigma(\tau)]$};
\draw[->] (s0)--node[above]{Same $\mathcal L$}(st);
\draw[->] (st)--node[above]{Fixed $\delta O$}(co);
\end{tikzpicture}
\caption{Steady-state evolution versus correlation calculation. Both rows use the same linear evolution operator but different initial matrices. The upper density matrix remains in steady state. In the lower row, $\sigma(0)$ is constructed only once at the start, then $\sigma$ evolves; at each delay, it is multiplied by the fixed $\delta O$ and the trace is taken. A delay-dependent correlation function does not require a time-dependent steady-state density matrix.}
\label{fig:ch2:qrt}
\end{figure}
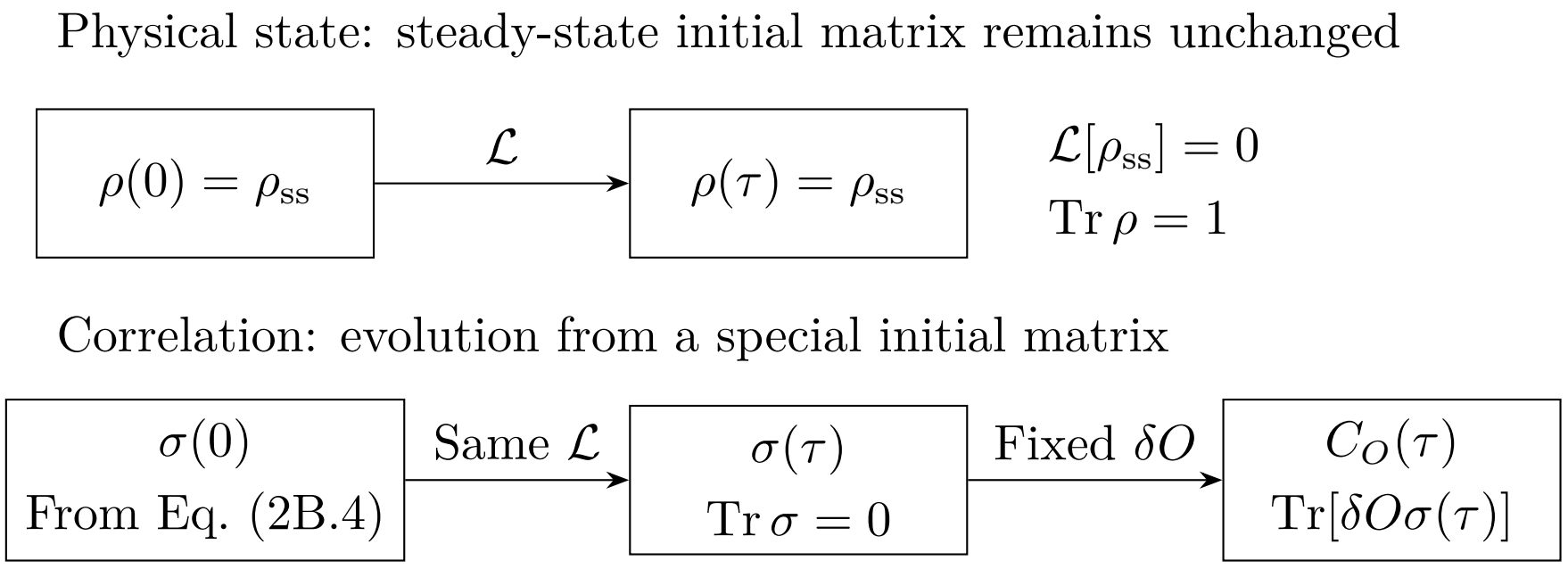

\subsection{From a Correlation Curve to a Noise Spectrum: Calculation Procedure}

The Fourier transform of the correlation function gives the noise spectrum. We use a one-sided symmetrized power spectral density in Hz, defined by
\begin{equation}
 S_O(f)=4\int_0^\infty C_O(\tau)\cos(2\pi f\tau)\,\dd\tau,
 \qquad \int_0^\infty S_O(f)\,\dd f=V_O.
 \label{extB:psd}
\end{equation}
The total spectral area equals the variance. The decay and oscillation timescales of the correlation determine how this area is distributed over frequency. For $N_{\mathrm a}$ independent atoms with identical responses, the correlation and spectrum of their average observable equal the single-atom results divided by $N_{\mathrm a}$.

For implementation, use $H_{\mathrm r}$ in the rotating reference of the main text, temporarily omit the tilde on the density matrix, and write the master equation as a function acting on any matrix $X$:
\begin{equation}
 \mathcal L[X]=-\frac{\ii}{\hbar}(H_{\mathrm r}X-XH_{\mathrm r})+\sum_k
 \left(L_kXL_k^\dagger-\frac12L_k^\dagger L_kX
                         -\frac12XL_k^\dagger L_k\right).
 \label{extB:matrixoperation}
\end{equation}
Then perform the following four steps.
\begin{enumerate}
 \item \textbf{Find the steady state.} Fix $H_{\mathrm r}$, $L_k$, and $O$ at the operating point, solve $\mathcal L[\rho_{\mathrm{ss}}]=0$, and impose $\Tr\rho_{\mathrm{ss}}=1$. This procedure assumes a stable, unique steady state.
 \item \textbf{Construct the initial matrix.} Calculate $\overline O$ and $\delta O$, then calculate $\sigma(0)$ once using Eq.~\eqref{extB:qrtinitial}.
 \item \textbf{Integrate and take the trace.} Starting from $\sigma(0)$, solve Eq.~\eqref{extB:qrtoutput} and record $C_O(\tau)$ at each delay. Use the full linear matrix operations and disable automatic normalization: the unit-trace constraint for $\rho$ must not be applied to $\sigma$.
 \item \textbf{Calculate the spectrum.} Apply the cosine integral to the recorded correlation curve. For uniform samples $\tau_n=n\Delta\tau$, the trapezoidal rule gives
 \[
 S_O(f_j)\simeq4\Delta\tau\sum_{n=0}^{K}w_nC_O(\tau_n)
                       \cos(2\pi f_j\tau_n),
 \]
  where $w_0=w_K=1/2$ and all other weights are 1. The time window should cover the correlation decay, and the step should resolve the fastest oscillation and decay. Halve the step and extend the window to check stability of the result over the frequency range of interest.
\end{enumerate}
Two direct checks are $C_O(0)=V_O$ and equality of the spectral integral to $V_O$. The first checks the initial matrix and readout; the second also checks frequency units, the one-sided-spectrum factor, and the integration range.

\subsection{Two-Level Example: A Fixed Steady State with Decaying Correlations}

Set $H_{\mathrm r}=0$ and use incoherent excitation and decay channels
\[
 L_\uparrow=\sqrt{\Gamma_\uparrow}\ket e\bra g,\qquad
 L_\downarrow=\sqrt{\Gamma_\downarrow}\ket g\bra e,
 \quad \Gamma_\uparrow=2000\,\mathrm{s}^{-1},\quad
 \Gamma_\downarrow=8000\,\mathrm{s}^{-1}.
\]
Measure excited-state population with $O=P_e=\operatorname{diag}(0,1)$. The preceding steps give
\begin{equation}
 \rho_{\mathrm{ss}}=\begin{pmatrix}0.8&0\\0&0.2\end{pmatrix},\quad
 \delta O=\begin{pmatrix}-0.2&0\\0&0.8\end{pmatrix},\quad
 \sigma(0)=\begin{pmatrix}-0.16&0\\0&0.16\end{pmatrix}.
 \label{extB:numericinitial}
\end{equation}
In this example, the diagonal elements of $\rho$ and $\sigma$ obey the same rate matrix, but have different initial values:
\[
 \frac{\dd}{\dd\tau}\begin{pmatrix}X_{gg}\\X_{ee}\end{pmatrix}
 =\begin{pmatrix}-2000&8000\\2000&-8000\end{pmatrix}
  \begin{pmatrix}X_{gg}\\X_{ee}\end{pmatrix}.
\]
Substituting $(0.8,0.2)^{\mathsf T}$ makes the right side zero, so the physical density matrix remains in steady state. Substituting $(-0.16,0.16)^{\mathsf T}$ gives a decaying solution, hence
\begin{equation}
 \begin{aligned}
 \sigma(\tau)&=\ee^{-\tau/\tau_{\mathrm c}}\sigma(0),\\
 C_{P_e}(\tau)&=\sigma_{ee}(\tau)=0.16\ee^{-\tau/\tau_{\mathrm c}},\\
 &\tau_{\mathrm c}=100\,\mu\mathrm{s},\quad \tau\geq0.
 \end{aligned}
 \label{extB:examplecorrelation}
\end{equation}
This distinguishes a fixed steady-state distribution from a delay-dependent correlation. For numerical evaluation, one may choose $\Delta\tau=1\,\mu\mathrm{s}$ and a maximum delay of $2\,\mathrm{ms}$, then check convergence. For the average population of $N_{\mathrm a}=10^4$ independent atoms, the result should be
\begin{equation}
 S_p(f)=\frac{6.4\times10^{-9}\,\mathrm{Hz}^{-1}}
 {1+(2\pi f\tau_{\mathrm c})^2},\qquad
 \int_0^\infty S_p(f)\,\dd f=1.6\times10^{-5}.
 \label{extB:lorentznoise}
\end{equation}
The single-measurement population standard deviation is $0.004$, and the low-frequency amplitude spectral density is $8.0\times10^{-5}\,\mathrm{Hz}^{-1/2}$. Slowing both transition rates while keeping the excitation probability fixed leaves the single-measurement variance unchanged but increases the correlation time, narrows the spectrum, and raises its low-frequency density.

\subsection{Converting Noise into Measurement Precision}

Finally, convert output fluctuations into uncertainty in the quantity being measured. Let that quantity be $x$ and the mean output $\overline y(x)$. Near an operating point with nonzero slope,
\begin{equation}
 K_x\simeq\frac{\overline y(x_0+\epsilon)-\overline y(x_0-\epsilon)}{2\epsilon},
 \qquad \delta x=\frac{\delta y}{|K_x|}.
 \label{extB:errorpropagation}
\end{equation}
Keep all other conditions identical in the two calculations and reduce $\epsilon$ to check slope stability. For independent repeated pulsed measurements, with interaction time $T$ and preparation/readout dead time $T_{\mathrm d}$ per cycle, a total time $\tau_{\mathrm{int}}$ permits approximately $M=\tau_{\mathrm{int}}/(T+T_{\mathrm d})$ cycles, giving $\delta x(\tau_{\mathrm{int}})=\delta x_{\mathrm{single}}/\sqrt M$. Samples in continuous measurements are generally correlated; the noise spectrum should be used to include bandwidth.

For the Rydberg receiver in the main text, consider the coherence component associated with absorption. Taking a real dipole moment and real probe amplitude (so $\Omega_p$ is real), one may choose in the rotating reference
\begin{equation}
 O_I=\frac{\ket1\bra2-\ket2\bra1}{2\ii},\qquad
 \overline O_I=\operatorname{Im}\rho_{21}^{\mathrm{ss}}.
 \label{extB:opticalobservable}
\end{equation}
Inserting the four-level Hamiltonian matrix, dissipation channels, and this $O_I$ into the preceding procedure gives $S_{O_I}$. In a linear readout model for a single component of independent, uniformly coupled atoms, let the optical conversion coefficient be $A(f)$. The atomic output noise is then $S_y^{\mathrm{at}}=|A|^2S_{O_I}/N_{\mathrm a}$. Calculating the field-to-output gain $G(2\pi f)$ of the main text at the same operating point gives
\begin{equation}
 S_E(f)=\frac{S_y(f)}{|G(2\pi f)|^2},\qquad
 (\delta E_{\mathrm{rms}})^2=\int_0^\infty |W(f)|^2S_E(f)\,\dd f.
 \label{extB:fieldnoise}
\end{equation}
Here $W$ is the filter applied to the equivalent input readout, and $f$ is the envelope or demodulation frequency. The gain must be nonzero throughout the frequency range used. At a zero-radio-frequency-field operating point without a local oscillator, a purely second-order response cannot be converted using this linear relation.

If the low-frequency equivalent input noise is approximately white, rectangular averaging of a constant signal over $T_{\mathrm{av}}$ has equivalent noise bandwidth $1/(2T_{\mathrm{av}})$, giving
\[
 \delta E_{\mathrm{rms}}\simeq\sqrt{\frac{S_E(0)}{2T_{\mathrm{av}}}},\qquad
 \eta_E\equiv\delta E_{\mathrm{rms}}\sqrt{T_{\mathrm{av}}}
 \simeq\sqrt{\frac{S_E(0)}2}.
\]
Including only the atomic contribution yields the atomic-noise-limited sensitivity of the chosen model. An actual photocurrent also involves photon counting, readout efficiency, and measurement backaction, which require an appropriate detection model. Noise power spectra may be added directly only when their contributions are uncorrelated.

\endgroup

\clearpage
\fancyhead[R]{\small Chapter 3}
\ctexset{chapter/format=\Large\normalfont,section/format=\large\normalfont,subsection/format=\normalsize\normalfont}
\captionsetup{labelfont=normalfont}

\chapter{Quantum Fields in Linear Media}
\label{chap:fieldquantization}

Chapter 2 treated the electromagnetic field as a prescribed classical drive and calculated atomic populations, coherences, and polarization. This explains absorption, saturation, and the radio-frequency response of Ryd\-berg atoms, but leaves unanswered questions about the statistics of the field itself: why can the same average electromagnetic power correspond to different photon-count fluctuations? Why does an idle port with no input signal still inject quantum noise into a system?

This chapter begins with a single lossless electromagnetic mode. We first express its energy in harmonic-oscillator form, then introduce quantization rules to establish the relations among photon number, field amplitude, and phase. We next discuss photodetection and homodyne readout, and finally extend the lossless-mode description to linear media with dispersion and absorption. Material response is represented mainly by a macroscopic dielectric function, without tracking individual atomic energy levels. Interactions in which both the field and few-level atoms are quantized are left to Chapter 4.

\section{An Electromagnetic Mode Is a Harmonic Oscillator}
\subsection{Starting with the Familiar LC Circuit}

The energy stored in an ideal LC circuit is the sum of its capacitive and inductive energies. Denoting the charge on a capacitor plate by $Q$ and the circuit current by $I=\dot Q$, we have
\begin{equation}
 W=\frac{Q^2}{2C}+\frac{LI^2}{2},\qquad
 L\ddot Q+\frac{Q}{C}=0,\qquad \omega_c=\frac{1}{\sqrt{LC}}.
 \label{eq:ch3:lc}
\end{equation}
Introducing the scaled coordinates $q=\sqrt L\,Q$ and $p=\dot q=\sqrt L\,I$, the stored energy and equations of motion become
\begin{equation}
 W=\frac12p^2+\frac12\omega_c^2q^2,
 \qquad \dot q=p,\qquad \dot p=-\omega_c^2q.
 \label{eq:ch3:osc}
\end{equation}
The variables $q$ and $p$ form a canonical pair. Here $p$ is a scaled generalized momentum, rather than a particle's mechanical momentum; both $p^2$ and $\omega_c^2q^2$ have units of joules. This scaling collects the circuit parameters into $\omega_c$ and gives the energy the standard harmonic-oscillator form.

Energy alternates between the capacitor and inductor while the total energy remains constant. An electromagnetic standing wave in a closed cavity has precisely the same property, except that the stored energy is distributed in three-dimensional space. In engineering electromagnetics, a cavity resonance is often represented by an equivalent lumped LC circuit. To make this correspondence precise, we need only specify the spatial normalization of the mode.

\subsection{Obtaining Mode Coordinates from Maxwell's Equations}

Consider an ideal closed cavity filled with a nonmagnetic, lossless medium whose dispersion is negligible over the frequency band of interest. The relative permittivity $\varepsilon_r(\mathbf r)$ is real and positive, the permeability is $\mu_0$, and the boundary is a perfect electric conductor (PEC). Choose a real transverse eigenmode function $\mathbf u(\mathbf r)$ satisfying
\begin{equation}
 \begin{aligned}
 \nabla\times\nabla\times\mathbf u&=\frac{\omega_c^2}{c^2}\varepsilon_r\mathbf u,\\
 \nabla\cdot(\varepsilon_r\mathbf u)&=0,\qquad
 \int_V\varepsilon_r\mathbf u^2(\mathbf r)\,\dd^3r=1.
 \end{aligned}
 \label{eq:ch3:mode}
\end{equation}
At the perfectly conducting boundary, $\mathbf n\times\mathbf u=0$. With the real permittivity and closed, lossless boundary conditions assumed here, the eigenmodes can be chosen as real vector functions. If degeneracy is present, a real orthogonal basis can be chosen within the degenerate subspace. Thus $\mathbf u^2=\mathbf u\cdot\mathbf u$. Under the normalization above, $\mathbf u$ has units of $\mathrm{m}^{-3/2}$.

In the absence of free charge, this eigenmode can be represented by a vector potential. Using the generalized Coulomb gauge $\nabla\cdot(\varepsilon_r\mathbf A)=0$ and setting the scalar potential to zero, we separate the spatial distribution from the time-dependent coordinate:
\begin{equation}
 \begin{aligned}
 \mathbf A(\mathbf r,t)&=\frac{q(t)}{\sqrt{\varepsilon_0}}\mathbf u(\mathbf r),\\
 \mathbf E(\mathbf r,t)&=-\frac{\dot q(t)}{\sqrt{\varepsilon_0}}\mathbf u(\mathbf r),\\
 \mathbf B(\mathbf r,t)&=\frac{q(t)}{\sqrt{\varepsilon_0}}\nabla\times\mathbf u(\mathbf r).
 \end{aligned}
 \label{eq:ch3:fieldsclassical}
\end{equation}
The spatial distribution is fixed entirely by $\mathbf u(\mathbf r)$; the only time-dependent unknown is the mode amplitude $q(t)$. Substituting $\mathbf A$ into the wave equation and using Eq.~\eqref{eq:ch3:mode} gives the harmonic-oscillator equation $\ddot q+\omega_c^2q=0$.

We now check the total electromagnetic energy stored in the cavity. Using the normalization in Eq.~\eqref{eq:ch3:mode}, the electric energy is
\[
 W_E=\frac12\int_V\varepsilon_0\varepsilon_r\mathbf E^2\dd^3r
 =\frac12\dot q^{\,2}.
\]
To calculate the magnetic energy, first take the dot product of the mode equation~\eqref{eq:ch3:mode} with $\mathbf u$ and integrate over the cavity. The vector identity
\[
 (\nabla\times\mathbf u)^2-\mathbf u\cdot(\nabla\times\nabla\times\mathbf u)
 =\nabla\cdot[\mathbf u\times(\nabla\times\mathbf u)],
\]
converts the divergence term into a surface integral. Since $\mathbf n\times\mathbf u=0$ at the PEC boundary, the boundary term vanishes, giving
\[
 \int_V(\nabla\times\mathbf u)^2\dd^3r
 =\int_V\mathbf u\cdot(\nabla\times\nabla\times\mathbf u)\dd^3r
 =\frac{\omega_c^2}{c^2}.
\]
Substituting $\varepsilon_0\mu_0=c^{-2}$, the magnetic energy and total energy reduce to
\begin{equation}
 \begin{aligned}
 W_B&=\frac{1}{2\mu_0}\int_V\mathbf B^2\dd^3r=\frac12\omega_c^2q^2,\\
 W&=W_E+W_B=\frac12p^2+\frac12\omega_c^2q^2,
 \end{aligned}
 \label{eq:ch3:modeenergy}
\end{equation}
where $p=\dot q$. This has exactly the same form as the LC circuit in Eq.~\eqref{eq:ch3:osc}, except that here $\frac12 p^2$ is the electric energy and $\frac12\omega_c^2 q^2$ is the magnetic energy. A different choice of coordinates can interchange the roles of these two terms without changing the oscillator physics.

Everything so far belongs to classical electromagnetics. The mode frequency $\omega_c$ and spatial distribution $\mathbf u(\mathbf r)$ are determined by Maxwell's equations and the boundary conditions, while the amplitudes $q,p$ can take continuous values. We now quantize only these two mode coordinates; under the assumptions of this section, the spatial mode remains the result of a classical calculation.

\section{From Mode Energy to Photon Number}
\subsection{Commutation Relations and Ladder Operators}

Quantization replaces the classical real variables $q$ and $p$ with Hermitian operators $\hat q$ and $\hat p$ acting on the quantum state of the mode, and requires them to satisfy the fundamental commutation relation:
\begin{equation}
 [\hat q,\hat p]\equiv\hat q\hat p-\hat p\hat q=\ii\hbar I,
 \qquad
 \hat H=\frac12\hat p^2+\frac12\omega_c^2\hat q^2.
 \label{eq:ch3:canonical}
\end{equation}
Here $I$ is the identity operator, which is often omitted below. This is the fundamental rule of canonical quantization adopted in this chapter; it cannot be derived from classical Maxwell equations. The commutation relation says that multiplying first by $\hat p$ and then by $\hat q$ differs from the reverse order by $\ii\hbar$. Chapter 2 already introduced noncommuting matrix multiplication ($[A,B]\neq 0$); the operators here involve the same kind of linear algebra.

Engineering electromagnetic analysis often combines two real oscillating quantities into a complex amplitude to simplify oscillation equations. Here we similarly combine $\hat q$ and $\hat p$ into dimensionless operators:
\begin{equation}
 \hat a=\sqrt{\frac{\omega_c}{2\hbar}}\hat q
       +\frac{\ii}{\sqrt{2\hbar\omega_c}}\hat p,
 \qquad
 \hat a^\dagger=\sqrt{\frac{\omega_c}{2\hbar}}\hat q
       -\frac{\ii}{\sqrt{2\hbar\omega_c}}\hat p.
 \label{eq:ch3:ladderdef}
\end{equation}
Because $\hat q$ and $\hat p$ are Hermitian (equal to their own conjugate transposes), $\hat a$ and $\hat a^\dagger$ are conjugate transposes of each other. Denote the two real coefficients by $A=\sqrt{\omega_c/(2\hbar)}$ and $B=1/\sqrt{2\hbar\omega_c}$ (noting that $AB=\frac{1}{2\hbar}$). Using $[\hat q,\hat p]=\ii\hbar$, their commutation relation follows directly:
\[
 \begin{aligned}
 [\hat a,\hat a^\dagger]
 &=(A\hat q+\ii B\hat p)(A\hat q-\ii B\hat p)\\
 &\quad-(A\hat q-\ii B\hat p)(A\hat q+\ii B\hat p)\\
 &=-2\ii AB[\hat q,\hat p]=1.
 \end{aligned}
\]
Expanding the product $\hat a^\dagger\hat a$ fully, the noncommuting cross terms give
\[
 \hat a^\dagger\hat a
 =A^2\hat q^2+B^2\hat p^2+\ii AB(\hat q\hat p-\hat p\hat q)
 =\frac{\hat H}{\hbar\omega_c}-\frac12.
\]
The mode energy operator, or Hamiltonian, can therefore be written as
\begin{equation}
 \hat n=\hat a^\dagger\hat a,
 \qquad \hat H=\hbar\omega_c\left(\hat n+\frac12\right),
 \qquad [\hat a,\hat a^\dagger]=1.
 \label{eq:ch3:hamilton}
\end{equation}
The constant $\frac12\hbar\omega_c$ in the energy formula, called the zero-point energy, arises precisely because $\hat q$ and $\hat p$ do not commute. Treating them as ordinary classical complex phasors would remove this term.

\subsection{Why Energy Increases in Discrete Steps}

Equation~\eqref{eq:ch3:hamilton} shows that finding the energy eigenstates of a mode amounts to finding the eigenstates of $\hat n=\hat a^\dagger\hat a$. Let a normalized eigenstate of $\hat n$ be $\ket{\nu}$, with eigenvalue $\nu$, so that $\hat n\ket{\nu}=\nu\ket{\nu}$. The quantity $\bra{\nu}\hat a^\dagger\hat a\ket{\nu}$ is the inner product of the column vector $\hat a\ket{\nu}$ with its own conjugate transpose, or its squared norm:
\[
 \nu=\bra{\nu}\hat a^\dagger\hat a\ket{\nu}
 =\|\hat a\ket{\nu}\|^2\geq0.
\]
Thus the eigenvalue $\nu$ cannot be negative. Also, $\hat a\hat a^\dagger-\hat a^\dagger\hat a=1$ gives $\hat n\hat a=\hat a^\dagger\hat a\hat a=(\hat a\hat a^\dagger-1)\hat a=\hat a(\hat n-1)$, or
\[
 [\hat n,\hat a]=-\hat a,\qquad
 \hat n(\hat a\ket{\nu})=(\nu-1)(\hat a\ket{\nu}).
\]
This means that, provided $\hat a\ket{\nu}$ is not the zero vector, applying $\hat a$ once to an eigenstate produces a new eigenstate whose eigenvalue is smaller by 1. If the initial value $\nu$ were not a nonnegative integer, repeated application of $\hat a$ would pass below 0 and produce a negative eigenvalue, contradicting nonnegativity. The descent must therefore stop exactly at $\nu=0$: the lowest state $\ket 0$ satisfies $\hat a\ket0=0$. It follows that the eigenvalues of $\hat n$ can only be the nonnegative integers $n=0,1,2,\ldots$.

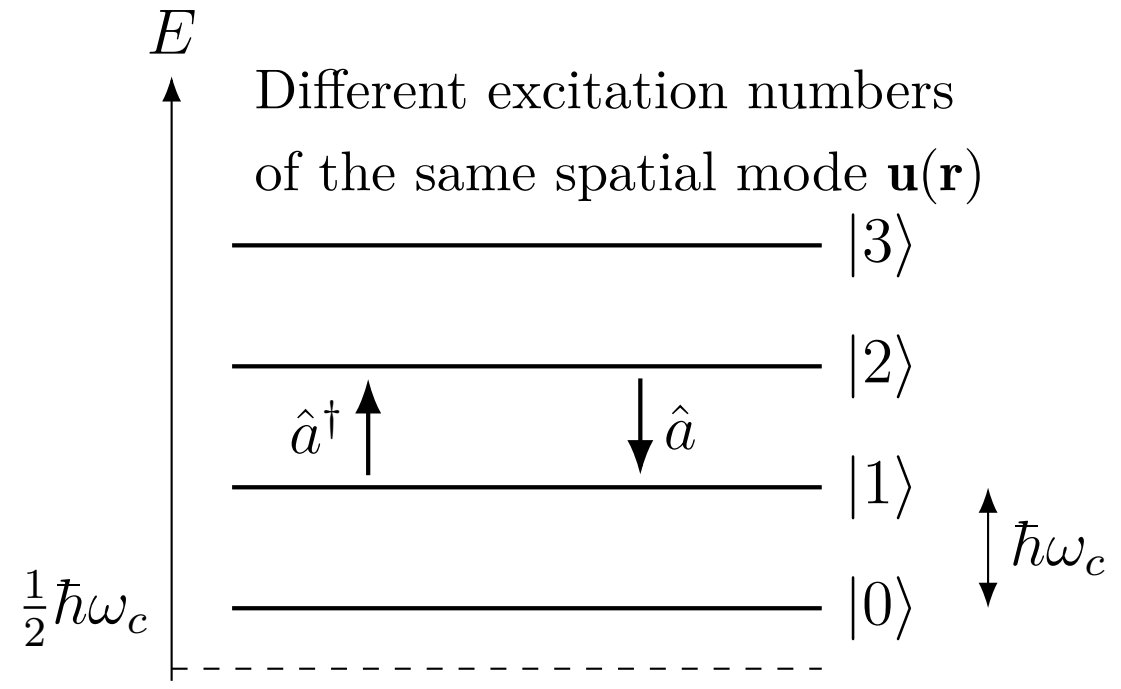
\begin{figure}[htbp]
\centering
\begin{tikzpicture}[x=1cm,y=.8cm,>=Latex]
 \draw[->] (0,0)--(0,5.0) node[above] {$E$};
 \foreach \n in {0,1,2,3} {
  \draw[thick] (.4,{\n+0.6})--(4.3,{\n+0.6});
  \node[right] at (4.3,{\n+0.6}) {$|\n\rangle$};
 }
 \draw[dashed] (0,0.1)--(4.3,.1);
 \node[left] at (0,.6) {$\frac12\hbar\omega_c$};
 \draw[->,thick] (1.3,1.7)--(1.3,2.5) node[midway,left] {$\hat a^\dagger$};
 \draw[->,thick] (3.1,2.5)--(3.1,1.7) node[midway,right] {$\hat a$};
 \draw[<->] (5.4,.6)--(5.4,1.6) node[midway,right] {$\hbar\omega_c$};
 \node[anchor=west,align=left,font=\small] at (.4,4.5) {Different excitation numbers\\of the same spatial mode $\mathbf u(\mathbf r)$};
\end{tikzpicture}
\caption{Energy levels of a single electromagnetic mode. All levels share the same spatial mode and differ in excitation number; adding one photon increases the energy by $\hbar\omega_c$.}
\label{fig:ch3:ladder}
\end{figure}

The transition coefficients between levels also follow directly from inner products. The squared norm of $\hat a\ket n$ is $\bra n\hat a^\dagger\hat a\ket n=n$, whereas that of $\hat a^\dagger\ket n$ is $\bra n\hat a\hat a^\dagger\ket n=\bra n(\hat n+1)\ket n=n+1$. After fixing the phases of the basis vectors, the ladder operators act as follows:
\begin{equation}
 \begin{gathered}
 \hat a\ket n=\sqrt n\ket{n-1},\qquad
 \hat a^\dagger\ket n=\sqrt{n+1}\ket{n+1},\\
 \hat n\ket n=n\ket n,\qquad
 E_n=\hbar\omega_c\left(n+\frac12\right).
 \end{gathered}
 \label{eq:ch3:ladderaction}
\end{equation}
Because $\hat a$ decreases the excitation number by 1 and $\hat a^\dagger$ increases it by 1, they are called the annihilation and creation operators, respectively; $\hat n$ is the photon-number operator. A ``photon'' is one quantized increment of excitation energy $\hbar\omega_c$ in this electromagnetic mode. The eigenstate $\ket n$ is called a number state or Fock state (see Figure~\ref{fig:ch3:ladder} for the energy levels). This photon concept applies equally to microwave cavities and lumped LC quantum circuits; it is not restricted to visible light.

In the basis $(\ket0,\ket1,\ket2,\ldots)$, $\hat a$ has nonzero entries only on the first superdiagonal, with successive values $1,\sqrt2,\sqrt3,\ldots$. For example, retaining only $\ket0$ through $\ket3$ gives the numerical matrices
\[
 a=\begin{pmatrix}0&1&0&0\\0&0&\sqrt2&0\\0&0&0&\sqrt3\\0&0&0&0\end{pmatrix},
 \qquad
 a^\dagger=\begin{pmatrix}0&0&0&0\\1&0&0&0\\0&\sqrt2&0&0\\0&0&\sqrt3&0\end{pmatrix},
\]
\[
 n=\operatorname{diag}(0,1,2,3).
\]
This truncation is convenient for numerical implementation, but the commutation relation is no longer exact at the highest level: retaining $0$ through $N$ gives $[a,a^\dagger]=I-(N+1)\ket N\bra N$. The populations of the highest few number states should therefore be sufficiently small, and convergence should be checked by increasing $N$.

\subsection{The Electric-Field Operator and Vacuum Fluctuations}

Inverting Eq.~\eqref{eq:ch3:ladderdef} gives
\[
 \hat q=\sqrt{\frac{\hbar}{2\omega_c}}(\hat a+\hat a^\dagger),\qquad
 \hat p=-\ii\sqrt{\frac{\hbar\omega_c}{2}}(\hat a-\hat a^\dagger).
\]
Substitution into the classical field expansion~\eqref{eq:ch3:fieldsclassical} yields
\begin{equation}
 \begin{aligned}
 \hat{\mathbf E}(\mathbf r)&=\ii\sqrt{\frac{\hbar\omega_c}{2\varepsilon_0}}
 \mathbf u(\mathbf r)(\hat a-\hat a^\dagger),\\
 \hat{\mathbf B}(\mathbf r)&=\sqrt{\frac{\hbar}{2\varepsilon_0\omega_c}}
 (\nabla\times\mathbf u)(\hat a+\hat a^\dagger).
 \end{aligned}
 \label{eq:ch3:fieldoperator}
\end{equation}
The spatial distribution is still the classical eigenmode $\mathbf u(\mathbf r)$, while the continuously variable mode amplitude has become an operator. Mean fields and variances are then obtained by taking expectation values in the field's quantum state, as in Chapter 2.

Choose a real unit direction $\mathbf e$ and define the vacuum field scale along it as
\[
 \mathcal E(\mathbf r)=\sqrt{\frac{\hbar\omega_c}{2\varepsilon_0}}
 \,\mathbf e\cdot\mathbf u(\mathbf r).
\]
In a number state $\ket n$, $\hat a\ket n$ is orthogonal to $\ket n$, so $\avg{\hat a}=0$. Similarly, $\avg{\hat a^\dagger}=0$, and the mean electric field is therefore zero. A number state has no defined mean-field phase, but this does not imply that the mean squared electric field vanishes. Expanding gives
\[
 \hat E^2=-\mathcal E^2
 (\hat a^2+\hat a^{\dagger2}-\hat a^\dagger\hat a-\hat a\hat a^\dagger).
\]
Here $\avg{\hat a^2}=\avg{\hat a^{\dagger2}}=0$, while $\avg{\hat a^\dagger\hat a}=n$ and $\avg{\hat a\hat a^\dagger}=n+1$, so
\begin{equation}
 \avg{\hat E}=0,\qquad
 \Var(E)=\avg{\hat E^2}-\avg{\hat E}^2=\mathcal E^2(2n+1).
 \label{eq:ch3:vacuum}
\end{equation}
Even for $n=0$, the variance remains $\mathcal E^2$. The state $\ket0$, called the vacuum state, contains no photons that can be annihilated (absorbed) from the mode. Nevertheless, it retains the zero-point energy $\frac12\hbar\omega_c$ and field fluctuations, called vacuum fluctuations. This result is the contribution of the selected single mode. Summing all frequencies at an ideal spatial point generally requires handling high-frequency divergences; real detection always has a finite spatial extent and frequency bandwidth.

This field scale can be obtained directly from engineering electromagnetic mode calculations. Define the effective mode volume along the field direction at position $\mathbf r_0$ by
\begin{equation}
 \begin{aligned}
 V_{\mathrm{eff}}(\mathbf r_0)
 &=\frac{\int_V\varepsilon_r(\mathbf r)|\mathbf E_{\mathrm{mode}}(\mathbf r)|^2\dd^3r}
 {\varepsilon_r(\mathbf r_0)|\mathbf e\cdot\mathbf E_{\mathrm{mode}}(\mathbf r_0)|^2},\\
 |\mathcal E(\mathbf r_0)|
 &=\sqrt{\frac{\hbar\omega_c}{2\varepsilon_0\varepsilon_r(\mathbf r_0)V_{\mathrm{eff}}(\mathbf r_0)}}.
 \end{aligned}
 \label{eq:ch3:veff}
\end{equation}
Any arbitrary scaling constant in the simulated eigenfield cancels between numerator and denominator. For example, in vacuum, $f_c=10\,\mathrm{GHz}$ and $V_{\mathrm{eff}}=1\,\mathrm{cm}^3$ give a vacuum electric-field standard deviation of approximately $6.1\times10^{-4}\,\mathrm{V/m}$. This is the fluctuation amplitude of a normalized mode, rather than the complex peak amplitude of a continuous wave.

\subsection{Multiple Modes and Multiple Photons Are Different Concepts}

A general electromagnetic field must be expanded in a set of eigenmodes $\mathbf u_m(\mathbf r)$. Under the lossless conditions above, orthonormality generalizes to $\int_V\varepsilon_r\mathbf u_m\cdot\mathbf u_n\dd^3r=\delta_{mn}$, and each mode has its own pair $q_m,p_m$. Cross terms vanish in the spatial integral, leaving a sum of independent harmonic-oscillator energies. Quantizing each mode gives
\begin{equation}
 \begin{gathered}
 [\hat a_m,\hat a_n^\dagger]=\delta_{mn},\qquad
 \hat H=\sum_m\hbar\omega_m\left(\hat a_m^\dagger\hat a_m+\frac12\right),\\
 \hat{\mathbf E}(\mathbf r)=\ii\sum_m\sqrt{\frac{\hbar\omega_m}{2\varepsilon_0}}
 \mathbf u_m(\mathbf r)(\hat a_m-\hat a_m^\dagger).
 \end{gathered}
 \label{eq:ch3:multimode}
\end{equation}
The mode index $m$ must be distinguished from the photon number $n_m$: $m$ labels different spatial, polarization, or frequency modes, determined by classical Maxwell equations, whereas $n_m$ specifies how many photons occupy mode $m$, as determined by the quantum state. A joint multimode number state is written $\ket{n_1,n_2,\ldots}$.

For example, the two-mode state $(\ket{1,0}+\ket{0,1})/\sqrt2$ contains only one photon in total; $\ket{1,1}$ contains two. The former can be produced by sending a single photon through an equal-power beam splitter, with the relative phase depending on the port phase convention. It satisfies $\avg{\hat a_1^\dagger\hat a_2}=1/2$: first-order coherence exists between the modes even though each has zero mean field. A zero mean field therefore does not preclude interference. Calculations should specify both the selected modes and their quantum state.

\section{Field Evolution and the Connection to Classical Amplitudes}
\subsection{Time-Dependent States and Time-Dependent Operators Are Equivalent}

As in Chapter 2, the quantum state can evolve according to the Schr\"odinger equation while operators remain fixed: this is the Schr\"odinger picture. For a lossless single mode, the evolution matrix is $U(t)=\exp(-\ii\hat Ht/\hbar)$, and the density matrix evolves as $\rho(t)=U(t)\rho(0)U^\dagger(t)$. Using the cyclic property of the trace, $\Tr(ABC)=\Tr(CAB)$, the mean field can be rewritten as
\[
 \avg{\hat E}(t)=\Tr[\hat E U(t)\rho(0)U^\dagger(t)]
 =\Tr[\underbrace{U^\dagger(t)\hat E U(t)}_{\hat E_H(t)}\rho(0)].
\]
Defining the time-dependent operator $\hat E_H(t)=U^\dagger(t)\hat E U(t)$ transfers the time dependence to the operator, while the state remains fixed at $\rho(0)$. This is the Heisenberg picture. The subscript $H$ indicates the picture and has a different meaning from the Hamiltonian $\hat H$.

Care is needed with the sign of the derivative. Since $\dot U=-(\ii/\hbar)\hat H U$ and $\dot U^\dagger=+(\ii/\hbar)U^\dagger\hat H$, differentiating $\hat a_H(t)=U^\dagger(t)\hat a U(t)$ gives the opposite sign to the density-matrix master equation $\dot\rho=-(\ii/\hbar)[\hat H,\rho]$ in Chapter 2:
\[
 \frac{\dd\hat a_H}{\dd t}=\dot U^\dagger\hat a U+U^\dagger\hat a\dot U=
 \frac{\ii}{\hbar}[\hat H,\hat a_H]
 =-\ii\omega_c\hat a_H.
\]
Here we used $[\hat n,\hat a]=-\hat a$. The free-evolution solution is therefore $\hat a_H(t)=\hat a\ee^{-\ii\omega_ct}$. To simplify notation, we omit the subscript $H$ below and take all explicitly time-dependent field operators to be in the Heisenberg picture. The electric-field operator becomes
\begin{equation}
 \begin{aligned}
 \hat{\mathbf E}(\mathbf r,t)&=\hat{\mathbf E}^{(+)}(\mathbf r,t)
 +\hat{\mathbf E}^{(-)}(\mathbf r,t),\\
 \hat{\mathbf E}^{(+)}(\mathbf r,t)&=\ii\sqrt{\frac{\hbar\omega_c}{2\varepsilon_0}}
 \mathbf u(\mathbf r)\hat a\ee^{-\ii\omega_ct}.
 \end{aligned}
 \label{eq:ch3:positive}
\end{equation}
Here $\hat{\mathbf E}^{(-)}=(\hat{\mathbf E}^{(+)})^\dagger$. Following the phasor convention of Chapter 2, the positive-frequency part contains $\ee^{-\ii\omega_c t}$ with the annihilation operator $\hat a$; the negative-frequency part is its conjugate transpose, containing $\ee^{+\ii\omega_c t}$ with the creation operator $\hat a^\dagger$. The positive and negative superscripts do not denote positive and negative energies. When calculating field statistics, the quantum state remains fixed at $\rho(0)$; free evolution must not be counted again in the density matrix.

\subsection{Why a Coherent State Resembles a Classical Harmonic Field}

A number state has a definite photon number but zero mean field. A phase-stable classical harmonic field is usually described by another type of state: an eigenstate of the annihilation operator $\hat a$, defined by
\begin{equation}
 \hat a\ket\alpha=\alpha\ket\alpha,
 \qquad \alpha\in\mathbb C.
 \label{eq:ch3:coherentdef}
\end{equation}
This is called a coherent state. The dimensionless complex number $\alpha=|\alpha|\ee^{\ii\phi}$ contains amplitude and phase information; it is not a photon number. Taking the conjugate transpose of Eq.~\eqref{eq:ch3:coherentdef} also gives $\bra\alpha\hat a^\dagger=\alpha^*\bra\alpha$, a relation used below to calculate expectation values.

To find this state, expand it in the number-state basis as $\ket\alpha=\sum_{n=0}^{\infty}c_n\ket n$. Applying $\hat a$ once and shifting the summation index by one gives
\[
 \hat a\ket\alpha=\sum_{n=0}^{\infty}\sqrt{n+1}\,c_{n+1}\ket n.
\]
Comparing each component with the right-hand side $\alpha\sum_nc_n\ket n$ gives the recurrence $c_{n+1}=\alpha c_n/\sqrt{n+1}$, or $c_n=\alpha^nc_0/\sqrt{n!}$. The normalization condition is
\[
 1=|c_0|^2\sum_{n=0}^{\infty}\frac{|\alpha|^{2n}}{n!}
 =|c_0|^2\ee^{|\alpha|^2}.
\]
Choosing the overall phase so that $c_0>0$, we obtain
\begin{equation}
 \begin{aligned}
 \ket\alpha&=\ee^{-|\alpha|^2/2}\sum_{n=0}^{\infty}
 \frac{\alpha^n}{\sqrt{n!}}\ket n,\\
 P_n&=|\braket{n}{\alpha}|^2
 =\ee^{-|\alpha|^2}\frac{|\alpha|^{2n}}{n!}.
 \end{aligned}
 \label{eq:ch3:coherentstate}
\end{equation}
The probability $P_n$ of measuring $n$ photons follows a Poisson distribution. Setting $\bar n=|\alpha|^2$ and writing $n/n!$ as $1/(n-1)!$ in the sum gives the mean photon number $\avg{\hat n}=\bar n=|\alpha|^2$. Similarly, $\avg{\hat n(\hat n-1)}=\bra\alpha\hat a^{\dagger2}\hat a^2\ket\alpha=|\alpha|^4=\bar n^2$, so
\begin{equation}
 \Var(n)=\avg{\hat n^2}-\avg{\hat n}^2=\bar n,\qquad \frac{\Delta n}{\bar n}=\frac{1}{\sqrt{\bar n}}.
 \label{eq:ch3:poisson}
\end{equation}
Here $\Delta n=\sqrt{\Var(n)}$ is the standard deviation. The photon number of a coherent state is not fixed. When $\bar n\gg1$, however, the relative fluctuation $1/\sqrt{\bar n}$ is small, and many measurements can be described by classical waves.

Taking the expectation value of Eq.~\eqref{eq:ch3:positive} and using $\avg{\hat a}=\bra\alpha\hat a\ket\alpha=\alpha$ gives the mean electric field of a coherent state:
\[
 \avg{\hat{\mathbf E}}(\mathbf r,t)
 =\ii\sqrt{\frac{\hbar\omega_c}{2\varepsilon_0}}\mathbf u(\mathbf r)
 \bigl(\alpha\ee^{-\ii\omega_ct}-\alpha^*\ee^{\ii\omega_ct}\bigr).
\]
Comparing with the classical monochromatic field $\mathbf E(\mathbf r,t)=\frac12(\mathbf E_0\ee^{-\ii\omega_ct}+\mathbf E_0^*\ee^{\ii\omega_ct})$ in Eq.~(2.12) of Chapter 2 gives the exact correspondence between the classical complex peak amplitude and the coherent-state parameter $\alpha$:
\begin{equation}
 \mathbf E_0(\mathbf r)=2\ii\sqrt{\frac{\hbar\omega_c}{2\varepsilon_0}}
 \mathbf u(\mathbf r)\alpha.
 \label{eq:ch3:classicalamplitude}
\end{equation}
Equation~\eqref{eq:ch3:classicalamplitude} relates the classical drive amplitude of Chapter 2 to the coherent-state parameter. In the linear model used here, the mean field of a coherent state satisfies classical Maxwell equations, while the electric-field variance remains $\mathcal E^2$. When $|\alpha|\gg1$ and atomic backaction on the drive source is negligible, the mode can usually be approximated as a prescribed classical drive. If a real radio-frequency source or laser has additional amplitude or phase noise, these statistics must also be included in the field state.

\subsection{Two Quadratures and the Thermal State}

Radio-frequency signals are often separated into in-phase and quadrature components, the I and Q channels. For a quantum mode, define the corresponding dimensionless quadrature operators:
\begin{equation}
 \hat X=\frac{\hat a+\hat a^\dagger}{2},\qquad
 \hat Y=\frac{\hat a-\hat a^\dagger}{2\ii},\qquad
 [\hat X,\hat Y]=\frac{\ii}{2}.
 \label{eq:ch3:quadratures}
\end{equation}
Both are Hermitian, and $\hat a=\hat X+\ii\hat Y$. In a coherent state, $\avg{\hat X}=\Re\alpha$ and $\avg{\hat Y}=\Im\alpha$. To find the fluctuations, expand $\hat X^2$:
\[
 \hat X^2=\frac14
 (\hat a^2+\hat a^{\dagger2}+\hat a^\dagger\hat a+\hat a\hat a^\dagger).
\]
Substituting $\avg{\hat a^2}=\alpha^2$, $\avg{\hat a\hat a^\dagger}=|\alpha|^2+1$, and related expressions, then subtracting the squared mean, gives $\Var(X)=1/4$; similarly, $\Var(Y)=1/4$.

The commutation relation $[\hat X,\hat Y]=\ii/2$ sets a lower bound on the measurement uncertainties of the two quadratures:
\begin{equation}
 \Delta X\,\Delta Y\geq\frac14.
 \label{eq:ch3:uncertainty}
\end{equation}
Both coherent and vacuum states have $\Delta X=\Delta Y=1/2$, exactly reaching the minimum uncertainty bound. This inequality follows directly from the nonnegativity of an operator inner product. For any real $\lambda$, define $\delta\hat X=\hat X-\avg{\hat X}$ and $\delta\hat Y=\hat Y-\avg{\hat Y}$. Since the squared norm of any state vector $\ket\phi=(\delta\hat X+\ii\lambda\delta\hat Y)\ket\psi$ is nonnegative,
\[
 \begin{aligned}
 0&\leq\avg{(\delta\hat X-\ii\lambda\delta\hat Y)
                   (\delta\hat X+\ii\lambda\delta\hat Y)}\\
 &=\Var(X)+\lambda^2\Var(Y)+\ii\lambda\avg{[\delta\hat X,\delta\hat Y]}\\
 &=\Var(X)+\lambda^2\Var(Y)-\frac{\lambda}{2}.
 \end{aligned}
\]
The quadratic polynomial on the right is nonnegative for every real $\lambda$, so its discriminant cannot be positive: $(-1/2)^2-4\Var(X)\Var(Y)\leq 0$, or $\Var(X)\Var(Y)\geq1/16$. Taking the square root gives Eq.~\eqref{eq:ch3:uncertainty}. Thus the in-phase and quadrature amplitudes of an electromagnetic field cannot both be determined with unlimited precision. This is an intrinsic statistical property of the quantum field, not additional noise from receiver electronics.

\begin{figure}[htbp]
\centering
\begin{tikzpicture}[font=\small,>=Stealth,scale=1.05]
\foreach \shift/\title in {0/{(a) Vacuum state},4/{(b) Coherent state},8/{(c) Thermal state}}{
 \begin{scope}[xshift=\shift cm]
 \node at (0.5,2.55) {\title};
 \draw[->,black!65] (-1,0)--(2,0) node[right]{$X$};
 \draw[->,black!65] (0,-0.95)--(0,1.7) node[above]{$Y$};
 \end{scope}
}
\draw[fill=black!10,fill opacity=0.6] (0,0) circle (0.5);
\draw[fill=black!10,fill opacity=0.6] (5,0.65) circle (0.5);
\draw[->] (4,0)--(5,0.65);
\fill (5,0.65) circle (1.3pt);
\node[anchor=west,font=\scriptsize] at (5.15,1.32) {$(\Re\alpha,\Im\alpha)$};
\draw[fill=black!10,fill opacity=0.6] (8,0) circle (0.85);
\node[align=center] at (0.5,-1.4) {$\avg{X}=\avg{Y}=0$\\$\Delta X=\Delta Y=1/2$};
\node[align=center] at (4.5,-1.4) {Mean amplitude set by $\alpha$\\$\Delta X=\Delta Y=1/2$};
\node[align=center] at (8.5,-1.4) {$\avg{X}=\avg{Y}=0$\\$\Delta X=\Delta Y>1/2$};
\end{tikzpicture}
\caption{Mean amplitudes and quadrature fluctuations for three field states. Circle centers indicate the means; the radius is one standard deviation of either quadrature. The thermal state is shown at nonzero temperature. A coherent state can have a nonzero mean amplitude but has the same fluctuation width as vacuum. These widths describe statistics from separate measurements of $X$ and $Y$ in different experiments; they do not imply that both can be measured precisely in one experiment or that every outcome lies inside the circle.}
\label{fig:ch3:quadratures}
\end{figure}
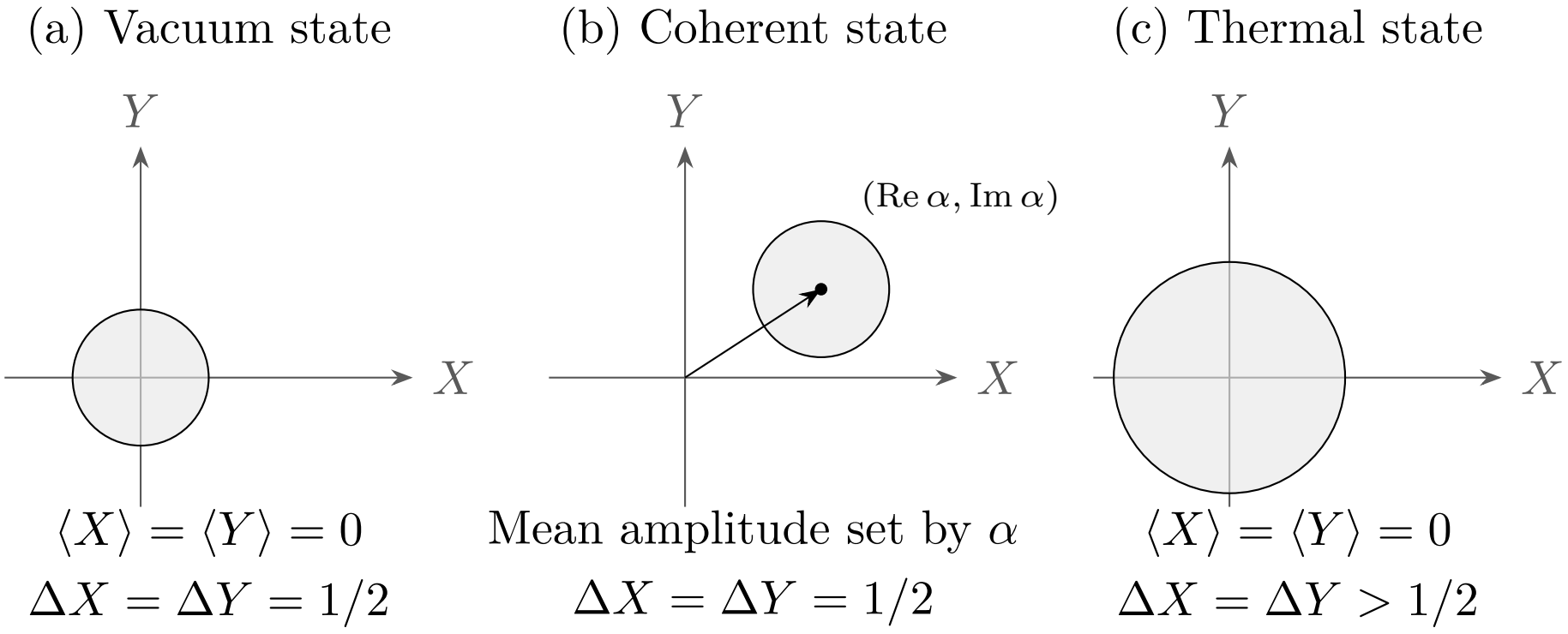

Unlike a coherent state with a defined phase, a single-mode thermal equilibrium state at absolute temperature $T$, such as blackbody radiation or a microwave thermal background, is an incoherent statistical mixture of number states $\ket n$. The energy spacing is $\hbar\omega_c$. Using the Boltzmann weight $\ee^{-n\hbar\omega_c/(k_BT)}$ from statistical physics and setting $z=\exp[-\hbar\omega_c/(k_BT)]<1$, its density matrix in the notation of Chapter 2 is
\begin{equation}
 \begin{aligned}
 \rho_{\mathrm{th}}&=(1-z)\sum_{n=0}^{\infty}z^n\ket n\bra n,\\
 \bar n_T\equiv\avg{\hat n}&=\frac{z}{1-z}
 =\frac{1}{\ee^{\hbar\omega_c/(k_BT)}-1}.
 \end{aligned}
 \label{eq:ch3:thermal}
\end{equation}
Here $k_B$ is Boltzmann's constant, and $\bar n_T$ is the Planck distribution for the thermal mean photon number. Normalization uses the geometric sum $\sum_{n=0}^\infty z^n=1/(1-z)$; successive differentiation of this series gives
\[
 \avg{\hat n}=\frac{z}{1-z}=\bar n_T,\qquad
 \avg{\hat n(\hat n-1)}=\frac{2z^2}{(1-z)^2}=2\bar n_T^2.
\]
The thermal photon-number variance is therefore $\Var(n)=\avg{\hat n^2}-\bar n_T^2=\bar n_T(\bar n_T+1)=\bar n_T+\bar n_T^2$, exceeding that of a coherent state with the same mean photon number by $\bar n_T^2$, due to the random Rayleigh fluctuations of the thermal field amplitude. Because $\rho_{\mathrm{th}}$ is diagonal, the mean electric field vanishes, and both quadrature variances are $\Var(X)=\Var(Y)=(2\bar n_T+1)/4$. Figure~\ref{fig:ch3:quadratures} compares the mean amplitudes and fluctuation widths of these three field states.

At $T=300\,\mathrm K$ and $f=10\,\mathrm{GHz}$, $\hbar\omega_c/(k_BT)\simeq1.60\times10^{-3}$, and the thermal mean photon number per mode is approximately $625$, so thermal fluctuations greatly exceed vacuum fluctuations. Microwave quantum experiments often use cryogenic environments with $k_BT\ll\hbar\omega_c$ to reduce thermal photon occupation.

At a wavelength of $780\,\mathrm{nm}$, room temperature gives $\hbar\omega_c/(k_BT)\simeq61.5$ and a thermal mean photon number of only about $2\times10^{-27}$. Thus, in the absence of applied light and with negligible stray background, the thermal state of this optical mode is already close to vacuum. This conclusion concerns only the thermal background; it does not imply that an optical fiber or free-space field carrying an optical signal must be in the vacuum state. Vacuum fluctuations remain even when the thermal photon number is nearly zero.

\section{From Field Operators to Practical Detection}
\subsection{Why Photon Counting Uses Positive- and Negative-Frequency Parts}

Like atomic absorption of a classical field in Chapter 2, absorption of electromagnetic energy by a detector, such as electrons in a photodiode or atoms, is driven by the electric-dipole interaction. We do not yet develop the detector's internal dynamics, but consider weak, near-resonant absorption by a detector initially in its ground state. Under the rotating-wave approximation, only a transition of the detector from its ground state to an excited state accompanied by annihilation of a field photon conserves energy. Thus, for an incident field in state $\ket\psi$, the final-state probability amplitude after one photoabsorption event is proportional to $\hat E^{(+)}(\mathbf r,t)\ket\psi$. Taking its squared norm, $\bra\psi\hat E^{(-)}\hat E^{(+)}\ket\psi$, or the trace for a statistical mixture, gives the instantaneous photon-count rate~\cite{ch3:glauber}:
\begin{equation}
 R(t)\propto\avg{\hat E^{(-)}(\mathbf r,t)\hat E^{(+)}(\mathbf r,t)}.
 \label{eq:ch3:count}
\end{equation}
The detection polarization has been selected here. The proportionality factor depends on the detector quantum efficiency, effective area, and response bandwidth.

For a single-mode field, Eq.~\eqref{eq:ch3:count} is proportional to $\avg{\hat a^\dagger\hat a}=\bar n$. Although vacuum has a nonzero $\avg{\hat E^2}$, it cannot supply one photon's energy to an ideal absorptive detector in its ground state and therefore produces no photon counts. Dark counts in real detectors arise from other processes.

Homodyne detection, discussed below, measures a field quadrature, and its output variance includes vacuum fluctuations. There is no contradiction: absorption counting and quadrature measurement correspond to different operator combinations. A discussion of quantum noise must first specify the detection method.

\subsection{What First- and Second-Order Correlations Tell Us}

Abbreviate the two sampled spacetime points $(\mathbf r_1,t_1)$ and $(\mathbf r_2,t_2)$ as $1$ and $2$. The first-order field correlation and its normalized form are defined by
\begin{equation}
 \begin{aligned}
 G^{(1)}(1,2)&=\avg{\hat E^{(-)}(1)\hat E^{(+)}(2)},\\
 g^{(1)}(1,2)&=\frac{G^{(1)}(1,2)}{\sqrt{G^{(1)}(1,1)G^{(1)}(2,2)}}.
 \end{aligned}
 \label{eq:ch3:g1}
\end{equation}
This has the same structure as a classical field cross-correlation. For a stationary field, the Fourier transform of $G^{(1)}(\tau)$ gives the corresponding ordered field spectrum, which cannot be used interchangeably with the symmetrized noise spectrum of Chapter 2.

Superposing two fields as $\hat E^{(+)}=\hat E_1^{(+)}+\ee^{\ii\phi}\hat E_2^{(+)}$ and substituting into the counting formula produces the cross term $2\Re[\ee^{\ii\phi}G^{(1)}(1,2)]$. For equal mean intensities, the fringe visibility is $|g^{(1)}|$; for unequal intensities, it also depends on their ratio. Interference requires correlation between the two fields, but does not require either field to have a nonzero mean electric field.

Now place photon counters at positions $1$ and $2$ and measure the joint probability of absorbing a photon at position 1 and another at position 2 simultaneously, or with a delay $\tau$. This requires two successive annihilation operations, $\hat E^{(+)}(2)\hat E^{(+)}(1)\ket\psi$, followed by taking the squared norm. The resulting second-order correlation is defined as
\begin{equation}
 \begin{split}
 G^{(2)}(1,2)&=\avg{\hat E^{(-)}(1)\hat E^{(-)}(2)
                    \hat E^{(+)}(2)\hat E^{(+)}(1)},\\
 g^{(2)}(1,2)&=\frac{G^{(2)}(1,2)}{G^{(1)}(1,1)G^{(1)}(2,2)}.
 \end{split}
 \label{eq:ch3:g2}
\end{equation}
Placing creation operators to the left and annihilation operators to the right is called normal ordering. Quantum operators generally do not commute, so normally ordered correlations cannot simply be replaced by intensity products in arbitrary order. All the normalized definitions above require a nonzero denominator.

For the same single mode at zero delay ($t_1=t_2$), spatial and field-scale factors cancel between numerator and denominator, giving
\[
 g^{(2)}(0)=\frac{\avg{\hat a^{\dagger2}\hat a^2}}{\avg{\hat a^\dagger\hat a}^2}.
\]
Using $\hat a^\dagger\hat a=\hat a\hat a^\dagger-1$, the numerator can be rewritten as
\[
 \begin{aligned}
 \hat a^{\dagger2}\hat a^2
 &=\hat a^\dagger(\hat a\hat a^\dagger-1)\hat a\\
 &=\hat n^2-\hat n=\hat n(\hat n-1).
 \end{aligned}
\]
Then, using $\Var(n)=\avg{\hat n^2}-\bar n^2$, we obtain
\begin{equation}
 g^{(2)}(0)=\frac{\avg{\hat n(\hat n-1)}}{\bar n^2}
 =1+\frac{\Var(n)-\bar n}{\bar n^2}.
 \label{eq:ch3:g2variance}
\end{equation}
If the mode initially contains $n$ photons, the first absorption has $n$ excitations available, and $n-1$ remain afterward. The probability weight for two absorptions therefore contains $n(n-1)$. The photon-number variances derived earlier give Table~\ref{tab:ch3:states}.

\begin{table}[htbp]
\centering\small
\caption{Photon statistics of the same detected electromagnetic mode in three typical quantum states. For vacuum, the denominator has $\bar n=0$, so $g^{(2)}$ is undefined.}
\label{tab:ch3:states}
\setlength{\tabcolsep}{4pt}
\renewcommand{\arraystretch}{1.35}
\begin{tabularx}{\linewidth}{@{}Yccc@{}}\toprule
Quantum state & \shortstack{Mean photon\\number\\$\bar n$}
 & \shortstack{Photon-number\\variance\\$\Var(n)$}
 & \shortstack{Second-order\\correlation\\$g^{(2)}(0)$}\\\midrule
Number state $\ket n$ ($n\geq1$) & $n$ & $0$ & $1-1/n$\\
Coherent state $\ket\alpha$ & $|\alpha|^2$ & $\bar n$ & $1$\\
Single-mode thermal state & $\bar n_T$ & $\bar n_T(\bar n_T+1)$ & $2$\\\bottomrule
\end{tabularx}
\end{table}

For example, at $\bar n=1$, a single-photon state, a coherent state, and a single-mode thermal state have the same mean energy. With the same detector coupling, their mean count rates are also equal, but their values of $g^{(2)}(0)$ are $0$, $1$, and $2$, respectively.
\begin{itemize}
 \item A coherent state has $g^{(2)}(0)=1$. An ideal coherent input produces Poisson counts.
 \item A single-mode thermal state has $g^{(2)}(0)=2$. Two detections are more likely to occur simultaneously or within a short delay; this is called photon bunching.
 \item A single-photon state has $g^{(2)}(0)=0$, because one photon cannot undergo two absorptions. More generally, $g^{(2)}(0)<1$ corresponds to $\Var(n)<\bar n$, called sub-Poissonian statistics.
\end{itemize}
For a nonnegative classical random intensity $I$, nonnegative variance implies $\avg{I^2}\geq\avg{I}^2$, so
\[
 g^{(2)}_{\mathrm{classical}}(0)=\frac{\avg{I^2}}{\avg{I}^2}=1+\frac{\Var(I)}{\avg{I}^2}\geq 1.
\]
Thus, within the corresponding detection model, $g^{(2)}(0)<1$ is a criterion for nonclassical photon statistics. A single value of $g^{(2)}(0)$, however, does not determine the single-photon preparation probability or state purity. For example, any mixture of vacuum and a nonzero single-photon component can have $g^{(2)}(0)=0$, even with a very small mean photon number.

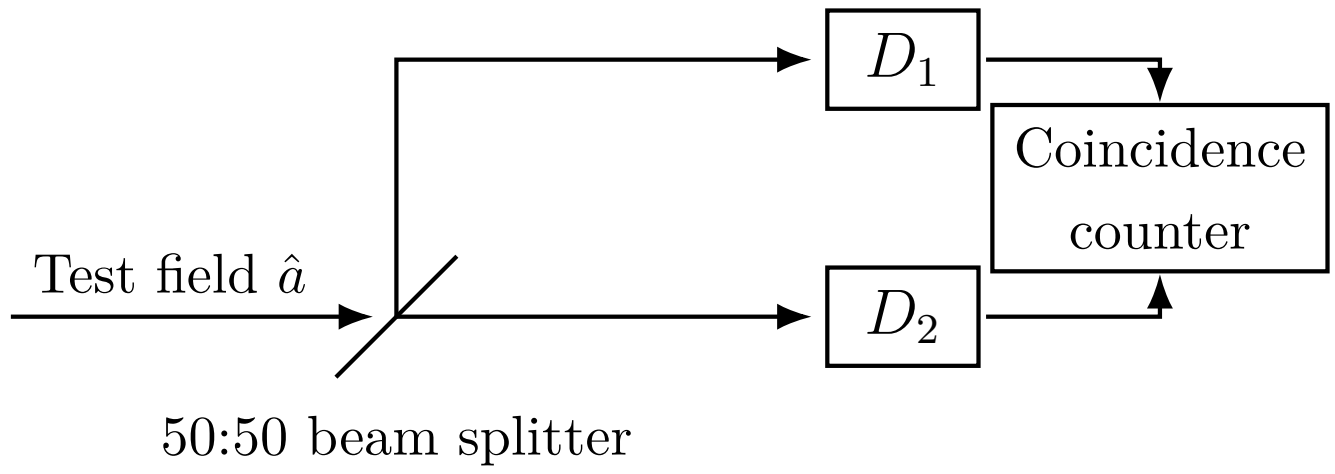
\begin{figure}[htbp]
\centering
\begin{tikzpicture}[>=Latex,thick]
 \draw[->] (0,0)--(2.4,0);
 \node[above,font=\small] at (1.05,0) {Test field $\hat a$};
 \draw (2.15,-.4)--(2.95,.4);
 \node[below,font=\small] at (2.55,-.5) {50:50 beam splitter};
 \draw[->] (2.55,0)--(5.3,0);
 \draw[->] (2.55,0)--(2.55,1.7)--(5.3,1.7);
 \node[draw,minimum width=1cm,minimum height=.65cm] at (5.9,0) {$D_2$};
 \node[draw,minimum width=1cm,minimum height=.65cm] at (5.9,1.7) {$D_1$};
 \node[draw,minimum height=1.1cm,align=center,font=\small] (counter) at (7.6,.85) {Coincidence\\counter};
 \draw[->] (6.45,0)--(7.6,0)--(counter.south);
 \draw[->] (6.45,1.7)--(7.6,1.7)--(counter.north);
\end{tikzpicture}
\caption{Hanbury Brown--Twiss (HBT) arrangement for measuring second-order correlations. Recording the trigger-time difference $\tau$ between the two detectors gives coincidence counts. A single photon passing through the beam splitter occupies a superposition of two modes, but absorption occurs in only one arm. The beam splitter does not duplicate the photon.}
\label{fig:ch3:hbt}
\end{figure}

The number of detected modes also matters. Collecting $M$ independent thermal modes of equal intensity gives $g^{(2)}(0)=1+1/M$. Finite detector time resolution further averages the bunching peak; the measured value depends on the field coherence time and detector response. Second-order correlations are often measured with the Hanbury Brown--Twiss (HBT) arrangement in Figure~\ref{fig:ch3:hbt}.

Further Reading 3B derives higher-order thermal-light correlations and discusses the effects of loss, background noise, and environments at different temperatures.

\subsection{Traveling Waves, Pulses, and Shot Noise}

For a cavity mode, $\bar n$ is the stored mean photon number and is dimensionless. For a traveling wave in a waveguide or free space, engineers are usually more concerned with power $P$, in $\mathrm W$, and mean photon flux, the number of photons crossing a section per second. For a narrowband optical or microwave beam centered at angular frequency $\omega_c$, the mean photon flux is $\Phi=P/(\hbar\omega_c)$, in $\mathrm s^{-1}$.

Under the narrowband-envelope and broadband-environment approximations, describe the traveling-wave envelope in a one-dimensional transmission line or collimated beam by a continuous-time bosonic operator $\hat b(t)$, with commutation relation
\begin{equation}
 [\hat b(t),\hat b^\dagger(t')]=\delta(t-t'),\qquad
 \avg{\hat b^\dagger(t)\hat b(t)}=\Phi(t).
 \label{eq:ch3:bflux}
\end{equation}
Since the Dirac delta $\delta(t-t')$ has units of $\mathrm s^{-1}$, $\hat b(t)$ has units of $\mathrm{s}^{-1/2}$. A pulse mode with normalized temporal envelope $u(t)$ can be described by the discrete pulse annihilation operator
\begin{equation}
 \int_{-\infty}^\infty|u(t)|^2\dd t=1,\qquad
 \hat a_u=\int_{-\infty}^\infty u^*(t)\hat b(t)\dd t,
 \qquad [\hat a_u,\hat a_u^\dagger]=1.
 \label{eq:ch3:temporal}
\end{equation}
Using the $\delta$-function integral in Eq.~\eqref{eq:ch3:bflux} gives $[\hat a_u,\hat a_u^\dagger]=\int|u(t)|^2\dd t=1$. Once the pulse envelope $u(t)$ is selected, the number states, coherent states, and photon-statistics results for a single mode apply to this traveling pulse.

We now derive the photodetector shot-noise power spectrum. Within a measurement window $T_m$, let an ideal steady coherent beam deliver a mean photon number $\bar N_{\mathrm{in}}=\Phi T_m$. If the detector quantum efficiency is $\eta_d$, each incident photon independently produces a photoelectron with probability $\eta_d$. Independent binomial sampling preserves a Poisson distribution, so the total photoelectron count $N_d$ in the window satisfies
\[
 \bar N_d=\eta_d\Phi T_m,\qquad \Var(N_d)=\bar N_d.
\]
With elementary charge $q_e=1.602\times 10^{-19}\,\mathrm C$, the mean photocurrent is $\bar I=q_e\bar N_d/T_m=q_e\eta_d\Phi$. Each measurement reports the current averaged over $T_m$, namely $I_m=q_eN_d/T_m$, whose variance is
\[
 \Var(I_m)=\frac{q_e^2\Var(N_d)}{T_m^2}
 =\frac{q_e^2\bar N_d}{T_m^2}
 =\frac{q_e\bar I}{T_m}.
\]
Compare this with the one-sided power-spectrum convention in Further Reading 2B: for a flat white-noise spectrum $S_I(f)$, rectangular averaging over $T_m$ gives an equivalent noise bandwidth $\Delta f=1/(2T_m)$ and a reading variance $\Var(I_m)=S_I\Delta f=S_I/(2T_m)$. Equating the two expressions gives the one-sided shot-noise power spectral density:
\begin{equation}
 S_I^{\mathrm{shot}}(f)=2q_e\bar I\qquad(\text{units: }\mathrm{A^2/Hz}).
 \label{eq:ch3:shot}
\end{equation}
Equation~\eqref{eq:ch3:shot} connects Poisson counting of coherent light to the shot-noise spectrum used in engineering. Atomic optical readout systems may also contain the atomic projection noise discussed in Chapter 2. A total noise model must propagate each noise source to the readout separately. Their power spectra can be added directly only when correlations are negligible and the same fluctuation has not been counted twice.

\subsection{Readout with a Local Oscillator and Phase Precision}

The mean reading of direct photodetection measures optical intensity and cannot determine the field phase by itself. Adding a local oscillator with known phase allows a selected quadrature to be read out. This is balanced homodyne detection, closely related to coherent radio-frequency demodulation. A homodyne measurement at one selected local-oscillator phase corresponds to one quadrature.

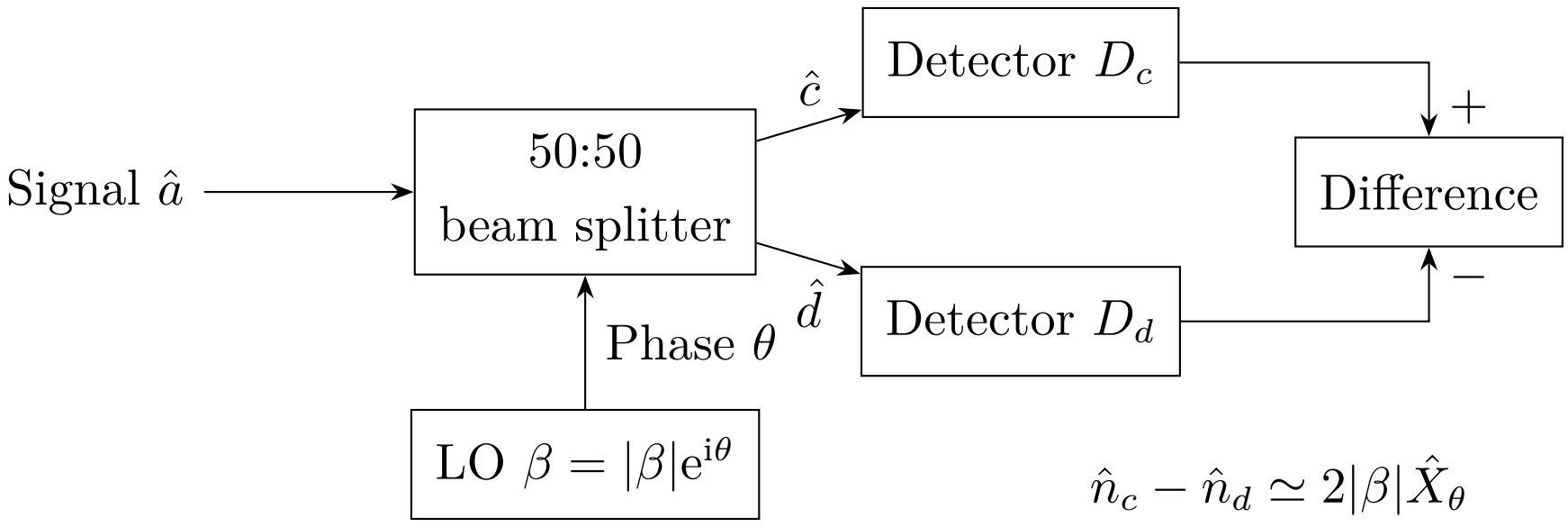
\begin{figure}[htbp]
\centering
\begin{tikzpicture}[font=\small,>=Stealth,
box/.style={draw,align=center,minimum height=0.8cm,inner sep=5pt}]
\node[box,minimum width=1.4cm] (bs) at (0,0) {50:50\\beam splitter};
\node[anchor=east] (in) at (-2.8,0) {Signal $\hat a$};
\node[box] (lo) at (0,-2) {LO $\beta=|\beta|\ee^{\ii\theta}$};
\node[box] (d1) at (3.2,0.95) {Detector $D_c$};
\node[box] (d2) at (3.2,-0.95) {Detector $D_d$};
\node[box] (diff) at (6.2,0) {Difference};
\draw[->] (in)--(bs);
\draw[->] (lo)--node[right]{Phase $\theta$}(bs);
\draw[->] (bs)--node[above]{$\hat c$}(d1);
\draw[->] (bs)--node[below]{$\hat d$}(d2);
\draw[->] (d1.east)-|node[pos=0.8,right]{$+$}(diff.north);
\draw[->] (d2.east)-|node[pos=0.8,right]{$-$}(diff.south);
\node[align=center] at (5.1,-2.15) {$\hat n_c-\hat n_d\simeq2|\beta|\hat X_\theta$};
\end{tikzpicture}
\caption{Balanced homodyne detection. The signal and strong local oscillator interfere at a beam splitter, and the two photon-number readings are subtracted. Changing the local-oscillator phase $\theta$ selects the measured quadrature. Compared with HBT detection, this arrangement adds a phase reference and measures a difference of readings rather than coincidences.}
\label{fig:ch3:homodyne}
\end{figure}

Send the weak signal mode $\hat a$ and a strong local-oscillator (LO) mode $\hat b$ at the same frequency into the two inputs of a balanced 50:50 beam splitter. With a port phase convention using real coefficients, the output modes are $\hat c=(\hat a+\hat b)/\sqrt2$ and $\hat d=(\hat b-\hat a)/\sqrt2$. Two identical photodetectors measure their photon numbers, and the readings are subtracted. The common local-oscillator intensity $\hat b^\dagger\hat b$ and signal intensity $\hat a^\dagger\hat a$ cancel exactly:
\[
 \hat n_c-\hat n_d=\hat c^\dagger\hat c-\hat d^\dagger\hat d
 =\hat a^\dagger\hat b+\hat b^\dagger\hat a.
\]
When the local oscillator is in a strong coherent state $\ket{\beta}$, with $\beta=|\beta|\ee^{\ii\theta}$ and $|\beta|\gg 1$, and its spatial and temporal mode matches the signal precisely, its operator in the leading term can be replaced by the classical complex amplitude $\hat b\simeq|\beta|\ee^{\ii\theta}$. The photoelectron-number difference then reduces to
\begin{equation}
 \begin{aligned}
 \hat n_c-\hat n_d&\simeq2|\beta|\hat X_\theta,\\
 \hat X_\theta&=\frac{\hat a\ee^{-\ii\theta}+\hat a^\dagger\ee^{\ii\theta}}{2}\\
 &=\hat X\cos\theta+\hat Y\sin\theta.
 \end{aligned}
 \label{eq:ch3:homodyne}
\end{equation}
Figure~\ref{fig:ch3:homodyne} shows the measurement arrangement. Choosing $\theta=0$ measures $\hat X$, while $\theta=\pi/2$ measures $\hat Y$. Increasing the local-oscillator amplitude increases the scale of the differential readout, raising quantum field fluctuations above the subsequent electronic noise floor, but does not reduce the signal's input-referred noise. For a coherent or vacuum signal, $\Var(X_\theta)=1/4$; other states require their own quantum-state variance.

As an example, consider the precision of a small-phase measurement on a coherent state using ideal homodyne readout. Let $\alpha=\sqrt{\bar n}\ee^{\ii\phi}$, with unknown phase $|\phi|\ll1$. Choosing $\theta=\pi/2$ gives
\[
 \avg{\hat Y}=\sqrt{\bar n}\sin\phi\simeq\sqrt{\bar n}\phi,
 \qquad \partial_\phi\avg{\hat Y}\simeq\sqrt{\bar n}.
\]
The coherent-state standard deviation is $\Delta Y=1/2$. Following Eq.~\eqref{eq:ramseynoise} in Chapter 2, divide the reading standard deviation by the response slope to obtain the phase uncertainty of one measurement:
\begin{equation}
 \Delta\phi=\frac{\Delta Y}{|\partial_\phi\avg{\hat Y}|}
 \simeq\frac{1}{2\sqrt{\bar n}}.
 \label{eq:ch3:phaseprecision}
\end{equation}
Here $\bar n$ is the mean photon number in the selected signal mode during the measurement, assuming unit detection efficiency, a sufficiently strong local oscillator, and good mode matching. Both $1/\sqrt{\bar n}$ and the independent-atom scaling $1/\sqrt{N_a}$ express the square-root improvement obtained by increasing independent statistical resources; the numerical coefficient depends on the measurement scheme.

A squeezed state can reduce one quadrature variance below vacuum while increasing the other to satisfy the uncertainty relation~\cite{ch3:gaussian}. Reducing $\Var(Y)$ while retaining the phase-response slope improves the precision above. Comparisons between schemes must also account for the energy used to prepare squeezing, loss, and detection efficiency. Squeezed light is already used to reduce readout quantum noise in LIGO gravitational-wave detection~\cite{ch3:ligo}.

\section{Why Lossy Systems Require Environmental Fields}

Chapter 2 incorporated environmental effects on quantum matter into a density-matrix master equation. Here we study quantum-field propagation through a prescribed linear medium. Material absorption and port leakage remove energy from the observed field and couple it to the environment. The attenuation obtained from a classical electromagnetic calculation must therefore be supplemented with environmental fields to determine the output quantum fluctuations.

For a broadband, approximately memoryless environment, the quantum Langevin equation~\eqref{eq:ch3:langevin} retains input noise explicitly; if only the intracavity state is needed, a master equation of the form used in Chapter 2 is sufficient. Their correspondence requires the same system partition, environment state, and approximations. The later treatment of dispersion and absorption with Green functions and noise currents retains the frequency dependence of material response and does not assume a memoryless environment in advance. The Markov approximation needs to be checked only when further reducing this description to constant decay rates. Further Reading 3A connects material oscillators, Green-function identities, and thermal-noise spectra.

\subsection{Checking an Attenuation Channel}

An attenuator with power transmittance $\eta\in(0,1)$ multiplies a classical complex amplitude by $\sqrt\eta$. If this rule is translated directly into the quantum operator relation $\hat a_{\mathrm{out}}=\sqrt\eta\,\hat a_{\mathrm{in}}$, then
\[
 [\hat a_{\mathrm{out}},\hat a_{\mathrm{out}}^\dagger]=\eta[\hat a_{\mathrm{in}},\hat a_{\mathrm{in}}^\dagger]=\eta\neq1.
\]
The output mode must satisfy $[\hat a_{\mathrm{out}},\hat a_{\mathrm{out}}^\dagger]=1$, which the expression above violates. Keeping only the input signal omits the field entering the output through coupling between the attenuator and its environment. Energy loss and noise injection must be treated within the same open-system model.

A consistent model represents the attenuator by a beam splitter of transmittance $\eta$. Introduce an independent bosonic operator $\hat v$ for the dissipative environment, satisfying $[\hat v,\hat v^\dagger]=1$ and commuting with the input signal operators. The complete input--output relation is
\begin{equation}
 \hat a_{\mathrm{out}}=\sqrt\eta\,\hat a_{\mathrm{in}}
 +\sqrt{1-\eta}\,\hat v.
 \label{eq:ch3:losschannel}
\end{equation}
Because input and environment operators commute, the output commutator is restored to $\eta+(1-\eta)=1$. This is consistent with the engineering thermal-noise relation for passive networks: for a matched attenuator in the classical limit $\hbar\omega\ll k_BT$, the environment's available output noise power over bandwidth $B$ is $(1-\eta)k_BTB$.

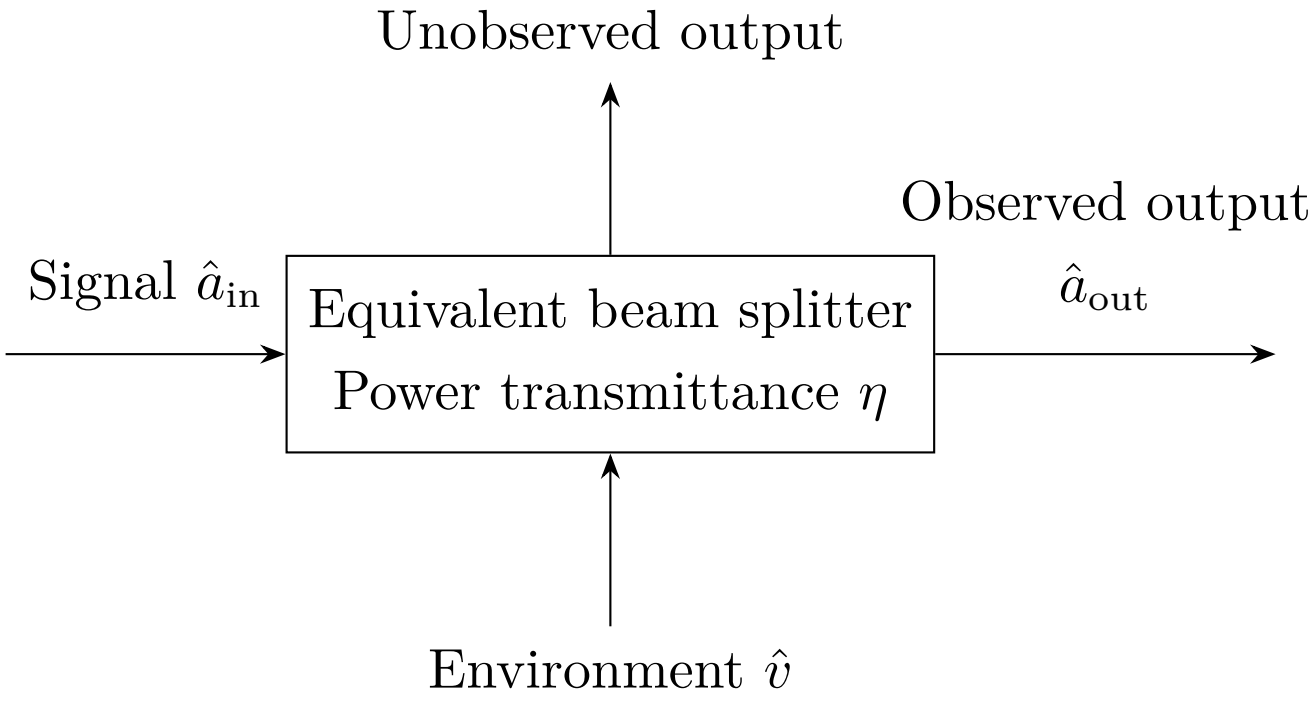
\begin{figure}[htbp]
\centering
\begin{tikzpicture}[font=\small,>=Stealth]
\node[draw,minimum width=2.1cm,minimum height=1.3cm,align=center] (bs) at (0,0) {Equivalent beam splitter\\Power transmittance $\eta$};
\draw[->] (-4,0)--node[midway,above=4pt]{Signal $\hat a_{\mathrm{in}}$}(bs.west);
\draw[->] (0,-1.8) node[below]{Environment $\hat v$}--(bs.south);
\draw[->] (bs.east)--node[midway,above=4pt,align=center]{Observed output\\$\hat a_{\mathrm{out}}$}(4.4,0);
\draw[->] (bs.north)--(0,1.8) node[above]{Unobserved output};
\node[align=center] at (0,-3.35) {$\hat a_{\mathrm{out}}=\sqrt\eta\,\hat a_{\mathrm{in}}+\sqrt{1-\eta}\,\hat v$\\[5pt]Zero-temperature environment: no thermal photons,\\but vacuum fluctuations remain};
\end{tikzpicture}
\caption{Equivalent beam-splitter model of an attenuation channel. Part of the signal enters an unobserved port, while the environmental field couples into the observed port. Observing the signal channel alone reveals loss; including the environment port restores the required output commutation relation. This is an equivalent model and does not require a physical beam-splitting interface inside the device.}
\label{fig:ch3:loss}
\end{figure}

Even when the environment is zero-temperature vacuum, $\hat v$ cannot be omitted. It adds no thermal photons to the output but still contributes to the quadrature variance. Figure~\ref{fig:ch3:loss} shows the signal and environment input ports.

Suppose that the initial signal and environment states are uncorrelated, and the environment is thermal with mean photon number $\bar n_v$, so that $\avg{\hat v}=0$ and $\Var(X_v)=(2\bar n_v+1)/4$. The output mean photon number and quadrature variance are then
\begin{equation}
 \begin{aligned}
 \avg{\hat n_{\mathrm{out}}}
 &=\eta\avg{\hat n_{\mathrm{in}}}+(1-\eta)\bar n_v,\\
 \Var(X_{\mathrm{out}})
 &=\eta\Var(X_{\mathrm{in}})+(1-\eta)\frac{2\bar n_v+1}{4}.
 \end{aligned}
 \label{eq:ch3:lossnoise}
\end{equation}
These expressions show how loss affects different quantum states:
\begin{itemize}
 \item For a zero-temperature vacuum environment ($\bar n_v=0$) and a coherent input ($\Var(X_{\mathrm{in}})=1/4$), the output variance after attenuation is $\frac14\eta+\frac14(1-\eta)=\frac14$. Thus a coherent state remains a pure coherent state $\ket{\sqrt\eta\alpha}$ under pure loss.
 \item With the environment still in vacuum, suppose the input quadrature variance is $0.1$ times the vacuum value. After $20\%$ power loss, the output variance is $0.8\times0.1+0.2=0.28$ times the vacuum value. Loss therefore weakens squeezing.
\end{itemize}

For a passive multiport network normalized to photon flux at one frequency, denote the input, output, and environment mode vectors by $\hat{\mathbf a}$, $\hat{\mathbf b}$, and $\hat{\mathbf v}$. Since all these ports have the same frequency, photon-flux normalization and classical power-wave normalization give the same scattering matrix $S$. With internal loss, the input--output relation must be written as
\begin{equation}
 \hat{\mathbf b}=S\hat{\mathbf a}+L\hat{\mathbf v},\qquad
 SS^\dagger+LL^\dagger=I.
 \label{eq:ch3:multiport}
\end{equation}
The commutation relations require $LL^\dagger=I-SS^\dagger$. This fixes the total weight of environmental coupling, but does not uniquely determine $L$. Nor does $S$ alone determine the actual noise spectrum: the temperatures and statistical states of the environment channels are also needed.

\subsection{Cavity Decay and Langevin Driving}

Damping of an intracavity field likewise requires environmental driving. If the mode energy decay rate is $\kappa$, the amplitude decay rate is $\kappa/2$, and the quality factor is $Q=\omega_c/\kappa$. If $\kappa$ includes internal loss and all port couplings, this $Q$ is the loaded quality factor.

Writing only $\dot{\hat a}=-(\ii\omega_c+\kappa/2)\hat a$ gives $\hat a(t)=\hat a(0)\ee^{-(\ii\omega_c+\kappa/2)t}$, whose equal-time commutator becomes $\ee^{-\kappa t}$ rather than remaining unity.

Under the Markov approximation of a memoryless, broadband stationary environment, the correct Heisenberg equation of motion must include driving by the port or environment traveling field $\hat b_{\mathrm{in}}(t)$:
\begin{equation}
 \begin{aligned}
 \dot{\hat a}&=-\left(\ii\omega_c+\frac\kappa2\right)\hat a
 +\sqrt\kappa\,\hat b_{\mathrm{in}}(t),\\
 [\hat b_{\mathrm{in}}(t),\hat b_{\mathrm{in}}^\dagger(t')]&=\delta(t-t').
 \end{aligned}
 \label{eq:ch3:langevin}
\end{equation}
This is the quantum Langevin equation. Here $\kappa$ has units of $\mathrm{s}^{-1}$ and $\hat b_{\mathrm{in}}(t)$ has units of $\mathrm{s}^{-1/2}$, so the drive $\sqrt\kappa\,\hat b_{\mathrm{in}}$ has the same dimensions as the time derivative $\dot{\hat a}$.

Writing the complex frequency as $s=\ii\omega_c+\kappa/2$, the general solution of the ordinary differential equation~\eqref{eq:ch3:langevin} is
\[
 \hat a(t)=\ee^{-st}\hat a(0)
 +\sqrt\kappa\int_0^t\ee^{-s(t-t')}\hat b_{\mathrm{in}}(t')\dd t'.
\]
Assume that the initial cavity operators at $t=0$ and the external field operators incident at $t>0$ are independent and commute. The $\delta(t'-t'')$ commutation relation then gives the equal-time commutator at any time:
\begin{equation}
 \begin{aligned}
 [\hat a(t),\hat a^\dagger(t)]
 &=\ee^{-\kappa t}[\hat a(0),\hat a^\dagger(0)]
 +\kappa\int_0^t\ee^{-\kappa(t-t')}\dd t'\\
 &=\ee^{-\kappa t}+(1-\ee^{-\kappa t})=1.
 \end{aligned}
 \label{eq:ch3:commcheck}
\end{equation}
The initial cavity mode's contribution to the commutator decreases with time, and the environmental input supplies exactly the missing amount. The damping coefficient $\kappa/2$ and input coupling coefficient $\sqrt\kappa$ therefore cannot be specified independently.

If only the intracavity state is needed, take the partial trace over the environment: sum the diagonal matrix elements in an environment basis while retaining the cavity state (see the definition in Section~\ref{ch4:sec_partialtrace}). Under the same Markov approximation, a zero-temperature environment gives a Lindblad master equation of the form used in Chapter 2, with jump operator $L_\kappa=\sqrt\kappa\,\hat a$, distinct from the noise coupling matrix $L$ in the multiport relation. A thermal environment gives two channels:
\[
 L_\downarrow=\sqrt{\kappa(\bar n_T+1)}\,\hat a,
 \qquad L_\uparrow=\sqrt{\kappa\bar n_T}\,\hat a^\dagger.
\]
The first decreases the cavity photon number; the second describes excitation of the cavity by the thermal environment. The Langevin equation retains the input field explicitly, whereas the master equation describes only the reduced cavity state. Both describe the same damping process.

\subsection{Port Input--Output Relations and Scattering Parameters}

Experiments usually drive a cavity through transmission-line, wave\-guide, or optical-fiber ports and measure the outgoing port waves. For a model with one coupling channel, using a port phase convention consistent with Eq.~\eqref{eq:ch3:langevin}, the input--output boundary relation is~\cite{ch3:inputoutput}
\begin{equation}
 \hat b_{\mathrm{out}}(t)=\hat b_{\mathrm{in}}(t)-\sqrt\kappa\,\hat a(t).
 \label{eq:ch3:inputoutput}
\end{equation}
The outgoing wave is the coherent sum of the directly reflected incident wave and the field leaking from the cavity. The minus sign depends on the port phase convention and must be consistent with the driving term in the Langevin equation. Solving the two equations together gives the scattering parameters measured in engineering practice.

Consider a single-port reflection cavity with external coupling rate $\kappa_{\mathrm{ex}}$ and intrinsic internal loss rate $\kappa_{\mathrm{i}}$, giving total decay rate $\kappa=\kappa_{\mathrm{ex}}+\kappa_{\mathrm{i}}$. The cavity Langevin equation and outgoing-port relation are
\begin{equation}
 \begin{aligned}
 \dot{\hat a}&=-\left(\ii\omega_c+\frac{\kappa_{\mathrm{ex}}+\kappa_{\mathrm{i}}}{2}\right)\hat a\\
 &\quad+\sqrt{\kappa_{\mathrm{ex}}}\,\hat b_{\mathrm{in}}(t)
 +\sqrt{\kappa_{\mathrm{i}}}\,\hat v_{\mathrm{in}}(t),\\
 \hat b_{\mathrm{out}}(t)&=\hat b_{\mathrm{in}}(t)-\sqrt{\kappa_{\mathrm{ex}}}\,\hat a(t).
 \end{aligned}
 \label{eq:ch3:s11time}
\end{equation}
Here $\hat v_{\mathrm{in}}$ is the input field of the internal-loss environment. Fourier transforming and replacing $\dd/\dd t$ by $-\ii\omega$ first gives
\[
 \hat a(\omega)=
 \frac{\sqrt{\kappa_{\mathrm{ex}}}\,\hat b_{\mathrm{in}}(\omega)
       +\sqrt{\kappa_{\mathrm{i}}}\,\hat v_{\mathrm{in}}(\omega)}
 {(\kappa_{\mathrm{ex}}+\kappa_{\mathrm{i}})/2-\ii(\omega-\omega_c)}.
\]
Substituting into the outgoing-port relation then yields
\begin{equation}
 \hat b_{\mathrm{out}}(\omega)=S_{11}(\omega)\hat b_{\mathrm{in}}(\omega)+L_{11}(\omega)\hat v_{\mathrm{in}}(\omega),
 \label{eq:ch3:s11freq}
\end{equation}
where the reflection coefficient $S_{11}(\omega)$ and internal-loss noise transmission coefficient $L_{11}(\omega)$ are
\begin{equation}
 \begin{aligned}
 S_{11}(\omega)&=\frac{(\kappa_{\mathrm{i}}-\kappa_{\mathrm{ex}})/2-\ii(\omega-\omega_c)}
 {(\kappa_{\mathrm{i}}+\kappa_{\mathrm{ex}})/2-\ii(\omega-\omega_c)},\\
 L_{11}(\omega)&=-\frac{\sqrt{\kappa_{\mathrm{ex}}\kappa_{\mathrm{i}}}}
 {(\kappa_{\mathrm{i}}+\kappa_{\mathrm{ex}})/2-\ii(\omega-\omega_c)}.
 \end{aligned}
 \label{eq:ch3:s11formula}
\end{equation}
The coefficient $S_{11}$ is identical to the classical single-port cavity reflection coefficient, while $L_{11}$ describes transfer from the internal-loss environment to the measurement port. Adding their squared magnitudes verifies $|S_{11}|^2+|L_{11}|^2=1$, consistent with Eq.~\eqref{eq:ch3:multiport}.

At resonance and critical coupling, $\kappa_{\mathrm{ex}}=\kappa_{\mathrm{i}}$, we have $S_{11}=0$ and $L_{11}=-1$. Exactly at this frequency, direct reflection of the incident wave cancels the cavity leakage, and the output is determined by the internal-loss environment. This frequency-domain result does not extend over the entire bandwidth. If the internal-loss environment is vacuum, the output still has quadrature fluctuations, but these do not produce additional thermal photon counts.

\section{Quantum Fields and Green Functions in Linear Media}
\subsection{Specifying the Scope of the Material Response}

The preceding mode expansion relies on a lossless, nondispersive medium and closed boundaries. Real structures often combine material absorption, dispersion, and radiative leakage, with resonances described by complex eigenfrequencies. Directly inserting $\omega_c-\ii\kappa/2$ into the Hamiltonian of a closed oscillator produces non-Hermitian evolution and does not give a model containing all quantum noise. As in port attenuation, environmental degrees of freedom must be included.

Macroscopic quantum electrodynamics offers a method suited to engineering electromagnetic calculations~\cite{ch3:scheel}: represent linear material response by a complex dielectric function, assign quantum noise currents to dissipative channels, and calculate the field with the classical dyadic Green function. This retains the familiar Maxwell solution procedure while accounting for quantum fluctuations.

To avoid cumbersome anisotropic tensor notation, this section considers local, isotropic, reciprocal, passive, nonmagnetic linear media:
\[
 \varepsilon_r(\mathbf r,\omega)=\varepsilon_r'(\mathbf r,\omega)
 +\ii\varepsilon_r''(\mathbf r,\omega),\qquad \varepsilon_r''(\mathbf r,\omega)\geq0\quad(\omega>0).
\]
The time convention remains $\ee^{-\ii\omega t}$, so a passive dissipative medium has imaginary part $\varepsilon_r''\geq 0$. The real part $\varepsilon_r'$ and imaginary part $\varepsilon_r''$ obey the Kramers--Kronig relations required by causality.

\subsection{From Noise Currents to Field Operators}

In classical frequency-domain electromagnetics, a prescribed external current density $\mathbf J(\mathbf r,\omega)$ produces an electric field satisfying the inhomogeneous vector wave equation obtained by eliminating the magnetic field:
\begin{equation}
 \left[\nabla\times\nabla\times
 -\frac{\omega^2}{c^2}\varepsilon_r(\mathbf r,\omega)\right]\mathbf E(\mathbf r,\omega)
 =\ii\mu_0\omega\mathbf J(\mathbf r,\omega).
 \label{eq:ch3:wavegreen}
\end{equation}
Define the inverse kernel of this linear wave operator as the classical dyadic Green function $\mathbf G(\mathbf r,\mathbf s,\omega)$, a $3\times 3$ matrix function satisfying the point-source equation
\begin{equation}
 \left[\nabla\times\nabla\times-\frac{\omega^2}{c^2}\varepsilon_r(\mathbf r,\omega)\right]
 \mathbf G(\mathbf r,\mathbf s,\omega)
 =\mathbf I\delta(\mathbf r-\mathbf s).
 \label{eq:ch3:greendef}
\end{equation}
The Green function satisfies outgoing radiation boundary conditions. Here $\mathbf I$ is the unit dyadic, and differentiation acts on the observation coordinate $\mathbf r$. In $\mathbf G$, column $k$ is the response to a unit point source at $\mathbf s$ oriented along coordinate direction $k$; its units are $\mathrm m^{-1}$. The field due to an arbitrary current distribution is
\[
 \mathbf E(\mathbf r,\omega)=\ii\mu_0\omega
 \int\mathbf G(\mathbf r,\mathbf s,\omega)\cdot\mathbf J(\mathbf s,\omega)\dd^3s.
\]

We now turn to the operator description. First specify the continuous-frequency expansion convention for the positive-frequency electric-field operator $\hat{\mathbf E}^{(+)}(\mathbf r,t)$:
\begin{equation}
 \begin{aligned}
 \hat{\mathbf E}^{(+)}(\mathbf r,t)
 &=\int_0^\infty\hat{\boldsymbol{\mathcal E}}(\mathbf r,\omega)
       \ee^{-\ii\omega t}\dd\omega,\\
 \hat{\mathbf E}(\mathbf r,t)
 &=\hat{\mathbf E}^{(+)}(\mathbf r,t)+\hat{\mathbf E}^{(-)}(\mathbf r,t).
 \end{aligned}
 \label{eq:ch3:fourier}
\end{equation}
Here $\hat{\boldsymbol{\mathcal E}}$ is the positive-frequency field spectral amplitude per unit angular frequency, with units $\mathrm{V\cdot m^{-1}\cdot s}$. There is no factor $1/(2\pi)$ before the integral; all coefficients below follow this convention.

To describe the quantum degrees of freedom of the material's dissipative channels at every position and frequency, introduce a continuous vector bosonic operator field $\hat{\mathbf f}(\mathbf s,\omega)$ with fundamental commutation relations
\begin{equation}
 \begin{aligned}
 &[\hat f_i(\mathbf s,\omega),\hat f_j^\dagger(\mathbf s',\omega')]\\
 &\qquad=\delta_{ij}\delta(\mathbf s-\mathbf s')\delta(\omega-\omega'),\\
 &[\hat f_i(\mathbf s,\omega),\hat f_j(\mathbf s',\omega')]=0.
 \end{aligned}
 \label{eq:ch3:bathcomm}
\end{equation}
Here $i,j\in\{x,y,z\}$ label Cartesian components. Equation~\eqref{eq:ch3:bathcomm} is the continuous space--frequency version of the discrete-mode relation $[\hat a_m,\hat a_n^\dagger]=\delta_{mn}$; hence $\hat f_i$ has units $\mathrm{m^{-3/2}\cdot s^{1/2}}$.

With these spectral and operator normalization conventions, macroscopic quantum electrodynamics gives the noise current~\cite{ch3:scheel}
\begin{equation}
 \hat{\boldsymbol{\mathcal J}}_N(\mathbf s,\omega)
 =\omega\sqrt{\frac{\hbar\varepsilon_0}{\pi}\varepsilon_r''(\mathbf s,\omega)}
 \,\hat{\mathbf f}(\mathbf s,\omega).
 \label{eq:ch3:noisecurrent}
\end{equation}
Its structure resembles $\sqrt\kappa\,\hat b_{\mathrm{in}}$: the noise-source amplitude is proportional to the square root of the dissipation strength. The normalization ensures that the resulting fields satisfy canonical commutation relations. Further Reading 3A explains how these environment operators relate to material oscillators.

Substituting the noise current into the classical Green-function solution and using $\mu_0\varepsilon_0=c^{-2}$ gives
\begin{equation}
 \hat{\boldsymbol{\mathcal E}}(\mathbf r,\omega)
 =\ii\sqrt{\frac{\hbar}{\pi\varepsilon_0}}\frac{\omega^2}{c^2}
 \int\dd^3s\,\sqrt{\varepsilon_r''(\mathbf s,\omega)}\,
 \mathbf G(\mathbf r,\mathbf s,\omega)\cdot\hat{\mathbf f}(\mathbf s,\omega).
 \label{eq:ch3:macfield}
\end{equation}
In Eq.~\eqref{eq:ch3:macfield}, geometry, scattering, and dispersion are described by the classical Green function, while quantum statistics come from the environment operators and their state. Commutation relations alone do not determine the noise level; the environment must also be specified as vacuum, thermal, or another state.

For an open-boundary problem with a finite lossy object in lossless free space, the radiation boundary at infinity is also a dissipation, or escape, channel. Mathematically, one can initially assign infinitesimal passive absorption $\varepsilon_r''\to 0^+$ to the entire infinite background, complete the integral over all space, and only then take the limit $\varepsilon_r''\to 0^+$. This automatically includes free-space vacuum fluctuations incident from infinity (see Further Reading 3A.2).

\subsection{Why Field Correlations Contain the Imaginary Part of the Green Function}

First suppose that the environment throughout space is in vacuum. Since $\hat f_i(\mathbf s,\omega)\ket0=0$, normally ordered field correlations vanish, whereas antinormally ordered correlations retain a vacuum contribution. Equation~\eqref{eq:ch3:bathcomm} gives
\[
 \avg{\hat f_k(\mathbf s,\omega)\hat f_{k'}^\dagger(\mathbf s',\omega')}
 =\delta_{kk'}\delta(\mathbf s-\mathbf s')\delta(\omega-\omega').
\]
Multiply component $i$ of Eq.~\eqref{eq:ch3:macfield} by the conjugate transpose of component $j$ and take the vacuum expectation value. The delta function $\delta(\mathbf s-\mathbf s')$ eliminates one spatial volume integral, giving
\[
 \begin{split}
 \avg{\hat{\mathcal E}_i(\mathbf r,\omega)
       \hat{\mathcal E}_j^\dagger(\mathbf r',\omega')}
 ={}&\frac{\hbar}{\pi\varepsilon_0}\frac{\omega^4}{c^4}\delta(\omega-\omega')\\
 &\times\int\dd^3s\,\varepsilon_r''(\mathbf s,\omega)
       \sum_kG_{ik}(\mathbf r,\mathbf s,\omega)G_{jk}^*(\mathbf r',\mathbf s,\omega).
 \end{split}
\]
The spatial integral simplifies using the following Green-function identity, derived in Further Reading 3A.2:
\begin{equation}
 \begin{aligned}
 &\frac{\omega^2}{c^2}\int\dd^3s\,\varepsilon_r''(\mathbf s,\omega)
 \sum_kG_{ik}(\mathbf r,\mathbf s,\omega)G_{jk}^*(\mathbf r',\mathbf s,\omega)\\
 &\qquad=\Im G_{ij}(\mathbf r,\mathbf r',\omega).
 \end{aligned}
 \label{eq:ch3:greenidentity}
\end{equation}
This identity is related to electromagnetic energy balance and reciprocity. When the two observation points coincide and the result is projected along a dipole direction, the imaginary part of the Green function corresponds to the total real power supplied by the point source. The next section uses this relation in numerical calculations. Open-space radiation must be included through the limiting procedure described above.

Substituting Eq.~\eqref{eq:ch3:greenidentity} and using $1/(\varepsilon_0c^2)=\mu_0$ gives
\begin{equation}
 \begin{aligned}
 &\avg{\hat{\mathcal E}_i(\mathbf r,\omega)
       \hat{\mathcal E}_j^\dagger(\mathbf r',\omega')}\\
 &\qquad=\frac{\hbar\mu_0}{\pi}\omega^2
 \Im G_{ij}(\mathbf r,\mathbf r',\omega)\delta(\omega-\omega').
 \end{aligned}
 \label{eq:ch3:vacgreen}
\end{equation}
For the reciprocal media considered here, $G_{ij}(\mathbf r,\mathbf r',\omega)=G_{ji}(\mathbf r',\mathbf r,\omega)$, so $\Im G_{ij}$ is simply the ordinary imaginary part of the matrix element.

Under the material and environment assumptions of this section, Eq.~\eqref{eq:ch3:vacgreen} reduces the spatial correlations of the vacuum field to the imaginary part of a classical Green function. Point-source responses from existing electromagnetic solvers can therefore also be used to calculate quantum-field correlations.

If the entire environment is in thermal equilibrium at temperature $T$, its correlations contain the thermal occupation number:
\[
 \begin{aligned}
 &\avg{\hat f_k^\dagger(\mathbf s,\omega)\hat f_{k'}(\mathbf s',\omega')}\\
 &\qquad=\bar n_T(\omega)\delta_{kk'}
 \delta(\mathbf s-\mathbf s')\delta(\omega-\omega').
 \end{aligned}
\]
The same steps give the field correlations for the two operator orderings:
\begin{equation}
 \begin{aligned}
 &\avg{\hat{\mathcal E}_i(\mathbf r,\omega)\hat{\mathcal E}_j^\dagger(\mathbf r',\omega')}\\
 &\quad=\frac{\hbar\mu_0}{\pi}\omega^2[\bar n_T(\omega)+1]
 \Im G_{ij}(\mathbf r,\mathbf r',\omega)\delta(\omega-\omega'),\\[3pt]
 &\avg{\hat{\mathcal E}_j^\dagger(\mathbf r',\omega')\hat{\mathcal E}_i(\mathbf r,\omega)}\\
 &\quad=\frac{\hbar\mu_0}{\pi}\omega^2\bar n_T(\omega)
 \Im G_{ij}(\mathbf r,\mathbf r',\omega)\delta(\omega-\omega').
 \end{aligned}
 \label{eq:ch3:thermalgreen}
\end{equation}
This is an electromagnetic fluctuation--dissipation relation. The first expression contains $\bar n_T+1$, the field factor for stimulated and spontaneous emission in atomic transitions; the second contains $\bar n_T$, the field factor for absorption from a thermal field. Their difference is independent of temperature and gives the commutation relation between the positive-frequency field spectrum and its conjugate. These are frequency-domain positive- and negative-frequency components, not directly the equal-time commutation relation of the total electric field.

\section{Calculations with Existing Electromagnetic Solvers}
\subsection{Normalizing Lossless Modes}

For a high-$Q$, weakly dispersive cavity that can be approximated as a closed lossless system, process the eigenmode solver results as follows:
\begin{enumerate}
 \item Find $\omega_c$ and the unnormalized field $\mathbf E_{\mathrm{mode}}$. For the closed cavity with real coefficients considered here, choose a real standing-wave basis. A nondegenerate mode can usually be made real by adjusting its overall phase; degenerate modes may also require linear combinations.
 \item Calculate
 \[
 N_u=\int_V\varepsilon_r|\mathbf E_{\mathrm{mode}}|^2\dd^3r,
 \qquad \mathbf u=\frac{\mathbf E_{\mathrm{mode}}}{\sqrt{N_u}}.
 \]
 In the chosen real basis, the squared magnitude equals the vector dotted with itself.
 \item Substitute $\mathbf u$ into Eq.~\eqref{eq:ch3:fieldoperator} or Eq.~\eqref{eq:ch3:veff} to find the vacuum field scale at the target position, then calculate means and fluctuations for the chosen quantum state.
 \item Obtain $\kappa=\omega_c/Q_L$ and $\kappa_{\mathrm{ex}}=\omega_c/Q_{\mathrm{ex}}$ from the loaded and external quality factors and insert them into the Langevin equation or Lindblad master equation. Retain separate coupling rates for multiple ports.
\end{enumerate}
With significant dispersion, stored energy also includes material polarization. In transparent or weakly lossy bands of a local, nonmagnetic medium, the commonly used Brillouin cycle-averaged energy formula is
\[
 \overline W=\frac14\int_V\left[
 \varepsilon_0\frac{\partial(\omega\varepsilon_r')}{\partial\omega}|\mathbf E_0|^2
 +\mu_0|\mathbf H_0|^2\right]\dd^3r,
\]
where $\mathbf E_0,\mathbf H_0$ are complex peak amplitudes. Mode normalization must be re-established from the total stored energy; replacing a single coefficient in the nondispersive formula is not universally valid. Green-function methods are preferable for strongly absorbing, low-$Q$, or open broadband structures.

\subsection{Obtaining Green Functions from Driven Sources}
\label{sec:ch3:dipolepower}

In lossy or open structures, the Green function can be extracted from a driven point-dipole response. One excitation at a specified frequency and direction gives the corresponding column or projection. The full dyadic requires three independent excitation directions, and broadband results require a frequency sweep.

At source point $\mathbf r_0$, place a test source with complex dipole moment $\mathbf p_0$. Its current density is $\mathbf J=-\ii\omega\mathbf p_0\delta(\mathbf r-\mathbf r_0)$. Substitution into the classical field integral gives
\begin{equation}
 \mathbf E_{\mathrm{dip}}(\mathbf r,\omega)
 =\mu_0\omega^2\mathbf G(\mathbf r,\mathbf r_0,\omega)\cdot\mathbf p_0.
 \label{eq:ch3:dipolegreen}
\end{equation}
Suppose the source lies in a local vacuum region at a finite distance from material interfaces. Along a real unit direction $\mathbf e$, take $\mathbf p_0=p_0\mathbf e$ with real $p_0$. The projected imaginary part at the source can be obtained from the self-field or the source output power:
\begin{equation}
 \begin{aligned}
 &\mathbf e\cdot\Im\mathbf G(\mathbf r_0,\mathbf r_0,\omega)\cdot\mathbf e\\
 &\quad=\frac{\Im[\mathbf e\cdot\mathbf E_{\mathrm{dip}}(\mathbf r_0,\omega)]}
 {\mu_0\omega^2p_0}
 =\frac{2P_{\mathrm{tot}}}{\mu_0\omega^3p_0^2}.
 \end{aligned}
 \label{eq:ch3:imGpower}
\end{equation}
Here we used the cycle-averaged power supplied outward by a point dipole,
\[
 P_{\mathrm{tot}}=\frac\omega2
 \Im[\mathbf p_0^*\cdot\mathbf E_{\mathrm{dip}}(\mathbf r_0,\omega)],
\]
which equals the sum of far-field radiated power and material absorption. Numerically, one can integrate the net outward power over a small closed surface surrounding the source in the lossless region, or calculate and add the far-field flux and material absorption separately. These are alternative procedures: adding absorption outside the small surface to its flux would double-count that power.

The real part of a point source's self-field in local vacuum is singular, whereas the projected imaginary part above is finite. A discretized calculation still requires source-model and convergence checks; when necessary, calculate the analytical free-space contribution separately from the scattered contribution. If an ideal point source is placed directly inside a local lossy continuum, the imaginary part of the self-field may also diverge, so the finite-value conclusion does not apply directly.

In free space (vacuum), the coincident-point imaginary part of the dyadic Green function has the exact analytical value
\[
 \Im\mathbf G_0(\mathbf r_0,\mathbf r_0,\omega)=\frac{k_0}{6\pi}\mathbf I,
 \qquad k_0=\frac{\omega}{c}.
\]
The local density of states (LDOS) describes the distribution of locally available modes. At a vacuum observation point $\mathbf r$, the electric LDOS projected along real unit direction $\mathbf e$ is defined as
\begin{equation}
 \mathcal D_{\mathbf e}(\mathbf r,\omega)
 =\frac{2\omega}{\pi c^2}\,
 \mathbf e\cdot\Im\mathbf G(\mathbf r,\mathbf r,\omega)\cdot\mathbf e.
 \label{eq:ch3:ldos}
\end{equation}
Its units are $\mathrm{m^{-3}\cdot s}$. This is a mode density weighted by the local electric field and direction, not an actual photon occupation number. In free space, the value for one direction is $\omega^2/(3\pi^2c^3)$, and the sum over three directions is $\omega^2/(\pi^2c^3)$. Surrounding structures change the Green function and therefore alter local field fluctuations and the radiation and absorption environment seen by a dipole.

To ensure reliable numerical results, perform four basic checks:
\begin{enumerate}
 \item Free-space benchmark: removing the structure should restore the analytical value of $\mathbf e\cdot\Im\mathbf G_0\cdot\mathbf e$, namely $\omega/(6\pi c)$, corresponding to the classical electric-dipole free-space radiated power $P_0=\frac{\mu_0\omega^4 p_0^2}{12\pi c}$.
 \item Reciprocity: in a reciprocal medium, exchanging source and observation points should give $\mathbf G(\mathbf r,\mathbf r_0,\omega)=\mathbf G^{\mathsf T}(\mathbf r_0,\mathbf r,\omega)$.
 \item Passivity: for a passive system without gain, the coincident-point projected imaginary part of the total Green function must be nonnegative, $\mathbf e\cdot\Im\mathbf G(\mathbf r_0,\mathbf r_0,\omega)\cdot\mathbf e\geq 0$, since total radiated and dissipated power is nonnegative. The scattered part $\Im\mathbf G_{\mathrm{sc}}$ obtained after subtracting free space may be negative, corresponding to radiation suppression by reflecting walls.
 \item Open boundaries and convergence: a PML represents radiation escaping toward the far field and cannot be treated directly as a physical material with a temperature. Vary the mesh, source discretization, and PML parameters to check convergence of the target projection and power.
\end{enumerate}

Eigenmode methods are suited to extracting field scales in approximately lossless cavities; driven Green-function methods are suited to field correlations in lossy open structures. Both retain classical electromagnetic algorithms for spatial calculations, while the quantum description supplies mode states and their statistics. The next section connects classical electromagnetic response to quantum coincidence counting through far-field scattering of two single photons.

\section{Scattering of Two Single Photons and Far-Field Correlations}
\label{sec:ch3:twophotons}

We have discussed photon counting and linear-system input--output relations separately. We now combine them in a scattering problem: two independently prepared single photons illuminate the same object, and coincidences are recorded in two far-field directions. The object may have arbitrary shape, but its response is assumed linear, time invariant, and passive; if it is lossy, environmental thermal photons are assumed negligible. The scattering amplitudes still come from classical Maxwell equations, while the quantum state determines how they are combined~\cite{ch3:huang}.

\subsection{Input State and Receiving Modes}

Choose two mutually orthogonal incident modes, represented by $\hat a_1,\hat a_2$. With one photon in each mode, the input state is
\begin{equation}
 \ket{\psi_{\mathrm{in}}}=\ket{1,1}
 =\hat a_1^\dagger\hat a_2^\dagger\ket{0,0}.
 \label{eq:ch3:twoinput}
\end{equation}
Here independence means that the input is a product of the two mode states. The beams may arrive from different directions, but their center frequencies, wavepackets, and polarization after detection selection should match so that the detection record cannot identify the photon's input of origin. Otherwise, interference between the two arrival processes is reduced.

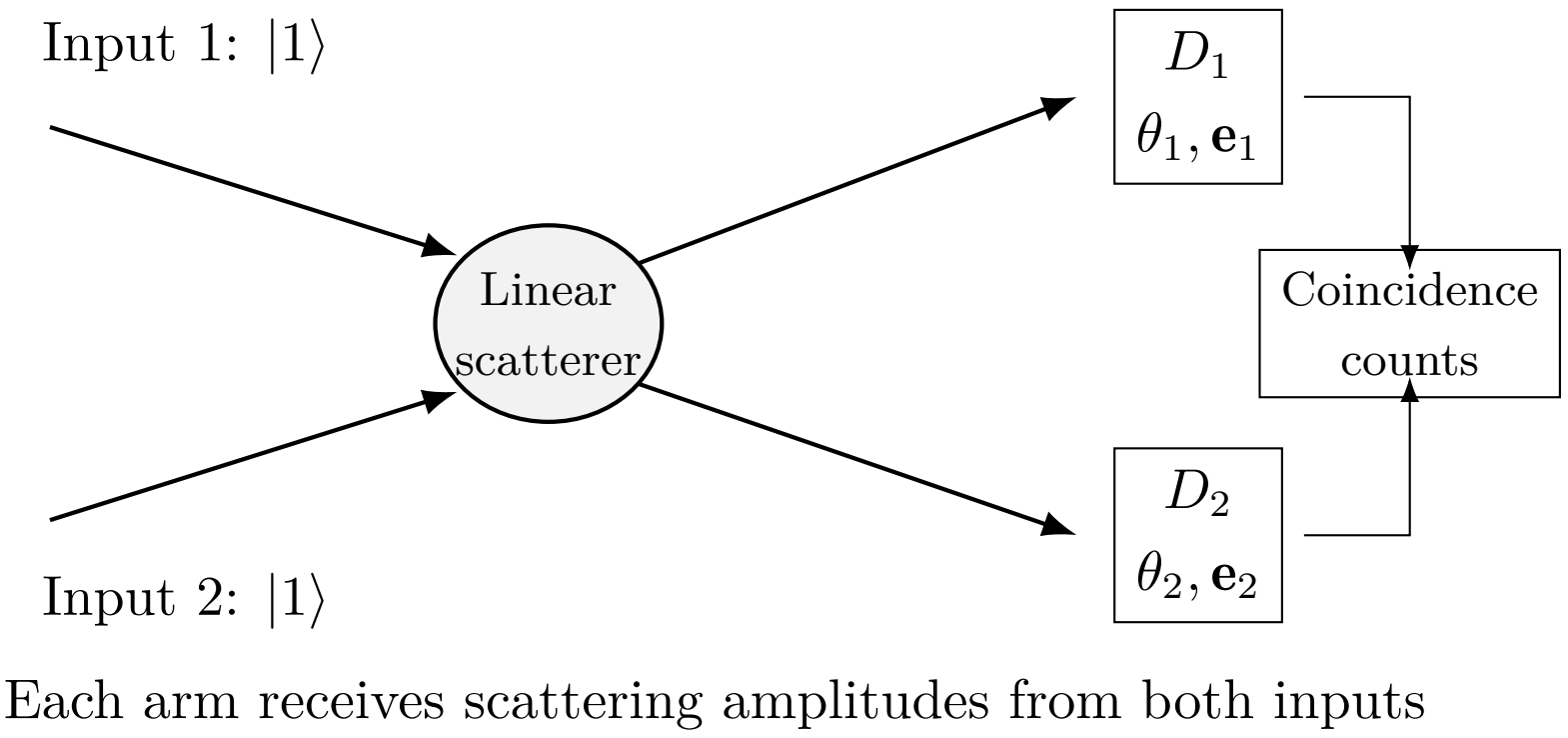
\begin{figure}[htbp]
\centering
\begin{tikzpicture}[>=Latex,font=\small]
 \draw[thick,fill=black!5] (3.5,0) ellipse (.75 and .65);
 \node[align=center,font=\footnotesize] at (3.5,0) {Linear\\scatterer};
 \draw[->,thick] (.2,1.3)--(2.9,.45);
 \draw[->,thick] (.2,-1.3)--(2.9,-.45);
 \node[above,align=center] at (1.1,1.5) {Input 1: $\ket1$};
 \node[below,align=center] at (1.1,-1.5) {Input 2: $\ket1$};

 \draw[->,thick] (4.1,.4)--(7,1.5);
 \draw[->,thick] (4.1,-.4)--(7,-1.4);
 \node[draw,align=center] at (7.8,1.5) {$D_1$\\$\theta_1,\mathbf e_1$};
 \node[draw,align=center] at (7.8,-1.4) {$D_2$\\$\theta_2,\mathbf e_2$};
 \draw[->] (8.5,1.5)--(9.2,1.5)--(9.2,.35);
 \draw[->] (8.5,-1.4)--(9.2,-1.4)--(9.2,-.35);
 \node[draw,align=center,font=\footnotesize] at (9.2,0) {Coincidence\\counts};
 \node[below,align=center] at (4.6,-2.2)
 {Each arm receives scattering amplitudes from both inputs};
\end{tikzpicture}
\caption{Two incident single photons and a far-field correlation measurement. The angle $\theta_j$ specifies a direction in the chosen observation plane, and $\mathbf e_j$ is the receiving polarization. Detection samples the field after spatial, polarization, and temporal selection.}
\label{fig:ch3:twoscatter}
\end{figure}

Each detection direction also requires a receiving aperture and polarization. First assume that each arm selects one normalized receiving mode, with mutually orthogonal output operators $\hat b_1,\hat b_2$, as shown in Figure~\ref{fig:ch3:twoscatter}. Under the narrowband approximation, the scattering response varies little over the photon bandwidth, giving
\begin{equation}
 \hat b_j=s_{j1}\hat a_1+s_{j2}\hat a_2+\hat v_j,
 \qquad j=1,2.
 \label{eq:ch3:twomap}
\end{equation}
The dimensionless complex transmission coefficient $s_{j\mu}$ connects input mode $\mu$ to receiving mode $j$. The operator $\hat v_j$ collects contributions from all other vacuum inputs and absorptive environments. These ensure the commutation relations of the complete outputs but contribute nothing to the normally ordered count averages considered here. The $2\times2$ submatrix formed from only two inputs and two outputs is generally not unitary because other scattering directions remain uncollected; it must not be forcibly normalized as a lossless two-port.

\subsection{How Two Probability Amplitudes Enter Coincidence Counts}

First calculate the mean photon numbers in the two arms. The input satisfies $\avg{\hat a_\mu^\dagger\hat a_\nu}=\delta_{\mu\nu}$, so
\begin{equation}
 \bar n_j=\avg{\hat b_j^\dagger\hat b_j}
 =|s_{j1}|^2+|s_{j2}|^2.
 \label{eq:ch3:twosingles}
\end{equation}
Thus the mean counts simply add the intensities from two independent classical excitations, with no first-order interference term between the inputs.

Coincidence counting instead requires successive annihilation of two photons. For a vacuum environment, retaining the terms acting on the occupied inputs gives
\[
 \begin{aligned}
 \hat b_2\hat b_1\ket{1,1}
 ={}&\bigl[s_{21}s_{11}\hat a_1^2
 +s_{22}s_{12}\hat a_2^2\\
 &\quad+(s_{11}s_{22}+s_{12}s_{21})\hat a_1\hat a_2\bigr]\ket{1,1}.
 \end{aligned}
\]
Each input contains only one photon, so $\hat a_1^2\ket{1,1}=\hat a_2^2\ket{1,1}=0$, while $\hat a_1\hat a_2\ket{1,1}=\ket{0,0}$. Hence
\begin{equation}
 \begin{aligned}
 G_{12}^{(2)}&\equiv\avg{\hat b_1^\dagger\hat b_2^\dagger\hat b_2\hat b_1}
 =\|\hat b_2\hat b_1\ket{1,1}\|^2\\
 &=|s_{11}s_{22}+s_{12}s_{21}|^2,\\[2pt]
 g_{12}^{(2)}&=\frac{|s_{11}s_{22}+s_{12}s_{21}|^2}
 {(|s_{11}|^2+|s_{12}|^2)(|s_{21}|^2+|s_{22}|^2)}.
 \end{aligned}
 \label{eq:ch3:twog2}
\end{equation}
For ideal counting over complete wavepackets, the total input photon number is two, so $G_{12}^{(2)}$ is the probability of detecting one photon in each arm. The normalization here uses mean photon numbers, not arbitrary trigger probabilities of detectors without photon-number resolution.

Writing $A=s_{11}s_{22}$ and $B=s_{12}s_{21}$ gives
\begin{equation}
 G_{12}^{(2)}=|A|^2+|B|^2+2\Re(AB^*).
 \label{eq:ch3:twopaths}
\end{equation}
The amplitude $A$ describes input 1 reaching detector 1 and input 2 reaching detector 2; $B$ describes the exchanged assignment. Both processes leave the same detection outcome, so their probability amplitudes are added before taking the squared magnitude. The two photons need not interact directly inside the object. If $A=-B\neq0$, coincidences vanish even though both single-arm counts can remain nonzero. This is Hong--Ou--Mandel (HOM) destructive interference~\cite{ch3:hom}.

If the photons are made distinguishable, for example by nonoverlapping arrival times, and counts cover the full wavepackets, the cross term vanishes and the reference value is $G_{12,\mathrm{dist}}^{(2)}=|A|^2+|B|^2$. Distinguishable and indistinguishable inputs should therefore be compared in the same apparatus. This measurement is a cross-correlation between two modes; $g_{12}^{(2)}<1$ by itself cannot be interpreted using the nonclassicality criterion for a single-mode autocorrelation. Equation~\eqref{eq:ch3:twog2} and the Cauchy--Schwarz inequality give $0\leq g_{12}^{(2)}\leq1$ for the ideal two-photon input here; any coincidence enhancement is relative to distinguishable inputs.

\subsection{An Angular Example with an Analytical Check}

Consider a small isotropic dielectric sphere in vacuum within the electric-dipole approximation. Two incident beams have equal field normalization at the sphere center: one propagates along $z$ and is polarized along $x$, and the other propagates along $x$ and is polarized along $y$. They excite electric dipoles along $x$ and $y$, respectively. In the $xy$ observation plane, take
\[
 \mathbf n(\theta)=(\cos\theta,\sin\theta,0),\qquad
 \mathbf e_\theta=(-\sin\theta,\cos\theta,0).
\]
Choose receiving polarization $\mathbf e_\theta$. The electric-dipole far field is proportional to $[I-\mathbf n\mathbf n]\mathbf p$. Projecting onto this polarization gives the transmission coefficients for the two excitations:
\begin{equation}
 s_{j1}=-c_j\sin\theta_j,\qquad
 s_{j2}=c_j\cos\theta_j.
 \label{eq:ch3:rayleighs}
\end{equation}
The coefficient $c_j$ includes the sphere polarizability, propagation distance, and receiving efficiency. At a given receiving direction, both inputs are projected onto the same polarization, so their different input polarizations no longer label their origins at detection. This example collects only scattered light and avoids the direct beams.

\begin{figure}[htbp]
\centering
\begin{tikzpicture}
 \begin{axis}[name=singles,width=6.25cm,height=5cm,
  xmin=30,xmax=150,ymin=0,ymax=1.14,
  xtick={30,90,150},ytick={0,.5,1},
  xlabel={$\theta\ ({}^\circ)$},ylabel={Relative mean count},
  tick label style={font=\footnotesize},label style={font=\small},
  title={(a) Single-arm counts},title style={font=\small},
  legend style={font=\scriptsize,draw=none,at={(.5,-.34)},anchor=north,legend columns=2},
  domain=30:150,samples=101]
  \addplot[black,thick] {sin(x)^2};\addlegendentry{Input 1}
  \addplot[black,thick,dashed] {cos(x)^2};\addlegendentry{Input 2}
  \addplot[black,thick,densely dotted] {1};\addlegendentry{Sum}
 \end{axis}
 \begin{axis}[at={(singles.east)},anchor=west,xshift=1.5cm,
  width=5.7cm,height=5cm,view={0}{90},
  xmin=30,xmax=150,ymin=30,ymax=150,
  xtick={30,90,150},ytick={30,90,150},
  xlabel={$\theta_1\ ({}^\circ)$},ylabel={$\theta_2\ ({}^\circ)$},
  tick label style={font=\footnotesize},label style={font=\small},
  title={(b) $g_{12}^{(2)}$},title style={font=\small},
  colormap={graymap}{gray(0cm)=(.15);gray(1cm)=(1)},
  point meta min=0,point meta max=1,
  colorbar,colorbar style={width=.18cm,ytick={0,.5,1},tick label style={font=\scriptsize}},
  domain=30:150,y domain=30:150,samples=35,samples y=35]
  \addplot3[surf,shader=interp] {sin(x+y)^2};
 \end{axis}
\end{tikzpicture}
\caption{Analytical results for a small sphere in the electric-dipole approximation. (a) The two inputs contribute $\sin^2\theta$ and $\cos^2\theta$ to the single-arm counts, summing to unity. (b) Angular correlation $g_{12}^{(2)}$; the coincidence probability includes a further factor $|c_1c_2|^2$. Dark lines mark cancellation of the two probability amplitudes. When the two angles coincide, light collected in that direction can be sent to two detectors through a 50:50 beam splitter.}
\label{fig:ch3:rayleighmap}
\end{figure}
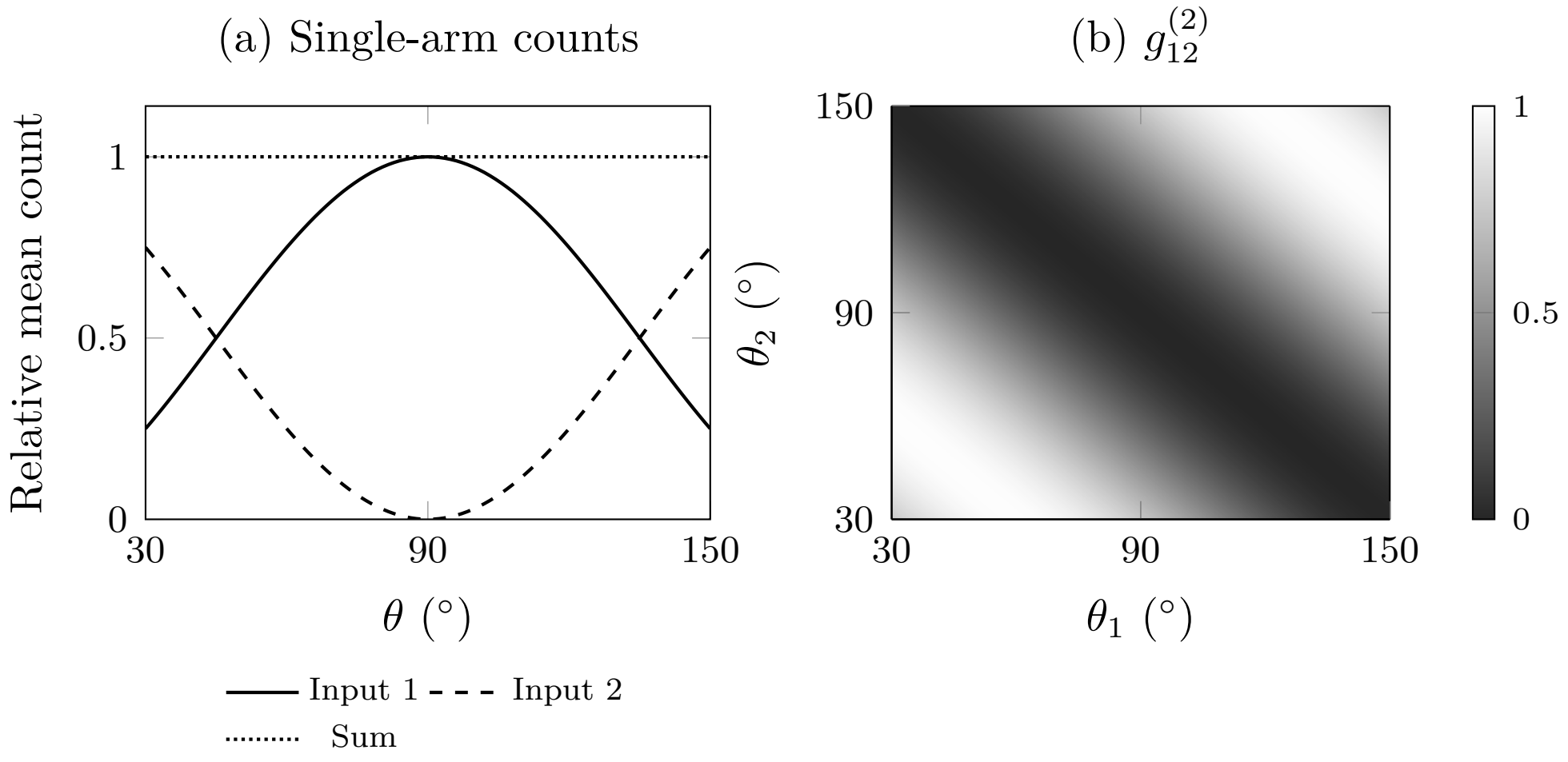

Substitution into Eqs.~\eqref{eq:ch3:twosingles} and~\eqref{eq:ch3:twog2} gives
\begin{equation}
 \begin{aligned}
 \bar n_j&=|c_j|^2,\\
 G_{12}^{(2)}&=|c_1c_2|^2\sin^2(\theta_1+\theta_2),\\
 g_{12}^{(2)}&=\sin^2(\theta_1+\theta_2).
 \end{aligned}
 \label{eq:ch3:rayleighg}
\end{equation}
If distance, receiving aperture, and other conditions are equal so that $|c_j|$ does not vary with scan angle, the single-arm mean counts are constant, but the two-photon coincidences have a clear angular structure, as shown in Figure~\ref{fig:ch3:rayleighmap}. For example, at $\theta_1=45^\circ$ and $\theta_2=135^\circ$, the two arrival amplitudes have equal magnitude and opposite phase, giving $g_{12}^{(2)}=0$; distinguishable inputs give $1/2$ at the same positions.

This example also illustrates a limitation of normalized correlations: the common sphere-polarizability factor cancels in $g_{12}^{(2)}$, so the plot alone cannot recover all material parameters of the sphere. Single-arm and absolute coincidence counts must also be recorded to retain information about scattering strength.

\subsection{Numerical Calculation and Engineering Uses}

For an arbitrary shape, the structure of Eq.~\eqref{eq:ch3:twog2} is unchanged; only the four complex transmission coefficients need numerical calculation. The procedure has three steps:
\begin{enumerate}
 \item Fix the spatial modes, frequency, and polarizations of the two incident beams and run one classical scattering calculation for each. Use the same power normalization for both excitations and retain the full complex fields; do not renormalize each radiation pattern to its own peak.
 \item Project each excitation's far field onto the receiving polarization and perform the mode-overlap integral to extract $s_{j\mu}$. Include the unscattered direct field if it lies within the receiving channel. Use the scattered field alone only when spatial selection excludes the direct beam.
 \item Scan $\theta_1,\theta_2$ and calculate $\bar n_1,\bar n_2,G_{12}^{(2)}$ and $g_{12}^{(2)}$ separately. The normalized correlation is undefined when a mean count is zero. Near zero, report the counts as well to avoid mistaking a low-count region for a useful operating point.
\end{enumerate}

For a narrow aperture and one receiving polarization, $s_{j\mu}$ differs from the corresponding classical complex far-field amplitude only by a factor common to that receiver. This factor cancels in $g^{(2)}$, so two electromagnetic solutions generate the entire angular correlation map. Absolute coincidence probabilities additionally require normalization of the input and receiving modes. Further Reading 3C treats finite bandwidth, unequal wavepackets, and multimode reception.

These calculations can guide designs that suppress or enhance coincidences between selected direction pairs. Geometry, resonances, and polarization conversion change the magnitudes and phases of $A$ and $B$, providing objective functions for designing two-photon correlations with dielectric scatterers and metasurfaces. For structural characterization, the phase combination $\arg(s_{11}s_{22}s_{12}^*s_{21}^*)$ in the correlation can add constraints absent from intensity measurements alone. However, classical coherent measurements may also access this phase information, so it does not by itself establish a quantum advantage.

For parameter estimation, compare errors at fixed incident photon number, measurement time, and detection loss. A zero or steep slope in a coincidence curve is only a candidate operating point; count probability and background noise must also be considered. The scatterer remains a linear medium throughout this chapter. Chapter 4 turns to joint quantum evolution of fields and few-level atoms, including changes caused by saturation and effective interactions between photons.

\begingroup
\renewcommand{\bibname}{References (Including Further Reading)}

\endgroup

\newpage
\begingroup
\setcounter{equation}{0}
\setcounter{subsection}{0}
\renewcommand{\theequation}{3A.\arabic{equation}}
\renewcommand{\thesubsection}{3A.\arabic{subsection}}
\renewcommand{\theHequation}{ch3.A.\arabic{equation}}
\renewcommand{\theHsubsection}{ch3.A.\arabic{subsection}}
\section*{Further Reading 3A\quad Dispersion, Dissipation, and Green-Function Identities}
\addcontentsline{toc}{section}{Further Reading 3A: Dispersion, Dissipation, and Green-Function Identities}

The main text related field operators to Green functions in lossy dispersive media. Here we add three details: which degrees of freedom are included in a quantized mode of a dispersive medium; how the Green-function integral identity is derived; and how field correlations are converted to the one-sided thermal-noise spectra commonly used in engineering.

\subsection{Polariton Modes from Mixing Fields and Matter}

In a dispersive medium, an electromagnetic wave exchanges energy with matter through driven oscillations of polarization charges. Begin with two coupled lossless harmonic oscillators representing an electromagnetic mode, with operator $\hat a$ and frequency $\omega_c$, and a material-polarization mode, with operator $\hat b$ and frequency $\omega_m$. In the near-resonant rotating-wave approximation, take real coupling strength $g$ and write the coupled-system Hamiltonian as
\begin{equation}
 \frac{\hat H}{\hbar}
 =\omega_c\hat a^\dagger\hat a+\omega_m\hat b^\dagger\hat b
 +g(\hat a^\dagger\hat b+\hat b^\dagger\hat a).
 \label{eq:ch3:polaritonham}
\end{equation}
Constant zero-point energies have been omitted. The operators $\hat a$ and $\hat b$ are independent bosonic operators, with $[\hat a,\hat a^\dagger]=[\hat b,\hat b^\dagger]=1$ and $[\hat a,\hat b]=0$.

The Heisenberg equations of motion are governed by the two-dimensional Hermitian matrix $\left(\begin{smallmatrix}\omega_c&g\\g&\omega_m\end{smallmatrix}\right)$. Setting its characteristic determinant to zero, $(\omega_c-\Omega)(\omega_m-\Omega)-g^2=0$, gives two new normal-mode eigenfrequencies:
\begin{equation}
 \Omega_\pm=\frac{\omega_c+\omega_m}{2}
 \pm\sqrt{\frac{(\omega_c-\omega_m)^2}{4}+g^2}.
 \label{eq:ch3:polaritonfreq}
\end{equation}
The new normal-mode operators $\hat p_\pm$ are orthogonal linear combinations of $\hat a$ and $\hat b$. At resonance, $\omega_c=\omega_m$, they are $\hat p_\pm=(\hat a\pm\hat b)/\sqrt2$, with eigenfrequencies $\omega_c\pm g$. A hybrid excitation formed from the field and material polarization is called a polariton. The two contributions have equal weight at resonance; away from resonance, their proportions vary with detuning.

Thus one excitation quantum $\hbar\Omega_\pm$ in a dispersive medium contains both field energy and material-polarization energy. Normalization cannot use the vacuum electric energy density $\varepsilon_0E^2/2$ alone.

If the polarization oscillator also couples to a continuum of environmental oscillators, energy can flow from the local mode into the environment. The Huttner--Barnett model diagonalizes the complete linear system of field, polarization, and reservoir, obtaining continuous bosonic degrees of freedom~\cite{ch3:huttner}. This provides an interpretation of $\hat{\mathbf f}(\mathbf s,\omega)$ in the main text. Eliminating the explicit material and reservoir variables leaves their effects encoded in a complex dielectric function and the corresponding noise current.

\subsection{The Inverse-Operator Identity and Open Radiation Boundaries}

Write the electromagnetic wave operator and its inverse as
\[
 \mathsf M=\nabla\times\nabla\times
 -\frac{\omega^2}{c^2}\varepsilon_r,\qquad \mathsf G=\mathsf M^{-1}.
\]
First work in a space where the inverse operator and its adjoint are both defined; boundary issues are addressed below. From $\mathsf G\mathsf M=\mathsf I$ and $\mathsf M^\dagger\mathsf G^\dagger=\mathsf I$, we obtain

\[
 \begin{aligned}
 \mathsf G(\mathsf M^\dagger-\mathsf M)\mathsf G^\dagger
 &=\mathsf G(\mathsf M^\dagger\mathsf G^\dagger)
 -(\mathsf G\mathsf M)\mathsf G^\dagger\\
 &=\mathsf G-\mathsf G^\dagger.
 \end{aligned}
\]
Define the generalized imaginary part $\operatorname{Im}_H\mathsf G=(\mathsf G-\mathsf G^\dagger)/(2\ii)$, which is itself Hermitian. Dividing the preceding equation by $2\ii$ gives
\begin{equation}
 \operatorname{Im}_{H}\mathsf G
 =\mathsf G\,\mathsf Q\,\mathsf G^\dagger,
 \qquad \mathsf Q\equiv\frac{\mathsf M^\dagger-\mathsf M}{2\ii}.
 \label{eq:ch3:inverseidentity}
\end{equation}
When the boundary terms vanish, the curl--curl operator equals its adjoint, so $\mathsf M^\dagger-\mathsf M$ retains only the imaginary part of the dielectric function:
\[
 \mathsf Q=-\frac{\omega^2}{c^2}
 \frac{\varepsilon_r^*-\varepsilon_r}{2\ii}
 =\frac{\omega^2}{c^2}\varepsilon_r''.
\]
Writing the operator products as spatial integrals and using reciprocity, $G_{ij}(\mathbf r,\mathbf r')=G_{ji}(\mathbf r',\mathbf r)$, reduces the generalized imaginary part to the ordinary imaginary part of each matrix element and gives Eq.~\eqref{eq:ch3:greenidentity} in the main text.

For a finite lossy object in lossless free space, the open radiation boundary cannot be neglected. The surface term from integration by parts retains a radiation contribution even as the boundary is moved to infinity. The generalized imaginary part then separates as
\begin{equation}
 \operatorname{Im}_{H}\mathsf G
 =\mathsf G\mathsf Q_{\mathrm{abs}}\mathsf G^\dagger
 +\mathsf R_{\mathrm{rad}},
 \label{eq:ch3:radiation}
\end{equation}
where $\mathsf Q_{\mathrm{abs}}$ describes material absorption and $\mathsf R_{\mathrm{rad}}$ the radiation-boundary contribution. The latter remains nonzero even without lossy material. For example, the free-space Green function has coincident-point imaginary part $\omega\mathbf I/(6\pi c)$.

An equivalent treatment retains infinitesimal absorption in the background. As long as absorption is nonzero, outgoing waves eventually decay at infinity, allowing the radiation contribution to be included in the all-space volume integral. Complete the integral before taking the background loss to zero. Setting the integrand to zero in vacuum beforehand would omit this contribution.

\subsection{From Field Correlations to Engineering Thermal-Noise Spectra}

At observation point $\mathbf r$ in vacuum, consider the total electric field along a real unit direction $\mathbf e$, namely $\hat E(t)=\mathbf e\cdot[\hat{\mathbf E}^{(+)}(\mathbf r,t)+\hat{\mathbf E}^{(-)}(\mathbf r,t)]$. In thermal equilibrium at temperature $T$, define the symmetrized time-correlation function used in Further Reading 2B:
\[
 C_E(\tau)=\frac12\avg{\hat E(t+\tau)\hat E(t)+\hat E(t)\hat E(t+\tau)}.
\]
Substitute the positive-frequency expansion~\eqref{eq:ch3:fourier} and thermal-equilibrium correlations~\eqref{eq:ch3:thermalgreen}, then use $\delta(\omega-\omega')$ to eliminate one frequency integral. Only cross terms between positive and negative frequencies remain, giving
\begin{equation}
 \begin{aligned}
 C_E(\tau)&=\int_0^\infty\mathcal A_E(\omega)\cos(\omega\tau)\dd\omega,\\
 \mathcal A_E(\omega)&=\frac{\hbar\mu_0}{\pi}\omega^2[2\bar n_T(\omega)+1]\\
 &\quad\times\mathbf e\cdot\Im\mathbf G(\mathbf r,\mathbf r,\omega)\cdot\mathbf e.
 \end{aligned}
 \label{eq:ch3:symcorr}
\end{equation}
Chapter 2 uses a one-sided spectrum per hertz, with convention $C_E(\tau)=\int_0^\infty S_E^{\mathrm{sym}}(f)\cos(2\pi f\tau)\dd f$. Setting $\omega=2\pi f$ gives $\dd\omega=2\pi\dd f$, so the spectral density must also be multiplied by $2\pi$. The resulting units are $\mathrm{(V/m)^2/Hz}$:
\begin{equation}
 \begin{aligned}
 S_E^{\mathrm{sym}}(f)&=2\pi\mathcal A_E(2\pi f)\\
 &=2\hbar\mu_0\omega^2[2\bar n_T(\omega)+1]\\
 &\quad\times\mathbf e\cdot\Im\mathbf G(\mathbf r,\mathbf r,\omega)\cdot\mathbf e,
 \qquad \omega=2\pi f.
 \end{aligned}
 \label{eq:ch3:sympsd}
\end{equation}

In the classical limit $\hbar\omega\ll k_BT$, $2\bar n_T+1\simeq2k_BT/(\hbar\omega)$, so
\begin{equation}
 S_E^{\mathrm{classical}}(f)\simeq4\mu_0k_BT\omega\,
 \mathbf e\cdot\Im\mathbf G(\mathbf r,\mathbf r,\omega)\cdot\mathbf e,
 \qquad\omega=2\pi f.
 \label{eq:ch3:nyquist}
\end{equation}
An electrically small receiving dipole that weakly perturbs the field connects this result to circuit-noise formulas. With direction $\mathbf e$ and effective length $\ell$, its open-circuit voltage is $V=\ell E$, giving
\[
 S_V(f)=\ell^2S_E^{\mathrm{classical}}(f).
\]
In the corresponding transmitting problem, the dipole moment and complex peak port current satisfy $I_0=-\ii\omega p_0/\ell$. Equation~\eqref{eq:ch3:imGpower} gives the resistance associated with radiation and absorption in the external medium:
\[
 \begin{aligned}
 R_{\mathrm{env}}&=\frac{2P_{\mathrm{tot}}}{|I_0|^2}\\
 &=\mu_0\omega\ell^2\,
 \mathbf e\cdot\Im\mathbf G(\mathbf r,\mathbf r,\omega)\cdot\mathbf e.
 \end{aligned}
\]
Substitution into the voltage spectrum yields
\[
 S_V(f)=4k_BTR_{\mathrm{env}}.
\]
This is the Johnson--Nyquist thermal-noise relation for this receiving model. All relevant radiative and absorptive environments are assumed to share one temperature, and the Green function is the background response before inserting the receiving dipole. Any conductor ohmic loss within the antenna requires its own resistance and noise contribution. If the parts have different temperatures, they cannot all be multiplied by a single $T$.
\endgroup

\vspace{1.5em}
\begingroup
\setcounter{equation}{0}
\setcounter{subsection}{0}
\renewcommand{\theequation}{3B.\arabic{equation}}
\renewcommand{\thesubsection}{3B.\arabic{subsection}}
\renewcommand{\theHequation}{ch3.B.\arabic{equation}}
\renewcommand{\theHsubsection}{ch3.B.\arabic{subsection}}
\section*{Further Reading 3B\quad Higher-Order Correlations and Nonequilibrium Thermal Radiation}
\addcontentsline{toc}{section}{Further Reading 3B: Higher-Order Correlations and Nonequilibrium Thermal Radiation}

A direct number-state density-matrix calculation for many modes can have very large dimension. For Gaussian states, one can instead propagate means and second moments, then use them to calculate higher-order correlations. We first specify when this simplification applies, then discuss field correlations and energy flow in environments at different temperatures.

\subsection{Fourth Moments of Gaussian States and the Siegert Relation}

For a set of modes $\hat a_i$, separate the mean amplitudes $\alpha_i=\avg{\hat a_i}$ and define $\delta\hat a_i=\hat a_i-\alpha_i$. A state is Gaussian if its quadrature Wigner distribution is a Gaussian function. Vacuum, coherent states, thermal states, and the usual squeezed states belong to this class~\cite{ch3:gaussian}.

These states remain Gaussian when transmitted through linear passive networks or evolved under quadratic Hamiltonians and Gaussian environments. An arbitrary statistical mixture, however, need not be Gaussian. For a Gaussian state, the means and the following two types of second moments determine all higher moments:
\begin{equation}
 N_{ij}=\avg{\delta\hat a_i^\dagger\delta\hat a_j},
 \qquad M_{ij}=\avg{\delta\hat a_i\delta\hat a_j}.
 \label{eq:ch3:NM}
\end{equation}
The matrix $N_{ij}$ describes mode occupations and ordinary cross-correlations, while $M_{ij}$ describes anomalous pair correlations. Coherent and ordinary thermal states have $M=0$; squeezed states generally have nonzero $M$. Apart from vacuum $\ket0$, number states with definite photon number are not Gaussian.

For zero-mean Gaussian fluctuations, Wick's theorem expresses a four-operator expectation as the sum of the three possible products of pair expectations. The original operator order within each pair is preserved. Thus
\begin{equation}
 \begin{aligned}
 &\avg{\delta\hat a_i^\dagger\delta\hat a_j^\dagger
        \delta\hat a_j\delta\hat a_i}\\
 &=\avg{\delta\hat a_i^\dagger\delta\hat a_i}
   \avg{\delta\hat a_j^\dagger\delta\hat a_j}\\
 &\quad+\avg{\delta\hat a_i^\dagger\delta\hat a_j}
        \avg{\delta\hat a_j^\dagger\delta\hat a_i}\\
 &\quad+\avg{\delta\hat a_i^\dagger\delta\hat a_j^\dagger}
        \avg{\delta\hat a_j\delta\hat a_i}\\
 &=N_{ii}N_{jj}+|N_{ij}|^2+|M_{ij}|^2.
 \end{aligned}
 \label{eq:ch3:wick}
\end{equation}
For nonzero mean amplitudes, first substitute $\hat a_i=\alpha_i+\delta\hat a_i$. Products of an odd number of zero-mean Gaussian fluctuations have zero expectation. Second-order terms follow directly from $N,M$, and fourth-order terms follow from Eq.~\eqref{eq:ch3:wick}.

For a zero-mean Gaussian field without anomalous correlations, $\alpha=0$ and $M=0$, only the first two terms remain. Dividing by $N_{ii}N_{jj}$ and normalizing according to the first-order correlation definition gives the Siegert relation:
\begin{equation}
 g^{(2)}(1,2)=1+|g^{(1)}(1,2)|^2.
 \label{eq:ch3:siegert}
\end{equation}
For example, if thermal light in the detected spatial and polarization mode has first-order time correlation $g^{(1)}(\tau)=\ee^{-|\tau|/\tau_c}\ee^{-\ii\omega_c\tau}$, then
\[
 g^{(2)}(\tau)=1+\ee^{-2|\tau|/\tau_c}.
\]
At zero delay, $g^{(2)}(0)=2$; at long delay it approaches unity. Under the Gaussian-statistics, spectral-shape, and detection-bandwidth conditions above, fitting an HBT correlation curve yields the coherence time $\tau_c$ and hence an estimate of the spectral linewidth.

Wick factorization requires a Gaussian state; the Siegert relation additionally requires zero mean and no anomalous correlations, so its scope is narrower. A single-photon state satisfies $\avg{\hat a^\dagger\hat a}=1$ and $\avg{\hat a^{\dagger2}\hat a^2}=0$, hence $g^{(2)}(0)=0$. Applying the Siegert relation to it gives an incorrect result.

\subsection{Effects of Pure Loss and Background Noise on Second-Order Correlations}

For a Gaussian input to a linear network, first and second moments can be propagated directly. Second moments alone cannot describe a non-Gaussian input, but normally ordered moments still obey a simple rule under pure loss. Consider Eq.~\eqref{eq:ch3:losschannel} with independent input and vacuum environment.

Since $\hat v\ket0_v=0$, every term containing environment operators has zero expectation after expanding the normally ordered products. Thus
\[
 \avg{\hat n_{\mathrm{out}}}=\eta\avg{\hat n_{\mathrm{in}}},\qquad
 \avg{\hat a_{\mathrm{out}}^{\dagger2}\hat a_{\mathrm{out}}^2}
 =\eta^2\avg{\hat a_{\mathrm{in}}^{\dagger2}\hat a_{\mathrm{in}}^2}.
\]
Substitution into the definition of $g^{(2)}(0)$ cancels the squared transmittance $\eta^2$ between numerator and denominator:
\[
 g^{(2)}_{\mathrm{out}}(0)=\frac{\eta^2\avg{\hat a_{\mathrm{in}}^{\dagger2}\hat a_{\mathrm{in}}^2}}{(\eta\avg{\hat n_{\mathrm{in}}})^2}=g^{(2)}_{\mathrm{in}}(0).
\]
As long as the output mean photon number is nonzero, pure loss leaves single-mode $g^{(2)}(0)$ unchanged. For example, attenuation turns a single photon into a mixture of vacuum and one photon. Two absorptions remain impossible, so $g^{(2)}(0)=0$, although the photon arrival probability is reduced.

Real measurements also include thermal background and detector dark counts. When the signal is attenuated but the background does not decrease by the same factor, background counts and their cross terms with the signal change the measured correlation, often weakening antibunching. Signal and background must then be modeled separately; Wick factorization cannot be applied directly to a non-Gaussian signal.

\subsection{Environments at Different Temperatures: Nonequilibrium Correlations and Radiative Energy Flow}

In thermal equilibrium, $\bar n_T(\omega)$ is the same everywhere and can be taken outside the spatial integral before applying the Green-function identity. This step fails when different regions have different temperatures. For example, an antenna receiving both cold-sky and ground radiation, or a cavity connected to ports at different temperatures, requires each environment's noise to be propagated separately before their total correlation is formed.

The calculation resembles noise analysis in a classical linear system. Let state vector $\hat{\mathbf x}$ be driven by environmental sources $\hat{\mathbf F}$:
\[
 \begin{aligned}
 \mathsf M(\omega)\hat{\mathbf x}(\omega)&=\hat{\mathbf F}(\omega),\\
 \hat{\mathbf x}(\omega)&=\mathsf R(\omega)\hat{\mathbf F}(\omega),
 \qquad \mathsf R=\mathsf M^{-1}.
 \end{aligned}
\]
Here $\mathsf R$ is the classical response matrix. To make the index ordering explicit, define the normally ordered source and state spectra by
\[
 \begin{aligned}
 \avg{\hat F_j^\dagger(\omega')\hat F_i(\omega)}
 &=2\pi\delta(\omega-\omega')D_{ij}(\omega),\\
 \avg{\hat x_j^\dagger(\omega')\hat x_i(\omega)}
 &=2\pi\delta(\omega-\omega')C_{ij}(\omega).
 \end{aligned}
\]
This uses a Fourier convention over the full real frequency axis, with $1/(2\pi)$ in the inverse transform. For independent sources, $\mathsf D$ is diagonal; a thermal channel's diagonal entry depends on its coupling strength and thermal occupation. Substituting the linear response term by term gives
\begin{equation}
 \mathsf C(\omega)=\mathsf R(\omega)\mathsf D(\omega)\mathsf R^\dagger(\omega).
 \label{eq:ch3:RDR}
\end{equation}
For a medium with local-equilibrium temperature $T(\mathbf s)$, using the positive-frequency expansion convention of Eq.~\eqref{eq:ch3:fourier} gives
\begin{equation}
 \begin{aligned}
 &\avg{\hat{\mathcal E}_j^\dagger(\mathbf r',\omega')\hat{\mathcal E}_i(\mathbf r,\omega)}\\
 &=\frac{\hbar}{\pi\varepsilon_0}\frac{\omega^4}{c^4}\delta(\omega-\omega')
 \int\dd^3s\,\varepsilon_r''(\mathbf s,\omega)\bar n_T(\mathbf s,\omega)\\
 &\quad\times\sum_kG_{ik}(\mathbf r,\mathbf s,\omega)G_{jk}^*(\mathbf r',\mathbf s,\omega).
 \end{aligned}
 \label{eq:ch3:noneqgreen}
\end{equation}
Each volume element's thermal-source strength is set by its loss and local temperature, then propagated to the observation point through the Green function. An open system must also include the external incident radiation environment, with its temperature or actual occupation specified. Integrating only thermal sources inside the finite object is insufficient.

As an example, consider a single-mode cavity coupled to two independent thermal environments. The channel decay rates are $\kappa_1,\kappa_2$, their thermal occupations are $\bar n_1,\bar n_2$, and the total decay rate is $\kappa=\kappa_1+\kappa_2$. In the narrowband Markov approximation, the mean photon number obeys
\begin{equation}
 \begin{aligned}
 \frac{\dd\avg{\hat n}}{\dd t}
 &=-\kappa\avg{\hat n}+\kappa_1\bar n_1+\kappa_2\bar n_2,\\
 \bar n_{\mathrm{ss}}
 &=\frac{\kappa_1\bar n_1+\kappa_2\bar n_2}{\kappa_1+\kappa_2}.
 \end{aligned}
 \label{eq:ch3:twobaths}
\end{equation}
The second expression follows by setting the time derivative in the first to zero. The same result can be obtained in the frequency domain in a rotating frame. Let $\nu=\omega-\omega_c$, with response $R(\nu)=(\kappa/2-\ii\nu)^{-1}$ and source spectrum $D=\kappa_1\bar n_1+\kappa_2\bar n_2$. The cavity mode's first-order normally ordered spectrum is
\[
 S_a^{\mathrm{ord}}(\nu)=|R(\nu)|^2 D(\nu)
 =\frac{\kappa_1\bar n_1+\kappa_2\bar n_2}{\nu^2+(\kappa/2)^2}.
\]
Here $S_a^{\mathrm{ord}}$ is a first-order field-amplitude correlation spectrum whose integral gives the mean photon number; it is not a photon-number fluctuation spectrum. Using
\[
 \int_{-\infty}^{\infty}\frac{\dd\nu}{\nu^2+(\kappa/2)^2}
 =\frac{2\pi}{\kappa}
\]
gives $\bar n_{\mathrm{ss}}=\int S_a^{\mathrm{ord}}(\nu)\dd\nu/(2\pi)$, in agreement with Eq.~\eqref{eq:ch3:twobaths}.

If $\bar n_1>\bar n_2$, a net energy flow from environment 1 to environment 2 persists even after the cavity mean photon number stops changing. The net power absorbed by environment 2 is
\begin{equation}
 \begin{aligned}
 P_{1\to2}&=\hbar\omega_c\kappa_2(\bar n_{\mathrm{ss}}-\bar n_2)\\
 &=\hbar\omega_c\frac{\kappa_1\kappa_2}{\kappa_1+\kappa_2}
 (\bar n_1-\bar n_2).
 \end{aligned}
 \label{eq:ch3:heatflux}
\end{equation}
Thus the resonance frequency and total linewidth determine only part of the response. Calculating field statistics and net energy flow in a system with multiple temperatures also requires the coupling rates and occupations of the individual environments.
\endgroup 

\newpage
\begingroup
\setcounter{equation}{0}
\setcounter{subsection}{0}
\renewcommand{\theequation}{3C.\arabic{equation}}
\renewcommand{\thesubsection}{3C.\arabic{subsection}}
\renewcommand{\theHequation}{ch3.C.\arabic{equation}}
\renewcommand{\theHsubsection}{ch3.C.\arabic{subsection}}
\section*{Further Reading 3C\quad Wavepackets and Reception in Two-Photon Scattering}
\addcontentsline{toc}{section}{Further Reading 3C: Wavepackets and Reception in Two-Photon Scattering}

The main text assumes narrow bandwidth, matched wavepackets, and single-mode reception. Here we examine how wavepacket overlap and the receiving method change coincidence counts, still using classical transmission coefficients as the basis of the calculation.

\subsection{Arrival-Time Difference and Interference Visibility}

Let the two inputs have normalized temporal envelopes $\xi_1(t)$ and $\xi_2(t)$, satisfying $\int|\xi_\mu(t)|^2\dd t=1$. Add a delay $\tau$ to input 2 and assume approximately constant scattering coefficients over the wavepacket bandwidth. If polarization selection removes information about the input of origin, the coincidence-detection probability amplitude at times $t_1,t_2$ is
\begin{equation}
 \begin{aligned}
 \Psi_{12}(t_1,t_2;\tau)
 ={}&A\xi_1(t_1)\xi_2(t_2-\tau)\\
 &+B\xi_2(t_1-\tau)\xi_1(t_2),
 \end{aligned}
 \label{eq:ch3:timeamplitude}
\end{equation}
where $A,B$ have the same meanings as in the main text. Integration over a window covering the entire pulse gives the coincidence probability
\begin{equation}
 \begin{aligned}
 P_{12}(\tau)&=\int\!\int|\Psi_{12}(t_1,t_2;\tau)|^2\dd t_1\dd t_2\\
 &=|A|^2+|B|^2+2\chi(\tau)\Re(AB^*),\\
 \chi(\tau)&=\left|\int\xi_1^*(t)\xi_2(t-\tau)\dd t\right|^2,
 \qquad 0\leq\chi\leq1.
 \end{aligned}
 \label{eq:ch3:overlap}
\end{equation}
Each squared term has unit time integral. The cross term separates into two conjugate overlap integrals, so its weight is $\chi$. Perfectly matched, synchronized wavepackets have $\chi=1$; nonoverlapping wavepackets have $\chi=0$. Thus independent single-photon sources do not require relative optical phase locking in the classical-field sense, but their spectra, arrival times, and distinguishability after detection selection must be controlled~\cite{ch3:partial}.

If $\Re(AB^*)<0$, scanning the delay produces a coincidence dip. Relative to the nonoverlapping reference $P_{\mathrm{dist}}=|A|^2+|B|^2$, its visibility is
\begin{equation}
 \mathcal V=\frac{P_{\mathrm{dist}}-P_{12}(0)}{P_{\mathrm{dist}}}
 =\frac{-2\chi(0)\Re(AB^*)}{|A|^2+|B|^2}.
 \label{eq:ch3:homvisibility}
\end{equation}
Even with complete wavepacket overlap, cancellation is incomplete unless the two scattering probability amplitudes have equal magnitude and opposite phase. Conversely, when the structure gives $A=-B$, visibility is limited directly by wavepacket overlap. For finite timing windows or time- or frequency-resolved detection, integrate Eq.~\eqref{eq:ch3:timeamplitude} with the actual detector response; the whole-pulse formula cannot be used universally.

\subsection{Frequency Response and Finite Receiving Apertures}

If the narrowband approximation fails, first propagate each input wavepacket through its classical frequency response. Define detuning $\nu=\omega-\omega_c$ and use $\xi_\mu(t)=\int\widetilde\xi_\mu(\nu)\ee^{-\ii\nu t}\dd\nu/(2\pi)$ to define
\begin{equation}
 \psi_{j\mu}(t)=\int\frac{\dd\nu}{2\pi}\,
 s_{j\mu}(\omega_c+\nu)\widetilde\xi_\mu(\nu)
 \ee^{-\ii\nu t}.
 \label{eq:ch3:filterwavepacket}
\end{equation}
The input delay can be included as a phase factor in $\widetilde\xi_2$. Then use
\[
 \Psi_{12}(t_1,t_2)
 =\psi_{11}(t_1)\psi_{22}(t_2)+\psi_{12}(t_1)\psi_{21}(t_2)
\]
to calculate the coincidence probability and normalize with $\bar n_j=\sum_\mu\int|\psi_{j\mu}(t)|^2\dd t$. Numerically, this means sweeping frequency to obtain transmission coefficients, multiplying by the input spectra, applying an inverse Fourier transform, and integrating over time.

A finite aperture requires distinguishing two receiving methods. If light couples coherently into one known mode, such as a single-mode fiber, first perform a complex-amplitude overlap integral of the far field and then use the two-mode formulas in the main text. If a detector receives several orthogonal modes without distinguishing them, sum their count probabilities instead. Denote the mode sets collected by the two arms by $\mathcal A,\mathcal B$. Under the matched-wavepacket conditions of the main text,
\begin{equation}
 \begin{aligned}
 \bar N_{\mathcal A}&=\sum_{j\in\mathcal A}
 (|s_{j1}|^2+|s_{j2}|^2),\\
 P_{\mathcal A\mathcal B}&=\sum_{j\in\mathcal A}\sum_{k\in\mathcal B}
 |s_{j1}s_{k2}+s_{j2}s_{k1}|^2,\\
 g_{\mathcal A\mathcal B}^{(2)}&=
 \frac{P_{\mathcal A\mathcal B}}{\bar N_{\mathcal A}\bar N_{\mathcal B}}.
 \end{aligned}
 \label{eq:ch3:bucket}
\end{equation}
Thus values of $g^{(2)}$ at different directions within an aperture cannot simply be arithmetically averaged. Angular, frequency, and temporal averaging all affect whether deep zeros survive. Thermal photons in the environment or detector dark counts require the corresponding background and cross terms. Only normally ordered contributions from a vacuum environment can be omitted directly.
\endgroup

\clearpage
\fancyhead[R]{\small Chapter 4}
\ctexset{chapter/format=\Large\normalfont,section/format=\large\normalfont,subsection/format=\normalsize\normalfont}
\captionsetup{labelfont=normalfont}

\chapter{Quantum Fields and Quantum Matter}
\label{ch4:chapter}

Chapter 2 treated the electromagnetic field as a prescribed classical drive and calculated atomic populations and coherences. Chapter 3 quantized the electromagnetic field and studied its propagation, loss, and detection in linear macroscopic structures. We now bring these two descriptions into one model: an atom can transfer its excitation energy to the electromagnetic field, and the field can in turn change the atom's quantum state. Field and matter evolve together as one coupled quantum system.

A direct physical question is this: what happens when an excited atom is placed in an empty cavity or in free space and the external excitation is switched off? In an ideal lossless single-mode cavity, the excitation can pass back and forth between the atom and the cavity mode. In a broadband, weakly coupled open environment, the excited state usually decays approximately exponentially, emitting a photon. Both behaviors arise from the same electric-dipole interaction; the difference lies in the available electromagnetic modes and their spectral distribution. A high $Q$ alone does not guarantee observable exchange: the coupling and loss rates must also be compared.

This chapter first develops the Jaynes--Cummings model of a single two-level atom coupled to a single cavity mode, then discusses spontaneous emission in open structures, cavity enhancement (the Purcell effect), and the quantum statistics of emitted photons. For engineering electromagnetic analysis, three questions must be distinguished: how strong is the single-photon coupling ($g$), into which channels does energy flow ($\kappa,\Gamma$), and what statistics does the light detected at a port exhibit ($g^{(2)}$)? Classical electromagnetic simulations of local field enhancement or a shortened radiative lifetime answer only part of the first two questions.

\section{How an Atom and a Field Form One System}
\subsection{Joint States and the Tensor-Product Space}

We retain the two atomic levels $\ket g$ (ground state) and $\ket e$ (excited state); the number states of the electromagnetic mode of interest are $\ket0,\ket1,\ldots$. To specify the states of the atom and field simultaneously, we pair their basis vectors and write
\begin{equation}
 \ket{g,n}\equiv\ket g\otimes\ket n,\qquad
 \ket{e,n}\equiv\ket e\otimes\ket n\qquad (n=0,1,2,\ldots).
 \label{ch4:product}
\end{equation}
The symbol $\otimes$ denotes the tensor product (also called the direct product, or the Kronecker product in matrix calculations). Physically, it simply combines the states of two subsystems into a joint basis vector. For example, $\ket{e,0}$ means ``the atom is excited and the cavity mode is in vacuum,'' whereas $\ket{g,1}$ means ``the atom is in the ground state and the cavity contains 1 photon.'' Although these two states have the same uncoupled energy when the atom and cavity are resonant (both are $\hbar\omega_0$, omitting the cavity zero-point energy), they are orthogonal states.

An arbitrary joint pure state is a linear superposition of these joint basis vectors:
\begin{equation}
 \ket\Psi=\sum_{n=0}^{\infty}
 \left(c_{g,n}\ket{g,n}+c_{e,n}\ket{e,n}\right),
 \quad \sum_{n=0}^{\infty}\left(|c_{g,n}|^2+|c_{e,n}|^2\right)=1.
 \label{ch4:jointstate}
\end{equation}
The squared magnitude of a probability amplitude gives a joint measurement probability. For example, $|c_{e,1}|^2$ is the probability of simultaneously finding the atom excited and 1 photon in the cavity. If the photon number is truncated to $n=0,\ldots,N$ in a numerical calculation, the joint state vector is a column vector of length $2(N+1)$.

Operators acting on the atom and field separately must also be extended to this joint space. Using the energy-level basis order of Chapter 2, $\ket g=\binom{1}{0}$ and $\ket e=\binom{0}{1}$, we define the atomic lowering operator $\hat\sigma_-$, raising operator $\hat\sigma_+$, and population-difference operator $\hat\sigma_z$:
\begin{equation}
 \begin{aligned}
 \hat\sigma_-&=\ket g\bra e=\begin{pmatrix}0&1\\0&0\end{pmatrix},\\
 \hat\sigma_+&=\hat\sigma_-^\dagger=\ket e\bra g=\begin{pmatrix}0&0\\1&0\end{pmatrix},\\
 \hat\sigma_z&=\ket e\bra e-\ket g\bra g=\begin{pmatrix}-1&0\\0&1\end{pmatrix}.
 \end{aligned}
 \label{ch4:atomicops}
\end{equation}
Here $\hat\sigma_z$ is the matrix $Z$ in Further Reading 2A.4 of Chapter 2.
(Some quantum mechanics textbooks order the basis as $(\ket e,\ket g)$, giving $\hat\sigma_z=\operatorname{diag}(1,-1)$. This book consistently orders levels from lower to higher energy, $(\ket g,\ket e)$. Keeping the operator definition $\ket e\bra e-\ket g\bra g$ in view is sufficient: the physical results are identical.) In the joint space, the full matrix forms of field and atomic operators are given by tensor products:
\[
 \hat a_{\rm joint}=I_{\rm a}\otimes\hat a,\qquad
 \hat\sigma_{-,\rm joint}=\hat\sigma_-\otimes I_{\rm f}.
\]
Here $I_{\rm a}$ is the $2\times 2$ atomic identity matrix, and $I_{\rm f}$ is the identity in the field space. $\hat a_{\rm joint}$ changes only the photon-number index $n$ on the right, whereas $\hat\sigma_{-,\rm joint}$ changes only the atomic-level index on the left:
\[
 \hat a\ket{e,n}=\sqrt n\ket{e,n-1},\qquad
 \hat\sigma_-\ket{e,n}=\ket{g,n},\qquad
 \hat\sigma_+\ket{g,n}=\ket{e,n}.
\]
Below, the identity operators $I_{\rm a},I_{\rm f}$ in the joint space are left implicit. Because the two types of operator act independently on different indices, operators belonging to the two different subspaces commute: $[\hat a,\hat\sigma_-]=[\hat a,\hat\sigma_+]=0$. Commutation relations within each subsystem must still be retained, such as $[\hat a,\hat a^\dagger]=I$ and $[\hat\sigma_+,\hat\sigma_-]=\hat\sigma_z$.

\subsection{Why the Atom and Field Cannot Always Have Separate States: Entanglement and the Partial Trace}
\label{ch4:sec_partialtrace}

If the atom independently occupies the state $\ket{\psi_{\rm a}}=c_g\ket g+c_e\ket e$ and the field independently occupies $\ket{\psi_{\rm f}}=\sum_n f_n\ket n$, their joint state is the tensor product $\ket\Psi=\ket{\psi_{\rm a}}\otimes\ket{\psi_{\rm f}}$. Its expansion coefficients must factorize as $c_{g,n}=c_gf_n$ and $c_{e,n}=c_ef_n$. Interaction generally causes the joint-state coefficients to lose this factorization.

Consider a superposition with a total of only 1 quantum of excitation:
\begin{equation}
 \ket\Psi=c_e\ket{e,0}+c_c\ket{g,1},\qquad
 |c_e|^2+|c_c|^2=1.
 \label{ch4:singleexcitation}
\end{equation}
The subscript $c$ denotes the cavity mode. If $c_e,c_c$ are both nonzero, this state cannot be written as a product of subsystem states. Suppose it could be written as $(u_g\ket g+u_e\ket e)\otimes(f_0\ket0+f_1\ket1)$. For the coefficients of $\ket{e,0}$ and $\ket{g,1}$ to be nonzero, all four quantities $u_e,f_0,u_g,f_1$ would have to be nonzero. But nonzero $\ket{g,0}$ and $\ket{e,1}$ components would then also appear, contradicting Eq.~\eqref{ch4:singleexcitation}. A joint pure state that cannot be factored into a product of two subsystem pure states is called an entangled state.

In this single-excitation state, finding the atom in the ground state implies that the cavity contains one photon; finding it excited implies that the cavity is in vacuum. This population correlation alone does not prove entanglement: an incoherent statistical mixture with the same proportions has the same population correlation. Equation~\eqref{ch4:singleexcitation} also retains a phase relation between its two components, which distinguishes it from a statistical mixture.

If an experiment measures only the atomic state, without measuring the cavity photon number, how should the atomic statistics be described? Starting from the joint density matrix $\rho=\ket\Psi\bra\Psi$, summing out the unmeasured photon-number index $n$ gives the reduced atomic density matrix $\rho_{\rm a}$:
\begin{equation}
 (\rho_{\rm a})_{ss'}=\sum_{n=0}^{\infty}\bra{s,n}\rho\ket{s',n},
 \qquad s,s'\in\{g,e\}.
 \label{ch4:partialtrace}
\end{equation}
This operation is called taking the partial trace over the field degrees of freedom, written $\rho_{\rm a}=\Tr_{\rm f}\rho$. Substituting Eq.~\eqref{ch4:singleexcitation} into Eq.~\eqref{ch4:partialtrace}, the orthogonality relation $\braket01=0$ makes all coherent cross terms containing $c_ec_c^*$ vanish in the sum. The reduced density matrices of the atom and field are therefore both diagonal:
\begin{equation}
 \begin{aligned}
 \rho_{\rm a}&=|c_c|^2\ket g\bra g+|c_e|^2\ket e\bra e
 =\begin{pmatrix}|c_c|^2&0\\0&|c_e|^2\end{pmatrix},\\
 \rho_{\rm f}&=|c_e|^2\ket0\bra0+|c_c|^2\ket1\bra1.
 \end{aligned}
 \label{ch4:reduced}
\end{equation}
When $c_e,c_c$ are both nonzero, the joint system is in a pure state, but observing either the atom or field alone gives a mixed state. Neither has off-diagonal coherence in the energy-level or number-state basis used here. The phase information has not disappeared from the joint state: it resides in atom--field cross correlations. Predicting subsequent exchange therefore generally requires retaining the joint state; $\rho_{\rm a}\otimes\rho_{\rm f}$ cannot be substituted for the original $\rho$.

\section{From the Dipole Interaction to Single-Mode Coupling Equations}
\subsection{What Determines the Coupling Coefficient \texorpdfstring{$g$}{g}?}

We use the electric-dipole approximation of Chapter 2: the atom is much smaller than the spatial scale over which the local electric field varies, so it is treated as a point dipole at $\mathbf r_0$. Its interaction Hamiltonian is
\begin{equation}
 \hat H_{\rm int}=-\hat{\mathbf d}\cdot\hat{\mathbf E}(\mathbf r_0),
 \qquad
 \hat{\mathbf d}=\mathbf d_{ge}\hat\sigma_-+
 \mathbf d_{eg}\hat\sigma_+.
 \label{ch4:dipole}
\end{equation}
Here $\mathbf d_{ge}=\bra g\hat{\mathbf d}\ket e$, $\mathbf d_{eg}=\mathbf d_{ge}^*$, and the two selected states are assumed to have no permanent electric dipole moment. In Chapter 2 the electric field was a prescribed classical function; here it is the operator derived in Chapter 3.

According to Eq.~(3.10) of Chapter 3, for a normalized real eigenmode $\mathbf u(\mathbf r)$ satisfying $\int_V\varepsilon_r\mathbf u^2\dd^3r=1$, the electric-field operator is
\begin{equation}
 \hat{\mathbf E}(\mathbf r)=\ii\boldsymbol{\mathcal E}(\mathbf r)
 (\hat a-\hat a^\dagger),\qquad
 \boldsymbol{\mathcal E}(\mathbf r)
 =\sqrt{\frac{\hbar\omega_c}{2\varepsilon_0}}\,\mathbf u(\mathbf r).
 \label{ch4:modefield}
\end{equation}
The vector $\boldsymbol{\mathcal E}(\mathbf r)$ is the vacuum field scale of this mode (units: $\mathrm{V/m}$). The scalar vacuum field scale in Chapter 3 is its projection along a specified real unit direction $\mathbf e$, namely $\mathcal E=\mathbf e\cdot\boldsymbol{\mathcal E}$. The standard deviation of the vacuum electric field along that direction is $|\mathcal E|$. Substituting Eq.~\eqref{ch4:modefield} into Eq.~\eqref{ch4:dipole}, we define the single-photon electric-dipole coupling coefficient (vacuum Rabi coupling rate):
\begin{equation}
 \begin{aligned}
 g&=-\frac{\ii}{\hbar}\,\mathbf d_{eg}\cdot\boldsymbol{\mathcal E}(\mathbf r_0),\\
 |g|&=\frac{|\mathbf d_{eg}\cdot\boldsymbol{\mathcal E}(\mathbf r_0)|}{\hbar}
 =\sqrt{\frac{\omega_c}{2\hbar\varepsilon_0}}\,
 |\mathbf d_{eg}\cdot\mathbf u(\mathbf r_0)|.
 \end{aligned}
 \label{ch4:gdefinition}
\end{equation}
$g$ has units of $\mathrm{s}^{-1}$ and is an amplitude coupling rate. $\hbar|g|$ is the magnitude of the interaction matrix element between $\ket{e,0}$ and $\ket{g,1}$. Stronger local mode fields and better alignment of the polarization with the transition dipole give stronger coupling. The mean electric field is zero in both vacuum and a single-photon state; $\boldsymbol{\mathcal E}$ must not be interpreted as a definite classical electric field present in either state.

For a single atom and a single mode, the reference phase of the atomic basis vectors can be chosen to make a nonzero $g$ real and positive. We first retain its complex form when writing the four interaction terms, then adopt the convention $g=|g|>0$. In multimode or multiatom problems, relative phases that cannot all be removed must be retained.

\subsection{Why Only Two Terms Are Retained: The Jaynes--Cummings Model}

Substituting $\hat{\mathbf d}$ and $\hat{\mathbf E}(\mathbf r_0)$ into Eq.~\eqref{ch4:dipole} and expanding the product gives four operator combinations:
\[
 \hat H_{\rm int}=\hbar\bigl(
 g\,\hat a\hat\sigma_++g^*\,\hat a^\dagger\hat\sigma_-
 -g^*\,\hat a\hat\sigma_--g\,\hat a^\dagger\hat\sigma_+
 \bigr).
\]
Consider their physical processes and phase evolution (under free evolution, $\hat a\propto\ee^{-\ii\omega_ct}$ and $\hat\sigma_+\propto\ee^{+\ii\omega_0t}$):
\begin{itemize}
 \item The first two terms are near-resonant exchange terms. $\hat a\hat\sigma_+$ annihilates 1 cavity photon while taking the atom from its ground state to its excited state (absorption); $\hat a^\dagger\hat\sigma_-$ returns the atom to its ground state while creating 1 cavity photon (emission). These terms evolve at the slowly varying difference frequency, $\ee^{\pm\ii(\omega_0-\omega_c)t}$, and approximately conserve the uncoupled energy near resonance ($\omega_0\simeq\omega_c$).
 \item The last two are counter-rotating terms. $\hat a\hat\sigma_-$ annihilates 1 photon and 1 atomic excitation simultaneously; $\hat a^\dagger\hat\sigma_+$ adds 1 photon and 1 atomic excitation simultaneously. These terms change the uncoupled excitation number and oscillate rapidly at the sum frequency, $\ee^{\pm\ii(\omega_0+\omega_c)t}$.
\end{itemize}
Provided that the coupling strength and detuning are much smaller than the carrier frequency, namely
\[
 g\sqrt{n+1},\ |\omega_0-\omega_c|\ll\omega_0+\omega_c,
\]
the rapidly oscillating last two terms contribute little on time scales long compared with the carrier period but short compared with the coupling dynamics. Neglecting them is the rotating-wave approximation (RWA). This does not deny energy conservation in the complete system; it omits processes far from resonance.

Omitting the constant zero-point energy, which does not affect the dynamics, and choosing the basis phases so that $g>0$ gives the Jaynes--Cummings (JC) model~\cite{ch4:jc}:
\begin{equation}
 \frac{\hat H}{\hbar}=\omega_c\hat a^\dagger\hat a
 +\omega_0\hat\sigma_+\hat\sigma_-
 +g(\hat a\hat\sigma_++\hat a^\dagger\hat\sigma_-).
 \label{ch4:jc}
\end{equation}
The first term is the cavity-photon energy, with $\ket g$ and the vacuum $\ket 0$ chosen as the energy reference. The second term, $\hbar\omega_0\hat\sigma_+\hat\sigma_-=\hbar\omega_0\ket e\bra e$, is the atomic excitation energy. The third describes one-for-one exchange of quantum excitations between the atom and cavity mode. When the coupling is no longer much smaller than the carrier frequency, the RWA, the level truncation, and the complete light--matter model must be reexamined; one cannot simply increase $g$ arbitrarily in the present model.

Equation~\eqref{ch4:jc} has an important conservation law. Define the total excitation-number operator (photon number plus atomic excitation number):
\[
 \hat N=\hat a^\dagger\hat a+\hat\sigma_+\hat\sigma_-.
\]
Using $[\hat a^\dagger\hat a,\hat a]=-\hat a$ and $[\hat\sigma_+\hat\sigma_-,\hat\sigma_+]=\hat\sigma_+$, we can show that $[\hat N,\hat a\hat\sigma_+]=0$, and hence
\[
 [\hat H,\hat N]=0.
\]
Thus the total excitation number is conserved in the undriven, lossless JC model. If there is initially only one excitation, for example $\ket{e,0}$, subsequent evolution involves only the two-dimensional subspace spanned by $\ket{e,0}$ and $\ket{g,1}$; it never enters $\ket{g,0}$ or the two-excitation subspace. The original infinite-dimensional problem becomes a $2\times2$ matrix problem.

\subsection{Connection to the Semiclassical Rabi Drive of Chapter 2}

Suppose the cavity contains a strong coherent field in the state $\ket\alpha$, with mean photon number $\bar n=|\alpha|^2\gg 1$, rather than a single photon. If exchanging a few photons does not appreciably change the drive over the time considered, and atom--field correlations are negligible, the field operator $\hat a$ in Eq.~\eqref{ch4:jc} can be replaced approximately by its classical expectation value $\alpha$. The interaction then reduces to an external drive acting only on the atom:
\[
 \hat H_{\rm int}\simeq\hbar g(\alpha\hat\sigma_++\alpha^*\hat\sigma_-).
\]
Comparing its amplitude with the semiclassical rotating-wave Hamiltonian $-\frac{\hbar}{2}(\Omega\hat\sigma_++\Omega^*\hat\sigma_-)$ of Chapter 2 gives the relation between the semiclassical Rabi frequency $|\Omega|$ and single-photon coupling rate $g$:
\begin{equation}
 |\Omega|=2g|\alpha|=2g\sqrt{\bar n}.
 \label{ch4:classicaldrive}
\end{equation}
Alternatively, substituting the coherent-state complex peak electric field $\mathbf E_0=2\ii\boldsymbol{\mathcal E}(\mathbf r_0)\alpha$ from Eq.~(3.18) of Chapter 3 into $\Omega=\mathbf d_{eg}\cdot\mathbf E_0/\hbar$ gives $\Omega=-2g\alpha$, whose magnitude agrees with Eq.~\eqref{ch4:classicaldrive}.

This semiclassical relation cannot be extrapolated to vacuum. For $\alpha=0$, the classical drive is indeed zero, but an excited atom still couples to the vacuum cavity mode because $\hat a^\dagger\ket0=\ket1$. The next section derives the corresponding oscillation frequency $2g$ directly from the joint-state equations.

\section{One Excitation Exchanged Between Atom and Cavity}
\subsection{Vacuum Rabi Oscillations: Solving the State Equations}

We now solve the time evolution from $\ket{\Psi(0)}=\ket{e,0}$: an excited atom and a cavity in vacuum. Since the total excitation number $N=1$ is conserved, the joint state at any time must have the form of Eq.~\eqref{ch4:singleexcitation}, $\ket{\Psi(t)}=c_e(t)\ket{e,0}+c_c(t)\ket{g,1}$. Using $\hat a^\dagger\hat\sigma_-\ket{e,0}=\ket{g,1}$ and $\hat a\hat\sigma_+\ket{g,1}=\ket{e,0}$, substituting $\ket{\Psi(t)}$ into the Schr\"odinger equation $\ii\hbar\partial_t\ket\Psi=\hat H\ket\Psi$ gives two first-order ordinary differential equations:
\begin{equation}
 \ii\frac{\dd}{\dd t}\begin{pmatrix}c_e\\c_c\end{pmatrix}
 =\begin{pmatrix}\omega_0&g\\g&\omega_c\end{pmatrix}
 \begin{pmatrix}c_e\\c_c\end{pmatrix}.
 \label{ch4:twoamplitudes}
\end{equation}
Write the atomic detuning relative to the cavity as $\delta=\omega_0-\omega_c$. Factoring out the common mean-frequency phase $\ee^{-\ii(\omega_0+\omega_c)t/2}$, and still denoting the amplitudes in the rotating frame by $c_e,c_c$, gives a symmetric form:
\begin{equation}
 \ii\frac{\dd}{\dd t}\begin{pmatrix}c_e\\c_c\end{pmatrix}
 =M\begin{pmatrix}c_e\\c_c\end{pmatrix},\qquad
 M=\begin{pmatrix}\delta/2&g\\g&-\delta/2\end{pmatrix}.
 \label{ch4:rotmatrix}
\end{equation}
Mathematically, Eq.~\eqref{ch4:rotmatrix} is identical to the coupled-mode equations for two coupled microwave resonators. The physical distinction is that $c_e(t)$ and $c_c(t)$ here are probability amplitudes of a joint quantum state. Their squared magnitudes $|c_e|^2$ and $|c_c|^2$ give the probabilities of finding the quantum excitation in the atom or cavity mode upon measurement.

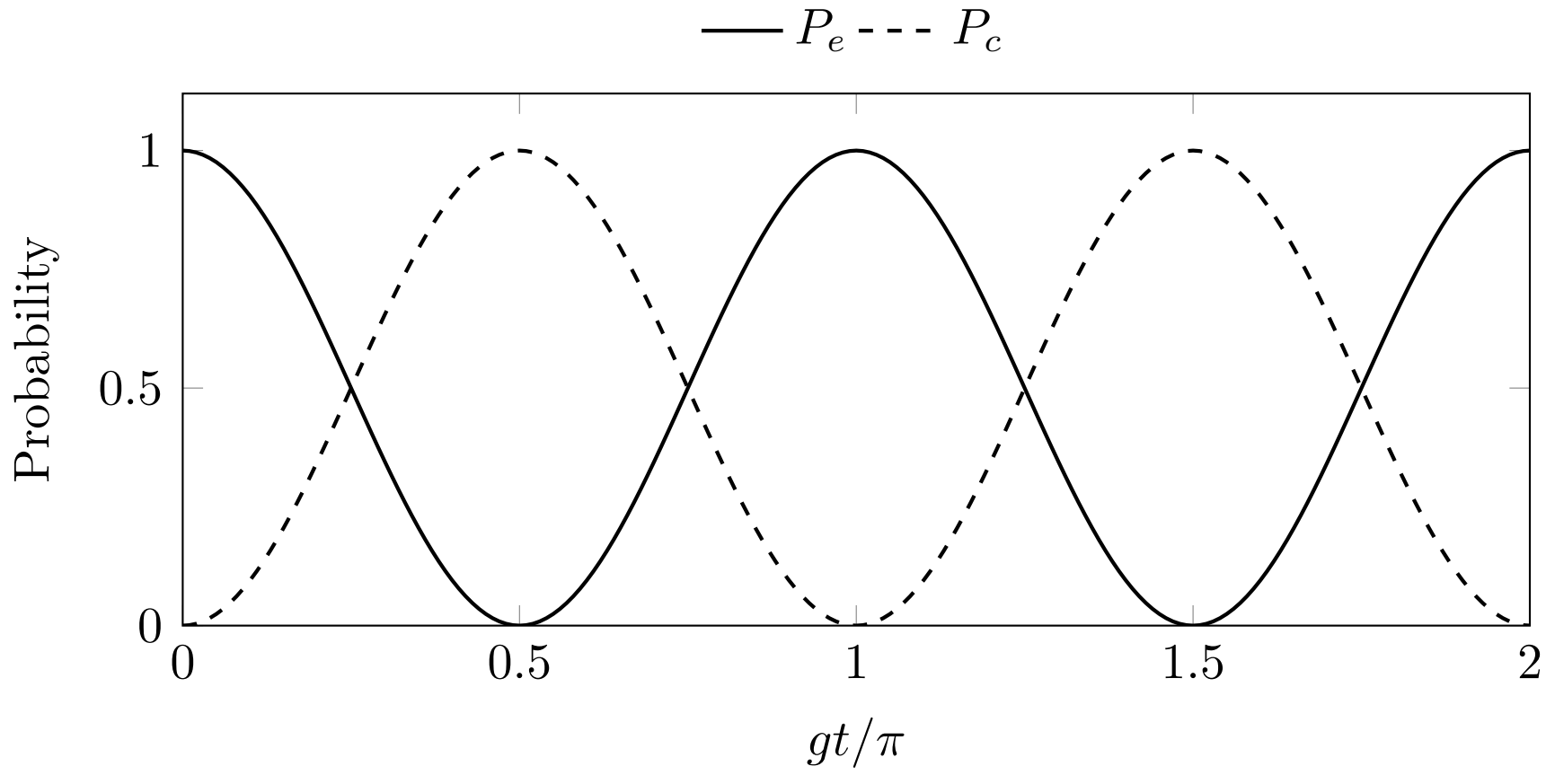
\begin{figure}[htbp]
\centering
\begin{tikzpicture}[font=\small]
\begin{axis}[width=11.5cm,height=5.5cm,xmin=0,xmax=2,ymin=0,ymax=1.12,
 xlabel={$gt/\pi$},ylabel={Probability},xtick={0,0.5,1,1.5,2},ytick={0,0.5,1},
 tick label style={font=\small},legend style={draw=none,at={(.5,1.04)},anchor=south,legend columns=2}]
\addplot[black,thick,domain=0:2,samples=201]{cos(deg(pi*x))^2};\addlegendentry{$P_e$}
\addplot[black,dashed,thick,domain=0:2,samples=201]{sin(deg(pi*x))^2};\addlegendentry{$P_c$}
\end{axis}
\end{tikzpicture}
\caption{Vacuum Rabi oscillations on resonance without loss (Eq.~\eqref{ch4:vacuumrabi}). At $gt/\pi=0$, the system is in $\ket{e,0}$; at $gt/\pi=0.5$, it has fully converted to the intracavity single-photon state $\ket{g,1}$. At intermediate times, such as $gt/\pi=0.25$, the atom and cavity field are entangled.}
\label{ch4:figexchange}
\end{figure}

Define the generalized single-photon vacuum Rabi frequency $\Omega_1=\sqrt{\delta^2+4g^2}$. Because $M$ satisfies $M^2=(\Omega_1^2/4)I$, its matrix exponential can be expanded exactly as $\ee^{-\ii Mt}=I\cos(\Omega_1t/2)-\ii(2M/\Omega_1)\sin(\Omega_1t/2)$. Applying it to the initial condition $\binom{c_e(0)}{c_c(0)}=\binom{1}{0}$ gives the analytical solution:
\begin{equation}
 \begin{aligned}
 c_e(t)&=\cos\frac{\Omega_1t}{2}
 -\ii\frac{\delta}{\Omega_1}\sin\frac{\Omega_1t}{2},\\
 c_c(t)&=-\ii\frac{2g}{\Omega_1}\sin\frac{\Omega_1t}{2},\\
 P_c(t)&=|c_c(t)|^2=\frac{4g^2}{\delta^2+4g^2}
 \sin^2\frac{\Omega_1t}{2},\qquad P_e(t)=1-P_c(t).
 \end{aligned}
 \label{ch4:exchange}
\end{equation}
Since this subspace contains at most 1 photon, $P_c(t)$ is also the mean cavity photon number $\avg{\hat a^\dagger\hat a}$. In particular, at exact atom--cavity resonance ($\delta=0$, $\Omega_1=2g$),
\begin{equation}
 \ket{\Psi(t)}=\cos(gt)\ket{e,0}-\ii\sin(gt)\ket{g,1},
 \quad P_e(t)=\cos^2(gt),\quad P_c(t)=\sin^2(gt).
 \label{ch4:vacuumrabi}
\end{equation}
At $t=0$, the energy is in the atom. At $t=\pi/(4g)$, the system reaches the maximally entangled state $(\ket{e,0}-\ii\ket{g,1})/\sqrt2$. At $t=\pi/(2g)$, the atom returns to its ground state with unit probability, creating exactly 1 photon in the cavity, $\ket{g,1}$. After another $\pi/(2g)$, the atom reabsorbs the photon and returns to $\ket{e,0}$. This exchange at angular frequency $2g$, with no external driving field and only a single excitation oscillating between the atom and an initially empty cavity mode, is called vacuum Rabi oscillation (Figure~\ref{ch4:figexchange}).

Throughout vacuum Rabi oscillation, $\avg{\hat a}=\avg{\hat\sigma_-}=0$, so both the mean electric field and mean dipole moment vanish, yet energy exchange continues. The joint state gives
\[
 \avg{\hat a^\dagger\hat\sigma_-}=c_ec_c^*
 =\ii\cos(gt)\sin(gt).
\]
The equation of motion then gives
\[
 \frac{\dd P_e}{\dd t}
 =-2g\operatorname{Im}\avg{\hat a^\dagger\hat\sigma_-}
 =-g\sin(2gt),
\]
in agreement with direct differentiation of $P_e=\cos^2(gt)$.

The exchange is controlled by atom--field cross correlations. Approximating $\avg{\hat a^\dagger\hat\sigma_-}$ by $\avg{\hat a^\dagger}\avg{\hat\sigma_-}$ would predict zero exchange. Semiclassical mean-field equations without additional spontaneous-emission or noise terms cannot produce vacuum Rabi oscillations from this initial state. The Bloch equations of Chapter 2, with empirical decay rates, can describe atomic population decay, but do not retain the atom--single-mode-field correlations present here.

\subsection{Dressed-State Energies and Vacuum Rabi Splitting}

This section discusses the coupled energy levels and resonant exchange. Readers interested in using frequency shifts at large detuning to read out the atomic state can turn next to Further Reading 4B.

\begin{figure}[htbp]
\centering
\begin{tikzpicture}[font=\small]
\begin{axis}[width=10.8cm,height=6cm,xmin=-4,xmax=4,ymin=-2.5,ymax=2.5,
 xlabel={$\delta/g$},ylabel={Shift from central frequency $/g$},xtick={-4,-2,0,2,4},ytick={-2,-1,0,1,2}]
\addplot[black!55,dashed,domain=-4:4]{x/2};
\addplot[black!55,dashed,domain=-4:4]{-x/2};
\addplot[black,thick,domain=-4:4,samples=121]{sqrt(x*x+4)/2};
\addplot[black,thick,domain=-4:4,samples=121]{-sqrt(x*x+4)/2};
\draw[<->] (axis cs:0,-1)--node[right]{$2$}(axis cs:0,1);
\end{axis}
\end{tikzpicture}
\caption{Avoided crossing of the single-excitation eigenfrequencies. The vertical axis subtracts the central frequency and is normalized by $g$; the separation of $2$ shown corresponds to an actual angular-frequency difference of $2g$. Dashed lines are the uncoupled atomic and cavity frequencies; solid lines are the two coupled eigenfrequency branches.}
\label{ch4:figanticross}
\end{figure}
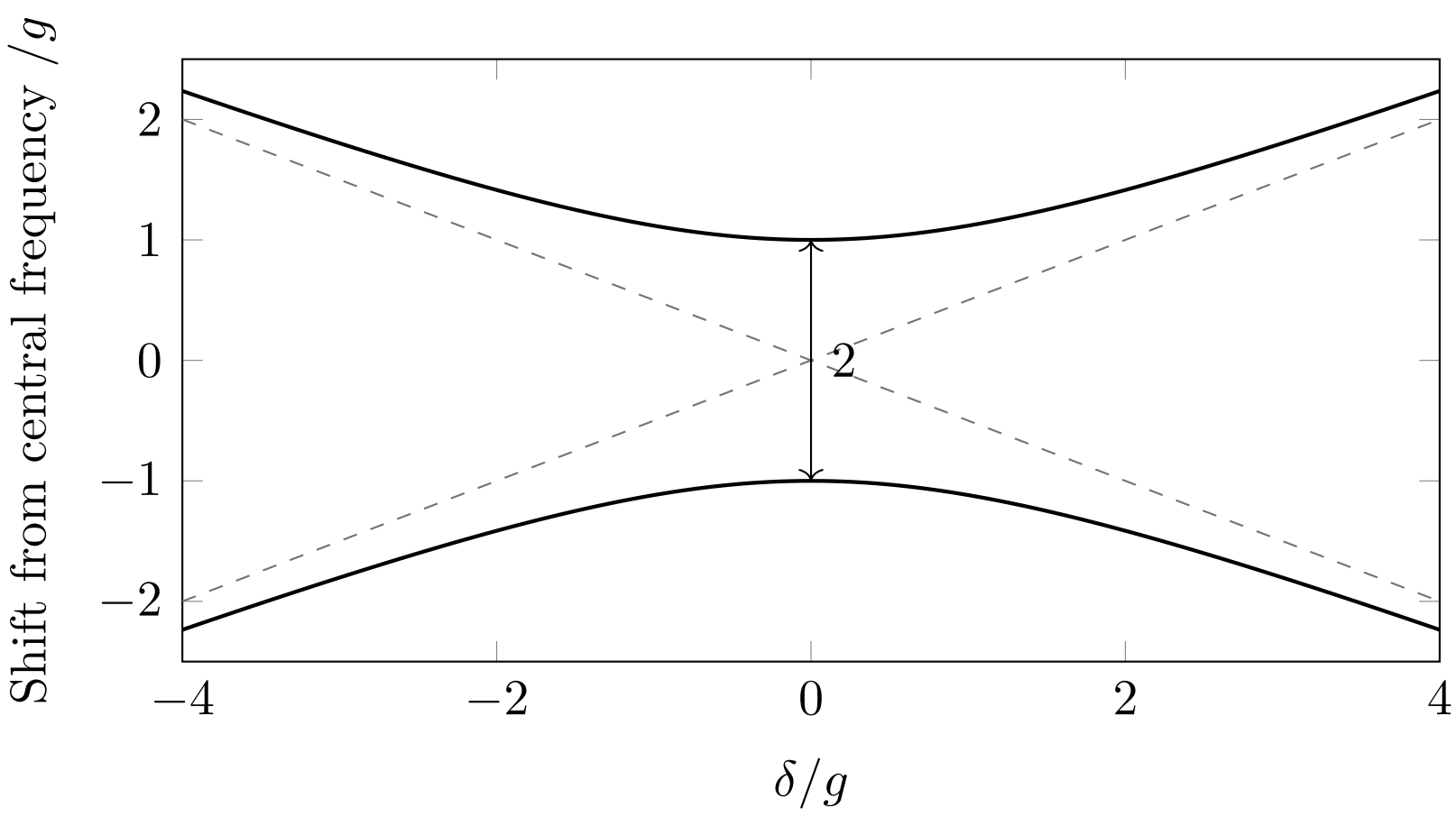

The eigenvalues of the $2\times 2$ coupling matrix in Eq.~\eqref{ch4:twoamplitudes} give the two coupled eigenenergies of the single-excitation subspace ($N=1$):
\begin{equation}
 E_{1,\pm}=\hbar\left[
 \frac{\omega_0+\omega_c}{2}
 \pm\frac12\sqrt{\delta^2+4g^2}\right].
 \label{ch4:dressedone}
\end{equation}
On resonance ($\delta=0$), the corresponding eigenstates are called dressed states. Their mixed field--matter structure resembles that of the polariton modes in Further Reading 3A, but the matter component here is a two-level atom:
\[
 \ket{1,\pm}=\frac{\ket{e,0}\pm\ket{g,1}}{\sqrt2},
 \qquad E_{1,\pm}=\hbar(\omega_c\pm g).
\]
The dipole coupling splits the originally degenerate states $\ket{e,0}$ and $\ket{g,1}$ into an upper and a lower eigenstate separated in frequency by $2g$, called vacuum Rabi splitting. Figure~\ref{ch4:figanticross} shows the avoided crossing of the energy levels as the detuning $\delta$ varies.

More generally, suppose the initial state is $\ket{e,n}$: the atom is excited, the cavity already contains $n$ photons, and the total excitation number is $m=n+1$. This state couples only to $\ket{g,n+1}$. Since $\hat a^\dagger\ket n=\sqrt{n+1}\ket{n+1}$ and $\hat a\ket{n+1}=\sqrt{n+1}\ket n$, in the $2\times 2$ matrix above we need only replace the off-diagonal entry $g$ by $g\sqrt{n+1}$. On resonance, the oscillation angular frequency and eigenenergies in the $m=n+1$ excitation subspace are therefore
\begin{equation}
 \Omega_{n+1}=2g\sqrt{n+1},\qquad
 E_{m,\pm}=m\hbar\omega_c\pm\hbar g\sqrt{m}\quad(m=n+1\geq 1).
 \label{ch4:enhancedrabi}
\end{equation}
The factor $\sqrt{n+1}$ has three implications:
\begin{enumerate}
 \item For $n=0$, $\Omega_1=2g\neq 0$: emission does not require a photon to be present initially in the mode. The energy comes from the excited atom.
 \item For $n\geq 1$, the squared magnitude of the transition matrix element is proportional to $(n+1)$. When the golden rule applies, the transition rate carries this factor as well: the ``$n$'' represents stimulated emission induced by existing photons, and the ``$1$'' represents spontaneous emission.
 \item The splitting varies as $\pm\hbar g\sqrt{m}$, that is, as $1,\sqrt2,\sqrt3,\ldots$. This nonlinear dependence makes the JC ladder anharmonic. In comparison, two coupled linear oscillators can still be transformed into two independent normal modes, each with equally spaced quantum energy levels.
\end{enumerate}

\section{From Single-Mode Exchange to Spontaneous Emission in Open Space}
\label{ch4:section_emission}
\subsection{Why Open Space Usually Produces Irreversible Decay}
\label{ch4:sec_spont}

In an ideal lossless single-mode cavity, energy remains within the atom and cavity mode and can be exchanged repeatedly. In open free space, the atom couples simultaneously to many propagating modes. The emitted field leaves the atom's vicinity, and its frequency components accumulate different phases. With a broadband environment, weak coupling, and no appreciable feedback, the atom usually exhibits approximately irreversible decay. An open waveguide with a reflecting termination, a band edge, or long-delay feedback need not satisfy these conditions.

We now express this picture mathematically. For the initial state $\ket{e,\{0\}}$, with the atom excited and all electromagnetic modes in vacuum, expand the single-excitation joint state as
\[
 \ket{\Psi(t)}=c_e(t)\ket{e,\{0\}}
 +\sum_\mu c_\mu(t)\ket{g,1_\mu}.
\]
Here $\ket{1_\mu}$ has 1 photon in mode $\mu$ and vacuum in every other mode. During a time interval over which depletion is negligible ($c_e(t)\simeq1$), transform to the rotating frames of the respective free evolutions (the interaction picture). The probability amplitude of mode $\mu$ then satisfies $\dot c_\mu(t)\simeq -\ii g_\mu^*\ee^{\ii(\omega_\mu-\omega_0)t}$. Integrating and taking the squared magnitude gives the probability at time $t$ of emission into mode $\mu$:
\begin{equation}
 |c_\mu(t)|^2\simeq |g_\mu|^2
 \frac{4\sin^2[(\omega_\mu-\omega_0)t/2]}
 {(\omega_\mu-\omega_0)^2}.
 \label{ch4:finiteemission}
\end{equation}
As a function of $\omega_\mu$, Eq.~\eqref{ch4:finiteemission} is centered at the atomic frequency $\omega_0$, has main-lobe width $4\pi/t$ between its first two zeros and peak height $|g_\mu|^2 t^2$, and follows a $\mathrm{sinc}^2$ profile. As the evolution time $t$ increases, this spectral window narrows, allowing less frequency mismatch, while its area grows in proportion to $t$. Using the distributional identity $\lim_{t\to\infty}\frac{4\sin^2(xt/2)}{tx^2}=2\pi\delta(x)$, summing the emission probabilities over all modes $\mu$ and dividing by the time $t$ gives the total spontaneous-emission rate (Fermi's golden rule):
\begin{equation}
 \begin{aligned}
 \Gamma&\simeq\frac1t\sum_\mu|c_\mu(t)|^2
 \simeq2\pi\sum_\mu|g_\mu|^2\delta(\omega_0-\omega_\mu)
 =2\pi J(\omega_0),\\
 J(\omega)&\equiv\sum_\mu|g_\mu|^2\delta(\omega-\omega_\mu).
 \end{aligned}
 \label{ch4:goldenrule}
\end{equation}
This limit is used within an intermediate time window: $t$ must be longer than the environment memory time but shorter than the time for substantial excitation depletion. The approximation $c_e\simeq1$ cannot be applied to physically infinite times. Here $J(\omega)$ is the dipole-coupling spectral density of the electromagnetic environment (units: $\mathrm{s}^{-1}$), and $\delta(\omega-\omega_\mu)$ is the Dirac $\delta$ function. Equation~\eqref{ch4:goldenrule} shows that the spontaneous-emission rate $\Gamma$ is not an intrinsic constant of an isolated atom. At the atomic transition frequency $\omega_0$, it is determined directly by the environment coupling spectral density $J(\omega_0)$.

If the environment spectrum $J(\omega)$ is sufficiently broad and flat near $\omega_0$, so that its correlation memory time is much shorter than $\Gamma^{-1}$, then in each infinitesimal interval $\dd t$ the atom loses excited-state population at the constant rate $\Gamma$: $\dd P_e=-\Gamma P_e\dd t$. Integration gives the familiar exponential decay law:
\begin{equation}
 P_e(t)=\ee^{-\Gamma t},\qquad
 |c_e(t)|=\ee^{-\Gamma t/2}.
 \label{ch4:exponential}
\end{equation}
This is the microscopic quantum origin of the dissipation rate $\Gamma$ in the optical Bloch equations and Lindblad master equation of Chapter 2. (If a narrow resonance or photonic bandgap makes $J(\omega)$ vary sharply near $\omega_0$, exponential decay fails; see the full integral-equation derivation in Further Reading 4A.)

\subsection{Expressing the Spontaneous-Emission Rate Through the Dyadic Green Function}
\label{ch4:sec_green}

Enumerating individual modes is usually inconvenient in open or lossy structures. Chapter 3 expressed the vacuum-field spectral correlations through the imaginary part of the dyadic Green function, allowing the coupling spectral density to be obtained directly from the classical electromagnetic response.

Let the atomic transition dipole be $\mathbf d=\mathbf d_{ge}$. Contracting the vacuum electric-field correlation spectrum $\frac{\hbar\mu_0}{\pi}\omega^2\operatorname{Im}\mathbf G(\mathbf r_0,\mathbf r_0,\omega)$ from Eq.~(3.47) of Chapter 3 with the two transition dipoles, and dividing by $\hbar^2$, expresses the coupling spectral density $J(\omega)$ directly through the imaginary part of the Green function:
\begin{equation}
 J(\omega)=\frac{\mu_0\omega^2}{\pi\hbar}
 \mathbf d^*\cdot\operatorname{Im}\mathbf G
 (\mathbf r_0,\mathbf r_0,\omega)\cdot\mathbf d.
 \label{ch4:spectralgreen}
\end{equation}
Substituting into $\Gamma=2\pi J(\omega_0)$ in Eq.~\eqref{ch4:goldenrule} gives the electromagnetic decay rate in a reciprocal, passive, nonmagnetic linear background under weak-coupling and short-memory conditions~\cite{ch4:dung}:
\begin{equation}
 \Gamma_{\rm em}=\frac{2\mu_0\omega_0^2}{\hbar}
 \mathbf d^*\cdot\operatorname{Im}\mathbf G
 (\mathbf r_0,\mathbf r_0,\omega_0)\cdot\mathbf d.
 \label{ch4:gammagreen}
\end{equation}
The emitter is assumed here to lie in a local vacuum region. The Green function includes only the background structure; the same atom must not be included again through a dielectric polarizability. If the emitter is embedded in a continuum medium, a microscopic local-field model must be specified separately.

Comparing Eq.~\eqref{ch4:gammagreen} with the projected electric local density of states defined by Eq.~(3.51) of Chapter 3, $\mathcal D_{\mathbf e}(\mathbf r_0,\omega_0)=\frac{2\omega_0}{\pi c^2}\mathbf e_d\cdot\operatorname{Im}\mathbf G(\mathbf r_0,\mathbf r_0,\omega_0)\cdot\mathbf e_d$, for a linearly polarized dipole $\mathbf d=|\mathbf d|\mathbf e_d$, gives another form of the spontaneous-emission rate:
\[
 \Gamma_{\rm em}=\frac{\pi\omega_0|\mathbf d|^2}{\varepsilon_0\hbar}\,\mathcal D_{\mathbf e}(\mathbf r_0,\omega_0).
\]
In free-space vacuum, substituting $\operatorname{Im}\mathbf G_0(\mathbf r_0,\mathbf r_0,\omega_0)=\frac{\omega_0}{6\pi c}\mathbf I$ from Chapter 3 and $\mu_0=1/(\varepsilon_0 c^2)$ gives the free-space spontaneous-emission rate (Einstein's $A$ coefficient):
\begin{equation}
 \Gamma_0=\frac{\omega_0^3|\mathbf d|^2}
 {3\pi\varepsilon_0\hbar c^3}.
 \label{ch4:gamma0}
\end{equation}
If lossy materials, such as a metal nanoantenna or a lossy substrate, surround the atom, the imaginary part $\operatorname{Im}\mathbf G$ includes both far-field radiation channels and channels in which material Ohmic loss converts energy into heat. The total decay rate therefore separates into
\begin{equation}
 \Gamma_{\rm em}=\Gamma_{\rm rad}+\Gamma_{\rm abs},\qquad
 \Gamma_{\rm tot}=\Gamma_{\rm em}+\Gamma_{\rm nr,int}.
 \label{ch4:ratedivision}
\end{equation}
Here $\Gamma_{\rm nr,int}$ is the emitter's internal nonradiative relaxation rate not already included in the background electromagnetic model. An important engineering point follows: placing an atom near a lossy metal surface can greatly shorten the total lifetime $1/\Gamma_{\rm tot}$ as $\operatorname{Im}\mathbf G$ increases sharply. But if the increase mainly enters material heating, $\Gamma_{\rm abs}$, a process called fluorescence quenching, the efficiency of radiated photons available for far-field collection, $\Gamma_{\rm rad}/\Gamma_{\rm tot}$, can deteriorate severely.

\subsection{Calculating \texorpdfstring{$\Gamma_{\rm em}$}{Γem} Directly from Classical Dipole Power}
\label{ch4:sec_power}

As shown by Eq.~\eqref{eq:ch3:imGpower} in Section~\ref{sec:ch3:dipolepower}, a point dipole at $\mathbf r_0$ with complex peak amplitude $\mathbf p_0$ supplies a total cycle-averaged real power $P_{\rm cl}=\frac{\omega_0}{2}\operatorname{Im}[\mathbf p_0^*\cdot\mathbf E_{\rm dip}(\mathbf r_0,\omega_0)]=\frac{\mu_0\omega_0^3}{2}\mathbf p_0^*\cdot\operatorname{Im}\mathbf G(\mathbf r_0,\mathbf r_0,\omega_0)\cdot\mathbf p_0$ in a classical electromagnetic simulation. Taking the test dipole directly as $\mathbf p_0=\mathbf d_{ge}$ and comparing with Eq.~\eqref{ch4:gammagreen} gives
\begin{equation}
 \begin{gathered}
 \Gamma_{\rm em}=\frac{4P_{\rm cl}(\mathbf p_0=\mathbf d_{ge})}{\hbar\omega_0},
 \qquad \frac{\Gamma_{\rm em}}{\Gamma_0}=\frac{P_{\rm cl}}{P_{\rm cl,0}},\\
 \frac{\Gamma_{\rm rad}}{\Gamma_0}=\frac{P_{\rm rad}}{P_{\rm cl,0}},
 \qquad \frac{\Gamma_{\rm abs}}{\Gamma_0}=\frac{P_{\rm abs}}{P_{\rm cl,0}}.
 \end{gathered}
 \label{ch4:classicalpower}
\end{equation}
Here $P_{\rm cl}$ is $P_{\rm tot}$ of Chapter 3, and $P_{\rm cl,0}$ is the power supplied by the same test dipole in free space, denoted $P_0$ in Chapter 3.

The factor $4$ comes from the two amplitude conventions. Classical power uses a complex peak dipole moment, whereas $\mathbf d_{ge}$ is a transition matrix element. Choosing the classical test source as $\mathbf p_0=2\mathbf d_{ge}$ increases its power fourfold, so we may also write $\Gamma_{\rm em}=P_{\rm cl}(2\mathbf d_{ge})/(\hbar\omega_0)$. This only calibrates an equivalent source: $2\mathbf d_{ge}$ is not the mean dipole moment of an excited atom. That mean is zero, and even a general two-level superposition satisfies $|\rho_{eg}|\leq1/2$.

Under these approximations and this source model, conventional electromagnetic solvers can calculate the ratios. Place the same test dipole at the same position, frequency, and orientation, and calculate the radiated power and material-absorption power in the structure, together with the free-space reference power. The radiative efficiency must still be calculated as $\Gamma_{\rm rad}/\Gamma_{\rm tot}$; any internal nonradiative processes of the emitter must be included separately. A PML only absorbs waves leaving the computational domain and must not be counted as absorption in the device material.

For multiple atoms, Green functions between different positions are also needed to describe coherent exchange and collective decay. Further Reading 4C develops the corresponding model.

\section{Exchange, Enhancement, and Photon Collection in a Lossy Cavity}
\label{ch4:sec_loss}
Cavity quantum electrodynamics (cavity QED) describes the dynamics of both the cavity field and quantum matter. The same resonant structure can also enter the atom's environmental description through a Green function; the difference is whether the cavity mode is retained explicitly. The mode should be retained when the intracavity state, atom--cavity correlations, or repeated excitation exchange is needed. If only the slower atomic evolution is of interest and the cavity response can adjust rapidly to it, the mode can be eliminated under suitable approximations and its effect represented by atomic decay and frequency shifts. Equation~\eqref{ch4:purcellrate} below gives the result in the single-excitation, fast-cavity limit without pure dephasing.

This choice also depends on detuning, the time scale of interest, and the frequency structure of the coupling spectrum; comparing $g$ with $\kappa$ alone is insufficient. If appreciable memory remains after cavity elimination, retain a memory kernel or reintroduce appropriate field degrees of freedom, as discussed in Further Reading 4A. In either description, each cavity channel must be counted only once. When the mode is explicit, its effect is described by coupling $g$ and cavity loss $\kappa$, while the atomic background decay $\Gamma_{\rm b}$ includes only the remaining channels.

\subsection{Assigning Each Loss Parameter: The Cavity QED Master Equation}

\begin{figure}[htbp]
\centering
\begin{tikzpicture}[font=\footnotesize,>=Stealth]
\node[draw,minimum width=2cm,minimum height=1.1cm,align=center] (atom) at (0,0) {Two-level atom\\$\omega_0$};
\node[draw,minimum width=2.4cm,minimum height=1.1cm,align=center] (cav) at (4.2,0) {Cavity mode $\hat a$\\$\omega_c$};
\draw[<->,thick] (atom)--node[above]{$g$}(cav);
\draw[->] (atom.south)--(0,-1.6) node[below,align=center]{Other atomic decay\\$\Gamma_{\rm b}$};
\draw[->] (cav.south)--(4.2,-1.6) node[below,align=center]{Other cavity loss\\$\kappa_{\rm i}$};
\draw[->,thick] (cav.east)--node[above=4pt,pos=0.53]{Target port $\kappa_{\rm ex}$}(8.1,0);
\draw[->] (atom.north east) to[out=70,in=110,looseness=4] node[above,align=center]{Pure dephasing $\gamma_\phi$} (atom.north west);
\end{tikzpicture}
\caption{Channels in an open single-atom, single-mode cavity QED system. $g$ describes reversible coherent exchange; $\Gamma_{\rm b}$ and $\kappa_{\rm i}$ are uncollected loss channels; $\kappa_{\rm ex}$ is the useful output into the target waveguide or detector. $\gamma_\phi$ denotes pure dephasing without population transfer.}
\label{ch4:figports}
\end{figure}
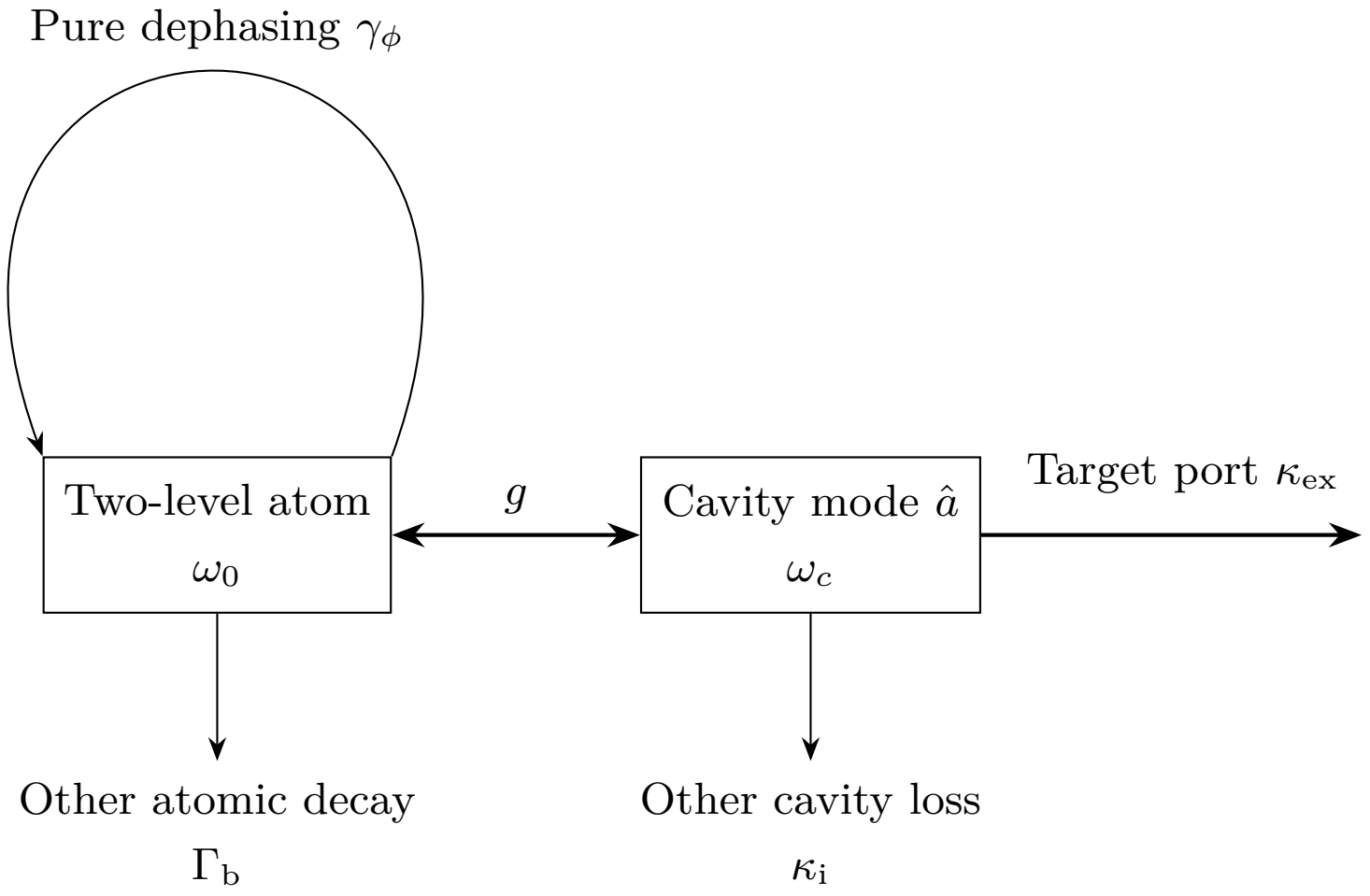

We now retain one principal cavity mode $\hat a$ explicitly and put the remaining loss channels that satisfy the Markov approximation into Lindblad dissipators. As shown in Figure~\ref{ch4:figports}, four rate parameters must be clearly distinguished:
\begin{itemize}
 \item $g$: the rate of coherent amplitude exchange between the atom and principal cavity mode (units: $\mathrm{s}^{-1}$).
 \item $\kappa=\kappa_{\rm ex}+\kappa_{\rm i}$: the total cavity photon-number (energy) decay rate, corresponding to the loaded quality factor $Q=\omega_c/\kappa$. Here $\kappa_{\rm ex}$ is the useful output coupling rate into the target measurement port, such as a coupled waveguide or fiber, and $\kappa_{\rm i}$ is the intrinsic internal loss rate from cavity-wall absorption, scattering, and similar processes.
 \item $\Gamma_{\rm b}$: atomic decay directly into free-space background radiation outside the cavity or through internal nonradiative processes, without passing through the principal cavity mode $\hat a$.
 \item $\gamma_\phi$: the atomic pure-dephasing rate.
\end{itemize}

Write the dissipative superoperator from Chapter 2 as $\Diss{L}\rho=L\rho L^\dagger-\frac12(L^\dagger L\rho+\rho L^\dagger L)$. In a zero-temperature environment, or vacuum at optical frequencies, the joint atom--cavity density matrix $\rho$ obeys
\begin{equation}
 \begin{aligned}
 \dot\rho={}&-\frac{\ii}{\hbar}[\hat H,\rho]
 +\Diss{\sqrt{\kappa_{\rm ex}}\hat a}\rho
 +\Diss{\sqrt{\kappa_{\rm i}}\hat a}\rho\\
 &+\Diss{\sqrt{\Gamma_{\rm b}}\hat\sigma_-}\rho
 +\Diss{\sqrt{\gamma_\phi/2}\hat\sigma_z}\rho.
 \end{aligned}
 \label{ch4:master}
\end{equation}
Thus $\Gamma_{\rm em}$ obtained from a total Green function containing the same cavity resonance cannot be used directly as $\Gamma_{\rm b}$ in Eq.~\eqref{ch4:master}.

Equation~\eqref{ch4:master} also assumes an approximately memoryless external environment and independently treatable channels. If microwave thermal occupation is appreciable, cavity loss must include downward and upward transitions, with operators $\sqrt{\kappa(\bar n_T+1)}\hat a$ and $\sqrt{\kappa\bar n_T}\hat a^\dagger$, respectively. The environment can no longer be treated as vacuum.

\subsection{Lossy Single-Excitation Dynamics and Conditions for Exchange}
\label{ch4:sec_damped}

Consider the undriven single-excitation problem with no pure dephasing ($\gamma_\phi=0$) and initial state $\ket{e,0}$. Once a cavity photon leaks out ($\kappa$) or atomic background decay occurs ($\Gamma_{\rm b}$), the system jumps to the zero-excitation ground state $\ket{g,0}$ and is not reexcited in this undriven model. We may therefore track the component in which no dissipative jump has yet occurred with the unnormalized single-excitation vector $\ket{\widetilde\Psi(t)}=c_e(t)\ket{e,0}+c_c(t)\ket{g,1}$. The full trace-preserving density matrix is then
\begin{equation}
 \rho(t)=\ket{\widetilde\Psi(t)}\bra{\widetilde\Psi(t)}
 +\bigl(1-|c_e(t)|^2-|c_c(t)|^2\bigr)\ket{g,0}\bra{g,0}.
 \label{ch4:nojumpstate}
\end{equation}
Substituting Eq.~\eqref{ch4:nojumpstate} into the master equation~\eqref{ch4:master} and moving to a frame rotating at the atomic frequency $\omega_0$ gives two first-order linear ordinary differential equations for the complex probability amplitudes:
\begin{equation}
 \frac{\dd}{\dd t}\begin{pmatrix}c_e\\c_c\end{pmatrix}
 =\begin{pmatrix}
 -\Gamma_{\rm b}/2&-\ii g\\
 -\ii g&-\kappa/2+\ii\delta
 \end{pmatrix}\begin{pmatrix}c_e\\c_c\end{pmatrix}\qquad (\delta=\omega_0-\omega_c).
 \label{ch4:dampedamplitudes}
\end{equation}
Multiply the first equation by $c_e^*$ and the second by $c_c^*$, add, and take twice the real part. The coupling terms in $g$ cancel exactly, giving the excitation-probability balance:
\begin{equation}
 \frac{\dd}{\dd t}(|c_e|^2+|c_c|^2)
 =-\Gamma_{\rm b}|c_e|^2-\kappa|c_c|^2.
 \label{ch4:balance}
\end{equation}

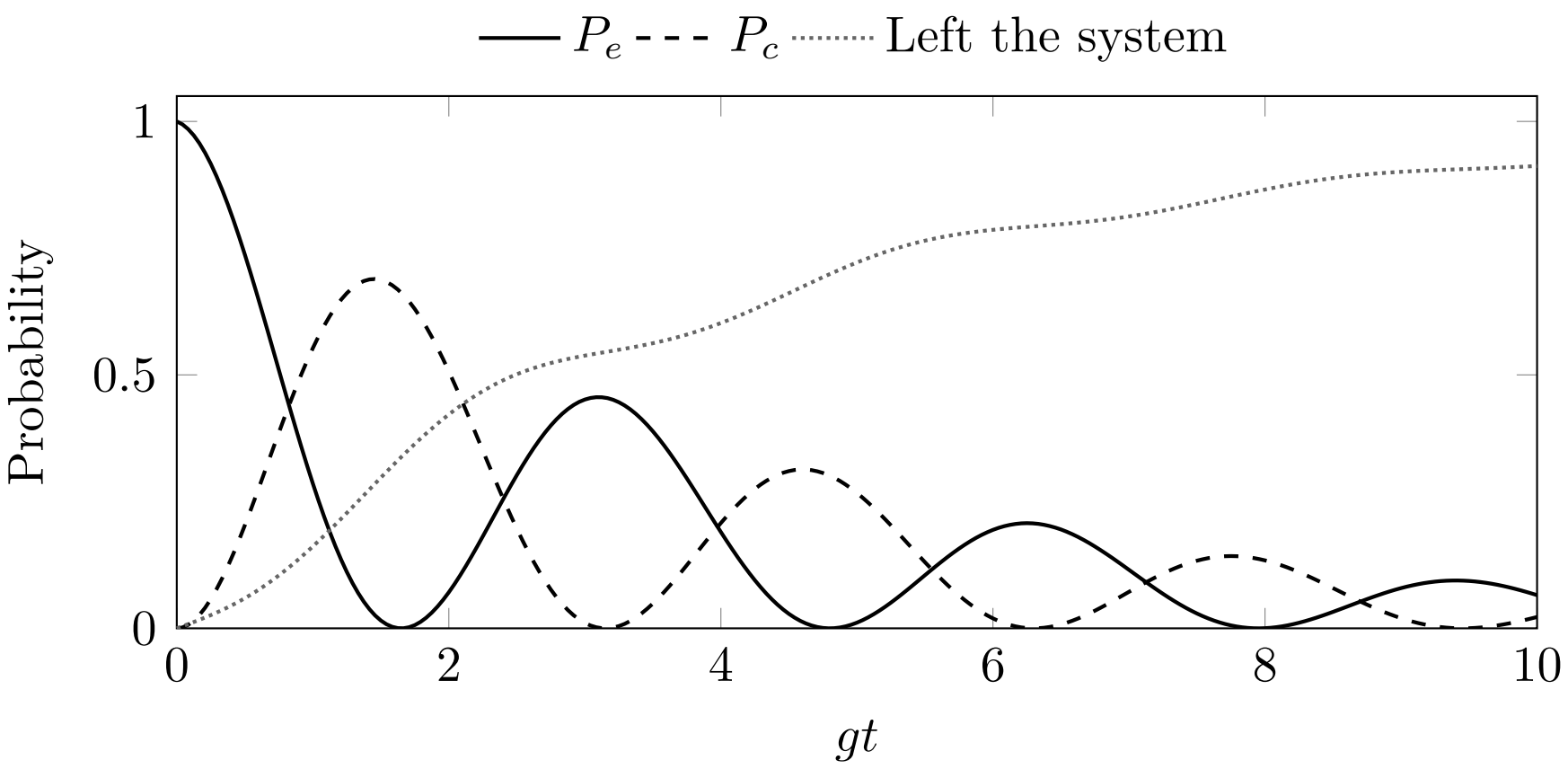
\begin{figure}[htbp]
\centering
\begin{tikzpicture}[font=\small]
\begin{axis}[width=11.6cm,height=5.5cm,xmin=0,xmax=10,ymin=0,ymax=1.05,
 xlabel={$gt$},ylabel={Probability},xtick={0,2,4,6,8,10},ytick={0,0.5,1},
 legend style={draw=none,at={(.5,1.03)},anchor=south,legend columns=3}]
\addplot[black,thick,domain=0:10,samples=241]{exp(-.25*x)*(cos(deg(sqrt(1-.075^2)*x))+.075/sqrt(1-.075^2)*sin(deg(sqrt(1-.075^2)*x)))^2};\addlegendentry{$P_e$}
\addplot[black,dashed,thick,domain=0:10,samples=241]{exp(-.25*x)/(1-.075^2)*sin(deg(sqrt(1-.075^2)*x))^2};\addlegendentry{$P_c$}
\addplot[black!60,densely dotted,thick,domain=0:10,samples=241]{1-exp(-.25*x)*((cos(deg(sqrt(1-.075^2)*x))+.075/sqrt(1-.075^2)*sin(deg(sqrt(1-.075^2)*x)))^2+sin(deg(sqrt(1-.075^2)*x))^2/(1-.075^2))};\addlegendentry{Left the system}
\end{axis}
\end{tikzpicture}
\caption{Single-excitation exchange with loss in the strong-coupling regime (Eq.~\eqref{ch4:dampedsolution}, with $\delta=0$, $\kappa/g=0.4$, $\Gamma_{\rm b}/g=0.1$, and $\gamma_\phi=0$). The atom and cavity undergo damped Rabi oscillations while the total excitation probability remaining in the system flows into external ports and the environment. The three curves sum identically to 1.}
\label{ch4:figdamped}
\end{figure}

On resonance ($\delta=0$), define the complex characteristic-root parameter
\[
 \nu=\sqrt{g^2-\left(\frac{\kappa-\Gamma_{\rm b}}{4}\right)^2},\qquad
 a_0=\frac{\kappa+\Gamma_{\rm b}}4.
\]
Solving Eq.~\eqref{ch4:dampedamplitudes} with $c_e(0)=1, c_c(0)=0$ gives the analytical solution:
\begin{equation}
 \begin{aligned}
 c_e(t)&=\ee^{-a_0t}
 \left[\cos(\nu t)+\frac{\kappa-\Gamma_{\rm b}}{4\nu}\sin(\nu t)\right],\\
 c_c(t)&=-\ii\frac g\nu\ee^{-a_0t}\sin(\nu t).
 \end{aligned}
 \label{ch4:dampedsolution}
\end{equation}
From $\nu$, the solution has oscillatory terms when $g>|\kappa-\Gamma_{\rm b}|/4$, and hyperbolic functions otherwise. At $\nu=0$, use the continuous limit $\sin(\nu t)/\nu\to t$. Oscillatory terms do not by themselves imply observable repeated exchange: the oscillation period must also be compared with the decay envelope (Figure~\ref{ch4:figdamped}).

A clear sufficient condition for multiple exchanges before substantial loss is $g\gg\kappa,\Gamma_{\rm b},\gamma_\phi$. Experimental criteria for strong coupling also depend on the measurement: no single inequality can replace both the condition for resolving time-domain exchange and that for resolving two spectral peaks. Another useful limit is a rapidly decaying cavity, $\kappa\gg g,\Gamma_{\rm b}$. After an initial transient of about $1/\kappa$, the cavity approximately follows the atom; the next subsection derives the resulting exponential decay.

Pure dephasing leaves the total excitation number unchanged but disrupts the phase relation between atom and field. For $\gamma_\phi\ne0$, the two-amplitude description is generally incomplete and the joint density matrix must be solved. Pure dephasing cannot simply be added to a population-loss rate.

\subsection{The Rapidly Decaying Cavity and the Purcell Enhancement Factor}

We now examine the rapidly decaying cavity limit, $\kappa\gg g,\Gamma_{\rm b}$. Because cavity damping is very fast, $c_c(t)$ follows changes in $c_e(t)$ almost instantaneously (adiabatic elimination). Setting $\dot c_c\simeq 0$ in the second equation of Eq.~\eqref{ch4:dampedamplitudes} gives the quasisteady relation
\[
 c_c(t)\simeq-\frac{\ii g}{\kappa/2-\ii\delta}\,c_e(t).
\]
Substituting this into the first equation of Eq.~\eqref{ch4:dampedamplitudes} and rationalizing the complex denominator gives
\[
 \begin{aligned}
 \dot c_e&\simeq-\left[\frac{\Gamma_{\rm b}}2+
 \frac{g^2}{\kappa/2-\ii\delta}\right]c_e\\
 &=-\left[\frac{\Gamma_{\rm b}+\Gamma_c(\delta)}2
 +\ii\delta\omega_{\rm cav}(\delta)\right]c_e.
 \end{aligned}
\]
The additional population decay rate and frequency shift introduced by the cavity are, respectively,
\begin{equation}
 \begin{aligned}
 \Gamma_c(\delta)&=\frac{g^2\kappa}{(\kappa/2)^2+\delta^2},\\
 \delta\omega_{\rm cav}(\delta)&=\frac{g^2\delta}{(\kappa/2)^2+\delta^2}.
 \end{aligned}
 \label{ch4:purcellrate}
\end{equation}
The first is Lorentzian and the second dispersive, analogous to the corrections to the real and imaginary parts of an input impedance produced by a coupled resonant circuit. Under this approximation, the atomic population decays at rate $\Gamma_{\rm b}+\Gamma_c$.

On atom--cavity resonance ($\delta=0$), the cavity-induced decay rate and dimensionless cooperativity $C$ are
\begin{equation}
 \Gamma_c(0)=\frac{4g^2}{\kappa},\qquad
 C\equiv\frac{\Gamma_c(0)}{\Gamma_{\rm b}}=\frac{4g^2}{\kappa\Gamma_{\rm b}}.
 \label{ch4:cooperativity}
\end{equation}
For the resonant fast-cavity limit without pure dephasing considered here, the cooperativity $C$ is the ratio of the atomic emission rate into the target cavity mode, $\Gamma_c(0)$, to the rate into all remaining background loss channels, $\Gamma_{\rm b}$.

For an isolated, weakly dissipative mode with negligible material dispersion, define the mode volume including projection along the dipole direction:
\[
 V_{\rm eff}=\frac{\int\varepsilon_r(\mathbf r)|\mathbf E_{\rm mode}(\mathbf r)|^2\dd^3r}
 {|\mathbf e_d\cdot\mathbf E_{\rm mode}(\mathbf r_0)|^2},
 \qquad \mathbf d=|\mathbf d|\mathbf e_d.
\]
Here we first take a real, linearly polarized transition dipole and an emitter in vacuum. The denominator already includes position and polarization matching to the mode.

Next, compare the resonant cavity-induced decay rate $\Gamma_c(0)=4g^2/\kappa$ with the free-space spontaneous-emission rate $\Gamma_0$ (Eq.~\eqref{ch4:gamma0}). Substituting $g^2=\frac{\omega_c|\mathbf d|^2}{2\hbar\varepsilon_0 V_{\rm eff}}$, $Q=\omega_c/\kappa$, and the resonant wavelength $\lambda_0=2\pi c/\omega_0$ gives the Purcell enhancement factor:
\begin{equation}
 F_c\equiv\frac{\Gamma_c(0)}{\Gamma_0}
 =\frac{3}{4\pi^2}\frac{\lambda_0^3}{V_{\rm eff}}Q,
 \qquad \lambda_0=\frac{2\pi c}{\omega_0}.
 \label{ch4:qv}
\end{equation}
Equation~\eqref{ch4:qv} describes the resonant contribution of a single cavity channel; the total decay still includes $\Gamma_{\rm b}$. Off resonance, the frequency factor in Eq.~\eqref{ch4:purcellrate} is also needed. Increasing $Q$ reduces $\kappa$, but once the fast-cavity approximation fails, $4g^2/\kappa$ can no longer represent a unidirectional emission rate. One must return to the coupled amplitudes or master equation. With strong dispersion, strong absorption, or overlapping modes, the simple lossless mode volume also ceases to apply; the full Green function and its coupling spectrum can be used instead. Appreciable environment memory additionally requires the dynamics described in Further Reading 4A.

\subsection{Entering the Cavity Is Not Reaching the Receiver: Optimizing Single-Photon Collection}

In practical single-photon sources and quantum sensors, we care not only about how fast the atom emits, but also about the fraction of photons that ultimately leave through the target coupling port $\kappa_{\rm ex}$. During undriven release of a single excitation, the instantaneous output photon flux at that port and the total collection efficiency (total probability of emission into the port) are
\begin{equation}
 \begin{aligned}
 \Phi_{\rm ex}(t)&=\kappa_{\rm ex}\avg{\hat a^\dagger\hat a}(t)
 =\kappa_{\rm ex}|c_c(t)|^2,\\
 \eta_{\rm ex}&=\int_0^\infty\Phi_{\rm ex}(t)\dd t
 =\kappa_{\rm ex}\int_0^\infty|c_c(t)|^2\dd t.
 \end{aligned}
 \label{ch4:collection}
\end{equation}

In the resonant fast-cavity limit without pure dephasing, the excitation enters the cavity mode with branching ratio $\beta_c=\frac{\Gamma_c}{\Gamma_{\rm b}+\Gamma_c}=\frac{C}{1+C}$, then leaves through the target port with fraction $\frac{\kappa_{\rm ex}}{\kappa}$. Thus $\eta_{\rm ex}\simeq\frac{\kappa_{\rm ex}}{\kappa}\frac{C}{1+C}$.

Without the fast-cavity approximation, but still within this chapter's JC model with Markovian losses, integrating Eq.~\eqref{ch4:dampedamplitudes} gives an exact total collection efficiency on resonance ($\delta=0$) and without pure dephasing ($\gamma_\phi=0$). Define $I_e=\int_0^\infty|c_e|^2\dd t$ and $I_c=\int_0^\infty|c_c|^2\dd t$. Integrating the probability balance~\eqref{ch4:balance} from $0$ to $\infty$ gives $\Gamma_{\rm b}I_e+\kappa I_c=1$. On the other hand, with $y=\operatorname{Im}(c_ec_c^*)$, Eq.~\eqref{ch4:dampedamplitudes} gives
\[
 \frac{\dd |c_c|^2}{\dd t}=-\kappa|c_c|^2+2gy,\qquad
 \dot y=g(|c_e|^2-|c_c|^2)
 -\frac{\kappa+\Gamma_{\rm b}}2y.
\]
Since both $|c_c|^2$ and $y$ vanish at the initial and final times, integrating the two equations separately from $0$ to $\infty$ gives $2g\int_0^\infty y\dd t=\kappa I_c$ and $g(I_e-I_c)=\frac{\kappa+\Gamma_{\rm b}}{2}\int_0^\infty y\dd t$. Eliminating $\int_0^\infty y\dd t$ and $I_e$ yields the exact expression for $\eta_{\rm ex}=\kappa_{\rm ex}I_c$:
\begin{equation}
 \begin{aligned}
 \eta_{\rm ex}&=\frac{\kappa_{\rm ex}}{\kappa+\Gamma_{\rm b}}
 \frac{4g^2}{4g^2+\kappa\Gamma_{\rm b}}\\
 &=\frac{\kappa_{\rm ex}}{\kappa+\Gamma_{\rm b}}\frac{C}{1+C},
 \qquad \kappa=\kappa_{\rm ex}+\kappa_{\rm i}.
 \end{aligned}
 \label{ch4:exactefficiency}
\end{equation}

\begin{figure}[htbp]
\centering
\begin{tikzpicture}[font=\small]
\begin{axis}[width=11.6cm,height=6cm,xmin=0,xmax=100,ymin=0,ymax=1.05,
 xlabel={$\kappa_{\rm ex}/(2\pi)\quad(\mathrm{MHz})$},ylabel={$\eta_{\rm ex}$},
 xtick={0,20,40,60,80,100},ytick={0,0.2,0.4,0.6,0.8,1}]
\addplot[black,thick,domain=0:100,samples=301]{x/(x+2)*400/(401+x)};
\addplot[only marks,mark=*,black,samples at={1,28.319604517,100}]{x/(x+2)*400/(401+x)};
\draw[black!55,densely dashed] (axis cs:28.319604517,0)--(axis cs:28.319604517,.870247777);
\node[anchor=south west] at (axis cs:30,.90) {Optimum $\eta_{\rm ex}\simeq0.870$};
\draw[->,black!60] (axis cs:32,.91)--(axis cs:28.319604517,.878);
\end{axis}
\end{tikzpicture}
\caption{Port collection efficiency versus output coupling rate, for $g/(2\pi)=10\,\mathrm{MHz}$ and $\Gamma_{\rm b}/(2\pi)=\kappa_{\rm i}/(2\pi)=1\,\mathrm{MHz}$, on resonance without pure dephasing. The curve is calculated from Eq.~\eqref{ch4:exactefficiency}; the optimum is approximately $28.3\,\mathrm{MHz}$, and the markers on either side correspond to $1$ and $100\,\mathrm{MHz}$. This is a model example, not a measured device curve.}
\label{ch4:figefficiency}
\end{figure}
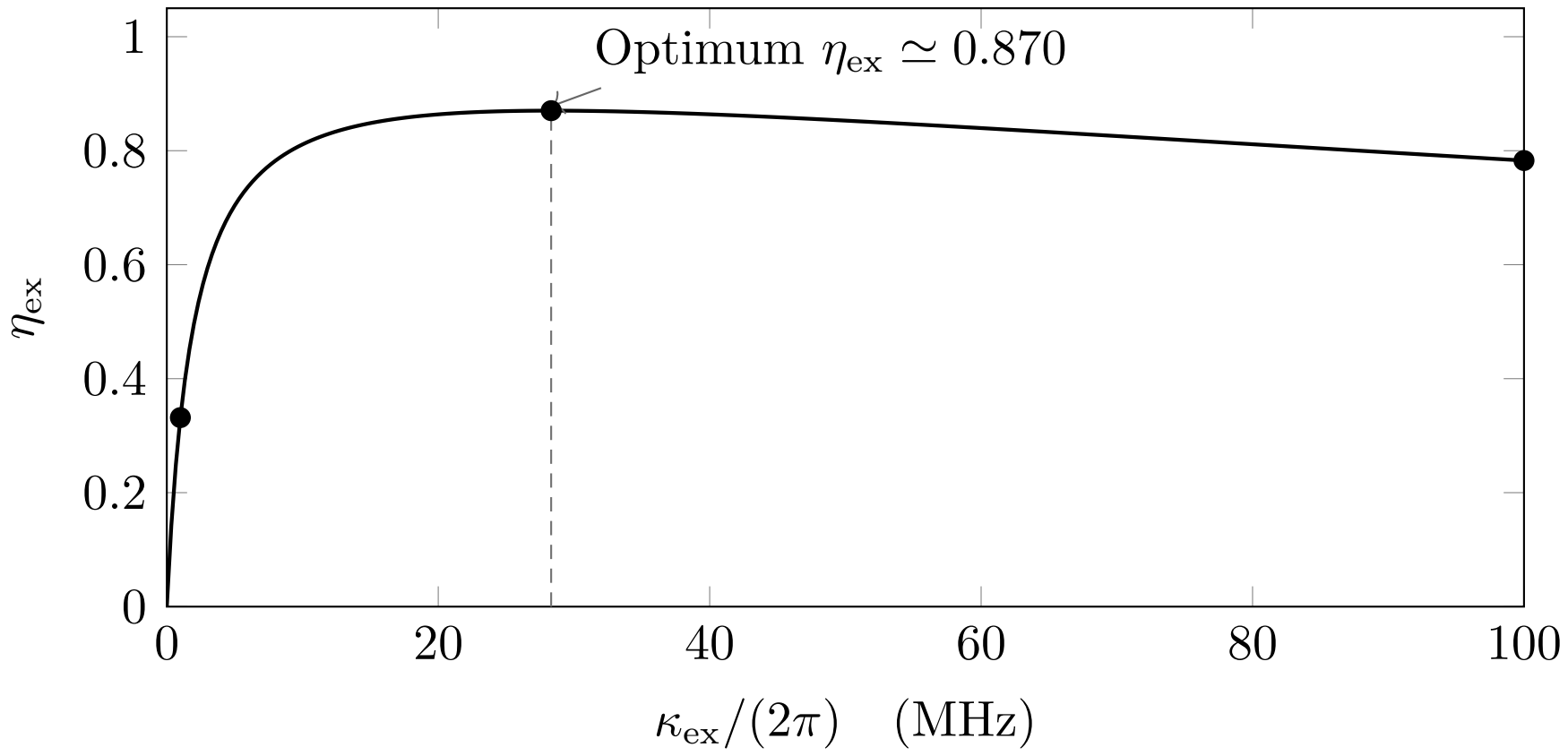

Equation~\eqref{ch4:exactefficiency} and Figure~\ref{ch4:figefficiency} reveal a central engineering design rule: the external port coupling rate $\kappa_{\rm ex}$ is optimal neither when made as large as possible nor when made as small as possible (highest $Q$):
\begin{itemize}
 \item If $\kappa_{\rm ex}$ is too small (weak output coupling), photons are trapped in the cavity and dissipated during repeated oscillations through intrinsic cavity loss $\kappa_{\rm i}$ and atomic background radiation $\Gamma_{\rm b}$. The first factor therefore tends to zero: $\frac{\kappa_{\rm ex}}{\kappa+\Gamma_{\rm b}}\to 0$.
 \item If $\kappa_{\rm ex}$ is too large (strong output coupling), excessive cavity damping reduces the cooperativity, $C=\frac{4g^2}{(\kappa_{\rm ex}+\kappa_{\rm i})\Gamma_{\rm b}}\to 0$. The effective cavity emission rate $4g^2/\kappa$ falls, and the excitation is more likely to be lost through atomic background channels.
 \item For fixed $g,\kappa_{\rm i}$ and $\Gamma_{\rm b}>0$, differentiating Eq.~\eqref{ch4:exactefficiency} with respect to $\kappa_{\rm ex}$ and setting the result to zero gives the optimal output coupling:
 \[
  \kappa_{\rm ex}^{\rm opt}=\sqrt{(\kappa_{\rm i}+\Gamma_{\rm b})\left(\kappa_{\rm i}+\frac{4g^2}{\Gamma_{\rm b}}\right)}.
 \]
 In the example of Figure~\ref{ch4:figefficiency}, $\kappa_{\rm ex}^{\rm opt}/(2\pi)=\sqrt{2\times 401}\approx 28.3\,\mathrm{MHz}$, giving a maximum collection efficiency of $87.0\%$.
\end{itemize}

\section{Driving, Detection, and Photon Correlations: Single-Photon Nonlinearity}
\label{ch4:section_driven}
\subsection{Preparing a Single-Photon Wavepacket on Demand}

The initial-state release process in Section~\ref{ch4:sec_loss} provides the basic prototype of a triggered single-photon source~\cite{ch4:reiserer}. An appropriate preparation pulse first places the atom in the excited state $\ket{e,0}$. The drive is then switched off, allowing this one quantum of excitation to enter the output waveguide through the cavity port $\kappa_{\rm ex}$. Without pure dephasing, the normalized temporal wavepacket of the output single-photon pulse, corresponding to $u(t)$ in Eq.~(3.27) of Chapter 3, follows directly from $c_c(t)$:
\begin{equation}
 \xi(t)=\frac{\sqrt{\kappa_{\rm ex}}\,c_c(t)}{\sqrt{\eta_{\rm ex}}},
 \qquad \int_0^\infty|\xi(t)|^2\dd t=1.
 \label{ch4:wavepacket}
\end{equation}
A common carrier phase is omitted here. In the ideal model of release after a single excitation, at most one photon is emitted per cycle. The target wavepacket mode is in the mixed state $\rho_{\rm out}=(1-\eta_{\rm ex})\ket0\bra0+\eta_{\rm ex}\ket{1_\xi}\bra{1_\xi}$. Its normalized second-order photon-number correlation is zero whenever $\eta_{\rm ex}>0$. This does not mean that a photon is received in every cycle: successful output still has probability $\eta_{\rm ex}$. If the preparation pulse continues after emission, it may reexcite the atom and produce several photons in one cycle. Pure dephasing also reduces the indistinguishability of wavepackets emitted in different cycles.

\subsection{Under Continuous Driving, First Identify the Detected Light}

Under continuous-wave excitation at angular frequency $\omega_L$, the system evolves to a nonequilibrium steady state and emits continuously. Define drive detunings from the atom and cavity as $\Delta_a=\omega_L-\omega_0$ and $\Delta_c=\omega_L-\omega_c$, respectively, so that $\Delta_c-\Delta_a=\omega_0-\omega_c=\delta$. Two common arrangements for continuous driving and detection must be distinguished carefully through their Hamiltonians and detection operators $\hat c$. Here $\Phi=\Tr(\hat c^\dagger\hat c\rho)$ has units of $\mathrm{s}^{-1}$, and $\hat c$ has units of $\mathrm{s}^{-1/2}$:
\begin{enumerate}
 \item Arrangement 1: drive the atom from the side and detect cavity transmission or atomic fluorescence. A classical field directly drives the atom with Rabi frequency $\Omega$. In a frame rotating at $\omega_L$, the effective Hamiltonian is
 \begin{equation}
  \begin{aligned}
  \frac{\hat H_r}{\hbar}={}&-\Delta_c\hat a^\dagger\hat a
  -\Delta_a\hat\sigma_+\hat\sigma_-\\
  &+g(\hat a\hat\sigma_++\hat a^\dagger\hat\sigma_-)
  -\frac12(\Omega\hat\sigma_++\Omega^*\hat\sigma_-).
  \end{aligned}
  \label{ch4:drivenhamiltonian}
 \end{equation}
 If output is detected at a cavity port $\kappa_{\rm ex}$ with no incident light, the output operator is $\hat c=\sqrt{\kappa_{\rm ex}}\hat a$. If instead atomic fluorescence is collected in free space without a cavity, away from the incident laser direction, with solid-angle collection rate $\Gamma_{\rm col}$, the output operator is $\hat c=\sqrt{\Gamma_{\rm col}}\hat\sigma_-$.
 \item Arrangement 2: inject coherent light $\beta_{\rm in}$ through a waveguide port and measure the port reflection or intracavity field. The incident photon flux is $|\beta_{\rm in}|^2$ (units: $\mathrm{s}^{-1}$). The drive now enters the cavity through the port. Set the side drive $\Omega=0$ in Eq.~\eqref{ch4:drivenhamiltonian} and add the port-drive term
 \[
  \hat H_{\rm port}=\ii\hbar\sqrt{\kappa_{\rm ex}}
  (\beta_{\rm in}\hat a^\dagger-\beta_{\rm in}^*\hat a).
 \]
 If there is an independent transmission port $\kappa_{\rm tr}$, replace the total cavity decay by $\kappa=\kappa_{\rm ex}+\kappa_{\rm tr}+\kappa_{\rm i}$ and include that channel in the master equation. The transmission detection operator is then $\hat c_{\rm tr}=\sqrt{\kappa_{\rm tr}}\hat a$. If reflection at the input port itself is measured, however, the Gardiner--Collett input--output relation, Eq.~(3.36) of Chapter 3, gives the reflected field as the coherent sum of direct reflection and cavity leakage:
 \[
  \hat c_{\rm out}=\beta_{\rm in}I-\sqrt{\kappa_{\rm ex}}\hat a.
 \]
 Reflected-light statistics must use the full $\hat c_{\rm out}$; intracavity correlations of $\hat a$ generally differ. The port dissipator in the master equation remains $\Diss{\sqrt{\kappa_{\rm ex}}\hat a}$. It cannot simply be replaced by the detection operator while leaving the Hamiltonian unchanged.
\end{enumerate}

\subsection{A Conditional-State Algorithm for the Steady-State Second-Order Correlation \texorpdfstring{$g^{(2)}(\tau)$}{g²(τ)}}
\label{ch4:sec_g2}

In a continuous steady state $\rho_{\rm ss}$ with nonzero mean photon flux, satisfying $\mathcal L[\rho_{\rm ss}]=0$, the delayed second-order correlation of the detected channel $\hat c$ is defined as
\begin{equation}
 g^{(2)}(\tau)=
 \frac{\avg{\hat c^\dagger(0)\hat c^\dagger(\tau)
 \hat c(\tau)\hat c(0)}_{\rm ss}}{\Phi_{\rm ss}^2},
 \qquad \Phi_{\rm ss}=\Tr(\hat c^\dagger\hat c\rho_{\rm ss}).
 \label{ch4:g2definition}
\end{equation}
How can the master equation directly evaluate this four-operator expectation value with delay $\tau$? Combining the Kraus measurement update of Further Reading 2A with the quantum regression theorem of Further Reading 2B gives a three-step procedure: update the state after detection, then evolve it~\cite{ch4:qutip}:
\begin{enumerate}
 \item Step 1: detection update at $\tau=0$. In the Markovian model, the quantum regression theorem and cyclicity of the trace write the numerator as $\Tr[\hat c^\dagger\hat c\,\ee^{\mathcal L\tau}(\hat c\rho_{\rm ss}\hat c^\dagger)]$. After the detector records the first photon at $\tau=0$, the system jumps (collapses) instantaneously from its steady state $\rho_{\rm ss}$ to the normalized conditional density matrix $\rho_{\rm det}(0)$:
 \begin{equation}
  \rho_{\rm det}(0)=
  \frac{\hat c\rho_{\rm ss}\hat c^\dagger}{\Tr(\hat c\rho_{\rm ss}\hat c^\dagger)}
  =\frac{\hat c\rho_{\rm ss}\hat c^\dagger}{\Phi_{\rm ss}}.
  \label{ch4:conditionalinitial}
 \end{equation}
 \item Step 2: evolve with the master equation for time $\tau$. Starting from $\rho_{\rm det}(0)$, use the original full Lindblad master equation to obtain $\rho_{\rm det}(\tau)$. Do not condition further on whether additional clicks occur in between. Using no-jump evolution instead would calculate a different quantity from the $g^{(2)}$ defined here.
 \item Step 3: read and normalize the conditional photon flux. At time $\tau$, the conditional photon flux is $\Phi(\tau\mid 0)=\Tr[\hat c^\dagger\hat c\rho_{\rm det}(\tau)]$. Divide it by the steady-state photon flux $\Phi_{\rm ss}$ to obtain $g^{(2)}(\tau)$:
 \begin{equation}
  \frac{\dd\rho_{\rm det}(\tau)}{\dd\tau}=\mathcal L[\rho_{\rm det}(\tau)],\qquad
  g^{(2)}(\tau)=
  \frac{\Tr[\hat c^\dagger\hat c\rho_{\rm det}(\tau)]}{\Phi_{\rm ss}}.
  \label{ch4:conditionalg2}
 \end{equation}
\end{enumerate}
$g^{(2)}(\tau)$ is the ratio of the conditional count rate, given a click at time zero, to the original steady-state count rate. The time-dependent operators in the definition use the Heisenberg picture. In the calculation, $\rho_{\rm ss}$ and $\hat c$ are fixed: construct the conditional initial state once, then evolve $\rho_{\rm det}$. The steady state itself does not change with $\tau$. This conditional matrix has unit trace and is positive semidefinite; its role differs from that of the traceless fluctuation auxiliary matrix in Further Reading 2B.

\subsection{Example: Antibunching in Resonance Fluorescence from One Two-Level Atom}

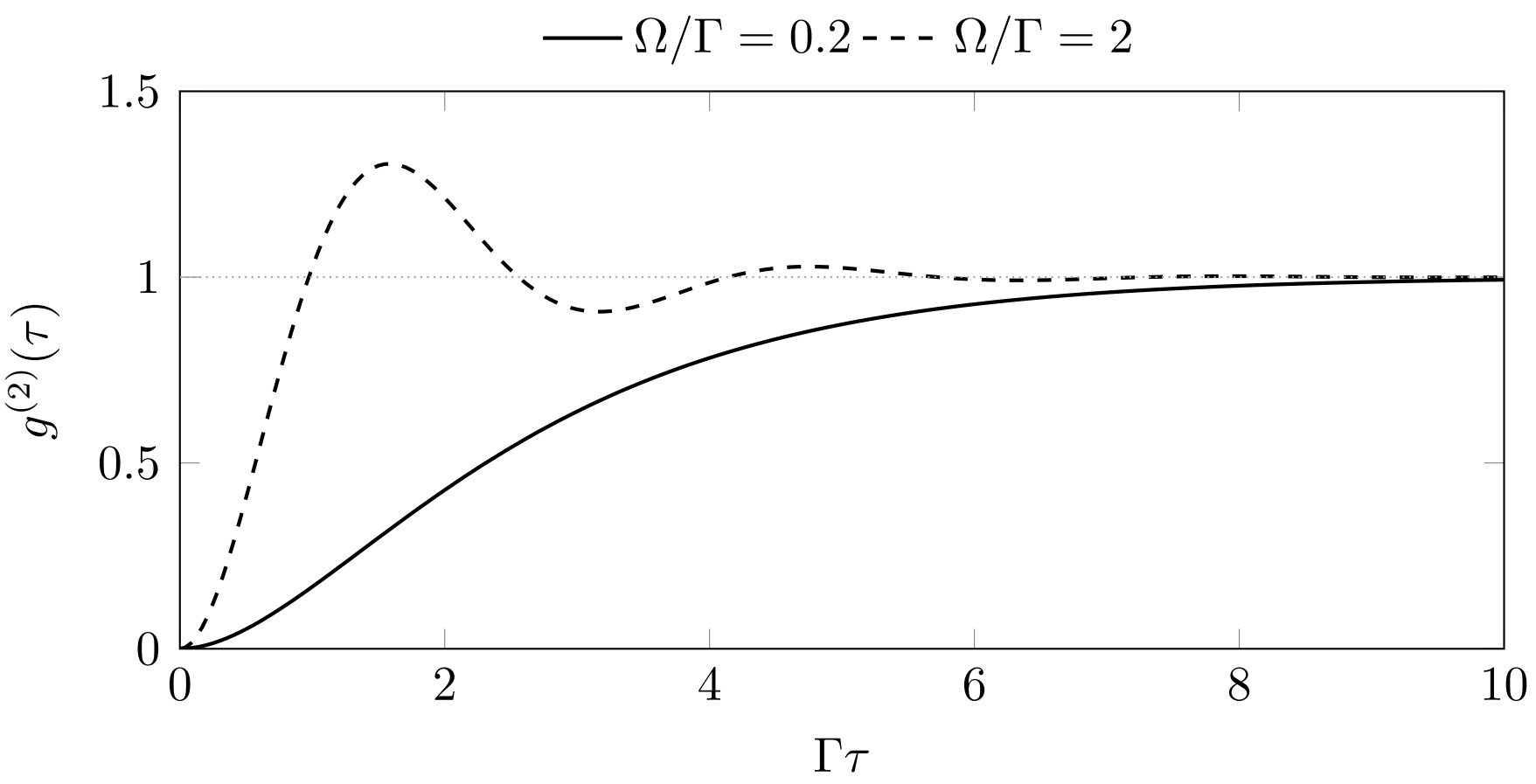
\begin{figure}[htbp]
\centering
\begin{tikzpicture}[font=\small]
\begin{axis}[width=11.6cm,height=5.8cm,xmin=0,xmax=10,ymin=0,ymax=1.5,
 xlabel={$\Gamma\tau$},ylabel={$g^{(2)}(\tau)$},xtick={0,2,4,6,8,10},ytick={0,0.5,1,1.5},
 legend style={draw=none,at={(.5,1.02)},anchor=south,legend columns=2}]
\addplot[black,thick,domain=0:10,samples=201]{1-exp(-.75*x)*((exp(.15*x)+exp(-.15*x))/2+5*(exp(.15*x)-exp(-.15*x))/2)};\addlegendentry{$\Omega/\Gamma=0.2$}
\addplot[black,dashed,thick,domain=0:10,samples=241]{1-exp(-.75*x)*(cos(deg(sqrt(4-.0625)*x))+.75/sqrt(4-.0625)*sin(deg(sqrt(4-.0625)*x)))};\addlegendentry{$\Omega/\Gamma=2$}
\addplot[black!35,densely dotted,domain=0:10]{1};
\end{axis}
\end{tikzpicture}
\caption{Second-order correlation $g^{(2)}(\tau)$ of resonance fluorescence from a single two-level atom (Eq.~\eqref{ch4:fluorescenceg2}, with ideal time resolution, no background, and no pure dephasing). For any nonzero drive, $g^{(2)}(0)=0$ at zero delay (photon antibunching). Strong driving reveals Rabi oscillations during reexcitation.}
\label{ch4:figg2}
\end{figure}

Apply the three-step procedure directly to the fluorescence correlation of a single two-level atom driven by a continuous laser in free space. Only side-scattered fluorescence is collected, so the detection operator is $\hat c=\sqrt{\Gamma_{\rm col}}\hat\sigma_-$.
\begin{itemize}
 \item Zero delay, $g^{(2)}(0)$: the square of a two-level lowering operator is identically zero, $\hat\sigma_-^2=\ket g\braket e g\bra e=0$, so at $\tau=0$,
 \begin{equation}
  g^{(2)}(0)=
  \frac{\avg{\hat\sigma_+^2\hat\sigma_-^2}_{\rm ss}}
  {\avg{\hat\sigma_+\hat\sigma_-}_{\rm ss}^2}=0.
  \label{ch4:antibunching}
 \end{equation}
 The reason is that detecting 1 fluorescence photon at $\tau=0$ collapses the conditional state~\eqref{ch4:conditionalinitial} immediately to the ground state, $\rho_{\rm det}(0)=\frac{\hat\sigma_-\rho_{\rm ss}\hat\sigma_+}{\bra e\rho_{\rm ss}\ket e}=\ket g\bra g$. An atom in its ground state cannot emit a second photon at that same instant: the external drive needs a finite time to reexcite it to $\ket e$. Consequently, $g^{(2)}(0)=0$ and $g^{(2)}(\tau)>g^{(2)}(0)$. This suppression of closely spaced emission events is called photon antibunching.
 \item Delayed recovery, $g^{(2)}(\tau)$: take a laser resonant with the atom ($\Delta_a=0$), positive real $\Omega$, and no pure dephasing. In the rotating frame, write the excited-state population as $p_e=\rho_{ee}$ and twice the imaginary part of the slowly varying coherence as $v=2\operatorname{Im}\tilde\rho_{eg}=2\operatorname{Im}s$. The optical Bloch equations of Chapter 2 then reduce to two real equations:
 \begin{equation}
  \begin{aligned}
  \dot p_e&=-\Gamma p_e+\frac\Omega2v,\\
  \dot v&=-\frac\Gamma2v+\Omega(1-2p_e),\qquad
  p_{e,\rm ss}=\frac{\Omega^2}{\Gamma^2+2\Omega^2}.
  \end{aligned}
  \label{ch4:fluorescencebloch}
 \end{equation}
 Detection at $\tau=0$ prepares the conditional initial ground state $\ket g\bra g$, giving $p_e(0)=0, v(0)=0$. Solving Eq.~\eqref{ch4:fluorescencebloch} from this initial condition yields the conditional excited-state population $p_e(\tau\mid\text{initial state }g)$. Substitution into Eq.~\eqref{ch4:conditionalg2} gives the correlation curve. The two real equations may also be integrated numerically:
 \begin{equation}
  g^{(2)}(\tau)=\frac{p_e(\tau\mid\text{initial state }g)}{p_{e,\rm ss}}.
  \label{ch4:fluorescenceg2}
 \end{equation}
\end{itemize}
As Figure~\ref{ch4:figg2} shows, under weak driving ($\Omega\ll\Gamma$), $g^{(2)}(\tau)$ rises monotonically from $0$ to $1$. Oscillatory terms appear for $\Omega>\Gamma/4$. With a sufficiently strong drive, the atom undergoes damped Rabi oscillations during reexcitation, making $g^{(2)}(\tau)$ oscillate about $1$ as it converges.

\subsection{Photon Blockade in a Cavity: The Difference from Linear Scattering in Chapter 3}

\begin{figure}[htbp]
\centering
\begin{tikzpicture}[font=\small,>=Stealth,x=1cm,y=0.72cm]
\draw[thick] (0,0)--(3.8,0) node[right]{$\ket{g,0}$};
\draw[thick] (0,3)--(3.8,3) node[right]{$\ket{1,-}$};
\draw[thick] (0,4.3)--(3.8,4.3) node[right]{$\ket{1,+}$};
\draw[thick] (0,6.6)--(3.8,6.6) node[right]{$\ket{2,-}$};
\draw[thick] (0,8.3)--(3.8,8.3) node[right]{$\ket{2,+}$};
\draw[densely dashed,black!55] (1.9,6)--(3.8,6);
\draw[->,thick] (0.8,0.12)--node[right]{$\hbar\omega_L$}(0.8,2.88);
\draw[->,thick,dashed] (2.5,3.12)--node[left,pos=.7]{$\hbar\omega_L$}(2.5,5.88);
\draw[<->] (3.5,6)--(3.5,6.6);
\node[anchor=west,align=left,font=\footnotesize,text width=4.2cm] at (6,5.15) {Second photon at the same frequency\\is detuned from the target level};
\draw[black!60] (3.6,6.3)--(5.3,6.3)--(5.3,5.15)--(5.8,5.15);
\end{tikzpicture}
\caption{Photon blockade in the JC ladder. The first lower-branch transition is resonant with the drive, whereas the second same-frequency excitation misses the next lower branch by energy $\hbar\Delta_{\rm anh}$. The dashed line marks the sum of two drive-photon energies. Vertical spacings are schematic; actual energies are given by Eq.~\eqref{ch4:jcladder}.}
\label{ch4:figblockade}
\end{figure}
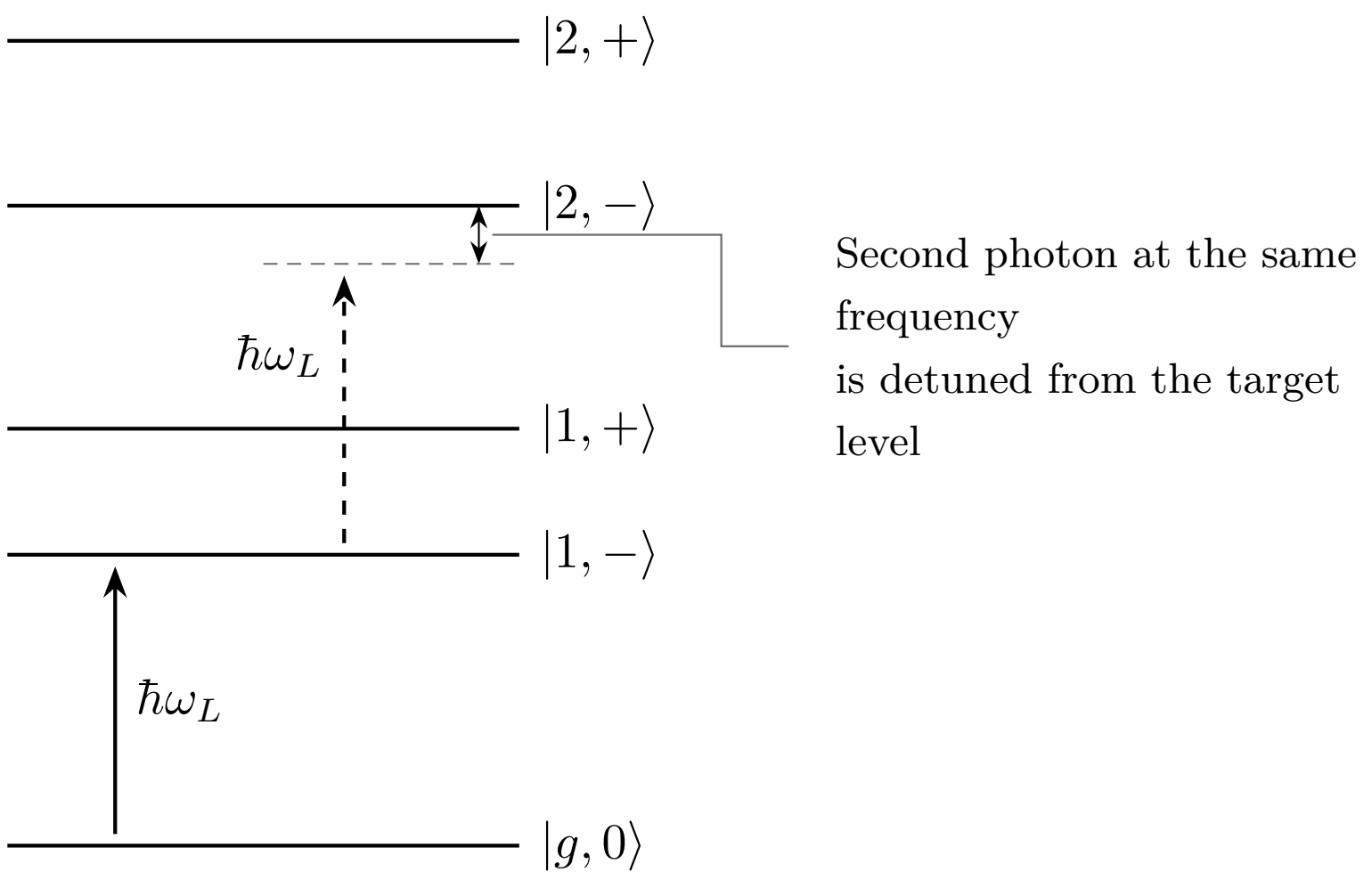

What happens when a two-level atom is placed in a strongly coupled cavity ($g\gg\kappa,\Gamma_{\rm b}$) illuminated through a port by a weak coherent laser? Equation~\eqref{ch4:enhancedrabi} shows that the dressed-state energy ladder of the resonant atom--cavity system is anharmonic, with unequal spacings:
\begin{equation}
 E_{m,\pm}=m\hbar\omega_c\pm\hbar g\sqrt m\qquad(m=1,2,\ldots).
 \label{ch4:jcladder}
\end{equation}
As Figure~\ref{ch4:figblockade} illustrates, tune the laser to the lower single-excitation state $\ket{1,-}$, so that $\omega_L=\omega_c-g$. The transition from the ground state to this lower branch is resonant with the incident light, although its excitation probability still depends on port coupling and loss. While the first photon remains in the cavity, a second photon at the same frequency, still with energy $\hbar\omega_L=\hbar(\omega_c-g)$, may attempt to excite the system to the lower two-excitation state $\ket{2,-}$. The required transition frequency, however, is $(E_{2,-}-E_{1,-})/\hbar=\omega_c-g(\sqrt2-1)$. The frequency mismatch, or anharmonic detuning, is
\begin{equation}
 \Delta_{\rm anh}=\frac{E_{2,-}-E_{1,-}}{\hbar}-\omega_L=g(2-\sqrt2)\approx 0.586g.
 \label{ch4:anharmonicity}
\end{equation}
When $\Delta_{\rm anh}$ exceeds the relevant linewidths and drive broadening, the second excitation is suppressed. A selected output channel may then satisfy $g^{(2)}(0)\ll1$, a phenomenon called conventional photon blockade~\cite{ch4:blockade}. This does not imply that the second photon is certainly reflected or that transmission has unit probability: reflection, transmission, and loss must still be calculated with a port model. Both photon blockade and Rydberg blockade use anharmonic levels to suppress multiple excitations, but their microscopic mechanisms differ: the former uses atom--cavity dressed levels, whereas the latter uses interactions between atoms.

The linear two-photon scattering of Chapter 3 combines fixed single-photon scattering amplitudes. Interference between indistinguishable paths can change coincidence counts even without effective photon--photon interactions. Here, the two-level atom saturates and the joint excitation spectrum is anharmonic, so fixed classical scattering coefficients generally cannot determine all two-photon processes.

Maxwell's equations remain applicable to the background structure in both cases. The difference is that this chapter also retains the joint state of a few matter levels and the field. If the input itself is a specified one- or two-photon wavepacket, its input modes must be described explicitly or a Fock-state input method used. A coherent drive with the same mean power is not a substitute.

\section{Putting the Model into Engineering Electromagnetic Calculations}
\subsection{Five Steps from Electromagnetic Simulation to a Fully Quantum Solution}

For a practical electromagnetic structure containing a quantum emitter---an atom, quantum dot, color center, or superconducting qubit---modeling and computation can follow this standard procedure:
\begin{enumerate}
 \item Extract classical electromagnetic parameters:\nopagebreak[4]
 \begin{itemize}
  \item If the background field is incorporated into the environment and the weak-coupling and Markov conditions of Section~\ref{ch4:section_emission} hold, use the point-dipole sources of Section~\ref{ch4:sec_power} and the power ratios in Eq.~\eqref{ch4:classicalpower} to calculate $\Gamma_{\rm rad},\Gamma_{\rm abs}$. With appreciable memory, retain the coupling spectrum over the relevant frequency band.
  \item If an isolated, weakly dissipative cavity mode with negligible material dispersion is retained explicitly, use an eigenmode solver to obtain $\omega_c$, the effective mode volume $V_{\rm eff}$, and the single-photon coupling rate $g$ (Eq.~\eqref{ch4:gdefinition}). Use port S parameters or eigenmode $Q$ values to separate $\kappa_{\rm ex}$ and $\kappa_{\rm i}$. More general structures require a field description consistent with their loss and dispersion.
 \end{itemize}
 \item Construct joint matrices. Choose a photon-number cutoff $N$ and retain $\ket0,\ldots,\ket N$. Use a matrix tensor-product function, such as \texttt{kron} in MATLAB or Python, to construct
 \[
  \hat a_{\rm joint}=I_2\otimes\hat a_N,\qquad
  \hat\sigma_{-,\rm joint}=\hat\sigma_-\otimes I_{N+1}.
 \]
 Then assemble the Hamiltonian matrix $\hat H$ and jump operators $L_k$. The joint matrix dimension is $2(N+1)$.
 \item Solve the dynamics or steady state. For transient pulses, solve the ordinary differential equation $\dot\rho=\mathcal L[\rho]$. For continuous-wave excitation, solve the linear algebraic system $\mathcal L[\rho_{\rm ss}]=0$ with the normalization constraint $\Tr\rho_{\rm ss}=1$. Use consistent time units. Values given as $g/(2\pi)$, $\kappa/(2\pi)$, or detunings in Hz must be multiplied by $2\pi$ and converted to consistent units. An inverse lifetime $1/T_1$ or $1/T_2$ is already a decay rate and must not be multiplied by $2\pi$ again.
 \item Calculate observables. Obtain the port photon flux from $\Phi=\Tr(\hat c^\dagger\hat c\rho)$ and the delayed second-order correlation $g^{(2)}(\tau)$ using the three-step procedure of Section~\ref{ch4:sec_g2}.
 \item Check physical consistency and numerical convergence. Verify trace, Hermiticity, positive semidefiniteness, and probability balance, then vary the time step, time window, and photon-number cutoff. For a continuously driven cavity, retaining only $\ket0,\ket1$ makes the truncated matrix satisfy $\hat a^2=0$ identically, producing an artificial zero second-order correlation. Allow at least two photons ($N\geq2$, meaning a cavity subspace of dimension at least 3 and a joint matrix of dimension at least 6), then increase $N$ further to check convergence. A small mean photon number does not replace a check of the second-order moment. For an undriven single-excitation initial-value problem, excitation-number conservation guarantees that the two-excitation subspace is never entered.
\end{enumerate}

\subsection{How Weak-Excitation Spectra Reduce to Classical Coupled-Mode Equations and S Parameters}
\label{ch4:sec_sparameters}

In an engineering experiment, one can first scan the port reflection spectrum $r(\omega_L)=S_{11}(\omega_L)$ with a weak continuous probe from a vector network analyzer or a tunable laser. For a very weak incident coherent field, $|\beta_{\rm in}|\to 0$, with mean cavity photon number much smaller than 1, the atom remains in $\ket g$ most of the time. Its mean population difference can therefore be approximated by the constant $\avg{\hat\sigma_z}\simeq -1$, and higher-order fluctuation correlations neglected by taking $\avg{\hat a\hat\sigma_z}\simeq\avg{\hat a}\avg{\hat\sigma_z}\simeq -\alpha$.

Denote the mean cavity amplitude by $\alpha=\avg{\hat a}$ and the slowly varying atomic dipole amplitude by $s=\avg{\hat\sigma_-}$. With coherent port driving $\hat H_{\rm port}$, take first moments of the master equation~\eqref{ch4:master}, use $[\hat\sigma_-,\hat a\hat\sigma_+]=-\hat a\hat\sigma_z$, and set $\avg{\hat a\hat\sigma_z}\simeq-\alpha$ in the expectation-value equations. This gives the linear coupled-mode equations:
\begin{equation}
 \begin{aligned}
 \dot\alpha&=\left(\ii\Delta_c-\frac{\kappa}{2}\right)\alpha-\ii g s
 +\sqrt{\kappa_{\rm ex}}\beta_{\rm in},\\
 \dot s&=(\ii\Delta_a-\gamma_2)s-\ii g\alpha,
 \qquad \gamma_2=\frac{\Gamma_{\rm b}}{2}+\gamma_\phi.
 \end{aligned}
 \label{ch4:linearresponse}
\end{equation}
Set the time derivatives to zero ($\dot\alpha=\dot s=0$), solve the second equation for $s=-\ii g\alpha/(\gamma_2-\ii\Delta_a)$, and substitute into the first. Combining this with the port reflection relation $\beta_{\rm out}=\beta_{\rm in}-\sqrt{\kappa_{\rm ex}}\alpha$ gives the complex reflection coefficient $r(\omega_L)=S_{11}(\omega_L)$ of a cavity containing one atom:
\begin{equation}
 \begin{aligned}
 \alpha&=\frac{\sqrt{\kappa_{\rm ex}}\beta_{\rm in}}
 {\kappa/2-\ii\Delta_c+g^2/(\gamma_2-\ii\Delta_a)},\\
 r(\omega_L)&\equiv\frac{\beta_{\rm out}}{\beta_{\rm in}}
 =1-\frac{\kappa_{\rm ex}}
 {\kappa/2-\ii\Delta_c+g^2/(\gamma_2-\ii\Delta_a)}.
 \end{aligned}
 \label{ch4:reflection}
\end{equation}
Setting $g$ to zero reduces Eq.~\eqref{ch4:reflection} to the empty-cavity reflection coefficient of Chapter 3. The atomic correction $\Sigma=g^2/(\gamma_2-\ii\Delta_a)$ has dimensions of a rate. Its real and imaginary parts modify the effective damping and frequency shift, respectively, acting like an additional load in the coupled-mode equations; it is not directly a circuit impedance measured in ohms.

Under resonant strong coupling, the real parts of the response poles lie approximately at $\Delta_c=\pm g$. Whether the measured spectrum shows two peaks or two dips also depends on port interference and loss. For example, an ideal single-port system without other losses can have $|r|=1$, with resonance information appearing mainly in the reflection phase.

The weak-excitation equations have the same form as classical linear coupled-mode equations, so spectral splitting alone does not establish nonclassical photon statistics. Power saturation can also occur in classical nonlinear systems. Verifying photon blockade requires calculating or measuring the second-order correlation of the relevant output; testing the JC ladder requires resolving its excitation-number-dependent anharmonic levels.

\begingroup
\renewcommand{\bibname}{References (Including Further Reading)}

\endgroup

\clearpage
\begingroup
\setcounter{equation}{0}
\setcounter{subsection}{0}
\renewcommand{\theequation}{4A.\arabic{equation}}
\renewcommand{\thesubsection}{4A.\arabic{subsection}}
\renewcommand{\theHequation}{ch4.A.\arabic{equation}}
\renewcommand{\theHsubsection}{ch4.A.\arabic{subsection}}
\section*{Further Reading 4A\quad Continuum Modes, Memory Effects, and Non-Markovian Evolution}
\addcontentsline{toc}{section}{Further Reading 4A: Continuum Modes, Memory Effects, and Non-Markovian Evolution}

Section~\ref{ch4:sec_spont} derived Fermi's golden rule using short-time perturbation theory and a $\mathrm{sinc}^2$ frequency window. Within the electric-dipole, two-level, rotating-wave, single-excitation model, this reading starts from the time-dependent Schr\"odinger equation and eliminates the continuum electromagnetic modes to derive an exact integro-differential equation with environment memory. It then explains when exponential decay with a constant rate (the Markov approximation) holds and when it fails.

\subsection{Why Eliminating Field Modes Produces a Convolution over the Past}

In the interaction picture, with each component rotating at its free frequency, the Schr\"odinger equations for the single-excitation state $\ket{\Psi(t)}=c_e(t)\ket{e,\{0\}}+\sum_\mu c_\mu(t)\ket{g,1_\mu}$ are
\begin{equation}
 \begin{aligned}
 \dot c_e(t)&=-\ii\sum_\mu g_\mu
 \ee^{-\ii(\omega_\mu-\omega_0)t}c_\mu(t),\\
 \dot c_\mu(t)&=-\ii g_\mu^*
 \ee^{\ii(\omega_\mu-\omega_0)t}c_e(t).
 \end{aligned}
 \label{ch4:Aamplitudes}
\end{equation}
Assume all electromagnetic modes are initially in vacuum, so that $c_\mu(0)=0$. Formally integrate the second equation in time:
\[
 c_\mu(t)=-\ii g_\mu^*\int_0^t
 \ee^{\ii(\omega_\mu-\omega_0)t'}c_e(t')\dd t'.
\]
Substitute this into the first equation. The two factors of $-\ii$ multiply to $-1$. Using the coupling spectral density $J(\omega)=\sum_\mu|g_\mu|^2\delta(\omega-\omega_\mu)$ from Eq.~\eqref{ch4:goldenrule} to replace the mode sum by a frequency integral gives an exact memory-kernel integro-differential equation for the atomic probability amplitude $c_e(t)$:
\begin{equation}
 \begin{aligned}
 \dot c_e(t)&=-\int_0^t F(t-t')c_e(t')\dd t',\\
 F(\tau)&=\sum_\mu|g_\mu|^2\ee^{-\ii(\omega_\mu-\omega_0)\tau}\\
 &=\int_0^\infty J(\omega)\ee^{-\ii(\omega-\omega_0)\tau}\dd\omega.
 \end{aligned}
 \label{ch4:Amemory}
\end{equation}
Here $F(\tau)$ is the memory kernel of the electromagnetic environment (units: $\mathrm{s}^{-2}$), the Fourier transform of the spectral density $J(\omega)$ determined by the imaginary part of the Green function in Eq.~\eqref{ch4:spectralgreen}. Equation~\eqref{ch4:Amemory} shows that the present rate of change $\dot c_e(t)$ generally depends on the entire past history $c_e(t')$: electromagnetic radiation emitted at $t'$ may be reflected by surrounding structures and return to act on the atom at $t$.

\subsection{When a Constant Decay Rate Is Sufficient: The Wigner--Weisskopf Approximation}

If the open environment spectrum $J(\omega)$ is very broad and smooth, its Fourier transform $F(\tau)$ is appreciable only within a very short environment correlation time $\tau_B$ and rapidly vanishes for $\tau\gg\tau_B$. Provided that the atom evolves slowly enough---over $\tau_B$, $c_e(t-\tau)\simeq c_e(t)$ barely changes---we can take $c_e(t)$ outside the integral and extend the upper limit to $\infty$:
\[
 \dot c_e(t)=-\int_0^t F(\tau)c_e(t-\tau)\dd\tau
 \simeq -c_e(t)\int_0^\infty F(\tau)\dd\tau.
\]
Using the generalized-function identity $\int_0^\infty\ee^{-\ii(\omega-\omega_0)\tau}\dd\tau=\pi\delta(\omega-\omega_0)-\ii\,\mathcal P\frac{1}{\omega-\omega_0}$, where $\mathcal P$ denotes the Cauchy principal value, gives a differential equation with constant coefficients:
\begin{equation}
 \dot c_e=-\left(\frac\Gamma2+\ii\Delta_{\rm L}\right)c_e,
 \quad \Gamma=2\pi J(\omega_0),\quad
 \Delta_{\rm L}=\mathcal P\int_0^\infty
 \frac{J(\omega)}{\omega_0-\omega}\dd\omega.
 \label{ch4:AMarkov}
\end{equation}
The real part gives the amplitude decay rate $\Gamma/2$, corresponding to population decay rate $\Gamma$; the imaginary part gives a frequency shift. The principal-value integral belongs to the present electric-dipole, rotating-wave, continuum-spectrum model and cannot give an absolute Lamb shift without a treatment of high frequencies. The main text's $\omega_0$ may be taken as the measurable transition frequency in a reference environment. When calculating structure-induced shifts, the reference contribution must be subtracted consistently.

A common scale criterion is $\Gamma\tau_B\ll1$, but it is not a general necessary-and-sufficient condition. Frequency shifts, band edges, initial transients, and delayed feedback must also be checked. Conversely, in a cavity with linewidth $\kappa$ and high $Q$, the memory time can reach $\tau_B\sim 1/\kappa$. When stronger coupling makes the decay rate comparable to or greater than $\kappa$, the condition $\Gamma\tau_B\ll 1$ fails completely and the cavity cannot be treated as a memoryless background.

\subsection{Treating Resonant Memory Kernels with Auxiliary Differential Equations}

If $J(\omega)$ contains pronounced narrow resonances, such as cavity or surface-plasmon modes, one can reduce the approximately $O(N_t^2)$ work needed for direct history convolution. Following the auxiliary differential equation (ADE) method used for dispersive media in finite-difference time-domain (FDTD) calculations, decompose the memory kernel, or fit it with vector fitting, as a sum of complex exponentials:
\[
 F(\tau)\simeq\sum_{j=1}^{M}A_j\ee^{-s_j\tau},\qquad
 \operatorname{Re}s_j>0.
\]
Defining auxiliary variables $z_j(t)=\int_0^t\ee^{-s_j(t-t')}c_e(t')\dd t'$ converts the memory equation~\eqref{ch4:Amemory}, within the chosen exponential-kernel model, into first-order ordinary differential equations:
\begin{equation}
 \dot c_e=-\sum_{j=1}^M A_jz_j,\qquad
 \dot z_j=c_e-s_jz_j,\qquad z_j(0)=0.
 \label{ch4:Aauxiliary}
\end{equation}
For example, approximate a narrowband cavity resonance by the Lorentzian spectrum
\[
 J(\omega)=\frac{g^2\kappa/(2\pi)}{(\kappa/2)^2+(\omega-\omega_c)^2}.
\]
For $\omega_c\gg\kappa$, when only the near-resonant envelope is of interest, the detuning integral can be extended over the full real axis, giving
\[
 F(\tau)\simeq g^2\ee^{-(\kappa/2-\ii\delta)\tau},\qquad \tau\geq0.
\]
The original positive-frequency integral in Eq.~\eqref{ch4:Amemory} is not exactly this single exponential: a narrowband extension has been made here. With $c_c=-\ii g z_1$, the auxiliary equations take the form of Eq.~\eqref{ch4:dampedamplitudes} for $\Gamma_{\rm b}=0$; adding the independent background term $-\Gamma_{\rm b}c_e/2$ recovers that equation. Thus retaining an explicit cavity mode can equivalently represent the corresponding single-exponential memory.

For a general fit, also check nonnegativity and causality of the coupling spectrum and compare with the original convolution. Requiring only that the fitted poles satisfy $\operatorname{Re}s_j>0$ is insufficient to ensure physically consistent quantum evolution.
\endgroup

\clearpage
\begingroup
\setcounter{equation}{0}
\setcounter{subsection}{0}
\renewcommand{\theequation}{4B.\arabic{equation}}
\renewcommand{\thesubsection}{4B.\arabic{subsection}}
\renewcommand{\theHequation}{ch4.B.\arabic{equation}}
\renewcommand{\theHsubsection}{ch4.B.\arabic{subsection}}
\section*{Further Reading 4B\quad Dispersive Coupling at Large Detuning and Quantum Nondemolition Readout}
\addcontentsline{toc}{section}{Further Reading 4B: Dispersive Coupling at Large Detuning and Quantum Nondemolition Readout}

When the atomic transition frequency $\omega_0$ is far detuned from the cavity resonance $\omega_c$, with $|\delta|=|\omega_0-\omega_c|\gg g\sqrt{n+1}$, energy mismatch suppresses real population transfer between atom and cavity mode. Their coupling nevertheless produces second-order shifts in each other's resonance frequencies. Measuring the cavity's microwave or optical reflection phase can use these shifts to read out the atomic energy-level state. This is a common readout method in circuit QED and cavity QED~\cite{ch4:blais}.

\subsection{Deriving the Dispersive Hamiltonian by Second-Order Perturbation and a Unitary Transformation}

First write the free atomic energy as $\hbar\omega_0\hat\sigma_z/2$, differing from the main text only by a constant. The single-excitation matrix method gives the exact eigenenergies in the subspace of total excitation number $n\geq1$, spanned by $\ket{e,n-1}$ and $\ket{g,n}$. At large detuning, $|\delta|\gg 2g\sqrt{n}$, expand the square root to second order as $\sqrt{\delta^2+4g^2n}\simeq|\delta|+\frac{2g^2n}{|\delta|}$. This uses the same reasoning as the far-detuned energy-shift calculation in Chapter 2:
\begin{itemize}
 \item The eigenstate dominated by $\ket{g,n}$, with the atom in its ground state $\sigma_z=-1$ and $n$ photons in the cavity, shifts from energy $n\hbar\omega_c-\frac12\hbar\omega_0$ to $E_{g,n}\simeq n\hbar(\omega_c-\chi)-\frac12\hbar\omega_0$.
 \item The eigenstate dominated by $\ket{e,n}$, with the atom excited, $\sigma_z=+1$, and $n$ cavity photons in the $m=n+1$ subspace, shifts from energy $n\hbar\omega_c+\frac12\hbar\omega_0$ to $E_{e,n}\simeq n\hbar(\omega_c+\chi)+\frac12\hbar(\omega_0+2\chi)$,
\end{itemize}
where the dispersive shift is defined as
\[
 \chi=\frac{g^2}{\delta}=\frac{g^2}{\omega_0-\omega_c}.
\]

The same result follows directly at the operator level to second order through a unitary change of basis (the Schrieffer--Wolff transformation). Write the JC Hamiltonian as $\hat H=\hat H_0+\hat V$, omitting the constant $\frac12\hbar\omega_0$ relative to the main text:
\[
 \hat H_0=\hbar\omega_c\hat n+\frac{\hbar\omega_0}{2}\hat\sigma_z,
 \qquad \hat V=\hbar g(\hat a\hat\sigma_++\hat a^\dagger\hat\sigma_-).
\]
Choose the small anti-Hermitian operator $S=\frac g\delta(\hat a\hat\sigma_+-\hat a^\dagger\hat\sigma_-)$, satisfying $S^\dagger=-S$ and $[S,\hat H_0]=-\hat V$. Transform $\hat H$ unitarily and expand to second order in $g/\delta$:
\[
 \begin{aligned}
 \hat H_{\rm eff}&=\ee^S\hat H\ee^{-S}\\
 &=\hat H_0+\underbrace{\hat V+[S,\hat H_0]}_{=0}
 +[S,\hat V]+\frac12[S,[S,\hat H_0]]+\cdots\\
 &=\hat H_0+\frac12[S,\hat V]+\cdots.
 \end{aligned}
\]
Using $[\hat a\hat\sigma_+,\hat a^\dagger\hat\sigma_-]=(\hat n+1)\ket e\bra e-\hat n\ket g\bra g=\hat n\hat\sigma_z+\frac12(\hat\sigma_z+I)$ and dropping the global constant $\frac12\hbar\chi I$ gives the effective dispersive Hamiltonian at large detuning:
\begin{equation}
 \frac{\hat H_{\rm eff}}\hbar
 =(\omega_c+\chi\hat\sigma_z)\hat n
 +\frac{\omega_0+\chi}{2}\hat\sigma_z,
 \qquad \chi=\frac{g^2}{\delta}.
 \label{ch4:Bdispersive}
\end{equation}

\subsection{Quantum Nondemolition Readout and Dephasing from Photon-Number Shot Noise}

The cross term $\hbar\chi\hat\sigma_z\hat n$ in Eq.~\eqref{ch4:Bdispersive} admits two complementary interpretations:

From the cavity's perspective, write the first term as $(\omega_c+\chi\hat\sigma_z)\hat n$. The cavity resonance frequency therefore depends on the atomic energy-level state:
\begin{itemize}
 \item For the atomic ground state $\ket g$ ($\sigma_z=-1$), the cavity resonance is $\omega_c-\chi$.
 \item For the excited state $\ket e$ ($\sigma_z=+1$), the cavity resonance is $\omega_c+\chi$.
\end{itemize}
The two cavity frequencies differ by $2|\chi|$; the shift direction is set by the sign of $\delta$. At a fixed probe frequency, balanced homodyne detection can compare the reflected-field phase or amplitude. Readout requires accumulating sufficient signal-to-noise ratio: a nonzero frequency shift does not imply instantaneous distinguishability.

In the ideal second-order model, $[\hat H_{\rm eff},\hat\sigma_z]=0$, so dispersive coupling itself does not change populations. This is the basis of approximate quantum nondemolition readout. Actual readout remains subject to relaxation, driving, and higher-order corrections. The limit $\delta\omega_{\rm cav}\to\chi$ in Eq.~\eqref{ch4:purcellrate} is instead the atomic frequency shift in an empty cavity, not either of the two atomic-state-dependent cavity shifts.

From the atom's perspective, absorb $\hbar\chi\hat\sigma_z\hat n$ into the second term, writing $\frac12(\omega_0+\chi+2\chi\hat n)\hat\sigma_z$. With $n$ photons in the cavity, the atomic transition angular frequency is therefore
\begin{equation}
 \omega_a(n)\simeq\omega_0+\chi+2\chi n.
 \label{ch4:Bstark}
\end{equation}
Here $\chi$ is the single-mode vacuum frequency correction, and $2\chi n$ is the photon-number-dependent AC Stark shift. Each added photon changes the transition frequency by $2\chi$. For an approximately coherent cavity field, $\Delta n=\sqrt{\bar n}$ corresponds to a frequency-fluctuation scale $2|\chi|\sqrt{\bar n}$. These fluctuations cause dephasing, but the dephasing rate also depends on the temporal correlation of the photon-number fluctuations and cannot be identified directly with this frequency scale. The outgoing light carries information about the atomic state and affects its coherent superposition even if that light is never read out.

Engineering designs for dispersive readout face two additional physical limits. First, the intracavity probe photon number must be much smaller than the critical photon number $n_{\rm crit}=\delta^2/(4g^2)$; otherwise, the second-order approximation fails. Second, for $|\delta|\gg g,\kappa$ and a sufficiently smooth background response, the atomic dressed state contains a cavity-photon component with weight approximately $(g/\delta)^2$. The atom therefore still undergoes residual Purcell decay through the cavity port at rate
\begin{equation}
 \Gamma_{\rm disp}\simeq\kappa\frac{g^2}{\delta^2}.
 \label{ch4:BPurcell}
\end{equation}
A Purcell filter changes the port frequency response to suppress loss at the atomic frequency while retaining output at the readout frequency. Weakly anharmonic superconducting devices also require multilevel corrections; the ideal two-level expression $\chi=g^2/\delta$ does not apply unconditionally.
\endgroup

\clearpage
\begingroup
\setcounter{equation}{0}
\setcounter{subsection}{0}
\renewcommand{\theequation}{4C.\arabic{equation}}
\renewcommand{\thesubsection}{4C.\arabic{subsection}}
\renewcommand{\theHequation}{ch4.C.\arabic{equation}}
\renewcommand{\theHsubsection}{ch4.C.\arabic{subsection}}
\section*{Further Reading 4C\quad Collective Effects Mediated by the Electromagnetic Field}
\addcontentsline{toc}{section}{Further Reading 4C: Collective Effects Mediated by the Electromagnetic Field}

The discussion of interatomic interactions in Chapter 2 treated dipole--dipole coupling in the near-field electrostatic approximation ($R\ll\lambda_0$), where it scales as $\propto R^{-3}$. For two or more atoms at arbitrary separations or in structures such as waveguides and cavities, the field radiated by one atom can drive another, producing coherent coupling. The atoms also radiate into a common open environment, giving collective dissipation through constructive interference (superradiance) or destructive interference (subradiance). This reading describes both effects with the dyadic Green function of Chapter 3.

\subsection{The Real Part of the Green Function Determines Coherent Exchange; Its Imaginary Part Determines Collective Decay}

Consider identical two-level atoms at $\mathbf r_1,\mathbf r_2$, both with transition angular frequency $\omega_0$ and real linearly polarized transition dipoles $\mathbf d_1,\mathbf d_2$. In a reciprocal background, eliminate the background field degrees of freedom using weak-coupling, near-resonance, and Markov approximations. The real and imaginary parts of the dyadic Green function $\mathbf G(\mathbf r_i,\mathbf r_j,\omega_0)$ then play distinct roles~\cite{ch4:dung,ch4:ficek}:

The real part $\operatorname{Re}\mathbf G$ gives the coherent exchange term:
\begin{equation}
 \begin{aligned}
 \hat H_{12}&=\hbar J_{12}
 (\hat\sigma_{1+}\hat\sigma_{2-}+\hat\sigma_{2+}\hat\sigma_{1-}),
 \\
 J_{12}&=-\frac{\mu_0\omega_0^2}{\hbar}
 \mathbf d_1\cdot\operatorname{Re}\mathbf G(\mathbf r_1,\mathbf r_2,\omega_0)
 \cdot\mathbf d_2.
 \end{aligned}
 \label{ch4:Cexchange}
\end{equation}
Check the near-field limit. At very short free-space separations, $k_0R\ll 1$ with $\mathbf R=\mathbf r_1-\mathbf r_2$, the quasistatic field of a point dipole gives $\mu_0\omega_0^2\operatorname{Re}\mathbf G_0(\mathbf r_1,\mathbf r_2,\omega_0)\simeq(3\hat{\mathbf R}\hat{\mathbf R}-\mathbf I)/(4\pi\varepsilon_0R^3)$. Substituting into Eq.~\eqref{ch4:Cexchange} recovers the near-field dipole coupling of Chapter 2:
\[
 \hbar J_{12}\simeq\frac{\mathbf d_1\cdot\mathbf d_2
 -3(\mathbf d_1\cdot\hat{\mathbf R})(\mathbf d_2\cdot\hat{\mathbf R})}
 {4\pi\varepsilon_0R^3}.
\]
In the far field ($k_0R\gg 1$) or in a waveguide, however, $\operatorname{Re}\mathbf G$ oscillates over long distances as $\cos(k_0R)/R$ in three dimensions or $\sin(k_{\rm wg}R)$ in a one-dimensional waveguide. The static $R^{-3}$ formula is no longer applicable.

The imaginary part $\operatorname{Im}\mathbf G$ gives the collective dissipation matrix:
\begin{equation}
 \Gamma_{ij}=\frac{2\mu_0\omega_0^2}{\hbar}
 \mathbf d_i\cdot\operatorname{Im}\mathbf G
 (\mathbf r_i,\mathbf r_j,\omega_0)\cdot\mathbf d_j
 \qquad(i,j\in\{1,2\}).
 \label{ch4:Cdecaymatrix}
\end{equation}
The diagonal entries $\Gamma_{11},\Gamma_{22}$ are the single-atom spontaneous-emission rates in Eq.~\eqref{ch4:gammagreen}. The off-diagonal entries $\Gamma_{12}=\Gamma_{21}$ describe quantum interference when the atoms emit into shared radiation channels. In classical power calculations, these correspond to cross terms in radiation or absorption from the two dipoles, related to the real part of their mutual impedance. The dissipative part of the two-atom density-matrix equation becomes
\begin{equation}
 \left.\dot\rho\right|_{\rm em}
 =\sum_{i,j=1}^{2}\Gamma_{ij}
 \left(\hat\sigma_{j-}\rho\hat\sigma_{i+}
 -\frac12\{\hat\sigma_{i+}\hat\sigma_{j-},\rho\}\right),
 \label{ch4:Ccollectivemaster}
\end{equation}
where $\{A,B\}=AB+BA$. Since total radiated and absorbed power in a passive environment is nonnegative, $\Gamma_{ij}$ must be positive semidefinite and must always satisfy $|\Gamma_{12}|\leq\sqrt{\Gamma_{11}\Gamma_{22}}$.

\subsection{Superradiant Bright States and Subradiant Dark States}

Assume equal individual electromagnetic decay rates, $\Gamma_{11}=\Gamma_{22}=\Gamma$. Diagonalizing the real symmetric decay matrix $\left(\begin{smallmatrix}\Gamma&\Gamma_{12}\\\Gamma_{12}&\Gamma\end{smallmatrix}\right)$ gives collective eigenrates $\Gamma_\pm=\Gamma\pm\Gamma_{12}$. Equation~\eqref{ch4:Ccollectivemaster} then becomes a standard Lindblad equation with two independent channels and collective jump operators
\begin{equation}
 L_\pm=\sqrt{\Gamma\pm\Gamma_{12}}\,
 \frac{\hat\sigma_{1-}\pm\hat\sigma_{2-}}{\sqrt2}.
 \label{ch4:Ccollapse}
\end{equation}
Consider the symmetric and antisymmetric entangled states sharing 1 excitation between the atoms:
\[
 \ket\pm=\frac{\ket{e,g}\pm\ket{g,e}}{\sqrt2}.
\]
Apply $L_\pm$ to these states. Since $L_+\ket-=0$ and $L_-\ket+=0$, including the dipole-exchange frequency shifts $\pm J_{12}$ of Eq.~\eqref{ch4:Cexchange} gives
\begin{itemize}
 \item The symmetric state $\ket+$ has transition frequency $\omega_0+J_{12}$ and electromagnetic decay rate $\Gamma_+=\Gamma+\Gamma_{12}$.
 \item The antisymmetric state $\ket-$ has transition frequency $\omega_0-J_{12}$ and electromagnetic decay rate $\Gamma_-=\Gamma-\Gamma_{12}$.
\end{itemize}
In free space, suppose the separation is much smaller than the wavelength, the atoms remain describable as two independent point dipoles, the coupling stays within this model's validity, and the dipole orientations are parallel. Then $\Gamma_{12}\to\Gamma$. The symmetric state $\ket+$ has twice the radiative rate, $\Gamma_+\to 2\Gamma$: the interference resembles an in-phase dipole array, but there is only one excitation in total. This is the superradiant bright state. The antisymmetric state $\ket-$ instead has a rate approaching zero, $\Gamma_-\to 0$, corresponding to complete far-field cancellation between opposite-phase dipoles and suppression of emission into the shared radiation channels. This is the subradiant dark state.

\subsection{Why \texorpdfstring{$N_a$}{Nₐ} Atoms Coupled to a Cavity Give a \texorpdfstring{$\sqrt{N_a}$}{√Nₐ} Collective Enhancement}

For $N_a$ two-level atoms with equal transition frequencies in one single-mode cavity, the interaction Hamiltonian is
\[
 \hat H_{\rm int}=\hbar\sum_{j=1}^{N_a}
 (g_j\hat a\hat\sigma_{j+}+g_j^*\hat a^\dagger\hat\sigma_{j-}).
\]
Denote the all-ground-state joint atomic state by $\ket G=\ket{g_1,g_2,\ldots,g_{N_a}}$, and the state in which only atom $j$ is excited by $\ket j=\hat\sigma_{j+}\ket G$. With 1 photon in the cavity, initially $\ket{G,1}$, one application of $\hat H_{\rm int}$ distributes that photon coherently among the atoms:
\begin{equation}
 \begin{aligned}
 \hat H_{\rm int}\ket{G,1}
 &=\hbar\left(\sum_{j=1}^{N_a}g_j\ket j\right)\otimes\hat a\ket1
 =\hbar g_{\rm eff}\ket{B,0},\\
 \ket B&\equiv\frac1{g_{\rm eff}}\sum_{j=1}^{N_a}g_j\ket j,
 \qquad g_{\rm eff}=\sqrt{\sum_{j=1}^{N_a}|g_j|^2}.
 \end{aligned}
 \label{ch4:Cbright}
\end{equation}
Conversely, the collective bright state $\ket{B,0}$ is taken by $\hat H_{\rm int}$ exactly back to $\hbar g_{\rm eff}\ket{G,1}$. The remaining $N_a-1$ atomic single-excitation combinations orthogonal to $\ket B$ all satisfy $\hat H_{\rm int}\ket{D_k,0}=0$. They do not couple directly to this cavity mode and are called collective dark states relative to that mode.

If all atoms have the same frequency, start in their ground states, and have negligible additional interatomic interactions, single-excitation evolution from $\ket{G,1}$ involves only $\ket{G,1}$ and $\ket{B,0}$. Within this subspace, it is equivalent to a two-amplitude problem with coupling $g_{\rm eff}$. For equal coupling magnitudes $|g_j|=g$,
\[
 g_{\rm eff}=g\sqrt{N_a}.
\]
This square root arises from normalization of the collective state and coherent coupling. Beyond the single-excitation subspace, a weakly excited ensemble may be approximated as a collective harmonic oscillator, but is not strictly equivalent to a two-level ``superatom.'' A small mean cavity photon number $\avg{\hat n}\ll N_a$ alone does not ensure that all atoms are weakly excited.

Practical collective enhancement is also limited by initial polarization, inhomogeneous broadening, dephasing, and mode overlap. Atomic frequency mismatch or additional interactions can mix bright and dark states, and dark states may radiate into other channels. For room-temperature microwave spin ensembles in particular, inserting the total particle number into $g\sqrt{N_a}$ does not establish strong coupling. The effective participating population difference and all loss linewidths must also be checked.
\endgroup

\clearpage
\fancyhead[R]{\small Chapter 5}
\ctexset{chapter/format=\Large\normalfont,section/format=\large\normalfont,subsection/format=\normalsize\normalfont}
\captionsetup{labelfont=normalfont}

\chapter{A Unified View and Future Directions}
\label{ch5:chapter}

The preceding chapters developed the theme of field--matter interaction. Chapter 1 established the descriptive framework; Chapters 2, 3, and 4 treated classical fields with quantum matter, quantum fields in linear media, and the joint evolution of quantum fields and quantum matter. This chapter returns to practical devices. It draws together the information retained by these models and how they answer engineering questions.

Waveguide transmission, saturated absorption in an atomic cell, and photon coincidences in two far-field directions require different unknowns and equations. An oversimplified model can omit the processes that determine the result; retaining too many degrees of freedom can make computation unnecessarily costly. We therefore consider which field and matter degrees of freedom to retain, how to connect electromagnetic and quantum solvers, and how to evaluate task performance. Research directions and a conceptual comparison conclude the chapter. Measurement precision and nonclassical indicators assess the outputs of these models and must be interpreted with their input, loss, and readout conditions.

\section{Choosing a Model for a Specific Problem}
\label{ch5:selection}
\subsection{Specify the Input State, Structure Response, and Measurement}

A conventional classical electromagnetic problem is usually specified by a structure and an excitation. Material parameters, geometry, ports, and incident waveforms define the calculation of scattering parameters or spatial fields. A quantum electromagnetic problem must also specify the input field state and the final measurement. These specifications are distinct from the spatial boundary conditions in Maxwell's equations.
\begin{itemize}
 \item For the same structure and exactly the same average incident power, a single-photon state $\ket1$, a coherent state $\ket\alpha$, and a thermal state $\rho_{\rm th}$ give very different photon-count fluctuations.
 \item For the same output quantum field, direct photodetection measures intensity (proportional to $\avg{\hat a^\dagger\hat a}$), balanced homodyne detection measures a quadrature (proportional to $\avg{\hat X_\theta}$), and two-detector coincidence measurements probe a quantity proportional to $\avg{\hat a_1^\dagger\hat a_2^\dagger\hat a_2\hat a_1}$. These are different observables.
\end{itemize}

Before building a model, state the computational task in a sentence containing all the necessary elements. For example: ``A single-photon wavepacket enters each of two input ports; calculate the coincidence probability in two far-field collection modes within a specified time window after scattering by a dielectric object.'' This specifies the input state, propagation structure, collection modes, and measurement. The vague request to ``calculate quantum scattering by an object'' does not determine a model.

For a linear, passive, time-invariant background structure, classical Maxwell equations still determine all spatial propagation. With single-frequency or narrowband photon-flux normalization, the multiport quantum input--output relation derived in Chapter 3 is
\begin{equation}
 \hat{\mathbf b}=S\hat{\mathbf a}+L\hat{\mathbf v},
 \qquad SS^\dagger+LL^\dagger=I.
 \label{ch5:network}
\end{equation}
Here $\hat{\mathbf a}$ and $\hat{\mathbf b}$ are column vectors of input and output annihilation operators, and $\hat{\mathbf v}$ contains dissipative environmental-channel operators. Classical electromagnetic software provides the scattering matrix $S(\omega)$. The relation $I-SS^\dagger=LL^\dagger$ allows an equivalent loss-channel coupling matrix to be constructed so that the outputs obey bosonic commutation relations. This factorization of $L$ is not unique. Output noise also depends on the environmental state and cannot be determined by $S$ alone.

The same classical scattering matrix $S$ propagates the mean field of a coherent state and determines single- or two-photon correlations, including the Hong--Ou--Mandel interference of Chapter 3. Under the linear, passive, time-invariant conditions above, the complex scattering matrix over the relevant frequencies and modes supplies propagation amplitudes. Input and environmental states, together with detection modes, determine the count statistics. Absence of few-level saturation alone is insufficient: parametric modulation or other nonlinear processes may also lie outside the network described by Eq.~\eqref{ch5:network}.

By contrast, a structure may contain saturable, few-level quantum matter: an atom, quantum dot, superconducting qubit, or strongly driven atomic ensemble. Exchanging an excitation with a photon can change the material state and hence the response seen by later photons. A fixed linear scattering matrix $S$ is then insufficient. The quantum-state evolution equations of Chapter 2 or 4 are needed.

\subsection{Connecting the Four Models to Classical Simulation}

Chapter 1 classified four descriptions according to the treatment of field and matter. After Chapters 2--4, we can identify their computational unknowns, suitable problems, and interface parameters from classical electromagnetic simulation, as in Table~\ref{ch5:models}. Here ``macroscopic media'' and ``linear media'' refer to materials whose linear response is described by permittivity $\varepsilon_r(\mathbf r,\omega)$ and permeability $\mu_r(\mathbf r,\omega)$. This does not deny the microscopic quantum origin of those parameters.

\begin{table}[htbp]
\centering\small
\renewcommand{\arraystretch}{1.22}\setlength{\tabcolsep}{4pt}
\caption{Unknowns, applications, and classical simulation interfaces for the four electromagnetic models used in this book.}
\label{ch5:models}
\begin{tabularx}{\textwidth}{@{}>{\raggedright\arraybackslash}p{2.1cm}>{\raggedright\arraybackslash}p{2.6cm}>{\raggedright\arraybackslash}p{4.0cm}Y@{}}
\toprule
Model & Main unknowns & Parameters from classical simulation & Typical questions\\
\midrule
1. Classical field + macroscopic medium (conventional electromagnetics) & Classical fields $\mathbf E(\mathbf r,t),\mathbf H(\mathbf r,t)$ and complex port amplitudes & Field distributions, eigenfrequency $\omega_c$, scattering matrix $S(\omega)$, antenna impedance & Wave propagation, impedance matching, antenna radiation patterns, radar cross section (RCS)\\
\addlinespace[3pt]
2. Classical field + quantum matter (Chapter 2: semiclassical) & Few-level density matrix $\rho(t)$ (population $p$, coherence $s$) and macroscopic polarization $\mathbf P$ & Local complex drive amplitude $\mathbf E_0(\mathbf r_0)$ at the atom and background field distribution & Rabi oscillations, Ramsey clocks, EIT windows, Rydberg-atom radio-frequency receivers, nonlinear saturable absorption\\
\addlinespace[3pt]
3. Quantum field + linear medium (Chapter 3: field quantization) & Field state, quadrature variance $\Var(X_\theta)$, correlations $g^{(1)},g^{(2)}$ & Normalized eigenmode $\mathbf u(\mathbf r)$, effective mode volume $V_{\rm eff}$, scattering matrix $S(\omega)$, imaginary part of the dyadic Green function $\operatorname{Im}\mathbf G$ & Vacuum fluctuations, shot noise and homodyne phase measurement, thermal/vacuum noise in lossy networks, far-field two-photon interference\\
\addlinespace[3pt]
4. Quantum field + quantum matter (Chapter 4: fully quantum) & Joint atom--field state, or an appropriate reduced density matrix and conditional state & Single-photon coupling $g$, external cavity coupling $\kappa_{\rm ex}$ and internal loss $\kappa_{\rm i}$, dipole power ratio $P_{\rm cl}/P_{\rm cl,0}$ (or $\mathbf G(\mathbf r_i,\mathbf r_j,\omega_0)$) & Vacuum Rabi splitting, spontaneous-emission lifetime and Purcell enhancement, triggered single photons, fluorescence antibunching and cavity photon blockade\\
\bottomrule
\end{tabularx}
\end{table}

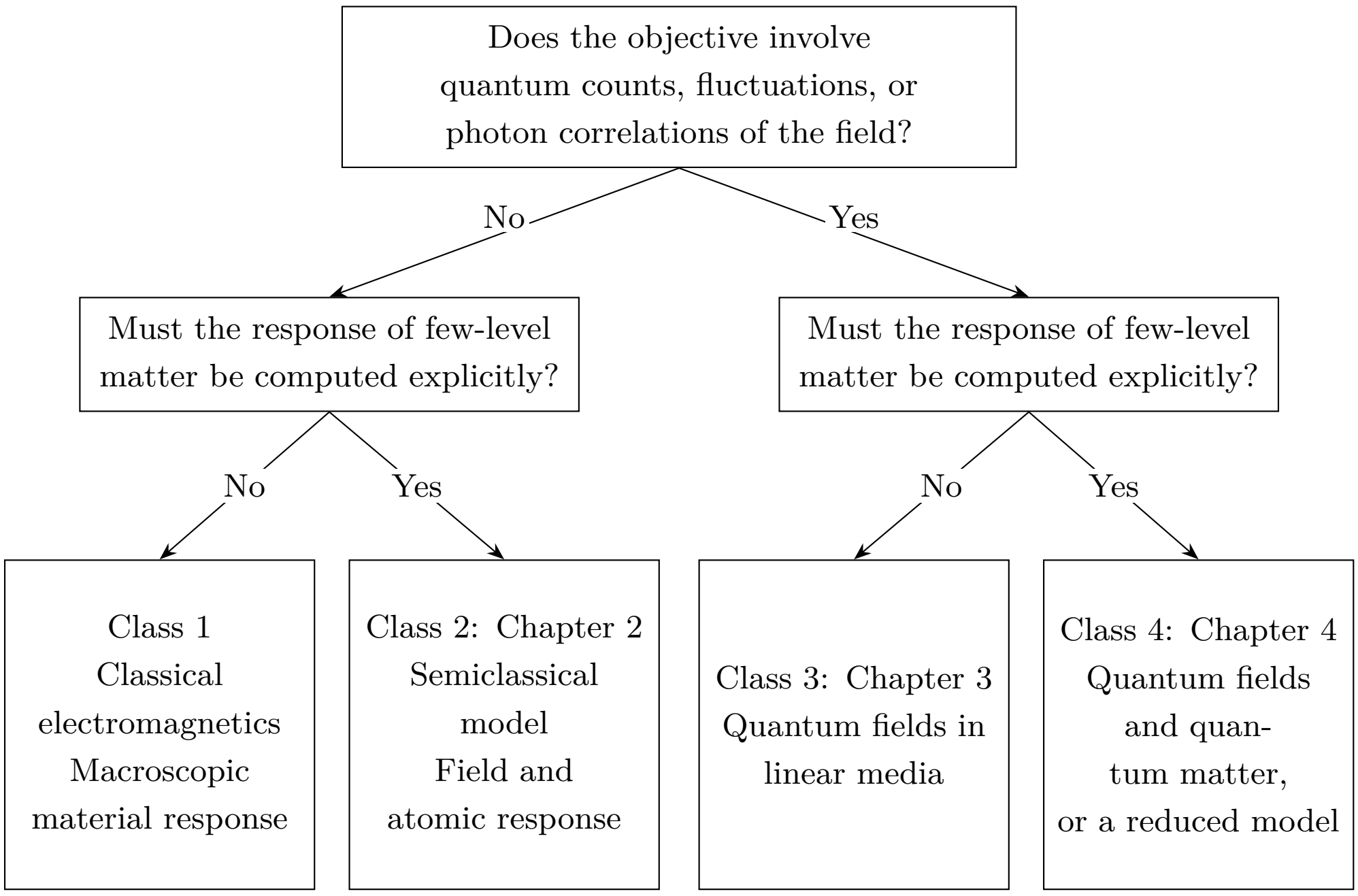
\begin{figure}[htbp]
\centering
\begin{tikzpicture}[font=\footnotesize,>=Stealth,
 box/.style={draw,align=center,inner sep=5pt},
 leaf/.style={draw,align=center,text width=2.8cm,minimum height=3.2cm,inner sep=3pt}]
\node[box,text width=6.2cm] (root) at (0,5.5) {Does the objective involve\\quantum counts, fluctuations, or\\photon correlations of the field?};
\node[box,text width=4.5cm] (left) at (-3.4,2.9) {Must the response of few-level matter be computed explicitly?};
\node[box,text width=4.5cm] (right) at (3.4,2.9) {Must the response of few-level matter be computed explicitly?};
\node[leaf] (a) at (-5.05,-.7) {Class 1\\Classical\\electromagnetics\\Macroscopic\\material response};
\node[leaf] (b) at (-1.7,-.7) {Class 2: Chapter 2\\Semiclassical model\\Field and atomic response};
\node[leaf] (c) at (1.7,-.7) {Class 3: Chapter 3\\Quantum fields in\\linear media};
\node[leaf] (d) at (5.05,-.7) {Class 4: Chapter 4\\Quantum fields\\and quantum matter,\\or a reduced model};
\draw[->] (root.south)--node[midway,above,fill=white,inner sep=1pt]{No}(left.north);
\draw[->] (root.south)--node[midway,above,fill=white,inner sep=1pt]{Yes}(right.north);
\draw[->] (left.south)--node[midway,fill=white,inner sep=1.5pt]{No}(a.north);
\draw[->] (left.south)--node[midway,fill=white,inner sep=1.5pt]{Yes}(b.north);
\draw[->] (right.south)--node[midway,fill=white,inner sep=1.5pt]{No}(c.north);
\draw[->] (right.south)--node[midway,fill=white,inner sep=1.5pt]{Yes}(d.north);
\end{tikzpicture}
\caption{An initial guide to model selection. Specify the desired output first, then decide which matter and field degrees of freedom to retain. Eliminating environmental modes still requires retaining noise and field-readout relations consistent with the decay.}
\label{ch5:choice}
\end{figure}

Figure~\ref{ch5:choice} gives a starting point for model selection. A few-level emitter need not couple strongly to a cavity to produce fluorescence statistics requiring a quantum description. Nor must all environmental photon modes be retained explicitly. Under suitable approximations, an atomic master equation gives the dynamics, while quantum input--output relations and emission operators give photon statistics. This reduced calculation still retains the quantum description needed for field measurements.

\subsection{Correct Mean Fields Do Not Guarantee Correct Higher-Order Correlations}

Section~\ref{ch4:sec_sparameters} showed that, in weak-excitation linear response, the mean cavity-port reflection coefficient $S_{11}(\omega_L)$ for a single atom has the classical coupled-mode form. Does this mean semiclassical mean-field equations can always replace the fully quantum equations?

For the single-atom, single-mode cavity of Chapter 4, write the mean cavity amplitude as $\alpha=\avg{\hat a}$, atomic dipole coherence as $s=\avg{\hat\sigma_-}$, and population difference as $z=\avg{\hat\sigma_z}$. Choose the phase convention in which $g$ is real and neglect thermal excitation. With driving only through the cavity port and no additional direct atomic drive, two exact first-moment equations are
\begin{equation}
 \begin{aligned}
 \dot s&=(\ii\Delta_a-\gamma_2)s
 +\ii g\avg{\hat a\hat\sigma_z},\\
 \dot z&=-\Gamma_{\rm b}(z+1)
 -4g\operatorname{Im}\avg{\hat a^\dagger\hat\sigma_-}.
 \end{aligned}
 \label{ch5:moment}
\end{equation}
Here $\gamma_2=\Gamma_{\rm b}/2+\gamma_\phi$. The joint expectation $\avg{\hat a\hat\sigma_z}$ in the first equation separates into the product of means and a correlation term:
\begin{equation}
 \avg{\hat a\hat\sigma_z}=\avg{\hat a}\avg{\hat\sigma_z}+C_{az}=\alpha z+C_{az},
 \qquad C_{az}\equiv\avg{\hat a\hat\sigma_z}-\alpha z.
 \label{ch5:cumulant}
\end{equation}
We retain the detuning convention $\Delta_a=\omega_L-\omega_0$ from Chapter 4. Equation~\eqref{ch5:cumulant} distinguishes two approximations.
\begin{enumerate}
 \item Mean-field factorization sets $C_{az}=0$ and replaces the joint expectation by $\alpha z$. It neglects this particular field--atom correlation. Under strong coherent driving it often describes the average response, but its accuracy must be checked at the operating point for the target quantity.
 \item Weak-excitation linearization further sets $z\simeq-1$, so the last term becomes $-\ii g\alpha$. This removes atomic saturation and cannot determine higher-order statistics such as photon blockade.
\end{enumerate}

Different observables involve different correlations. For undriven vacuum Rabi exchange starting from $\ket{e,0}$, for example, $\alpha=s=C_{az}=0$ can hold throughout, while energy still passes between atom and cavity. The cavity photon-number equation is
\begin{equation}
 \frac{\dd\avg{\hat n}}{\dd t}
 =-\kappa\avg{\hat n}
 +2g\operatorname{Im}\avg{\hat a^\dagger\hat\sigma_-}.
 \label{ch5:numberexchange}
\end{equation}
The excited-state population is $p_e=(z+1)/2$, with exchange term $-2g\operatorname{Im}\avg{\hat a^\dagger\hat\sigma_-}$. This is equal and opposite to the exchange term in the photon-number equation. Without driving or loss, adding the equations gives conservation of $\avg{\hat n}+p_e$. A zero mean field does not justify setting the exchange correlation to zero, nor can all quantum effects be assigned to $C_{az}$ alone.

After selecting a model, check three scales. Does the drive populate truncated levels or multiphoton states? Is the eliminated environment's correlation time much shorter than the retained system's evolution time? Can the detector resolve the fast processes of interest? A high-$Q$ cavity retained explicitly as part of the system can still have a Markovian external environment. Eliminating that narrowband cavity as well may instead produce appreciable memory in the reduced atomic dynamics. Approximation validity depends on the system--environment partition.

\section{From Equations to Computable Results}
\label{ch5:calculation}
\subsection{Connecting Electromagnetic Meshes and Quantum Hilbert Spaces}

After choosing the field and matter degrees of freedom, the next step is to formulate solvable equations. An electromagnetic mesh and a quantum state space serve different purposes. Finite-element or FDTD calculations supply spatial fields; a small set of modes and parameters relevant to emission, transmission, or measurement then defines the quantum model.

Suppose $K$ electromagnetic modes are retained explicitly, with mode $k$ truncated to photon numbers $0,\ldots,N_k$. There are also $M$ matter subsystems, with $d_j$ levels in subsystem $j$. The tensor-product rule gives the joint state-space dimension
\begin{equation}
 d=\prod_{j=1}^{M}d_j\prod_{k=1}^{K}(N_k+1).
 \label{ch5:dimension}
\end{equation}
The state vector has $d$ complex components, the density matrix has $d^2$ entries, and the vectorized evolution matrix is $d^2\times d^2$.

One two-level atom and a cavity truncated at $N_1=5$ give $d=12$, so the density matrix has only $144$ entries and is readily solved directly. If $10^6$ discrete field degrees of freedom were independent oscillators with two number states each, the unreduced tensor-product dimension would be $2^{10^6}\simeq10^{301030}$. This estimates the cost of storing a general joint state. It does not imply that every multimode quantum calculation requires that size: a fixed small excitation number, Gaussian states, symmetry, or low-rank methods can greatly reduce the unknowns.

The quantum basis dimension therefore cannot be equated with the Maxwell mesh size. The number of quantum modes depends on the target observable, bandwidth, and coupling spectrum. Convergence should be checked by increasing that number.

For three typical tasks, the solvers can exchange the following data:
\begin{enumerate}
 \item Multiport linear photonic networks (Chapter 3): an electromagnetic frequency sweep supplies the complex $N\times N$ scattering matrix $S(\omega)$. Quantum calculations use $S(\omega)$ to transform quadrature covariance matrices or compute multiphoton coincidence amplitudes.
 \item Strongly coupled cavity-QED and circuit-QED systems (Sections~\ref{ch4:sec_loss} and~\ref{ch4:section_driven}): electromagnetic calculations supply the main-mode frequency $\omega_c$, normalized mode field and effective mode volume $V_{\rm eff}$ (combined with dipole magnitude and orientation to obtain $g$), external coupling $\kappa_{\rm ex}$, and internal loss $\kappa_{\rm i}$. The quantum solver assembles the Jaynes--Cummings master equation in a space of dimension $d=2(N+1)$.
 \item Spontaneous emission and multi-atom dipole interactions in complex micro- and nanostructures (Section~\ref{ch4:section_emission} and Further Reading 4A and 4C): driven point-dipole simulations supply $\mathbf G(\mathbf r_i,\mathbf r_j,\omega)$. For a local total decay rate, the total power ratio $P_{\rm cl}/P_{\rm cl,0}$ for the same test dipole can also be used. Quantum calculations convert these into spontaneous-decay rates $\Gamma_{ij}$, the dipole coupling $J_{12}$, or a non-Markovian memory kernel $F(\tau)$.
\end{enumerate}

Check units and physical assignments when transferring parameters. If $g/(2\pi)=10\,\mathrm{MHz}$ is reported, use $g=2\pi\times10^7\,\mathrm{s}^{-1}$ in the equations. A lifetime $T_1$ gives $\Gamma=1/T_1$, without another factor of $2\pi$. Reported linewidths may be full widths or half widths at half maximum; check their definitions. Distinguish the cavity photon-number decay rate $\kappa$ from the amplitude decay rate $\kappa/2$.

Partition cavity modes and background loss as in Section~\ref{ch4:sec_loss}. Section~\ref{ch4:sec_power} gives the amplitude calibration and radiative/absorptive decomposition for converting classical dipole power into decay rates. These local power ratios cannot replace the complex Green function between different emitter positions or determine coherent dipole coupling by themselves.

\subsection{Solve the Quantum State, Then Extract the Measured Observable}

Under appropriate Markov and coarse-graining approximations, the reduced density matrix $\rho(t)$ of the selected subsystem obeys the common Lindblad form
\begin{equation}
 \dot\rho=\mathcal L[\rho]
 =-\frac{\ii}{\hbar}[\hat H,\rho]
 +\sum_k\left(L_k\rho L_k^\dagger
 -\frac12\{L_k^\dagger L_k,\rho\}\right).
 \label{ch5:master}
\end{equation}
Each jump operator $L_k$ includes the square root of its channel's decay rate and has units of $\mathrm{s}^{-1/2}$. For continuous-wave driving with a time-independent generator in the chosen rotating frame and a unique stable steady state, set $\dot\rho=0$ and impose unit trace. This gives a linear system. Further Reading 5B explains matrix vectorization; software interfaces are described in the QuTiP documentation~\cite{ch5:qutip}:
\[
 \mathcal L[\rho_{\rm ss}]=0,\qquad \Tr\rho_{\rm ss}=1.
\]

After finding $\rho(t)$ or $\rho_{\rm ss}$, select the observable corresponding to the actual detector:
\begin{itemize}
 \item Single-time means: the excited-state population is $\Tr(\ket e\bra e\rho)$, the mean cavity photon number is $\Tr(\hat a^\dagger\hat a\rho)$, and the target-port photon flux is $\Phi=\Tr(\hat c^\dagger\hat c\rho)$. With a directly reflected coherent background $\beta_{\rm in}$, the phase convention of Chapter 4 gives $\hat c=\beta_{\rm in}I-\sqrt{\kappa_{\rm ex}}\hat a$.
 \item Two-time spectra and coincidences: Further Reading 2B gives the auxiliary matrix and Fourier-transform steps for the symmetrized fluctuation spectrum $S_O(f)$. Section~\ref{ch4:sec_g2} gives conditional-state propagation for the delayed photon correlation $g^{(2)}(\tau)$. The former auxiliary matrix has zero trace; the latter detection-conditioned state has unit trace. Their initial conditions are not interchangeable.
\end{itemize}

\subsection{Pulsed Single Photons, Continuous Photon Streams, and Detection Corrections}
\label{ch5:protocol}

If a complete wavepacket contains at most one photon per preparation, it cannot produce two ideal absorptive detection events. The coincidence probability within that pulse is therefore zero, giving $g^{(2)}_{\rm pulse}[0]=0$ when the mean count is nonzero. The bracketed zero labels coincidences integrated over the full counting window of the same pulse. If re-excitation can occur during preparation of a real pulsed source, the multiphoton probability must also be calculated. A two-level emitter alone does not guarantee at most one photon per trial.

Figure 4.6 shows single-atom fluorescence under continuous laser driving. After emission, the atom returns to its ground state and can be excited again. It therefore emits many photons over a long interval. Its steady-state correlation is
\begin{equation}
 g^{(2)}(\tau)=\frac{\Phi(\tau\mid0)}{\Phi_{\rm ss}}.
 \label{ch5:conditional}
\end{equation}
Here $\Phi(\tau\mid0)$ is the conditional count rate at delay $\tau$ after a photon is detected at time zero. Immediately after ideal two-level fluorescence emission, the atom has not yet been re-excited, so $g^{(2)}(0)=0$. With the unique stable steady state assumed here, the conditional rate returns to its steady value at delays well beyond the relaxation times, giving $g^{(2)}\to1$. Widely separated counts become uncorrelated; this does not mean the fluorescence becomes a coherent state.

Detector timing jitter and finite counting windows smooth the curve. Let $h(\tau)$ be the normalized relative-time response, with $\int h(\tau)\dd\tau=1$. Neglecting background and detector dead time, the measured steady-state correlation is
\begin{equation}
 g^{(2)}_{\rm meas}(\tau)-1
 =\int_{-\infty}^{\infty}h(\tau-u)
 [g^{(2)}(u)-1]\dd u.
 \label{ch5:timing}
\end{equation}
If the antibunching dip is narrower than the instrument response, the measured zero-delay value exceeds the ideal value. The dip and oscillation timescales follow from the actual dynamics, generally involving drive, relaxation, and dephasing. They cannot all be replaced by one lifetime.

Background counts also reduce correlation contrast. Assume statistically independent signal and background whose counts add in the selected modes and time window, a Poisson background, and equal signal fractions at the two HBT detectors. Incoherent stray light or detector dark counts may provide such a background if these conditions hold. Incoherence alone does not imply Poisson statistics: single-mode thermal light has $g^{(2)}_{\rm bg}(0)=2$, so the following result with $g^{(2)}_{\rm bg}=1$ does not apply directly. Denote the signal and background rates by $\Phi_{\rm sig}$ and $\Phi_{\rm bg}$, and define
\[
 q_s=\frac{\Phi_{\rm sig}}{\Phi_{\rm sig}+\Phi_{\rm bg}}.
\]
Coincidences contain signal--signal, signal--background, and background--background contributions. Their normalized weights are $q_s^2$, $2q_s(1-q_s)$, and $(1-q_s)^2$. Temporarily neglecting temporal smoothing gives
\begin{equation}
 \begin{aligned}
 g^{(2)}_{\rm meas}(\tau)
 &=q_s^2g^{(2)}_{\rm sig}(\tau)+2q_s(1-q_s)+(1-q_s)^2\\
 &=1+q_s^2[g^{(2)}_{\rm sig}(\tau)-1].
 \end{aligned}
 \label{ch5:bgcorrection}
\end{equation}
If background accounts for $10\%$ of all counts, then $q_s=0.9$. Even for $g^{(2)}_{\rm sig}(0)=0$, the measured value is $1-0.9^2=0.19$. Unequal signal fractions require replacing $q_s^2$ by $q_{s,1}q_{s,2}$. With timing jitter, convolve the correlation deviation using Eq.~\eqref{ch5:timing}.

This correction does not apply to a background that adds coherently to the signal in the same mode. It also omits detector dead time and afterpulsing. If these effects matter, model the actual output operator or detection process. Detector dead time is the interval after a click during which the detector cannot respond. Record it separately from preparation and reset overhead between experimental trials.

\subsection{Three Levels of Numerical Validation}

\begin{enumerate}
 \item Electromagnetic model. Check convergence with mesh, domain size, and boundaries. A passive structure with consistent port normalization should satisfy $I-SS^\dagger\succeq0$. For a real unit direction $\mathbf e$, the coincident-point free-space Green function should approach $\mathbf e\cdot\operatorname{Im}\mathbf G_0(\mathbf r_0,\mathbf r_0,\omega)\cdot\mathbf e=\omega/(6\pi c)$. Treat point-source singularities consistently with the chosen Green-function definition.
 \item Quantum dynamics. Check $\Tr\rho=1$, $\rho=\rho^\dagger$, and $\rho\succeq0$, with errors decreasing as numerical tolerances are tightened. Removing dissipation conserves only quantities consistent with the Hamiltonian; bare energy and total excitation number generally need not be conserved under driving. Increase the cavity photon-number cutoff and check $\avg{\hat n}$ and $\avg{\hat a^{\dagger2}\hat a^2}$ separately for convergence.
 \item Physical limits and statistics. Taking $g\to0$ should recover the corresponding independent subsystems. Weak-excitation first-order response should agree with linear coupled-mode theory. For a steady state with decaying correlations, $g^{(2)}(\tau)$ should return to the uncorrelated baseline. Under the one-sided spectral convention of Chapter 2, $\int_0^\infty S_O(f)\dd f=C_O(0)$ should hold. Finite time windows and frequency ranges require integration-truncation checks.
\end{enumerate}

A small equation residual does not establish that the physical model is accurate enough. Vary the approximation level and compare with solvable limits or experiments under the same input and measurement conditions.

\section{How Should Quantum Advantage Be Assessed?}
\label{ch5:advantage}

The preceding sections addressed model selection and solution. We now consider how to evaluate those results. Atomic readout in Chapter 2, field-state propagation in Chapter 3, and photon emission in Chapter 4 serve different tasks, but each requires a statement of response, noise, efficiency, and resources. The precision relations below serve those tasks. Further Reading 5A derives precision bounds from complete measurement probabilities.

\subsection{Quantum Matter, Nonclassical Phenomena, and Task Performance}

Device assessment should distinguish three claims: quantum matter such as atomic levels serves as the sensing element; nonclassical statistics are observed; or performance improves for a defined task at fixed resources. These claims require different evidence.

For example, $g^{(2)}(0)<1$ in an ideal single-mode or stationary autocorrelation, or quadrature noise below the vacuum baseline, can test the corresponding nonclassicality. A HOM experiment also requires specified input and detection conditions: an arbitrary correlation dip between two ports is not necessarily nonclassical evidence. Even verified nonclassicality must be followed by a calculation of its practical benefit for the measurement task.

Atomic levels provide stable frequency references and specific field-coupling mechanisms, but external fields, collisions, and other effects can still shift the operating frequency. Comparing receiver sensitivity with a conventional antenna and low-noise amplifier requires the same task and an accounting of atomic noise, probe-light noise, collection region, bandwidth, and auxiliary equipment. Similarly, small $g^{(2)}(0)$ indicates relative suppression of multiphoton events. It does not by itself specify the single-photon probability, wavepacket purity, or usable flux.

\subsection{Response Slope, Noise Spectrum, and Measurement Time}
Sensitivity assessment must include response slope, noise, and measurement time together~\cite{ch5:degen}.

\label{ch5:precisionsection}

Let $x$ be the quantity to be measured and $y$ the random output of one trial. Near the operating point $x_0$, expand the mean response as
\[
 \overline y(x)\simeq\overline y(x_0)+K(x-x_0),
 \qquad K=\left.\frac{\partial\overline y}{\partial x}\right|_{x_0}.
\]
For $K\ne0$, the sample mean $\overline y_M$ of $M$ independent readings defines a local estimator
\[
 \widehat x=x_0+\frac{\overline y_M-\overline y(x_0)}{K}.
\]
If a single reading has variance $\sigma_y^2$, then $\Var(\overline y_M)=\sigma_y^2/M$, giving
\begin{equation}
 \delta x\equiv\sqrt{\Var(\widehat x)}
 =\frac{\sigma_y}{|K|\sqrt M}.
 \label{ch5:error}
\end{equation}
This assumes a locally linear response and negligible estimator bias. The standard deviation equals the root-mean-square error only when the estimator is unbiased. Correlated trials also invalidate direct use of the independent-sample $1/\sqrt M$ scaling.

An experimental cycle $T_{\rm c}$ includes preparation, interaction, readout, and reset. Let the total integration time be $\tau_{\rm int}$. With many repetitions, $M\simeq\tau_{\rm int}/T_{\rm c}$, and the time-normalized sensitivity is
\begin{equation}
 \eta_x\equiv\delta x\sqrt{\tau_{\rm int}}
 =\frac{\sigma_y\sqrt{T_{\rm c}}}{|K|}.
 \label{ch5:sensitivity}
\end{equation}
Its units are $[x]\sqrt{\mathrm s}$, often written as $[x]/\sqrt{\mathrm{Hz}}$. If preparation, readout, and reset do not accumulate the phase of interest, their combined overhead may be called dead time. Do not add these durations again if they are already included in $T_{\rm c}$.

For continuous steady-state readout, let $K(f)$ be the small-signal transfer function and $S_y(f)$ the one-sided output-noise spectrum. The input-referred spectrum is
\begin{equation}
 S_x(f)=\frac{S_y(f)}{|K(f)|^2},\qquad
 S_x^{1/2}(f)=\frac{\sqrt{S_y(f)}}{|K(f)|}.
 \label{ch5:inputnoise}
\end{equation}
If response and noise are approximately flat at low frequencies and the averaging time $T$ greatly exceeds the correlation time, rectangular averaging without dead time gives $\delta x^2\simeq S_x(0)/(2T)$. Thus $\eta_x\simeq\sqrt{S_x(0)/2}$. The factor $1/2$ follows from the one-sided spectrum and the rectangular average's equivalent noise bandwidth $1/(2T)$. Spectral and filtering conventions therefore matter.

Local field enhancement fits this framework as well. In the chosen in-phase readout component, suppose a structure amplifies both the signal $x$ and external noise $n_{\rm ext}$ in the same input mode by $F>0$. Let $n_{\rm add}$ denote independent added readout noise. Use the local linear model
\[
 y=K_0F(x+n_{\rm ext})+n_{\rm add},
\]
with both noise terms having zero mean. Then
\begin{equation}
 \sigma_{x,\rm eq}^2
 =\frac{\Var(y)}{|K_0F|^2}
 =\sigma_{\rm ext}^2+
 \frac{\sigma_{\rm add}^2}{|K_0F|^2}.
 \label{ch5:fieldenhancement}
\end{equation}
If $K_0$ and $\sigma_{\rm add}$ do not vary with $F$, dominant internal readout noise allows an approximately $F$-fold reduction in the input-referred noise standard deviation. Once external noise in the same mode dominates, increasing $F$ no longer improves the noise-limited signal-to-noise ratio. This is a consequence of referring noise through the signal path. It does not imply that room-temperature microwave reception is always limited by external thermal noise.

A real thermal background may occupy many directions and modes. The enhancement structure may also change bandwidth and the atomic operating point, and add its own loss noise. Each noise source then requires a new transfer calculation to the output. A single field-enhancement factor cannot be applied to all noise sources.

\subsection{Independent and Entangled Atoms: Time and Dephasing}

Suppose each atom accumulates phase $\phi=\Delta T$ during free evolution for time $T$, where $\Delta$ is the angular-frequency offset to be estimated. Near the chosen operating point, the Ramsey excitation probability for independent atoms is
\[
 p(\phi)=\frac{1+\mathcal V_{\rm ind}\sin\phi}{2},
 \qquad 0<\mathcal V_{\rm ind}\leq1.
\]
Here $\mathcal V_{\rm ind}$ is the fringe visibility, corresponding to $C$ in Eq.~\eqref{eq:ramsey} of Chapter 2.
For $N_{\rm a}$ independent atoms, the total excitation count satisfies
\[
 \bar n_e=N_{\rm a}p,\qquad
 \Var(n_e)=N_{\rm a}p(1-p).
\]
At $\phi=0$, the standard deviation is $\sqrt{N_{\rm a}}/2$ and the response slope is $\mathcal V_{\rm ind}N_{\rm a}/2$. The single-trial phase uncertainty is therefore
\begin{equation}
 \delta\phi_{\rm ind}=\frac1{\mathcal V_{\rm ind}\sqrt{N_{\rm a}}}.
 \label{ch5:sql}
\end{equation}
At ideal unit visibility, this becomes the independent-atom standard quantum limit $1/\sqrt{N_{\rm a}}$ for this protocol.

Now suppose a multi-atom GHZ state can be prepared:
\[
 \ket{\Psi_{\rm GHZ}}
 =\frac{\ket{g}^{\otimes N_{\rm a}}+\ket{e}^{\otimes N_{\rm a}}}{\sqrt2}.
\]
The fully excited component acquires a factor $\ee^{\ii N_{\rm a}\phi}$, so its relative phase is $N_{\rm a}\phi$. One readout applies a phase-controlled $\pi/2$ analysis pulse to all atoms, measures their levels individually, and determines the parity of the total excitation count. Choose the analysis-pulse phase so that the operating point lies at the maximum fringe slope. Label one parity outcome 1 and the other 0, giving
\[
 p_{\rm ent}(\phi)=\frac{1+\mathcal V_{\rm ent}\sin(N_{\rm a}\phi)}2.
\]
At the zero-phase operating point, this $0/1$ reading has variance $1/4$ and probability slope $\mathcal V_{\rm ent}N_{\rm a}/2$, so
\begin{equation}
 \delta\phi_{\rm ent}=\frac1{\mathcal V_{\rm ent}N_{\rm a}}.
 \label{ch5:entprecision}
\end{equation}
Labeling parity directly as $\pm1$ gives variance 1 and doubles the slope, leaving the precision unchanged. The ideal result has Heisenberg scaling, $1/N_{\rm a}$. It requires the entangled state and collective readout. The shorter fringe period also requires a sufficiently narrow prior phase range.

For equal $T$, atom number, and total experimental time, include each protocol's cycle duration. The repeated-measurement error ratio is
\begin{equation}
 \frac{\delta\phi_{\rm ent}(\tau_{\rm int})}
 {\delta\phi_{\rm ind}(\tau_{\rm int})}
 =\frac{\mathcal V_{\rm ind}}{\mathcal V_{\rm ent}\sqrt{N_{\rm a}}}
 \sqrt{\frac{T_{{\rm c},{\rm ent}}}{T_{{\rm c},{\rm ind}}}}.
 \label{ch5:timefair}
\end{equation}
Using ideal independent atoms as the reference, take $N_{\rm a}=100$ and $\mathcal V_{\rm ent}=0.2$. Equal cycle durations give an error ratio of $0.5$. If the entangled protocol takes four times as long per cycle, the ratio returns to 1. This assumes the prescribed readout occurs in every trial. If preparation relies on post\-selection, its success probability must also be included.

Dephasing changes the useful phase-accumulation time. Consider only independent Markovian pure dephasing at equal rates for all atoms, with $\gamma_\phi>0$. Chapter 2 gives $C=\ee^{-\gamma_2T}$; for pure dephasing, $\gamma_2=\gamma_\phi$, hence
\[
 \mathcal V_{\rm ind}=\ee^{-\gamma_\phi T},\qquad
 \mathcal V_{\rm ent}=\ee^{-N_{\rm a}\gamma_\phi T}.
\]
If each protocol optimizes $T$ separately, compare estimates of the same frequency $\Delta$, not phase errors obtained at different $T$. Neglect dead time and assume ideal preparation and readout. Using $\delta\Delta=\delta\phi/T$ and $M\simeq\tau_{\rm int}/T$ gives
\begin{equation}
 \begin{aligned}
 \delta\Delta_{\rm ind}^2&=
 \frac{\ee^{2\gamma_\phi T}}{N_{\rm a}T\tau_{\rm int}},\\
 \delta\Delta_{\rm ent}^2&=
 \frac{\ee^{2N_{\rm a}\gamma_\phi T}}{N_{\rm a}^2T\tau_{\rm int}}.
 \end{aligned}
 \label{ch5:dephasingcomparison}
\end{equation}
Minimizing each expression over $T$ yields $T_{\rm ind}=1/(2\gamma_\phi)$ and $T_{\rm ent}=1/(2N_{\rm a}\gamma_\phi)$. Both minimum frequency variances are $2\mathrm e\gamma_\phi/(N_{\rm a}\tau_{\rm int})$. In this simple model, faster dephasing of the entangled state cancels the improved ideal single-trial phase scaling. The conclusion depends on the noise model and protocol; it does not establish that entangled sensing can never offer an advantage~\cite{ch5:escher}.

\subsection{How Transmission and Detection Loss Affect Nonclassicality and Performance}
\label{ch5:loss}

Real electromagnetic systems have waveguide loss, mode mismatch, and finite detector quantum efficiency. Let $\eta\in(0,1)$ be the total power transmittance from generation to conversion into a photoelectron, including collection and detection efficiency. With a vacuum environment and no dark counts or thermal background, each photon independently survives this pure-loss channel with probability $\eta$. Equation~\eqref{eq:ch3:losschannel} of Chapter 3 gives $\hat a_{\rm out}=\sqrt\eta\hat a_{\rm in}+\sqrt{1-\eta}\hat v$. Compare the effects of loss on two quantum indicators.

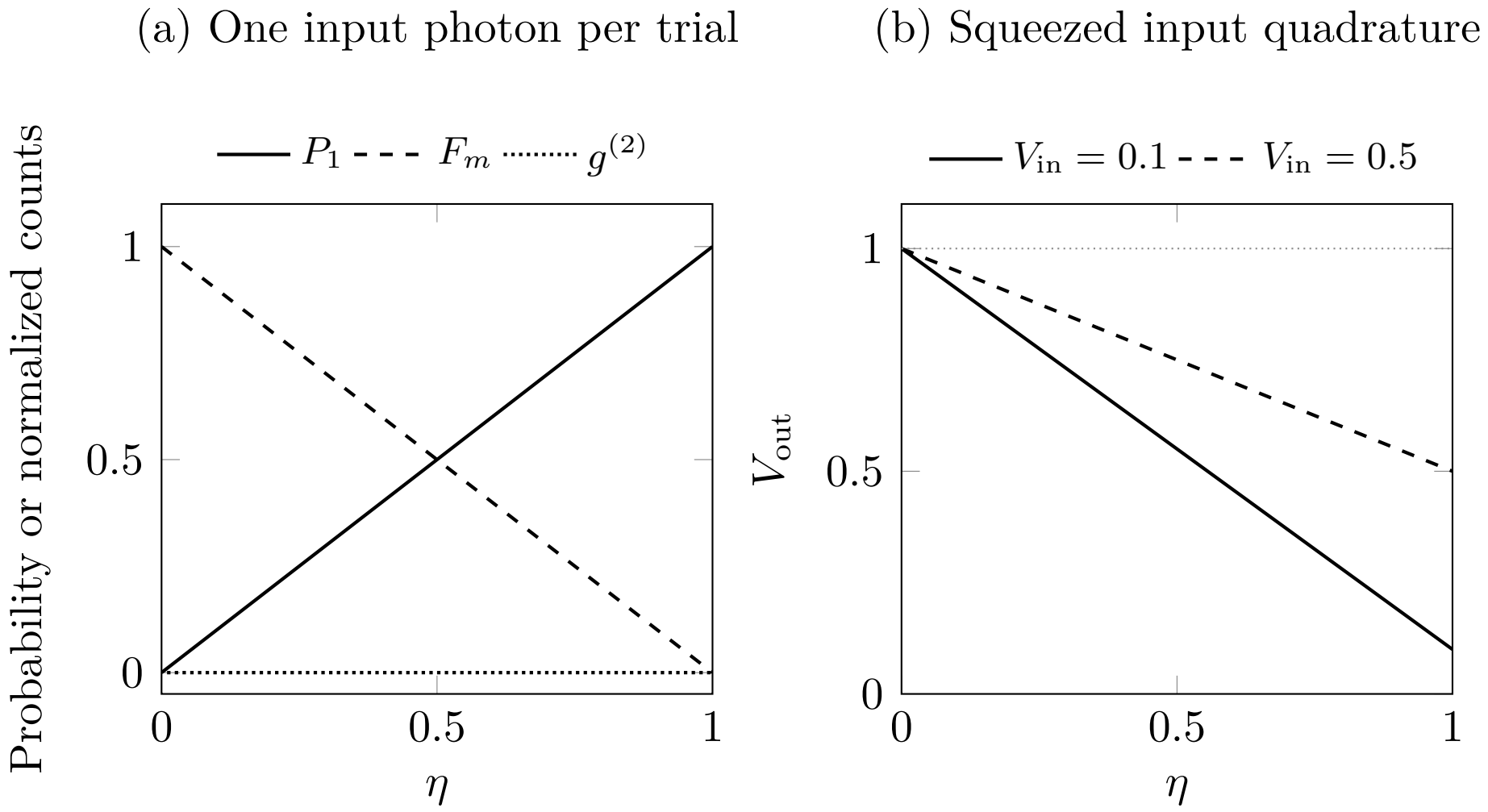
\begin{figure}[htbp]
\centering
\begin{tikzpicture}[font=\small]
\begin{axis}[name=left,width=6.1cm,height=5.6cm,xmin=0,xmax=1,ymin=-.05,ymax=1.1,
 xlabel={$\eta$},ylabel={Probability or normalized counts},xtick={0,.5,1},ytick={0,.5,1},
 title={(a) One input photon per trial},title style={yshift=25pt,font=\small},
 legend style={draw=none,font=\footnotesize,at={(.5,1.02)},anchor=south,legend columns=3}]
\addplot[black,thick,domain=0:1]{x};\addlegendentry{$P_1$}
\addplot[black,dashed,thick,domain=0:1]{1-x};\addlegendentry{$F_m$}
\addplot[black,densely dotted,thick,domain=.01:1]{0};\addlegendentry{$g^{(2)}$}
\end{axis}
\begin{axis}[at={(left.east)},anchor=west,xshift=1.55cm,width=6.1cm,height=5.6cm,
 xmin=0,xmax=1,ymin=0,ymax=1.1,xlabel={$\eta$},ylabel={$V_{\rm out}$},xtick={0,.5,1},ytick={0,.5,1},
 title={(b) Squeezed input quadrature},title style={yshift=25pt,font=\small},
 legend style={draw=none,font=\footnotesize,at={(.5,1.02)},anchor=south,legend columns=2}]
\addplot[black,thick,domain=0:1]{1-.9*x};\addlegendentry{$V_{\rm in}=0.1$}
\addplot[black,dashed,thick,domain=0:1]{1-.5*x};\addlegendentry{$V_{\rm in}=0.5$}
\addplot[black!45,densely dotted,domain=0:1]{1};
\end{axis}
\end{tikzpicture}
\caption{Pure loss for two different input states. (a) Single-photon transmission probability decreases, while $g^{(2)}$ remains zero for $\eta>0$. At $\eta=0$, there are no counts, so neither $F_m$ nor $g^{(2)}$ is defined; the endpoints indicate limits. (b) Squeezed noise returns toward the vacuum baseline $V=1$. The two panels use different input states.}
\label{ch5:lossfig}
\end{figure}

1. Photon-count fluctuations (the Fano factor) and second-order correlation $g^{(2)}$:
Let the input photon number within a pulse window be the random variable $n$. Binomial thinning with transmittance $\eta$ gives a detected output count $m$. For fixed input $n$, the conditional mean and variance are $\mathbb E[m\mid n]=\eta n$ and $\Var(m\mid n)=\eta(1-\eta)n$. Averaging over the input distribution and applying the law of total variance gives
\begin{equation}
 \begin{aligned}
 \bar m&=\eta\bar n,\\
 \Var(m)&=\mathbb E[\Var(m\mid n)]
 +\Var(\mathbb E[m\mid n])
 =\eta(1-\eta)\bar n+\eta^2\Var(n).
 \end{aligned}
 \label{ch5:thinning}
\end{equation}
The Fano factor $F_n\equiv\Var(n)/\bar n$ measures departure from Poisson photon-number fluctuations. Poissonian coherent light has $F_n=1$, an ideal definite-number state has $F_n=0$, and sub-Poissonian light has $F_n<1$. Dividing Eq.~\eqref{ch5:thinning} by $\bar m=\eta\bar n$ gives the linear relation
\begin{equation}
 F_m=1-\eta+\eta F_n.
 \label{ch5:fano}
\end{equation}
Meanwhile, the second factorial moment satisfies $\mathbb E[m(m-1)]=\eta^2\mathbb E[n(n-1)]$. For a nonzero mean input count and $\eta>0$, the normalized second-order correlation obeys
\begin{equation}
 g^{(2)}_{\rm out}
 =\frac{\mathbb E[m(m-1)]}{\bar m^2}
 =\frac{\eta^2\mathbb E[n(n-1)]}{(\eta\bar n)^2}
 =g^{(2)}_{\rm in}.
 \label{ch5:g2loss}
\end{equation}
For exactly one input photon per trial ($\rho_{\rm in}=\ket1\bra1$, $\bar n=1, F_n=0, g^{(2)}_{\rm in}=0$), a loss channel of efficiency $\eta$ produces a vacuum--single-photon mixture:
\begin{equation}
 \rho_{\rm out}=(1-\eta)\ket0\bra0+\eta\ket1\bra1.
 \label{ch5:singleloss}
\end{equation}
As Figure~\ref{ch5:lossfig}(a) shows, for any $\eta>0$ the normalized correlation remains $g^{(2)}_{\rm out}=0$. Yet the single-photon delivery probability falls to $P_1=\eta$, and the Fano factor becomes $F_m=1-\eta$. As $\eta\to0$, the limit $F_m\to1$ means only that the variance-to-mean ratio approaches the Poisson baseline. The output remains a vacuum--single-photon mixture, not Poissonian light. A single-photon source must therefore be characterized by more than $g^{(2)}(0)=0$. Report the end-to-end single-photon probability, repetition rate, and multiphoton probability; two-photon interference also requires indistinguishability.

2. Noise reduction in a quadrature-squeezed state:
Use the quadrature definition $\hat X_\theta=(\hat a\ee^{-\ii\theta}+\hat a^\dagger\ee^{\ii\theta})/2$ from Chapter 3. Relative to the vacuum variance $1/4$, define the normalized variance $V\equiv 4\Var(X_\theta)$. Vacuum and coherent states have $V=1$. A squeezed state measured along its squeezed quadrature has $V_{\rm in}=\ee^{-2r}<1$, where $r>0$ is the squeezing parameter. The pure-loss transformation of Chapter 3 gives, for the same quadrature,
\[
 \hat X_{\theta,\rm out}=\sqrt\eta\hat X_{\theta,\rm in}
 +\sqrt{1-\eta}\hat X_{\theta,v}.
\]
The input and environmental vacuum are uncorrelated, so the cross-covariance vanishes, and $\Var(X_{\theta,v})=1/4$. Taking the variance and multiplying by $4$ gives
\begin{equation}
 V_{\rm out}=\eta V_{\rm in}+1-\eta.
 \label{ch5:quadratureloss}
\end{equation}
Equation~\eqref{ch5:quadratureloss} shows that the loss term $1-\eta$ sets a floor on the output variance~\cite{ch5:caves}; see Figure~\ref{ch5:lossfig}(b). At $\eta=0.5$, arbitrarily strong input squeezing can only approach half the vacuum variance, or about $3.01\,\mathrm{dB}$ of noise reduction. For $V_{\rm in}=0.1$, corresponding to $10\,\mathrm{dB}$ of input squeezing, $V_{\rm out}=0.55$. The output noise reduction is $-10\log_{10}0.55\simeq2.6\,\mathrm{dB}$ relative to vacuum. This directly represents a signal-to-noise improvement only when the compared schemes have equal signal-response slopes.

These count relations assume independent loss with equal survival probability for every photon. If a device selectively changes spatial modes, spectra, or temporal profiles, or adds thermal background or dark counts, recalculate for the actual modes and time window. The invariance of $g^{(2)}$ cannot then be assumed.

\FloatBarrier
\subsection{Resource Accounting for Postselection and Scattering Zeros}
\label{ch5:postsection}

Near an interference dark port, the output probability $p$ may be small while its relative rate of change $(\partial_xp)/p$ is large. This ratio is not an estimation precision. Count fluctuations, success probability, and total experimental time must also be included.

With success probability $p_s$ per trial, $M$ trials yield only $Mp_s$ successful samples on average. For parameter-independent selection, multiply the information per successful sample by $p_s$ to obtain information per attempted trial. If $p_s$ varies with the parameter, the success and failure counts also carry information. Further Reading 5A derives this decomposition and uses $p(x)=\sin^2x$ to illustrate a divergent relative slope with finite total Fisher information. Whether information is finite in general depends on the probability model and its regularity conditions, not merely on the presence of a dark port.

The two-photon scattering dips of Chapter 3 and the photon blockade of Chapter 4 should be assessed together with singles, unnormalized coincidence rates, and detection efficiency. Comparisons must also fix the prior range, total measurement time, and input resources, while specifying preparation, control, local-oscillator, and reset overhead.

\section{Future Directions in Quantum Electromagnetics}

The limits of the preceding models suggest directions for further work. Complex structures require more suitable field degrees of freedom. Complete measurement systems must connect material response and readout. Appreciable propagation delays require environmental memory, and device optimization must connect electromagnetic parameters to final task performance.
\label{ch5:outlook}
\subsection{Quasinormal-Mode Quantization and Few-Mode Descriptions of Open Structures}

Connecting Maxwell solutions for complex structures to quantum dynamics requires a manageable and physically consistent interface. For open resonators with appreciable radiation leakage or material absorption, ordinary real-frequency normalization for lossless cavity modes cannot be carried over directly.

Quasinormal-mode (QNM) quantization~\cite{ch5:franke} and constructions of a few interacting lossy modes from classical coupling spectra~\cite{ch5:medina} offer two concrete routes. They retain the relevant spectral structure over the required bandwidth and include those modes in an open-system calculation. A finite-mode fit generally has error; it is not an exact finite-dimensional replacement for every continuum. Interference between resonances also cannot always be represented by uncoupled Lorentzian peaks.

Validation must go beyond a single classical spectrum. It should include different dipole positions and orientations, radiative and absorptive contributions, time-domain dynamics, and output statistics. Multiple emitters require a spectral-density matrix; a scalar spectrum at one position is usually insufficient. If the enlarged system still has a short-memory environment, its master equation can be used. The explicit modes can then store and return memory from the emitter subsystem's perspective.

\subsection{Complete Measurement Systems Linking Atoms, Optics, and Radio-Frequency Readout}

Rydberg-atom receivers, atomic magnetometers, and solid-state color-center sensors must connect the measured field, sensing medium, control light, and detector. Their modeling can be divided into three parts.
\begin{itemize}
 \item Electromagnetic front end: calculate how the cell, resonant enhancement structure, or waveguide transfers signal fields and environmental noise, retaining spatial distributions and mode normalization.
 \item Atomic medium: include multilevel response, thermal motion, transit, collisions, and dipole interactions as required by the operating conditions. Velocity averaging, spatial transport, and many-body correlations require their corresponding variables; they cannot all be reduced to one empirical linewidth.
 \item Optical and electronic readout: calculate probe propagation through the medium, detector response, and noise. When quantum backaction matters, retain the relevant joint dynamics or equivalent noise correlations.
\end{itemize}
Together, these parts should give the signal transfer function $K(f)$ and input-referred field-noise spectrum $S_E(f)$ at the same operating point. Sensitivity, bandwidth, dynamic range, and spatial resolution can then be compared. The calculation also identifies whether improvement comes from field enhancement, atomic response, or readout.

\subsection{Waveguide Quantum Electrodynamics and Delayed Feedback}

Emitters coupled to the same waveguide can exchange excitations through guided modes, producing collective radiation and coherent coupling. Guided propagation reduces free-space geometric spreading. Propagation loss, mode coupling, and decoherence still limit the interaction distance.

For emitter separation $L$ and group velocity $v_g$ in the relevant band, the propagation delay is approximately $\tau_d=L/v_g$. A reduced master equation that neglects delay may fail when this time approaches an important system timescale, or when reflections return previously emitted fields to an emitter. Delay equations, explicit propagating modes, or other memory-kernel methods can be selected for the problem and checked for approximation error.

These models concern on-chip quantum networks, long-range emitter coupling, and multiphoton-state generation. Direction-dependent chiral coupling also requires suitable mode polarization and transition selection rules; propagation delay alone does not imply it. A further limitation concerns two-time correlations: when reduced dynamics has appreciable memory, the original Markovian quantum-regression procedure cannot be retained without validation.

\subsection{Inverse Design of Electromagnetic Structures for Measurement Tasks}

Quantum devices still require mode control, matching, low loss, and appropriate field enhancement, but the design objective should target the final task. Examples include increasing single-photon output while limiting multiphoton probability, or reducing the estimation variance of Eq.~\eqref{ch5:error} at fixed resources. With geometry and material parameters $\boldsymbol\theta$ as design variables, write
\begin{equation}
 \begin{aligned}
 &\mathop{\rm minimize}_{\boldsymbol\theta}\quad
 \delta x^2(\boldsymbol\theta)\quad\text{or}\quad-\eta_{\rm ex}(\boldsymbol\theta),\\
 &\text{subject to Maxwell's equations, quantum dynamics,}\\
 &\text{and resource and fabrication constraints.}
 \end{aligned}
 \label{ch5:design}
\end{equation}
These are examples of two different objectives. Single-photon output efficiency also requires a specified preparation and collection protocol. Normalized $g^{(2)}$ cannot be used directly as the absolute probability of a multiphoton event.

Design parameters first change mode fields, scattering matrices, and Green functions, then coupling, loss, and quantum states, and finally measurement outcomes. For a linear steady-state readout, the adjoint equation in Further Reading 5B reduces the cost of repeated state-derivative calculations. Derivatives of electromagnetic parameters with respect to geometry still require the Maxwell solver, its adjoint, or another method. Objectives involving transient collection efficiency, correlations, or noise spectra also require differentiating their propagation and normalization steps.

Recheck the model as the structure changes during optimization. Nearby resonances may, for example, invalidate an earlier single-mode approximation. Verify the predicted improvement in the updated model and examine fabrication errors and operating-point drift.

\clearpage
\section{Connecting Engineering and Quantum Electromagnetics}
\label{ch5:summary}

Table~\ref{ch5:rosetta} summarizes several useful connections developed in this book. A quantum description continues to use the fields, boundaries, modes, and ports of classical electromagnetics, while adding states, operators, and measurement probabilities. Each equation in the table has its own conditions of validity. In particular, lossless single-mode formulas cannot be applied directly to arbitrary lossy, dispersive structures.

\begingroup
\small
\setlength{\tabcolsep}{4pt}
\renewcommand{\arraystretch}{1.27}
\begin{longtable}{@{}>{\raggedright\arraybackslash}p{2.65cm}>{\raggedright\arraybackslash}p{4.35cm}>{\raggedright\arraybackslash}p{\dimexpr\textwidth-7cm-4\tabcolsep\relax}@{}}
\caption{Connections between engineering electromagnetics and quantum descriptions.}\label{ch5:rosetta}\\
\toprule
Engineering concept & Quantum description and basic equations & Conditions and computational considerations\\
\midrule
\endfirsthead
\multicolumn{3}{l}{Table~\thetable\quad (continued)}\\
\toprule
Engineering concept & Quantum description and basic equations & Conditions and computational considerations\\
\midrule
\endhead
\midrule
\multicolumn{3}{r}{Continued on next page}\\
\endfoot
\bottomrule
\endlastfoot
LC resonant circuit or cavity eigenmode
& Quantum harmonic oscillator:\newline
$\hat H=\hbar\omega_c(\hat a^\dagger\hat a+\tfrac12)$\newline
$[\hat a,\hat a^\dagger]=1$
& Classical field equations determine the spatial profile of an ideal independent mode; its amplitude becomes an operator. Adding one photon increases the energy by $\hbar\omega_c$. A lossy open structure requires compatible descriptions of the modes and environment.\\
\addlinespace[4pt]
Mode energy normalization and effective mode volume
& Local vacuum field scale:\newline
$|\mathcal E(\mathbf r_0)|=\sqrt{\dfrac{\hbar\omega_c}{2\varepsilon_0\varepsilon_r(\mathbf r_0)V_{\rm eff}}}$\newline
$|g|=|\mathbf d_{eg}\cdot\boldsymbol{\mathcal E}(\mathbf r_0)|/\hbar$
& This expression applies to a lossless, nondispersive single mode with the corresponding local mode-volume definition and $\varepsilon_r(\mathbf r_0)>0$. The rms vacuum field of this mode along the local polarization is $|\mathcal E|$; it is not the total fluctuation of all electromagnetic vacuum modes.\\
\addlinespace[4pt]
Complex amplitude of a coherent carrier
& Coherent state: $\hat a\ket\alpha=\alpha\ket\alpha$\newline
$\bar n=|\alpha|^2$, $\Var(n)=\bar n$\newline
$\mathbf E_0=2\ii\boldsymbol{\mathcal E}\alpha$
& The last equation uses the phase convention $\hat{\mathbf E}^{(+)}=\ii\boldsymbol{\mathcal E}\hat a\ee^{-\ii\omega_ct}$ adopted in this book, with $\mathbf E_0$ the complex peak amplitude. A real source may also contain thermal and technical noise. The presence of a carrier alone does not establish a pure coherent state.\\
\addlinespace[4pt]
In-phase and quadrature components; I/Q readout
& $\hat X=(\hat a+\hat a^\dagger)/2$\newline
$\hat Y=(\hat a-\hat a^\dagger)/(2\ii)$\newline
$\Delta X\Delta Y\geq1/4$
& Balanced homodyne detection selects one quadrature. Simultaneous readout of both quadratures must include the added noise introduced by the specific detection scheme.\\
\addlinespace[4pt]
Lossy passive network and thermal noise
& $\hat{\mathbf b}=S\hat{\mathbf a}+L\hat{\mathbf v}$\newline
$SS^\dagger+LL^\dagger=I$
& Environment operators preserve the output commutation relations; the environment state determines thermal noise. Even a vacuum environment allows loss to bring squeezed noise back toward the vacuum baseline. Classical thermal-noise relations also require the appropriate temperature and port normalization.\\
\addlinespace[4pt]
Susceptibility and Lorentzian response
& Two-level steady-state susceptibility:\newline
$\chi\propto\dfrac{-\Delta+\ii\gamma_2}{\Delta^2+\gamma_2^2}\dfrac1{1+\mathcal S}$
& Here $\Delta=\omega-\omega_0$ and $\mathcal S$ is the saturation parameter. A closed two-level model gives a Lorentzian response in weak fields. Saturation reduces the absorptive response per unit field strength; multilevel coherent processes can also produce EIT.\\
\addlinespace[4pt]
Total power delivered by a classical point dipole; Green function
& Local decay rate:\newline
$\Gamma_{\rm em}/\Gamma_0=P_{\rm cl}/P_{\rm cl,0}$\newline
For absolute values and amplitude calibration, see\newline
Eqs.~\eqref{ch4:gammagreen} and \eqref{ch4:classicalpower}.
& Use the linear reciprocal background, local vacuum, and weak-coupling Markov conditions of Chapter 4. Evaluate $\mathbf G$ at the source position and transition frequency. $P_{\rm cl}$ is the cycle-averaged total power for a complex peak dipole moment, including radiation and material absorption; these should be recorded separately.\\
\addlinespace[4pt]
Coupled-circuit splitting and cavity enhancement
& Resonant lossless JC model:\newline
single-excitation energy splitting $2\hbar g$\newline
Resonant fast-cavity, weak-coupling limit:\newline
$\Gamma_{\rm c}(0)\simeq4g^2/\kappa$
& The mean weak-field response can have the same form as classical coupled-mode theory. Resolving spectral peaks also depends on loss and detection. The JC model provides anharmonicity through its $\sqrt m$ level structure; photon blockade further requires suitable driving and sufficiently narrow linewidths.\\
\end{longtable}
\endgroup

Field--matter interaction is the organizing theme of this book. Chapter 2 retains the quantum state of matter; Chapter 3 retains the quantum state of the field in a linear medium; Chapter 4 further retains their joint evolution and correlations. For a specific task, first determine which information affects the result, then choose and solve the corresponding model and check the approximations, measurement conditions, and resources. For researchers in engineering electromagnetics, spatial field solutions, port definitions, and error checks remain the foundation of reliable results. A quantum description brings excitation exchange and measurement statistics into the same calculation.

\begingroup
\renewcommand{\bibname}{References (Including Further Reading)}

\endgroup

\clearpage
\begingroup
\setcounter{equation}{0}\setcounter{subsection}{0}
\renewcommand{\theequation}{5A.\arabic{equation}}
\renewcommand{\thesubsection}{5A.\arabic{subsection}}
\renewcommand{\theHequation}{ch5.A.\arabic{equation}}
\renewcommand{\theHsubsection}{ch5.A.\arabic{subsection}}
\section*{Further Reading 5A\quad From Counting Probabilities to Parameter Estimation Precision}
\addcontentsline{toc}{section}{Further Reading 5A: From Counting Probabilities to Parameter Estimation Precision}

Section~\ref{ch5:precisionsection} estimates measurement precision by dividing the standard deviation by the response slope. When a measurement has several output channels, or when data are selected by a success criterion, it is more convenient to calculate Fisher information from the complete outcome probabilities. Here we consider a single real parameter $x$ and derive a lower bound on the variance of a locally unbiased estimator.

\subsection{Classical Fisher Information and the Cram\'{e}r--Rao Bound}

Suppose each experimental trial has several mutually exclusive discrete outcomes $r$. For example, $r$ may denote the number of photons received by a detector or the energy level occupied by an atom. When the parameter to be estimated is $x$, outcome $r$ occurs with probability $p_r(x)$, with normalization $\sum_r p_r(x)=1$. For outcomes with nonzero probabilities that vary smoothly with $x$, the classical Fisher information per measurement is defined as
\begin{equation}
 F(x)\equiv\sum_r p_r(x)
 \left[\frac{\partial\ln p_r(x)}{\partial x}\right]^2
 =\sum_r\frac{[\partial_xp_r(x)]^2}{p_r(x)}.
 \label{ch5:Afisher}
\end{equation}
$F(x)$ has dimensions of $x^{-2}$. It is the sum of the squared relative slope for each possible outcome $r$, weighted by its probability $p_r(x)$.

We now connect this quantity to a lower bound on the variance of an unbiased estimator. Assume that the support of the probability distribution is unchanged near the operating point and that summation and differentiation can be interchanged. For $M$ independent experimental trials, the recorded outcome sequence is $\mathbf r=(r_1,\ldots,r_M)$, with joint probability distribution $P(\mathbf r\mid x)=\prod_{j=1}^{M}p_{r_j}(x)$. Define the derivative of the log likelihood, or score function, as $U(\mathbf r)=\partial_x\ln P(\mathbf r\mid x)=\sum_{j=1}^{M}\partial_x\ln p_{r_j}(x)$.
Differentiating the normalization $\sum_{\mathbf r}P(\mathbf r\mid x)=1$ with respect to $x$ gives the mean $\mathbb E[U]=\sum_{\mathbf r}P(\partial_x\ln P)=\partial_x\sum_{\mathbf r}P=0$. Independence of the trials then gives the variance of $U$ as $\mathbb E[U^2]=M F(x)$.
Now take any unbiased estimator $\widehat x(\mathbf r)$, satisfying $\mathbb E[\widehat x]=x$. Differentiating the identity $\sum_{\mathbf r}[\widehat x(\mathbf r)-x]P(\mathbf r\mid x)=0$ with respect to $x$ yields
\[
 \sum_{\mathbf r}[\widehat x(\mathbf r)-x]\partial_x P(\mathbf r\mid x)-\sum_{\mathbf r}P(\mathbf r\mid x)=0
 \implies \mathbb E\bigl[(\widehat x-x)U\bigr]=1.
\]
Applying the Cauchy--Schwarz inequality $\bigl(\mathbb E[(\widehat x-x)U]\bigr)^2\leq \mathbb E[(\widehat x-x)^2]\,\mathbb E[U^2]=\Var(\widehat x)\cdot MF(x)$ to the left-hand side gives the Cram\'{e}r--Rao inequality:
\begin{equation}
 \Var(\widehat x)\geq\frac1{MF(x)}\iff
 \delta x\geq\frac{1}{\sqrt{MF(x)}}.
 \label{ch5:Acrb}
\end{equation}
This bound applies to locally unbiased estimation under the regularity conditions above. It does not guarantee that a particular algorithm attains the bound with a finite sample. Zero-probability points, parameter boundaries, and identifiability problems require separate analysis.

In particular, suppose a measurement has only two outcomes, such as an atom in the excited state $r=1$ or ground state $r=0$. This is a Bernoulli measurement with probabilities $p(x)$ and $1-p(x)$. Substitution into Eq.~\eqref{ch5:Afisher} immediately gives
\begin{equation}
 F(x)=\frac{(\partial_xp)^2}{p}+\frac{(-\partial_xp)^2}{1-p}
 =\frac{(\partial_xp)^2}{p(1-p)}.
 \label{ch5:Abernoulli}
\end{equation}
Substituting Eq.~\eqref{ch5:Abernoulli} into Eq.~\eqref{ch5:Acrb}, with the single-trial variance $\sigma_y^2=p(1-p)$ and slope $K=\partial_x p$, gives the bound $\frac{1}{\sqrt{MF}}=\frac{\sqrt{p(1-p)}}{|\partial_x p|\sqrt{M}}$. This is the same local precision as Eq.~\eqref{ch5:error} in the main text and the independent-atom projection-noise result in Chapter 2.

\subsection{Why Postselection and Dark Interference Ports Cannot Evade the Information Bound}

In a coincidence-counting or postselection experiment, let the total probability that a trial passes the selection criterion be $p_s(x)$. Such a trial is called a success, for example when coincident photons are detected. A failure, such as photon loss or a missing trigger, has probability $p_{\rm fail}(x)=1-p_s(x)$. Conditional on success, the resolved outcomes $r$ have probabilities $q_r(x)$, with $\sum_r q_r(x)=1$. The complete unconditional distribution for each trial is therefore $p_r(x)=p_s(x)q_r(x)$ together with $p_{\rm fail}(x)=1-p_s(x)$.

Insert this complete distribution into Eq.~\eqref{ch5:Afisher}. Using $\partial_x(p_s q_r)=q_r\partial_x p_s+p_s\partial_x q_r$ and the normalization derivative $\sum_r\partial_xq_r=0$, which eliminates the cross term $2(\partial_xp_s)\sum_r\partial_xq_r=0$, the total Fisher information separates exactly into two contributions:
\begin{equation}
 F_{\rm all}(x)=p_s(x)F_{\rm cond}(x)
 +\frac{[\partial_xp_s(x)]^2}{p_s(x)[1-p_s(x)]},
 \qquad F_{\rm cond}(x)\equiv\sum_r\frac{[\partial_xq_r(x)]^2}{q_r(x)}.
 \label{ch5:Apostselection}
\end{equation}
The first term is the information in the successful subsample, weighted by its probability of occurrence. The second is the information carried by the relative frequencies of success and failure. Here $0<p_s<1$ is assumed, and all failures are grouped into one outcome. If different failure outcomes can be resolved, retaining the full record may provide more information. Reporting only the conditional information $F_{\rm cond}$ while omitting the success probability is insufficient.

Consider a dark-port probability $p(x)=\sin^2x$. Near $x\to0$, we have $p\simeq x^2$, and the relative slope $\partial_x\ln p=2\cot x\simeq2/x$ becomes large. Yet at an operating point with $0<p<1$, the complete binary Fisher information is
\[
 F(x)=\frac{(2\sin x\cos x)^2}{\sin^2x\cos^2x}=4.
\]
The information does not diverge as the dark point is approached. Exactly at $x=0$, direct substitution gives $0/0$, and the measurement cannot distinguish $x$ from $-x$. The limit from neighboring points therefore cannot be treated unconditionally as an attainable precision.

Probability normalization must also be preserved when background is included. Suppose the background independently produces a click within a detection gate with probability $0<\epsilon_{\rm bg}<1$, and the readout distinguishes only a click from no click. A click from either source counts as a success, so the total probability is
\begin{equation}
 p_{\rm click}(x)=1-[1-\sin^2x](1-\epsilon_{\rm bg})
 =\epsilon_{\rm bg}+(1-\epsilon_{\rm bg})\sin^2x.
 \label{ch5:Adarkbackground}
\end{equation}
This expression never exceeds one over the entire parameter range. At a nonsingular operating point, substitution into the binary-information formula gives
\begin{equation}
 F_{\rm click}(x)
 =\frac{4(1-\epsilon_{\rm bg})\sin^2x}
 {\epsilon_{\rm bg}+(1-\epsilon_{\rm bg})\sin^2x}
 \simeq\frac{4(1-\epsilon_{\rm bg})}{\epsilon_{\rm bg}}x^2.
 \label{ch5:Adarkinformation}
\end{equation}
The final approximation applies sufficiently close to zero. The information then tends to zero: background retains click fluctuations while the first-order slope of the signal probability vanishes. This conclusion is specific to the stated detection model; it does not imply that all dark-port schemes lack practical value.

\subsection{The State-Dependent Upper Bound: Quantum Fisher Information}

In a quantum model, let the density matrix $\rho(x)$ depend on the parameter to be measured. A generalized measurement is described by effect matrices $\{E_r\}$ satisfying $E_r\succeq 0$ and $\sum_r E_r=I$. Examples include quadrature measurements and photon-number projective measurements $E_n=\ket n\bra n$. If the measurement setting is fixed and the parameter enters the probabilities through the state, the Born rule gives $p_r(x)=\Tr[E_r\rho(x)]$.

Braunstein and Caves showed~\cite{ch5:braunstein} that, over all measurement schemes $\{E_r\}$ allowed by quantum mechanics, the classical Fisher information has an upper bound determined solely by the quantum state $\rho(x)$. This is the quantum Fisher information ($F_Q$):
\begin{equation}
 F(x)\leq F_Q[\rho(x)],\qquad
 \delta x\geq\frac{1}{\sqrt{MF(x)}}\geq\frac{1}{\sqrt{MF_Q[\rho(x)]}}.
 \label{ch5:Aqcrb}
\end{equation}
For pure-state evolution $\ket{\psi(x)}=\ee^{-\ii x\hat G}\ket{\psi_0}$ generated by a fixed Hermitian operator, $F_Q$ is four times the variance of $\hat G$: $F_Q=4\Var(\hat G)$. The inequality separates quantum-sensor design into two steps: prepare a state carrying substantial parameter information under the available resources, then choose a realizable readout to extract it. The error bound still requires the regularity and local-unbiasedness conditions above. Practical loss, restricted readout, and the prior parameter range must also be included.
\endgroup

\clearpage
\begingroup
\setcounter{equation}{0}\setcounter{subsection}{0}
\renewcommand{\theequation}{5B.\arabic{equation}}
\renewcommand{\thesubsection}{5B.\arabic{subsection}}
\renewcommand{\theHequation}{ch5.B.\arabic{equation}}
\renewcommand{\theHsubsection}{ch5.B.\arabic{subsection}}
\section*{Further Reading 5B\quad Vectorizing the Steady-State Master Equation and Computing Adjoint Sensitivities}
\addcontentsline{toc}{section}{Further Reading 5B: Vectorizing the Steady-State Master Equation and Computing Adjoint Sensitivities}

In electromagnetic inverse design, such as metasurface topology optimization or waveguide shape optimization, we often first solve a linear system $A\mathbf e=\mathbf b$. The adjoint variable method then obtains the objective gradient with respect to thousands of geometric parameters through just one additional adjoint solve. With a prescribed Hamiltonian matrix and dissipation parameters, a finite-dimensional Lindblad master equation is also linear in the density matrix. If these coefficients depend on the state through a self-consistent field, the coupled equations require a separate linearization. This reading gives the full matrix--vector form of the master equation and derives adjoint sensitivity equations for quantum steady-state observables.

\subsection{Column-Stacking the Density Matrix: The Kronecker-Product Form}

Let the Hilbert-space dimension be $d$, so that the density matrix $\rho$ is a $d\times d$ complex matrix. Under column-stacking vectorization, the columns of $\rho$ are placed end to end to form a $d^2\times 1$ column vector $\mathbf q=\operatorname{vec}(\rho)$. For a two-level atom, with $d=2$ and basis ordered as $(\ket g,\ket e)$, for example,
\[
 \rho=\begin{pmatrix}\rho_{gg}&\rho_{ge}\\\rho_{eg}&\rho_{ee}\end{pmatrix}
 \implies
 \mathbf q=\operatorname{vec}(\rho)=(\rho_{gg},\rho_{eg},\rho_{ge},\rho_{ee})^{\mathsf T}.
\]
The superscript $\mathsf T$ denotes transpose without complex conjugation. This agrees with MATLAB column-stacking. In NumPy, use \texttt{reshape(-1, order='F')} explicitly, because the default order is row-major. For a matrix product $A\rho B$, column-stacking obeys the Kronecker-product identity
\begin{equation}
 \operatorname{vec}(A\rho B)
 =(B^{\mathsf T}\otimes A)\operatorname{vec}(\rho)
 =(B^{\mathsf T}\otimes A)\mathbf q.
 \label{ch5:Bvec}
\end{equation}
Apply Eq.~\eqref{ch5:Bvec} term by term to the Lindblad master equation, Eq.~\eqref{ch5:master}, noting that $H\rho=H\rho I$, $\rho H=I\rho H$, and $(L_k^\dagger)^{\mathsf T}=L_k^*$. The result is the first-order linear matrix differential equation $\dot{\mathbf q}=\mathsf L\mathbf q$, with the $d^2\times d^2$ Liouvillian superoperator matrix $\mathsf L$ given explicitly by
\begin{equation}
 \begin{aligned}
 \mathsf L={}&-\frac{\ii}{\hbar}
 \bigl(I\otimes H-H^{\mathsf T}\otimes I\bigr)\\
 &+\sum_k\left[L_k^*\otimes L_k
 -\frac12 I\otimes(L_k^\dagger L_k)
 -\frac12(L_k^\dagger L_k)^{\mathsf T}\otimes I\right].
 \end{aligned}
 \label{ch5:Bliouvillian}
\end{equation}
Here $H$ and $L_k$ denote matrices in the chosen basis, and $\mathsf L$ has units of $\mathrm{s}^{-1}$. In code, \texttt{kron} can assemble $\mathsf L$ according to Eq.~\eqref{ch5:Bliouvillian}. If $\mathsf L$ is time independent, the evolution is $\mathbf q(t)=\ee^{\mathsf L t}\mathbf q(0)$. For large problems, one usually computes the action of the matrix exponential on a vector or integrates directly. If $\mathsf L(t)$ varies with time, this constant-matrix exponential generally cannot be used directly.

\subsection{Removing the Singularity of \texorpdfstring{$\mathsf L\mathbf q=\mathbf 0$}{Lq=0} with a Rank-One Trace Constraint}

Finding a continuous-wave steady state means solving the homogeneous linear system $\mathsf L\mathbf q=\mathbf 0$. Because the master equation preserves the trace, $\frac{\dd}{\dd t}\Tr\rho=0$, the matrix $\mathsf L$ is necessarily singular, with at least one zero eigenvalue. Its inverse $\mathsf L^{-1}$ therefore cannot be formed directly.

Assume a unique stable steady state. To combine the steady-state equation $\mathsf L\mathbf q=\mathbf 0$ and trace normalization $\Tr\rho=1$ into an invertible inhomogeneous system $A\mathbf q=\mathbf u$, define the vectorized identity $\boldsymbol\ell=\operatorname{vec}(I_d)$. The trace of any density matrix is the inner product $\boldsymbol\ell^\dagger\mathbf q=\sum_{i=1}^d\rho_{ii}=\Tr\rho$, while trace preservation, $\Tr(\mathcal L[\rho])=0$, is equivalent to the left-null-vector relation
\[
 \boldsymbol\ell^\dagger\mathsf L=\mathbf 0^{\mathsf T}.
\]
Introduce a fixed positive reference rate $\gamma_{\rm ref}>0$ so that $\mathsf L/\gamma_{\rm ref}$ is dimensionless, and take the maximally mixed state vector $\mathbf u=\operatorname{vec}(I_d/d)$, which satisfies $\boldsymbol\ell^\dagger\mathbf u=\Tr(I_d/d)=1$. Construct the modified full-rank $d^2\times d^2$ matrix $A$ and linear system
\begin{equation}
 A\equiv\frac{\mathsf L}{\gamma_{\rm ref}}+\mathbf u\boldsymbol\ell^\dagger,
 \qquad A\mathbf q=\mathbf u.
 \label{ch5:Bnormalized}
\end{equation}
Premultiplication by $\boldsymbol\ell^\dagger$, using $\boldsymbol\ell^\dagger\mathsf L=0$ and $\boldsymbol\ell^\dagger\mathbf u=1$, gives $\boldsymbol\ell^\dagger\mathbf q=1$. Substitution back into Eq.~\eqref{ch5:Bnormalized} cancels $\mathbf u$ on both sides, leaving $\mathsf L\mathbf q=0$. The equation thus enforces both stationarity and unit trace.

Under the assumed unique stable steady state, the rank-one term removes the null space along the steady-state direction and makes $A$ invertible. In practice, solve the linear system, for example with \texttt{q = A \textbackslash{} u} in MATLAB, rather than forming $A^{-1}$ explicitly. If several steady states exist, this added term does not make the solution unique. Near very slow relaxation, also check conditioning and numerical error.

\subsection{One Adjoint Solve for the Gradients of All Design Parameters}

Let the real-valued inverse-design objective be a steady-state observable $J(\boldsymbol\theta)=\Tr[W(\boldsymbol\theta)\rho_{\rm ss}(\boldsymbol\theta)]$, with a Hermitian observation operator $W=W^\dagger$. For example, for the steady-state photon flux leaving a cavity with no direct incident background, take $W=\hat c^\dagger\hat c=\kappa_{\rm ex}\hat a^\dagger\hat a$. This differs from the time-integrated output probability of a pulsed source. With $\mathbf w=\operatorname{vec}(W)$ and the trace identity $\Tr(W^\dagger\rho)=\operatorname{vec}(W)^\dagger\operatorname{vec}(\rho)$, the objective becomes the vector inner product
\[
 J(\boldsymbol\theta)=\mathbf w^\dagger(\boldsymbol\theta)\mathbf q(\boldsymbol\theta).
\]
We seek the derivatives of $J$ with respect to $M_{\rm p}$ real design parameters $\theta_j$, with $j=1,\ldots,M_{\rm p}$. Differentiate $A(\boldsymbol\theta)\mathbf q(\boldsymbol\theta)=\mathbf u$ in Eq.~\eqref{ch5:Bnormalized} with respect to $\theta_j$. Keep the basis, dimension $d$, and reference rate fixed, so that $\mathbf u$ is unchanged. The state derivative then satisfies
\[
 A\,\frac{\partial\mathbf q}{\partial\theta_j}
 =-\left(\frac{\partial A}{\partial\theta_j}\right)\mathbf q
 \implies
 \frac{\partial\mathbf q}{\partial\theta_j}=-A^{-1}\left(\frac{\partial A}{\partial\theta_j}\right)\mathbf q.
\]
Solving for $\partial\mathbf q/\partial\theta_j$ separately for each parameter $\theta_j$ would require $M_{\rm p}$ linear solves, which can be costly. The key step in the adjoint method is to solve once for an adjoint vector $\boldsymbol\lambda$ at the current design point and for the same scalar objective, then reuse it for every parameter direction:
\begin{equation}
 A^\dagger\boldsymbol\lambda=\mathbf w
 \iff \boldsymbol\lambda^\dagger=\mathbf w^\dagger A^{-1}.
 \label{ch5:Badjoint}
\end{equation}
Insert Eq.~\eqref{ch5:Badjoint} into the total derivative $\frac{\partial J}{\partial\theta_j}=\operatorname{Re}\bigl[(\partial_{\theta_j}\mathbf w)^\dagger\mathbf q+\mathbf w^\dagger(\partial_{\theta_j}\mathbf q)\bigr]$. The state derivative $\partial\mathbf q/\partial\theta_j$ is eliminated, giving the adjoint gradient formula for the quantum master equation:
\begin{equation}
 \frac{\partial J}{\partial\theta_j}
 =\operatorname{Re}\left[
 \left(\frac{\partial\mathbf w}{\partial\theta_j}\right)^\dagger\mathbf q
 -\boldsymbol\lambda^\dagger
 \left(\frac{\partial A}{\partial\theta_j}\right)\mathbf q\right]
 \qquad(j=1,2,\ldots,M_{\rm p}).
 \label{ch5:Bgradient}
\end{equation}
For a fixed design point and one scalar objective, only one forward steady-state solve and one adjoint solve are required. The inner products are then evaluated for each parameter. The first term in Eq.~\eqref{ch5:Bgradient} retains changes in the readout itself: when $\kappa_{\rm ex}$ changes, for example, $W=\kappa_{\rm ex}\hat a^\dagger\hat a$ changes as well.

The saving comes from avoiding repeated state-derivative solves. Each $\partial_{\theta_j}A$ and $\partial_{\theta_j}\mathbf w$ must still be evaluated. If they arise from geometry changes, derivatives from the electromagnetic solver must also be included. A different objective or an updated design point generally requires new solves. Correlation functions, transient objectives, and nonlinear readouts also require derivatives of the relevant intermediate quantities.

When implementing adjoint optimization, the analytic resonant two-level steady state from Chapter 4 provides a useful check. Set pure dephasing to zero, choose the excited-state population operator $W=\ket e\bra e$, and differentiate with respect to the drive Rabi frequency $\theta=\Omega$. The analytic population and its derivative with respect to $\Omega$ are
\begin{equation}
 p_{e,\rm ss}=\frac{\Omega^2}{\Gamma^2+2\Omega^2},\qquad
 \frac{\partial p_{e,\rm ss}}{\partial\Omega}
 =\frac{2\Omega\Gamma^2}{(\Gamma^2+2\Omega^2)^2}.
 \label{ch5:Bcheck}
\end{equation}
At $\Omega=\Gamma$, for example, the steady-state population is $1/3$ and the exact derivative is $\frac{2}{9\Gamma}$. Check that Eq.~\eqref{ch5:Bgradient} agrees with both this analytic value and a central finite difference before increasing model complexity. Each new model should still be checked for state residuals, truncation, and gradient convergence.
\endgroup

\clearpage
\def\AuthorPublicationAppendixEmbedded{1}
\ifdefined\AuthorPublicationAppendixEmbedded
  \let\AuthorPubBegin\relax
  \let\AuthorPubEnd\relax
\else
  \documentclass[UTF8,fontset=fandol,zihao=-4,openany,oneside]{ctexbook}
  \usepackage[paperwidth=176mm,paperheight=250mm,top=21mm,bottom=21mm,left=21mm,right=19mm,headheight=15pt,headsep=6mm]{geometry}
  \usepackage{enumitem,fancyhdr}
  \usepackage[unicode,hidelinks]{hyperref}
  \hypersetup{pdftitle={Quantum Electromagnetics: Books, Chapters, and Papers by the Author},pdfauthor={Wei E. I. Sha}}
  \ctexset{chapter={format=\Large\normalfont,beforeskip=3pt,afterskip=22pt},section={format=\large\normalfont,beforeskip=19pt,afterskip=10pt},subsection={format=\normalsize\normalfont,beforeskip=13pt,afterskip=7pt}}
  \setlength{\parindent}{2em}
  \setlength{\parskip}{2pt}
  \linespread{1.22}
  \pagestyle{fancy}
  \fancyhf{}
  \fancyhead[L]{\small Quantum Electromagnetics}
  \fancyhead[R]{\small Appendix: Author Publications}
  \fancyfoot[C]{\small\thepage}
  \renewcommand{\headrulewidth}{0.35pt}
  \fancypagestyle{plain}{\fancyhf{}\fancyfoot[C]{\small\thepage}\renewcommand{\headrulewidth}{0pt}}
  \def\AuthorPubBegin{\begin{document}}
  \def\AuthorPubEnd{\end{document}}
\fi
\AuthorPubBegin
\begingroup
\setlength{\emergencystretch}{3em}
\clubpenalty=10000
\widowpenalty=10000
\raggedbottom
\urlstyle{same}
\providecommand{\AuthorPubSubsection}[1]{\par\ifdim\dimexpr\pagegoal-\pagetotal\relax<180pt\relax\newpage\fi\subsection*{#1}}
\lefthyphenmin=3\righthyphenmin=3
\providecommand{\AuthorPubDOI}[1]{\par\nobreak\noindent DOI: \url{https://doi.org/#1}}
\chapter*{Appendix\quad Books, Chapters, and Papers by the Author on Quantum Electromagnetics}
\addcontentsline{toc}{chapter}{Appendix\quad Books, Chapters, and Papers by the Author on Quantum Electromagnetics}
\markboth{Books and Papers by the Author on Quantum Electromagnetics}{Books, Chapters, and Papers by the Author on Quantum Electromagnetics}
\ifdefined\fancyhead\fancyhead[R]{\small Appendix: Author Publications}\fi

This appendix lists the author's books, book chapters, and journal papers on quantum electromagnetics and related topics. They cover field quantization, vacuum fluctuations, semiclassical coupled calculations, spontaneous emission, photon correlations, Rydberg-atom sensing, and quantum transport. Quantum-inspired optimization relevant to engineering electromagnetics is listed separately. These studies use quantum or statistical-physics models to construct classical computational algorithms; they do not amount to computing with quantum hardware.

\section*{Books and Book Chapters}

\AuthorPubSubsection{Related Books}
\begin{enumerate}[label={[B\arabic*]},leftmargin=3.2em,labelsep=0.6em,itemsep=9pt,topsep=5pt]
\item \begin{minipage}[t]{\linewidth}\raggedright Zhixiang Huang, Wei Sha, and Xianliang Wu. Symplectic Algorithms in Electromagnetic Computation: Theory and Applications. Beijing: Science Press, 2014. ISBN: 978-7-03-041083-2. [In Chinese; title translated.]
\par Chapter 8, ``Application of symplectic algorithms to the time-dependent Schr\"odinger equation,'' is directly relevant to numerical quantum dynamics.
\end{minipage}

\end{enumerate}

\AuthorPubSubsection{Book Chapters}
\begin{enumerate}[label={[B\arabic*]},start=2,leftmargin=3.2em,labelsep=0.6em,itemsep=9pt,topsep=5pt]
\item \begin{minipage}[t]{\linewidth}\raggedright Wei E. I. Sha and Qun Ren. Quantum effects on Maxwell's equations. In: Er-Ping Li and Zhong Lin Wang. \emph{New Advances in Maxwell's Equations and Applications}. Springer, 2024, Chapter 2: 29--73. (The year follows the publisher's recommended citation; first published online in February 2025.)
\AuthorPubDOI{10.1007/978-3-031-75786-0_2}
\end{minipage}

\item \begin{minipage}[t]{\linewidth}\raggedright Zhixiang Huang, Ming Fang, Jianwei You, Guoda Xie, and Wei E. I. Sha. Size-scale effects on Maxwell's equations. In: Er-Ping Li and Zhong Lin Wang. \emph{New Advances in Maxwell's Equations and Applications}. Springer, 2024, Chapter 3: 75--126. (The year follows the publisher's recommended citation; first published online in February 2025.)
\AuthorPubDOI{10.1007/978-3-031-75786-0_3}
\end{minipage}

\item \begin{minipage}[t]{\linewidth}\raggedright Ming Fang, Henglei Du, Chengpu Liu, Wei E. I. Sha, and Zhixiang Huang. Computational methods and techniques for nonlinear optics. In: Nicolae C. Panoiu (ed.). \emph{Fundamentals and Applications of Nonlinear Nanophotonics}. Elsevier, 2024, Chapter 12: 441--471. (The book was first published in 2023; the chapter bibliographic year is 2024.)
\AuthorPubDOI{10.1016/B978-0-323-90614-2.00004-3}
\end{minipage}

\item \begin{minipage}[t]{\linewidth}\raggedright Zhixiang Huang, Guoda Xie, Xingang Ren, and Wei E. I. Sha. The symplectic FDTD method for Maxwell and Schrödinger equations. In: Qiang Ren, Su Yan, and Atef Z. Elsherbeni (eds.). \emph{Advances in Time-Domain Computational Electromagnetic Methods}. Wiley-IEEE Press, 2022, Chapter 15: 557--609. (The chapter was first published in November 2022; the book copyright year is 2023. The chapter record gives pp. 557--609; the main text starts on p. 559.)
\AuthorPubDOI{10.1002/9781119808404.ch15}
\end{minipage}

\item \begin{minipage}[t]{\linewidth}\raggedright Xiaoyan Y. Z. Xiong and Wei E. I. Sha. The FDTD method: essences, evolutions, and applications to nano-optics and quantum physics. In: Sarhan M. Musa (ed.). \emph{Computational Nanotechnology Using Finite Difference Time Domain}. CRC Press, 2014, Chapter 2: 37--82. (First released in December 2013; the first-edition copyright year, 2014, is used here.)
\AuthorPubDOI{10.1201/b16319-2}
\end{minipage}

\end{enumerate}

\clearpage
\section*{Journal Papers}
\AuthorPubSubsection{Quantum Electromagnetic Theory, Green Functions, and Vacuum Fluctuations}
\begin{enumerate}[label={[J\arabic*]},start=1,leftmargin=3.2em,labelsep=0.6em,itemsep=9pt,topsep=5pt]
\item \begin{minipage}[t]{\linewidth}\raggedright Wei E. I. Sha, Zhihao Lan, Menglin L. N. Chen, Yongpin P. Chen, and Sheng Sun. Spin and orbital angular momenta of electromagnetic waves: from classical to quantum forms. \emph{IEEE Journal on Multiscale and Multiphysics Computational Techniques}, 2024, 9: 113--117.
\AuthorPubDOI{10.1109/JMMCT.2024.3370729}

\end{minipage}

\item \begin{minipage}[t]{\linewidth}\raggedright Wei Sha, Lei Ying, and Gaobiao Xiao. Green functions in electromagnetics. Wuli, 2024, 53(1): 43--49. [In Chinese; title translated.]
\AuthorPubDOI{10.7693/wl20240105}

\end{minipage}

\item \begin{minipage}[t]{\linewidth}\raggedright Tian Xia, Phillip R. Atkins, Wei E. I. Sha, and Weng C. Chew. Casimir force: vacuum fluctuation, zero-point energy, and computational electromagnetics. \emph{IEEE Antennas and Propagation Magazine}, 2021, 63(5): 14--24.
\AuthorPubDOI{10.1109/MAP.2021.3073118}

\end{minipage}

\item \begin{minipage}[t]{\linewidth}\raggedright Wei Sha. Computational electromagnetics at micro- and nanoscales: perspectives and future directions. Chinese Journal of Radio Science, 2020, 35(2): 242--251. [In Chinese; title translated.]
\AuthorPubDOI{10.13443/j.cjors.2019083102}

\end{minipage}

\item \begin{minipage}[t]{\linewidth}\raggedright Weng Cho Chew, Wei E. I. Sha, and Qi I. Dai. Green's dyadic, spectral function, local density of states, and fluctuation dissipation theorem. \emph{Progress In Electromagnetics Research}, 2019, 166: 147--165.
\AuthorPubDOI{10.2528/PIER19111801}

\end{minipage}

\item \begin{minipage}[t]{\linewidth}\raggedright Weng Cho Chew, Aiyin Y. Liu, Carlos Salazar-Lazaro, Dong-Yeop Na, and Wei E. I. Sha. Hamilton equations, commutator, and energy conservation. \emph{Quantum Reports}, 2019, 1(2): 295--303.
\AuthorPubDOI{10.3390/quantum1020027}

\end{minipage}

\item \begin{minipage}[t]{\linewidth}\raggedright Wei E. I. Sha, Aiyin Y. Liu, and Weng Cho Chew. Dissipative quantum electromagnetics. \emph{IEEE Journal on Multiscale and Multiphysics Computational Techniques}, 2018, 3: 198--213.
\AuthorPubDOI{10.1109/JMMCT.2018.2881691}

\end{minipage}

\item \begin{minipage}[t]{\linewidth}\raggedright Weng Cho Chew, Aiyin Y. Liu, Carlos Salazar-Lazaro, and Wei E. I. Sha. Quantum electromagnetics: a new look---Part I. \emph{IEEE Journal on Multiscale and Multiphysics Computational Techniques}, 2016, 1: 73--84.
\AuthorPubDOI{10.1109/JMMCT.2016.2617018}

\end{minipage}

\item \begin{minipage}[t]{\linewidth}\raggedright Weng Cho Chew, Aiyin Y. Liu, Carlos Salazar-Lazaro, and Wei E. I. Sha. Quantum electromagnetics: a new look---Part II. \emph{IEEE Journal on Multiscale and Multiphysics Computational Techniques}, 2016, 1: 85--97.
\AuthorPubDOI{10.1109/JMMCT.2016.2614800}

\end{minipage}

\item \begin{minipage}[t]{\linewidth}\raggedright Phillip R. Atkins, Qi I. Dai, Wei E. I. Sha, and Weng Cho Chew. Casimir force for arbitrary objects using the argument principle and boundary element methods. \emph{Progress In Electromagnetics Research}, 2013, 142: 615--624.
\AuthorPubDOI{10.2528/PIER13082105}
\end{minipage}

\end{enumerate}

\AuthorPubSubsection{Semiclassical Coupling and Numerical Methods for Quantum Dynamics}
\begin{enumerate}[label={[J\arabic*]},start=11,leftmargin=3.2em,labelsep=0.6em,itemsep=9pt,topsep=5pt]
\item \begin{minipage}[t]{\linewidth}\raggedright Guoda Xie, Wangchao Hu, Guilin Hou, Xingang Ren, Wei E. I. Sha, and Zhixiang Huang. Construction of an efficient CFS-PML technique based on the vector-scalar potential formulation and its application in quantum-electromagnetic multiphysics simulations. \emph{Computer Physics Communications}, 2027, 330: 110427. (Published online; the volume year is 2027.)
\AuthorPubDOI{10.1016/j.cpc.2026.110427}

\end{minipage}

\item \begin{minipage}[t]{\linewidth}\raggedright Guoda Xie, Zhixiang Huang, Jian Wei You, Zhihao Lan, Nicolae C. Panoiu, and Wei E. I. Sha. Universal vector--scalar potential framework for inhomogeneous electromagnetic system and its application in semiclassical quantum electromagnetics. \emph{IEEE Transactions on Plasma Science}, 2021, 49(11): 3459--3471.
\AuthorPubDOI{10.1109/TPS.2021.3117607}

\end{minipage}

\item \begin{minipage}[t]{\linewidth}\raggedright Ming Fang, Zhixiang Huang, Wei E. I. Sha, and Costas M. Soukoulis. Modelling of the fluctuation and coherent dynamics in active metamaterial devices. \emph{IEEE Transactions on Nanotechnology}, 2021, 20: 543--551.
\AuthorPubDOI{10.1109/TNANO.2021.3092059}

\end{minipage}

\item \begin{minipage}[t]{\linewidth}\raggedright Guoda Xie, Zhixiang Huang, Ming Fang, and Wei E. I. Sha. Simulating Maxwell--Schrödinger equations by high-order symplectic FDTD algorithm. \emph{IEEE Journal on Multiscale and Multiphysics Computational Techniques}, 2019, 4: 143--151.
\AuthorPubDOI{10.1109/JMMCT.2019.2920101}

\end{minipage}

\item \begin{minipage}[t]{\linewidth}\raggedright Jing Shen, Wei E. I. Sha, Xiaojing Kuang, Jinhua Hu, Zhi-Xiang Huang, and Xian-Liang Wu. Symplectic pseudospectral time-domain scheme for solving time-dependent Schrödinger equation. \emph{Progress In Electromagnetics Research M}, 2018, 66: 109--118.
\AuthorPubDOI{10.2528/PIERM18010808}

\end{minipage}

\item \begin{minipage}[t]{\linewidth}\raggedright Yongpin P. Chen, Wei E. I. Sha, Lijun Jiang, Min Meng, Yu Mao Wu, and Weng Cho Chew. A unified Hamiltonian solution to Maxwell--Schrödinger equations for modeling electromagnetic field-particle interaction. \emph{Computer Physics Communications}, 2017, 215: 63--70.
\AuthorPubDOI{10.1016/j.cpc.2017.02.006}

\end{minipage}

\item \begin{minipage}[t]{\linewidth}\raggedright Christopher Jayun Ryu, Aiyin Y. Liu, Wei E. I. Sha, and Weng Cho Chew. Finite-difference time-domain simulation of the Maxwell--Schrödinger system. \emph{IEEE Journal on Multiscale and Multiphysics Computational Techniques}, 2016, 1: 40--47.
\AuthorPubDOI{10.1109/JMMCT.2016.2605378}

\end{minipage}

\item \begin{minipage}[t]{\linewidth}\raggedright Jing Shen, Wei E. I. Sha, Zhixiang Huang, Mingsheng Chen, and Xianliang Wu. High-order symplectic FDTD scheme for solving a time-dependent Schrödinger equation. \emph{Computer Physics Communications}, 2013, 184(3): 480--492.
\AuthorPubDOI{10.1016/j.cpc.2012.09.032}
\end{minipage}

\end{enumerate}

\AuthorPubSubsection{Spontaneous Emission, Photon Correlations, and Light--Matter Interaction}
\begin{enumerate}[label={[J\arabic*]},start=19,leftmargin=3.2em,labelsep=0.6em,itemsep=9pt,topsep=5pt]
\item \begin{minipage}[t]{\linewidth}\raggedright Jun-Yong Yan, Paul C. A. Hagen, Hang-Yu Ge, Fang-Yuan Li, Hans-Georg Babin, Wei E. I. Sha, Bo Wang, Zhi-hua Gan, Andreas D. Wieck, Arne Ludwig, Chao-Yuan Jin, Vollrath M. Axt, Da-Wei Wang, Moritz Cygorek, and Feng Liu. Robust biexciton preparation for entangled photon-pair generation by single-shot Floquet driving. \emph{Nature Communications}, 2026, 17(1): 9745.
\AuthorPubDOI{10.1038/s41467-026-77221-9}

\end{minipage}

\item \begin{minipage}[t]{\linewidth}\raggedright Zi-Yang Xu, Wei E. I. Sha, Zhong-Jiang Tang, and Hang Xie. Anomalous inverse Faraday effect for graphene quantum dots in optical vortices. \emph{Physical Review B}, 2026, 114(15): 154430.
\AuthorPubDOI{10.1103/jzck-plc9}

\end{minipage}

\item \begin{minipage}[t]{\linewidth}\raggedright Chengnian Huang, Hangyu Ge, Yijia Cheng, Zi He, Feng Liu, and Wei E. I. Sha. Modeling of far-field quantum coherence by dielectric bodies based on the volume integral equation method. \emph{IEEE Transactions on Antennas and Propagation}, 2026, 74(7): 6983--6993.
\AuthorPubDOI{10.1109/TAP.2026.3683145}

\end{minipage}

\item \begin{minipage}[t]{\linewidth}\raggedright Yu-Tong Wang, Qi-Hang Ye, Jun-Yong Yan, Yufei Qiao, Yu-Xin Liu, Yong-Zheng Ye, Chen Chen, Xiao-Tian Cheng, Chen-Hui Li, Zi-Jian Zhang, Cheng-Nian Huang, Yun Meng, Kai Zou, Wen-Kang Zhan, Chao Zhao, Xiaolong Hu, Clarence Augustine T. H. Tee, Wei E. I. Sha, Zhixiang Huang, Huiyun Liu, Chao-Yuan Jin, Lei Ying, and Feng Liu. Moiré cavity quantum electrodynamics. \emph{Science Advances}, 2025, 11(21): eadv8115.
\AuthorPubDOI{10.1126/sciadv.adv8115}

\end{minipage}

\item \begin{minipage}[t]{\linewidth}\raggedright Chen Chen, Jun-Yong Yan, Hans-Georg Babin, Jiefei Wang, Xingqi Xu, Xing Lin, Qianqian Yu, Wei Fang, Run-Ze Liu, Yong-Heng Huo, Han Cai, Wei E. I. Sha, Jiaxiang Zhang, Christian Heyn, Andreas D. Wieck, Arne Ludwig, Da-Wei Wang, Chao-Yuan Jin, and Feng Liu. Wavelength-tunable high-fidelity entangled photon sources enabled by dual Stark effects. \emph{Nature Communications}, 2024, 15(1): 5792.
\AuthorPubDOI{10.1038/s41467-024-50062-0}

\end{minipage}

\item \begin{minipage}[t]{\linewidth}\raggedright Siyuan Huang, Yongpin Chen, Yang (Michael) Yang, and Wei E. I. Sha. Accurate optical optimization of light-emitting diodes with consideration of coupling between Purcell factor and transmittance coefficient. \emph{Optics Express}, 2022, 30(14): 24544--24552.
\AuthorPubDOI{10.1364/OE.462536}

\end{minipage}

\item \begin{minipage}[t]{\linewidth}\raggedright Yongpin P. Chen, Wei E. I. Sha, Lijun Jiang, and Jun Hu. Graphene plasmonics for tuning photon decay rate near metallic split-ring resonator in a multilayered substrate. \emph{Optics Express}, 2015, 23(3): 2798--2807.
\AuthorPubDOI{10.1364/OE.23.002798}

\end{minipage}

\item \begin{minipage}[t]{\linewidth}\raggedright Yongpin P. Chen, Wei E. I. Sha, Wallace C. H. Choy, Lijun Jiang, and Weng Cho Chew. Study on spontaneous emission in complex multilayered plasmonic system via surface integral equation approach with layered medium Green's function. \emph{Optics Express}, 2012, 20(18): 20210--20221.
\AuthorPubDOI{10.1364/OE.20.020210}

\end{minipage}

\item \begin{minipage}[t]{\linewidth}\raggedright Peng-fei Qiao, Wei E. I. Sha, Wallace C. H. Choy, and Weng Cho Chew. Systematic study of spontaneous emission in a two-dimensional arbitrary inhomogeneous environment. \emph{Physical Review A}, 2011, 83(4): 043824.
\AuthorPubDOI{10.1103/PhysRevA.83.043824}
\end{minipage}

\end{enumerate}

\AuthorPubSubsection{Rydberg-Atom Sensing and Reception}
\begin{enumerate}[label={[J\arabic*]},start=28,leftmargin=3.2em,labelsep=0.6em,itemsep=9pt,topsep=5pt]
\item \begin{minipage}[t]{\linewidth}\raggedright Yang Yan, Xinyi Y. I. Xu, Jinpeng Yuan, Wei E. I. Sha, Feng Liu, Zhihao Lan, Kang Liu, Zi He, Andrea Massa, Lirong Wang, Erping Li, Liantuan Xiao, and Suotang Jia. A Rydberg atomic synthetic-aperture radar imaging system approaching the Rayleigh criterion. \emph{Nature Sensors}, 2026, 1(11): 989--997.
\AuthorPubDOI{10.1038/s44460-026-00129-w}

\end{minipage}

\item \begin{minipage}[t]{\linewidth}\raggedright Wenjie Ding, Xinyi Y. I. Xu, Guoda Xie, Xingang Ren, Yingsong Li, Wei E. I. Sha, and Zhixiang Huang. Efficient simulation of Doppler-broadened Rydberg quantum receivers via Monte Carlo wave functions. \emph{IEEE Journal on Multiscale and Multiphysics Computational Techniques}, 2026, 11: 379--388.
\AuthorPubDOI{10.1109/JMMCT.2026.3707907}

\end{minipage}

\item \begin{minipage}[t]{\linewidth}\raggedright Zhida Wang, Yang Yan, Xinyi Y. I. Xu, Wei E. I. Sha, Xinyan Liang, Jinpeng Yuan, Lirong Wang, Liantuan Xiao, and Suotang Jia. Receiving amplitude-phase shift keying signals in microwave fields with a Rydberg atomic receiver. \emph{Chinese Physics B}, 2026, 35(8): 083203.
\AuthorPubDOI{10.1088/1674-1056/ae37fb}

\end{minipage}

\item \begin{minipage}[t]{\linewidth}\raggedright Shuai S. A. Yuan, Xinyi Y. I. Xu, Jinpeng Yuan, Guoda Xie, Chongwen Huang, Xiaoming Chen, Zhixiang Huang, and Wei E. I. Sha. Electromagnetic modeling and capacity analysis of Rydberg atom-based MIMO system. \emph{IEEE Antennas and Wireless Propagation Letters}, 2025, 24(7): 1839--1843.
\AuthorPubDOI{10.1109/LAWP.2025.3549313}

\end{minipage}

\item \begin{minipage}[t]{\linewidth}\raggedright Xinyi Y. I. Xu, Guoda Xie, Jinlou Ma, Lei Ying, Jinpeng Yuan, Zhixiang Huang, and Wei E. I. Sha. Fast simulation for interacting four-level Rydberg atoms: electromagnetically induced transparency and Autler--Townes splitting. \emph{Optics Express}, 2024, 32(12): 21755--21766.
\AuthorPubDOI{10.1364/OE.523897}
\end{minipage}

\end{enumerate}

\AuthorPubSubsection{Quantum Transport and Related Device Models}
\begin{enumerate}[label={[J\arabic*]},start=33,leftmargin=3.2em,labelsep=0.6em,itemsep=9pt,topsep=5pt]
\item \begin{minipage}[t]{\linewidth}\raggedright Liang Tian, Wei E. I. Sha, Hao Xie, Dongxue Liu, Tian-Ge Sun, Yin-Shui Xia, and Wenchao Chen. Carrier transport simulation methods for electronic devices with coexistence of quantum transport and diffusive transport. \emph{Journal of Applied Physics}, 2024, 135(22): 225703.
\AuthorPubDOI{10.1063/5.0209479}

\end{minipage}

\item \begin{minipage}[t]{\linewidth}\raggedright Hang Xie, Yu Zhang, Yanho Kwok, and Wei E. I. Sha. Charge separation and dissipation in molecular wires under a light radiation. \emph{Organic Electronics}, 2018, 58: 94--104.
\AuthorPubDOI{10.1016/j.orgel.2018.03.040}

\end{minipage}

\item \begin{minipage}[t]{\linewidth}\raggedright Zi Shuai Wang, Wei E. I. Sha, and Wallace C. H. Choy. Exciton delocalization incorporated drift-diffusion model for bulk-heterojunction organic solar cells. \emph{Journal of Applied Physics}, 2016, 120(21): 213101.
\AuthorPubDOI{10.1063/1.4970958}

\end{minipage}

\item \begin{minipage}[t]{\linewidth}\raggedright Hang Xie, Feng Jiang, and Wei E. I. Sha. Numerical methods for spin-dependent transport calculations and spin bound states analysis in Rashba waveguides. \emph{Computer Physics Communications}, 2016, 198: 118--127.
\AuthorPubDOI{10.1016/j.cpc.2015.09.008}
\end{minipage}

\end{enumerate}

\AuthorPubSubsection{Quantum-Inspired Optimization and Related Computation}
The following papers study optimization inspired by quantum annealing and computational methods based on statistical physics, with applications to metasurface design and communications signal processing.
\begin{enumerate}[label={[J\arabic*]},start=37,leftmargin=3.2em,labelsep=0.6em,itemsep=9pt,topsep=5pt]
\item \begin{minipage}[t]{\linewidth}\raggedright Yutong Jiang, Zhe Zheng, Shuai S. A. Yuan, Hao Xue, Long Li, and Wei E. I. Sha. Quantum-inspired optimization for 2-bit quantized phase-only metasurfaces with near-field multiobjective control. \emph{IEEE Transactions on Microwave Theory and Techniques}, 2025, 73(12): 9886--9901.
\AuthorPubDOI{10.1109/TMTT.2025.3619563}

\end{minipage}

\item \begin{minipage}[t]{\linewidth}\raggedright Qing-Guo Zeng, Xiao-Peng Cui, Xian-Zhe Tao, Jia-Qi Hu, Shi-Jie Pan, Wei E. I. Sha, and Man-Hong Yung. High-order-modulation large MIMO detector based on physics-inspired methods. \emph{Physical Review Applied}, 2025, 24(3): 034027.
\AuthorPubDOI{10.1103/zmgv-m2ql}

\end{minipage}

\item \begin{minipage}[t]{\linewidth}\raggedright Shuai S. A. Yuan, Yutong Jiang, Ziyi Zhang, Jia Nan Zhang, Feng Liu, Jian Wei You, and Wei E. I. Sha. Quantum annealing-inspired optimization for space--time coding metasurface. \emph{IEEE Transactions on Antennas and Propagation}, 2025, 73(9): 6512--6524.
\AuthorPubDOI{10.1109/TAP.2025.3573526}
\end{minipage}

\end{enumerate}

\clearpage
\endgroup
\AuthorPubEnd

\end{document}